\RequirePackage{silence}
\documentclass[fleqn,usenatbib,useAMS]{mnras} 
\usepackage{graphicx}
\usepackage{amsmath}

\usepackage{amsfonts}
\usepackage{float}
\usepackage{bm}
\usepackage[perpage,symbol*]{footmisc}
\usepackage{ae,aecompl}
\usepackage{array}
\usepackage{soul}
\usepackage{mathtools}
\usepackage{multirow}

\newcommand{\appropto}{\mathrel{\vcenter{
  \offinterlineskip\halign{\hfil$##$\cr 
    \propto\cr\noalign{\kern2pt}\sim\cr\noalign{\kern-2pt}}}}}

\newcommand{\ssim}{\,{\sim}\,} 

\DeclareRobustCommand{\perthousand}{%
  \ifmmode
    \text{\textperthousand}%
  \else
    \textperthousand
  \fi}

\title[Distinguishing Standard from Modified Gravity in the Local Group and beyond]{Distinguishing Standard from Modified Gravity in the Local Group and beyond} 

\author[Indranil Banik, Supervisor: Hongsheng Zhao]{Indranil Banik$^{1}$\thanks{Email: \href{mailto:ib45@st-andrews.ac.uk}{ib45@st-andrews.ac.uk} (Indranil Banik)\newline $~~~~~~~~~~~~~~$ \href{mailto:hz4@st-andrews.ac.uk}{hz4@st-andrews.ac.uk} (Hongsheng Zhao)}, Supervisor: Hongsheng Zhao$^{1}$\\
$^{1}$Scottish Universities Physics Alliance, University of Saint Andrews, North Haugh, Saint Andrews, Fife, KY16 9SS, United Kingdom}

\pubyear{2018}
\begin{document}
\label{firstpage}
\pagerange{\pageref{firstpage}--\pageref{lastpage}}

\maketitle

\begin{abstract}

The works in this portfolio test the hypothesis that it is not possible to extrapolate the Newtonian inverse square law of gravity from Solar System to galaxy scales. In particular, I look into various tests of Modified Newtonian Dynamics (MOND), which posits a modification below a very low acceleration threshold. Although discrepancies with Newtonian dynamics are indeed observed, they can usually be explained by invoking an appropriate distribution of invisible mass known as dark matter (DM). This leads to the standard cosmological paradigm, $\Lambda$CDM. I consider how it may be distinguished from MOND using collision velocities of galaxy clusters, which should sometimes be much faster in MOND. I focus on measuring these velocities more accurately and conclude that this test ought to be feasible in the near future.

For the time being, I look at the much nearer and more accurately observed Local Group (LG) of galaxies. Its main constituents $-$ the Milky Way (MW) and Andromeda (M31) $-$ should have undergone a past close flyby in MOND but not in $\Lambda$CDM. The fast MW-M31 relative motion around the time of their flyby would have allowed them to gravitationally slingshot any passing LG dwarf galaxies out at high speed. I consider whether there is any evidence for such high-velocity galaxies (HVGs). Several candidates are found in two different $\Lambda$CDM models of the LG, one written by a founding figure of the paradigm. The properties of these HVGs are similar to what might be expected in MOND, especially their tendency to lie close to a plane. Being more confident of its validity, I then used MOND to determine the escape velocity curve of the MW over the distance range 8$-$50 kpc, finding reasonable agreement with the latest observations. I finish by discussing possible future directions for MOND research.

\end{abstract}

\nokeywords


\section{Introduction and Context}
\label{Introduction}

This PhD is about testing the fundamental physical laws governing the Universe. Currently understood physics is capable of explaining a huge variety of observations. However, the great pillars of modern physics $-$ general relativity and quantum mechanics $-$ are difficult to reconcile \citep[e.g.][]{Carlip_2001}. As nature must be internally consistent, it follows that there must be systems whose behaviour can't be accurately predicted using our current incomplete understanding of physics.

Fortunately, there are indeed a few known examples of such systems, some of which were discovered almost a century ago. In particular, the dynamics of galaxies and galaxy clusters require at least one new fundamental assumption \citep{Zwicky_1937, Rubin_Ford_1970}. Before considering these observations and why they imply new physics, I will first briefly review our current understanding of the physics which is thought to govern such systems.

The main force at play here is gravity. Although this is intrinsically weaker than the other fundamental forces, it is cumulative in the sense that masses always attract each other. It is the very strength of the electromagnetic force which prevents significant separation of charge and thus ensures this force only acts over short distances in practice. The strong and weak nuclear forces are limited in their range, thus leaving gravity as the dominant influence known to act on large astronomical systems. It is precisely this kind of system which I will focus on, looking in particular at galaxies but also galaxy groups and clusters.

\newpage
\subsection{Gravity in the Solar System}

One of the most important advances in our understanding of celestial dynamics came about when Johannes Kepler discovered the eponymous laws of planetary motion in the early 1600s, benefiting from earlier observations by Tycho Brahe. Based on the assumption that planets orbit around the Sun, these empirical laws provided the first indication that Solar System dynamics could be understood with a few simple rules. In particular, Kepler's Third Law $-$ that the orbital period scales as the $\frac{3}{2}$ power of orbital semi-major axis $-$ later proved crucial. Four centuries later, this law is currently at the heart of how data from the Kepler telescope is used to infer orbital properties of exoplanets \citep{Borucki_1997}.

Kepler soon realised that this law also applied to the moons of Jupiter, albeit with a different normalisation. Once Newton discovered his laws of motion, consistency with Kepler's Third Law immediately showed that there must be a force towards the Sun with magnitude decreasing as $\frac{1}{r^2}$. This also provided a good explanation for the shapes of cometary orbits (especially Halley's Comet) and for how the Moon remains bound to the Earth. Thus, he soon realised that all massive objects must exert a force towards them in the same way as the Sun, Jupiter and Earth. This gave rise to what we now call Newton's Universal Law of Gravitation. In general, it states that 
\begin{eqnarray}
	\bm{g}_{_N} ~=~ -\sum_i{\frac{GM_i \left( \bm{r} - \bm{r}_i \right) }{{| \bm{r} - \bm{r}_i |}^3}}
	\label{Newtonian_gravity}
\end{eqnarray}

$\bm{g}_{_N}$ is the predicted acceleration of an object at position $\bm{r}$ due to the gravity of other masses $M_i$ located at positions $\bm{r}_i$. The $-$ sign indicates that gravity is attractive. For over two centuries, celestial motions indicated that $\bm{g}_{_N} = \bm{g}$ to within observational uncertainties, where $\bm{g}$ is the actual gravitational field inferred in some less model-dependent way (often based on planetary trajectories). Thus, Equation \ref{Newtonian_gravity} ruled the heavens until the Industrial Revolution.

Eventually, improved technology enabled more precise observations which highlighted tensions with the theory. These came to a head with the observation of how much light is deflected by the gravitational field of the Sun $-$ the deflection is twice the prediction of Newtonian dynamics \citep{Eddington_1919}.  Apparently, Newton's laws do not work for relativistic particles like photons. However, there are small deviations even for non-relativistic planets. This is now known to underlie the `anomalous' precession of Mercury's orbital perihelion by an extra 43'' every century. The historical attempt to explain this small but statistically significant discrepancy is excellently reviewed in \citet{Ruskin_2017}, highlighting several analogies with the ongoing missing mass vs modified gravity debate at the heart of this thesis.

In addition to observational discrepancies, it was also necessary to reconcile Newtonian gravity with the Special Theory of Relativity \citep{Einstein_1905}. The latter precluded instantaneous action at a distance, even though this is how the former works. These issues were resolved using the General Theory of Relativity \citep{Einstein_1915}. With this in hand, it seemed that all Solar System motions could be adequately explained.

General Relativity often yields very similar predictions to Newtonian gravity. This is especially true in galaxies, which are the focus of this article. Thus, I will treat the predictions of the latter as equal to those of the former (except when discussing gravitational lensing). This is because Newtonian gravity is much simpler to handle than General Relativity. Although the latter is important for cosmological-scale problems, I do not directly address such large scales. In particular, I only consider systems whose mean density is much larger than that of the Universe as a whole, making the system effectively decoupled from the large-scale cosmic expansion \citep[Hubble flow,][]{Hubble_1929}.

\subsection{Gravity beyond the Solar System $-$ the discovery of acceleration discrepancies}

In the past century, observations beyond the Solar System became increasingly accurate. These showed a remarkable phenomenon that I shall call `acceleration discrepancies'. Although it is currently not possible to directly observe accelerations in systems much larger than the Solar System, one can reasonably assume that a star in a rotating disk galaxy has a centripetal acceleration of $\frac{v^2}{r}$. This requires careful observation of the galaxy to be sure that it really is a rotating disk. Fortunately, this can be confirmed with only minimal assumptions based on the line of sight (`radial') velocity of its different parts. These motions cause a Doppler shift in the wavelengths of spectral lines that are nowadays measurable using integral field unit spectroscopy. In the case of the Large Magellanic Cloud, the rotation of the galaxy can be seen directly using proper motions \citep{Van_der_Marel_2014}. In this way, it is possible to obtain an observational estimate of the acceleration that makes few assumptions, especially with regards to the gravity theory.

As we do have such theories, this opens the possibility of testing them much more thoroughly using the latest Galactic and extragalactic observations. To be useful, theories of gravitation need to predict the acceleration based on the mass distribution (e.g. using a procedure similar to Equation \ref{Newtonian_gravity}). With some assumptions, we can convert observed light into an idea of how the mass is distributed in a particular system, thus determining the expected $\bm{g} \left( \bm{r} \right)$.

When Newtonian gravity is used to do this, the predicted acceleration often falls far short of the observed value. An early example of this was in the Coma Cluster of galaxies, where Fritz Zwicky found the need for ${\sim 100 \times}$ more matter than suggested by its observed brightness \citep{Zwicky_1937}. It was eventually realised that much of this mass exists as hot gas, which is in fact very bright $-$ but in X-rays, inaccessible to observations by ground-based telescopes \citep{Sarazin_1986}.

Another major acceleration discrepancy was found in the rotation curves of disk galaxies. Beyond the majority of their visible matter, Equation \ref{Newtonian_gravity} implies the rotation speed should decrease as $v_c \appropto \frac{1}{\sqrt{r}}$, the rotation curve version of Kepler's Third Law. However, observed rotation curves tend to remain flat out to large distances. An early indication of this came from the nearest large external galaxy, Andromeda \citep[M31,][]{Babcock_1939}. Later work confirmed that M31 indeed rotates much faster in its outer parts than can be expected on the basis of its visible mass \citep{Rubin_Ford_1970}. This was later confirmed with radio observations of the 21 cm hyperfine transition of neutral hydrogen \citep{Roberts_1975}. Such observations also indicated flat rotation curves for several other galaxies \citep{Rogstad_1972}. Radio observations were important because they extend out to much larger radii than optical measurements. This is due to star formation requiring a threshold gas density. At sufficiently large distances, the exponentially declining gas density \citep{Freeman_1970} falls below this threshold, leaving the outer parts of galaxies with very few stars.

If rotation curves are flat, the discrepancy with a Keplerian decline should become more pronounced at a larger distance. This should be easily detectable when considering the forces \emph{between} galaxies rather than the internal forces within them. In this regard, an important constraint is provided by the dynamics of the Milky Way (MW) and M31, galaxies which are ${\ssim 0.8}$ Mpc apart \citep{McConnachie_2012}. The basic idea is that they must have started receding from each other shortly after the Big Bang. However, they are presently approaching each other at $\sim$110 km/s, as inferred from the observed radial velocity of M31 \citep{Slipher_1913} corrected for the motion of the Sun within the MW \citep{Schmidt_1958}. Therefore, the gravitational attraction between the galaxies must have been strong enough to turn their initial recession around \citep{Kahn_Woltjer_1959}.\footnote{Deviations of velocity $\bm{v}$ from a pure Hubble expansion ($\bm{v} \equiv H \bm{r}$) are called peculiar velocities ($\bm{v}_{pec} \equiv \bm{v} - H \bm{r}$), where $\bm{r}$ is used for position and $H \equiv \frac{\dot{a}}{a}$ is the logarithmic time derivative of the cosmic scale-factor $a \left( t \right)$.} Using this constraint (known as the timing argument), it was found that the total mass in the MW \& M31 needed to be ${\ssim 4 \times}$ the observed matter in them.

Although this result was surprising, there were other reasons to suppose that the MW \& M31 are more massive than might be expected from imaging of their visible disks. With the advent of computers, \citet{Hohl_1971} used $N$-body simulations to show that self-gravitating disks are unstable, rapidly becoming dynamically hot (i.e. developing non-circular motions comparable to the circular rotation speed). The instability develops over only a few orbital periods, whereas the Universe is ${\ssim 40\times}$ older than the orbital period of the Sun \citep{McMillan_2017}. Thus, observed spiral galaxies can't be self-gravitating and must be surrounded by a dynamically hot halo. Moreover, this halo has to dominate the mass of the galaxy. As no such component is seen, \citet{Ostriker_Peebles_1973} suggested that it is dark.

\subsection{The Massive Compact Halo Object hypothesis}
\label{MACHO}

One possibility for this dark matter (DM) was a large number of as yet undetected very faint stars or stellar remnants around each galaxy \citep{Carr_1994}. This theory of massive compact halo objects (MACHOs) could be tested using gravitational microlensing searches \citep{Kerins_1995}. The basic idea is that a massive object would occasionally appear to pass very close to a star on our sky \citep{Refsdal_1966}. This alignment would cause the foreground mass to gravitationally deflect light from the background star, which would therefore appear to brighten and then fade.\footnote{There would be two apparent images of the star, but in microlensing these are $-$ by definition $-$ unresolved. If they are resolved, then the lensing is said to be `strong'.}

The obvious problem with searching for such microlensing events is that the true luminosity of a star can change. Generally, this would be associated with a change in its temperature. This would alter the colour of the star $\equiv$ the ratio of its fluxes in two different wavebands. However, gravitational microlensing equally affects photons of all wavelengths. This allows us to distinguish between microlensing and intrinsic variability by observing in two or more wavelength bands \citep{Paczynski_1986}.

Using these ideas, the EROS collaboration conducted a careful search for microlensing events. This involved continuous monitoring of 7 million stars in two fields of view towards the Large and Small Magellanic Clouds over a period of 6.7 years. Instead of the ${\ssim 39}$ events expected under the MACHO hypothesis, only 1 candidate event was found \citep{EROS_2007}. Similar results had already been reached several years earlier \citep{MACHO_2000}. As a result, it has become clear that MACHOs almost certainly do not have enough mass to account for the acceleration discrepancies in our Galaxy or to stabilise its disk if it obeys Newtonian dynamics.

\subsection{Non-baryonic dark matter}

This leads to several possibilities, none of which are based solidly on existing laws of physics. The most popular idea is to maintain the assumption of a large amount of mass in the outskirts of galaxies. This DM hypothesis is one of the key pillars of the currently prevailing cosmological paradigm \citep[$\Lambda$CDM,][]{Ostriker_Steinhardt_1995}. Cold gas in the amounts required would easily be detected and would in any case likely have clumped into MACHOs, contradicting microlensing observations (Section \ref{MACHO}). Although a small amount of hot gas is expected and has indeed been detected around the MW \citep{Nicastro_2016}, this can't constitute all of the DM. Thus, one needs to assume that the DM is not composed of baryons at all.

This leads to the present situation where no known fundamental particle has the properties required of the DM. Thus, it is thought to consist of an undiscovered stable particle, or at least one with a decay time longer than the age of the Universe \citep[e.g.][and references therein]{Steigman_1985}. The leading contender is a weakly interacting massive particle \citep[WIMP,][]{Griest_1993}, though a much lower mass axion could also work \citep{Kamionkowski_1998}.

Multi-decade searches for a WIMP have now ruled out a substantial part of the parameter space that was thought to be feasible before the searches started \citep[e.g.][]{Ackerman_2015, LUX_2016, PandaX_2016}. Moreover, an important motivation for the WIMP hypothesis is that nature might respect a new fundamental symmetry called supersymmetry \citep{Jungman_1996}. This predicts a plethora of new particles. However, recent null results from the Large Hadron Collider argue against the simplest forms of supersymmetry \citep{ATLAS_Collaboration_2015}.

Less attention has been paid to the possibility of axion DM, though this has started to change recently due to null detections of WIMPs \citep{Baer_2015}. Axions may be easier to search for as they interact with a strong magnetic field \citep{Sikivie_1983}. As neutron stars indeed have very strong magnetic fields, this has allowed some constraints to be placed on axion properties if they are ubiquitous enough to comprise the DM \citep{Berenji_2016}. Although a promising start, this leaves open most of the axion mass range calculated by \citet{Borsanyi_2016}. In fact, this range is difficult to probe by the Axion Dark Matter Experiment \citep{Duffy_2006}, one of the longest-running searches for axion DM. Thus, the acceleration discrepancies may yet be resolved using axions. Until then, it is prudent to consider other possibilities.

\subsection{Modified Newtonian Dynamics}
\label{MOND_introduction}

Just as we do not yet have a complete understanding of particle physics, so also we do not yet understand gravity. Therefore, another possibility is to suppose that the acceleration discrepancies are caused by a breakdown of Newtonian gravity in the relevant systems. This rather old idea was alluded to by \citet{Zwicky_1937} in the same paper that first reported significant acceleration discrepancies. There, Zwicky suggested that the inverse square law of gravity might break down at large distances.

As more observational data was gathered, certain patterns in the acceleration discrepancy became apparent $-$ where it appeared and where it did not. In this respect, a crucial discovery was the Tully-Fisher Relation concerning the dynamics of spiral galaxies \citep{Tully_Fisher_1977}. Eventually, Mordehai Milgrom realised that the important physical parameter is not the size of a system but the typical acceleration within it \citep{Milgrom_1983}. If modified gravity is the answer, then Newtonian gravity needs to break down below an acceleration scale $a_{_0}$. This theory of Modified Newtonian Dynamics (MOND) assumes that the gravitational field strength $g$ at distance $r$ from an isolated point mass $M$ transitions from the usual inverse square law (Equation \ref{Newtonian_gravity}) at short range to
\begin{eqnarray}
	g ~=~ \frac{\sqrt{GMa_{_0}}}{r} ~~~\text{for } ~r \gg \sqrt{\frac{GM}{a_{_0}}}
	\label{Deep_MOND_limit}
\end{eqnarray}

$a_{_0}$ is a fundamental acceleration scale of nature which must have an empirical value close to $1.2 \times {10}^{-10}$ m/s$^2$ to match galaxy rotation curves \citep{McGaugh_2011}.

In the 1990s, another unexplained acceleration was observed $-$ that of the whole Universe. Instead of slowing down due to the attractive effect of gravity, the cosmic scale-factor $a \left( t \right)$ seemed to be speeding up \citep[$\overset{..}{a} > 0$,][]{Riess_1998}. This could be fit into the context of General Relativity by reintroducing the cosmological constant term $\Lambda$, a direct coupling between the metric and Ricci curvature tensors. This `dark energy' can be viewed as a uniform energy density fundamental to the fabric of spacetime itself. Considering the behaviour of quantum systems, this makes some sense $-$ such systems have a zero point energy due to inherent uncertainty in field strengths and their time derivatives (e.g. in the position and velocity of a particle). Thus, a pendulum can never be exactly at the bottom and have zero velocity. Consequently, the energy of the pendulum must be slightly above the classical minimum.

Similarly, an apparently empty region of spacetime must have some value for quantities such as the electric field strength. Although it might be 0 classically, this is no longer feasible quantum mechanically $-$ there must be some uncertainty. For this reason, it is possible that spacetime itself has a minimum (zero-point) energy density associated with it $-$ a cosmic ground state.

This quantum-mechanical phenomenon seems to be having a significant effect on the expansion rate history of the Universe. This raises the question of whether there are other circumstances in which quantum effects might force us to revise our classical (non-quantum) expectations for the motions of astrophysical objects. A possibly useful analogy could be drawn with a gas $-$ at high enough temperatures, it behaves classically. However, quantum effects become important at low temperatures, when large-scale properties such as the heat capacity start to behave differently. Instead of being temperature-independent, this decreases with temperature due to the `freezing out' of quantised degrees of freedom that ultimately underlie heat capacity. A rough estimate of when this occurs (i.e. when equipartition of energy breaks down) can be found by equating the classical result for the mean energy of each particle with the Fermi energy.

The details of how quantum mechanics works with gravity are still unclear. Classically, the energy density in a gravitational field is given by
\begin{eqnarray}
	u ~=~ -\frac{g^2}{8 \pi G}
	\label{Classical_energy_density}
\end{eqnarray}

In a remarkable coincidence called the cosmic coincidence of MOND, $a_{_0}$ is comparable to the value of $g$ at which this equation yields an energy density equal in magnitude to the dark energy density $u_{_\Lambda} = \rho_{_\Lambda} c^2$ implied by the accelerating expansion of the Universe \citep{Riess_1998}. Thus,
\begin{eqnarray}
	\frac{g^2}{8\rm{\pi}G} ~<~ u_{_\Lambda} ~~\Leftrightarrow~~ g ~\la~ 2\rm{\pi}a_{_0}
	\label{Threshold_acceleration}
\end{eqnarray}

This strongly suggests that MOND is simply an empirical way of capturing deviations from classical gravity which arise due to quantum effects \citep{Milgrom_1999}. After all, $u_{_\Lambda}$ is likely a quantum mechanical effect $-$ if it dominates the energy density in a particular region of spacetime, then quantum gravity effects could well be important. Assuming that our classical gravity theories are only approximations to the true quantum gravity theory underlying nature, it would not be surprising if our existing theories failed at accelerations ${\la a_{_0}}$ but worked at higher accelerations. Indeed, there are some specific suggestions for how quantum gravity might work which yield MOND-like behaviour at low accelerations \citep[e.g.][]{Pazy_2013, Verlinde_2016, Smolin_2017}.

MOND was originally formulated as a non-relativistic theory, only to be applied in systems where Newtonian gravity and General Relativity meant much the same thing. This covers internal motions of galaxies and should cover forces between nearby galaxies. To describe these situations, a modified version of the usual Poisson Equation of Newtonian gravity is used \citep{Bekenstein_Milgrom_1984}.
\begin{eqnarray}
	\label{Modified_Poisson_Equation}
	\nabla \cdot \left[ \mu \left(g \right) \bm{g} \right] &=& -4 \pi G \rho ~~\text{  where} \\
	g &\equiv & \left| \bm{g} \right| ~~\text{  and} \\
	\mu \left( g \right) &=& \frac{g}{g + a_{_0}}
\end{eqnarray}

Here, I used the simple interpolating function $\mu \left( g \right)$ to capture how nature transitions between the Newtonian ($g \gg a_{_0}$) and deep-MOND ($g \ll a_{_0}$) regimes \citep{Famaey_Binney_2005}. This fairly gradual transition works very well with high-precision kinematic data from our own Galaxy \citep{Iocco_Bertone_2015} and from a large sample of ${\ssim 6000}$ elliptical galaxies probing accelerations up to ${\ssim 30 a_{_0}}$ \citep[][figure 2]{Chae_2017}. In spherical symmetry, it implies that the true gravity $g$ can be obtained from the Newtonian gravity $g_{_N}$ using
\begin{eqnarray}
	g ~=~ \frac{g_{_N}}{2} ~+~ \sqrt{ \left( \frac{g_{_N}}{2}\right)^2 + g_{_N} a_{_0}}
	\label{Simple_mu}
\end{eqnarray}


\begin{figure}
	\centering
		\includegraphics[width = 8.5cm] {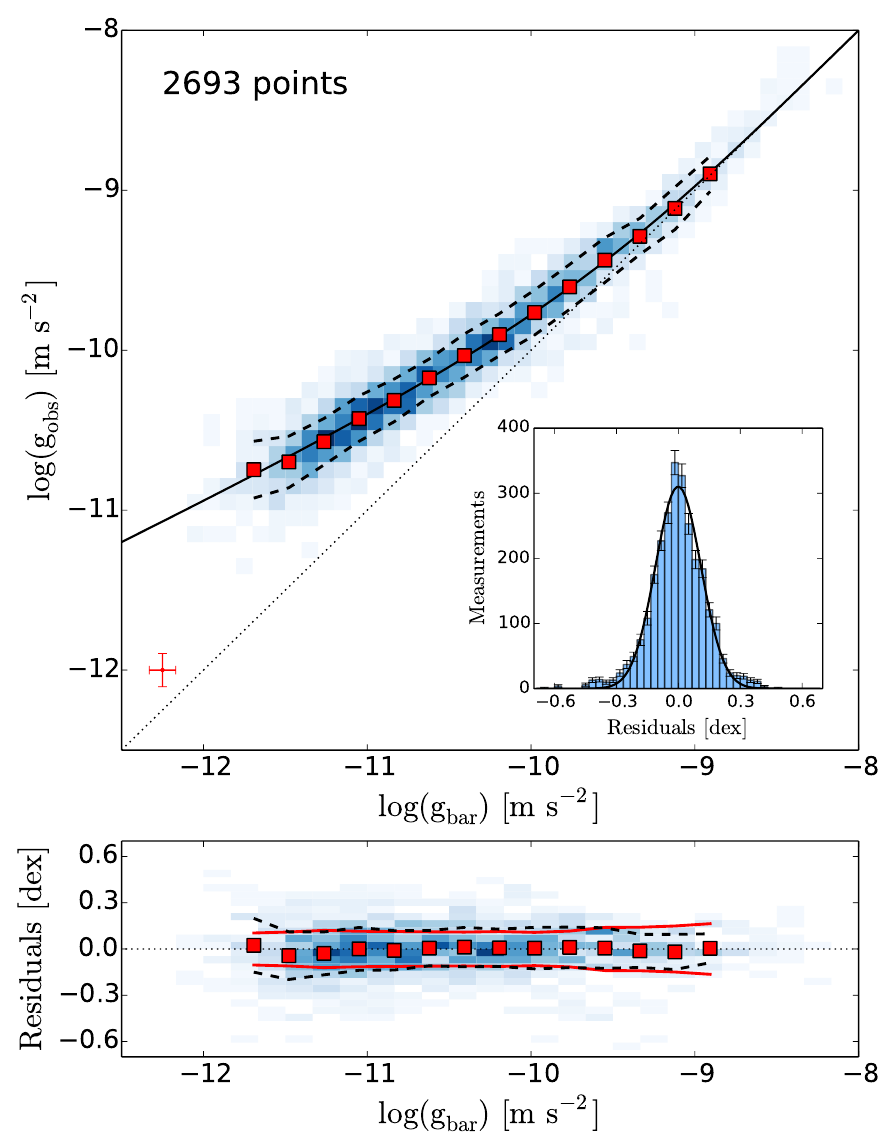}
		\caption{The relation between the actual acceleration $g_{_{obs}}$ in rotationally supported disk galaxies and the prediction $g_{_{bar}}$ of Newtonian gravity using their visible matter. Data from 153 galaxies (yielding 2,693 data points) are shown here as a 2D histogram, with darker shades of blue used to represent more common bins. The typical error budget is shown in red towards the bottom left. Figure from \citet{McGaugh_Lelli_2016}.}
\label{Acceleration_discrepancy_acceleration_relation}
\end{figure}

MOND has been remarkably successful at predicting rotation curves of disk galaxies merely by applying Equation \ref{Modified_Poisson_Equation} to their distribution of visible mass \citep{Famaey_McGaugh_2012}. This works because there is a very tight correlation between the actual accelerations in such galaxies and the predictions of Newtonian gravity over ${\ssim 5}$ orders of magnitude (dex) in mass ($10^7 M_\odot - 10^{12}M_\odot$) and ${\ssim 2}$ dex in surface brightness (Figure \ref{Acceleration_discrepancy_acceleration_relation}). This radial acceleration relation (RAR) is underpinned by mass estimates based on near-infrared photometry collected with the Spitzer Space Telescope \citep{SPARC}, taking advantage of reduced variability in stellar mass-to-light ($M/L$) ratios at these wavelengths \citep{Bell_de_Jong_2001, Norris_2016}. The kinematics are estimated using only the most reliable rotation curves \citep[][section 3.2.2]{SPARC}. The tightness of the RAR in the face of observational uncertainties is perhaps the clearest indication yet that our current understanding of gravity breaks down at very low acceleration.

In this context, it would be easy to explain why Newtonian gravity is off by the same factor close to a low mass galaxy and far from a more massive galaxy, as long as ${g}_{_N}$ is equal at both positions. However, in a DM context, this requires a tight correlation between each galaxy's rotation curve shape, DM halo scale radius and mass such that $\la 10^{-5}$ of the available phase space volume is actually filled \citep{Salucci_2007}. Those authors noted that ``theories of the formation of spirals do not trivially imply the existence of such a surface that underlies the occurrence of a strong dark-luminous coupling.'' More recent investigations continue to have difficulty explaining such correlations using collisionless DM, the standard version of $\Lambda$CDM \citep{Salucci_2017}.

Such correlations are intrinsic to MOND, which therefore predicts a global scaling relation between the asymptotic rotation curve $v_{_f}$ of a galaxy and its mass $M$. Beyond the bulk of the visible light from any galaxy, it can be well-approximated as a point mass in the deep-MOND regime ($g \ll a_{_0}$), thus making Equation \ref{Deep_MOND_limit} valid. As the centripetal acceleration is $\frac{v^2}{r}$ and $g$ is also $\propto \frac{1}{r}$, the rotation curve of every galaxy should eventually go flat at a level
\begin{eqnarray}
	v_{_f} &=& \sqrt{rg} \\
	    &=& \sqrt[4]{GMa_{_0}}
	\label{BTFR}
\end{eqnarray}

\begin{figure}
	\centering
		\includegraphics[width = 8.5cm] {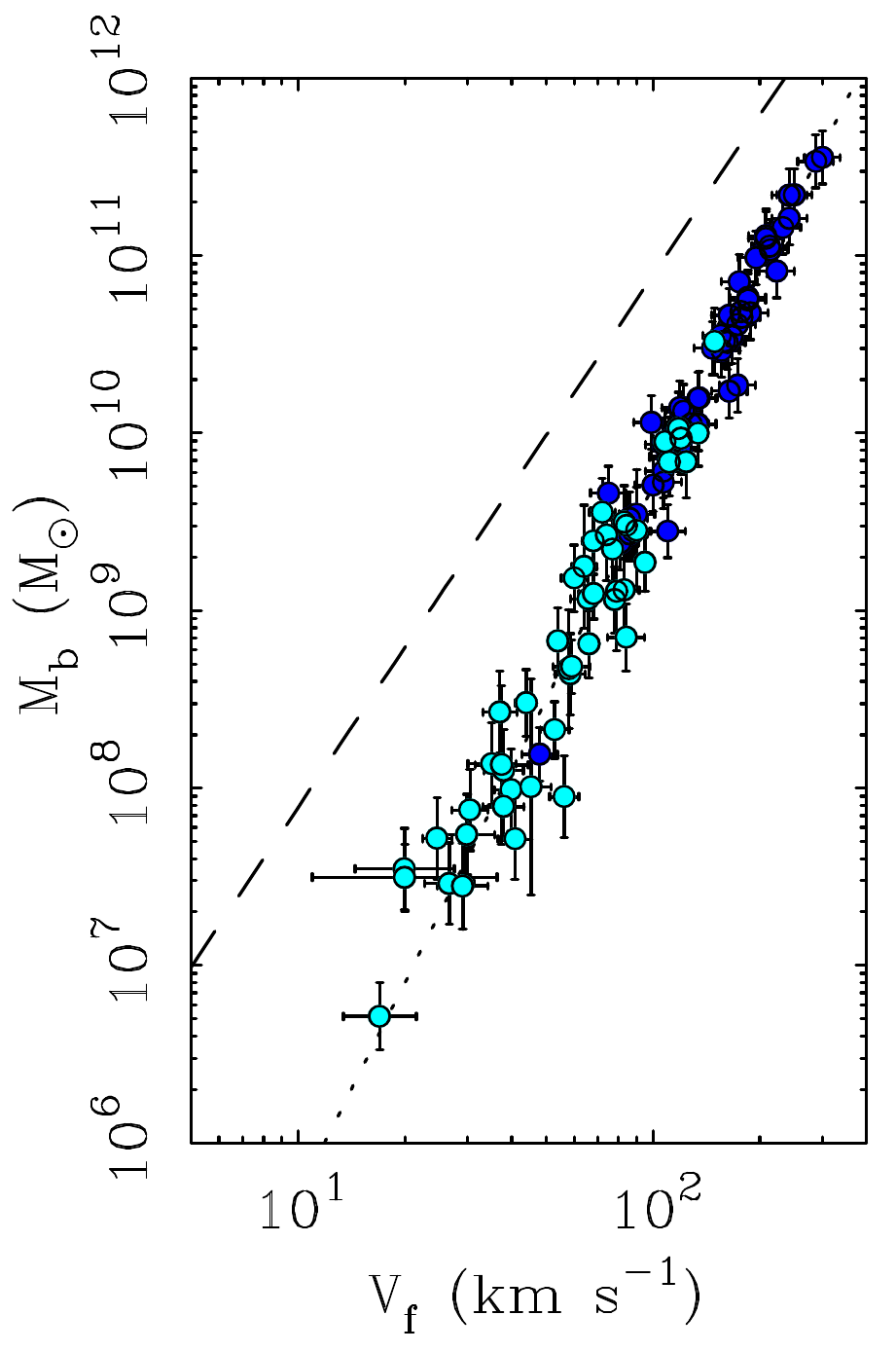}
		\caption{Asymptotic rotation velocity $v_{_f}$ as a function of the total (stellar plus gas) baryonic mass $M_b$ for ${\ssim 80}$ galaxies. The dashed line shows the expected trend using Newtonian gravity and a constant ratio of 5:1 between dark matter and baryons, the ratio required by $\Lambda$CDM for the Universe as a whole \citep{Planck_2015}. The dotted line shows the prediction of MOND (Equation \ref{BTFR}). Dark blue points are star-dominated galaxies while light-blue points are gas-dominated ones. Figure from \citet{Famaey_McGaugh_2012}.}
	\label{BTFR_diagram}
\end{figure}

Observed rotation curves are indeed asymptotically flat, with $v_{_f}$ related to the baryonic mass of a galaxy according to the empirical Baryonic Tully-Fisher Relation (BTFR). Figure 3 of \citet{Famaey_McGaugh_2012} $-$ reproduced in my Figure \ref{BTFR_diagram} $-$ shows that it agrees very well with Equation \ref{BTFR}, the MOND prediction for the BTFR. In the context of $\Lambda$CDM, this agreement is puzzling. It implies a baryon:DM ratio that varies between galaxies and is often much lower than in the Universe as a whole \citep[$\sim$5,][]{Planck_2015}. This is only possible if a substantial fraction of the baryons have been lost from most spiral galaxies $-$ loss of the sub-dominant baryons would not much affect the total mass and rotation curve, but it would reduce the baryon:DM ratio. Recent work on heavy element abundances in galaxies strongly argues against such large amounts of baryons being lost \citep{Rachel_2015}.

Supposing nonetheless that this is possible, the DM would evolve quite differently to the baryons as DM can't radiate. Thus, while supernovae (SNe) can heat up gas and eject it from a galaxy, the DM could not be directly heated by radiation from SNe in this way. As events like SNe are to some extent stochastic, one expects different relative amounts of baryons and DM in different systems.

As an example, I consider the gas fractions of galaxies with equal baryonic mass. Gas-rich and gas-poor galaxies must have evolved differently. The gas-poor galaxy most likely had much more SNe and ejected a larger fraction of its baryons. However, as the DM component dominates at large radii, $v_{_f}$ must be a property of its distribution. Therefore, one expects the galaxy with the lower gas fraction to have started out with more baryons i.e. in a more massive DM halo. This would imply a higher $v_{_f}$ than for the gas-rich galaxy. Yet, there is no correlation between the gas fraction and how far off a galaxy is from satisfying Equation \ref{BTFR}, i.e. the BTFR residual \citep[][figure 4]{Lelli_2017}. This figure also shows that the BTFR residual is uncorrelated with galaxy mass, size and surface density (only the last is a direct consequence of how MOND works as surface densities are related to accelerations, but total mass by itself need not be).



The environment of each galaxy should also play some role. But even in the same conditions, it is inevitable that star formation and SNe feedback is somewhat stochastic, especially in dwarf galaxies. This makes it all the more surprising that there is no evidence for any deviation from Equation \ref{BTFR} over ${\ssim 5}$ orders of magnitude in baryonic mass and a similar range in surface density. In fact, observations constrain any possible intrinsic scatter to ${\la 0.05}$ dex (12\%). Nearly 40 years after Equation \ref{BTFR} was first proposed \citep{Milgrom_1983}, it has remained consistent with rotation curve observations.

A relation roughly like the RAR should arise in $\Lambda$CDM because lower mass DM halos have shallower gravitational potential wells. This should make it easier for baryons to be ejected via energetic processes like SNe feedback. Still, the tightness of the observed RAR is difficult to explain in this way \citep{Desmond_2016, Desmond_2017}. Some attempts have been made to do so \citep[e.g.][]{Keller_Wadsley_2017}, but so far these have investigated only a very small range of galaxy masses and types. In these limited circumstances, there does seem to be a tight correlation of the sort observed. However, a closer look reveals that several other aspects of the simulations are inconsistent with observations \citep{Milgrom_2016}. For example, the rotation curve amplitudes are significantly overestimated in the central regions \citep[][figure 4]{Keller_Wadsley_2016}. This issue was recently revisited by \citet{Tenneti_2018}, who found that it was possible to get a tight RAR but with the wrong low-acceleration behaviour ($g \propto {g_{_N}}^{0.7}$ rather than the observed ${g_{_N}}^{0.5}$) and too high a transition acceleration above which $g \to g_{_N}$.

\begin{figure}
	\centering
		\includegraphics[width = 8.5cm] {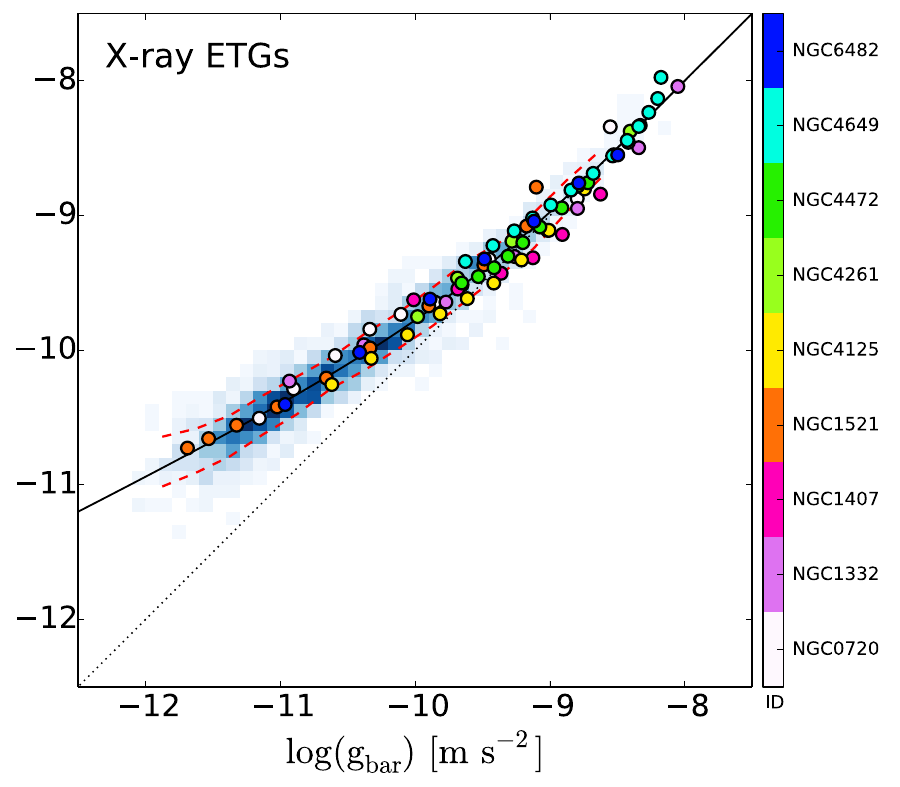}
		\caption{Similar to Figure \ref{Acceleration_discrepancy_acceleration_relation} but showing results for elliptical galaxies as coloured dots, with different colours used to represent different galaxies. The data is based on Chandra and XMM-Newton X-ray observations. The relation for spiral galaxies (Figure \ref{Acceleration_discrepancy_acceleration_relation}) is shown in the background using various shades of blue (more frequent regions shown in a darker shade). Figure from \citet{Lelli_2017}, which shows that similar results are obtained using the 21 cm neutral hydrogen line (see their figure 8).}
	\label{X_ray_combined_RAR}
\end{figure}

Although the disk galaxy RAR may eventually be accounted for in $\Lambda$CDM, it has recently become clear that elliptical galaxies follow the same RAR as spirals (Figure \ref{X_ray_combined_RAR}). This poses additional problems for $\Lambda$CDM because feedback would almost certainly work quite differently in spiral and elliptical galaxies, surely leaving them with different proportions of dark and visible matter.

Galactic gravitational fields can also be probed based on how they deflect light from background galaxies. The lensing is said to be strong if it leads to two distinct images. MOND and General Relativity are expected to have the same relation between gravitational field and light deflection \citep{Chiu_2006}, at least in the tensor-vector-scalar (TeVeS) relativistic extension of MOND \citep{Bekenstein_2004}. With this assumption, \citet{Tian_2017} showed that MOND works rather well with the Sloan Lensing Advanced Camera for Surveys strong lens sample of elliptical galaxies \citep{Bolton_2008}.

If the background galaxy is not multiply imaged but only appears mildly distorted, then this is termed weak gravitational lensing. It has recently become possible to detect weak lensing by galaxies using stacking techniques \citep{Brimioulle_2013}. As this requires a large number of background galaxies to detect, it can't probe too close to the lensing galaxy (within ${\ssim 50}$ kpc). However, it can probe much further out than rotation curves, reaching a typical distance of 250 kpc. Thus, weak lensing and rotation curves are complementary probes of galactic gravitational fields.

50 kpc is much larger than the typical extent of galaxies and their MOND radii \citep[e.g. the MOND radius of the MW is 9 kpc while its disk scale length is 2.15 kpc,][]{Bovy_2013}. Thus, the MOND-predicted gravitational field in the relevant region can be well approximated by Equation \ref{Deep_MOND_limit}. This leads to the simple prediction that a mass $M$ deflects any sufficiently distant ray of light through an angle
\begin{eqnarray}
	\Delta \phi ~=~ \frac{2 \pi \sqrt{GMa_{_0}}}{c^2} ~=~ 2\pi\left(\frac{v_{_f}}{c} \right)^2
	\label{Light_deflection_MOND}
\end{eqnarray}

\citet{Milgrom_2013} showed that this rather simple expectation is consistent with the weak lensing data collected by \citet{Brimioulle_2013}. Importantly, Equation \ref{Light_deflection_MOND} works around both spiral and elliptical galaxies if we assume that foreground galaxies with redder colours are typically ellipticals which have a higher $M/L$ than the bluer spirals.

Individual rotation curves often reveal additional insights. This is especially true for low surface brightness galaxies, which exhibit larger acceleration discrepancies. For example, NGC 1560 has large discrepancies at all radii \citep{Broeils_1992}. It has a pronounced dip in its baryonic radial density profile at ${\ssim 5}$ kpc. In $\Lambda$CDM, this would not affect the overall density of matter very much, because the vast majority of the matter must be dark. As the DM can't radiate and cool, it would not have settled into a disk (unlike the baryons). In any case, the DM needs to remain in a spheroidal halo to explain the stability of the observed disk \citep{Ostriker_Peebles_1973}.

Being dynamically hot, the DM would have a smooth distribution that is hardly affected by a sharp dip in the surface density of the sub-dominant baryons. As this smooth DM component needs to dominate the mass of the galaxy, one might expect the resulting gravitational field to be smooth, yielding a smooth rotation curve. However, there is a sharp dip in the rotation curve corresponding to that in the distribution of baryons \citep{Gentile_2010}.

NGC 1560 is just one example of features in the rotation curve corresponding to features in the underlying distribution of baryons. In fact, the correspondence is almost one-to-one \citep{Sancisi_2004}. While unsurprising when the baryons dominate the gravitational field, this is surprising when the smooth DM component is supposedly dominant.

There are other hints that the gravitational field in low surface brightness galaxies is actually dominated by their baryonic disk. Such galaxies often show spiral structure \citep[e.g.][]{McGaugh_1995}. This is probably the result of a self-gravitating instability in the disk \citep{Lin_Shu_1964}. However, the disk can't be self-gravitating if the mass of the galaxy is dominated by DM, even in the inner parts. This suggests that the large acceleration discrepancies in such systems are caused by matter distributed within the baryonic disk.

In MOND, the stability problem of disk galaxies is resolved by modifying the gravitational field equation to Equation \ref{Modified_Poisson_Equation} \citep{Brada_Milgrom_1998}. Roughly speaking, this is because $\bm{g}$ is sub-linearly affected by the underlying matter density (in Equation \ref{Deep_MOND_limit}, $g \propto \sqrt{M}$ rather than the Newtonian scaling $g \propto M$). Thus, an enhancement to the density has a smaller effect on $\bm{g}$, thereby limiting the tendency of material to be attracted to the overdensity and enhance it further. This makes it more likely that other mechanisms (e.g. pressure) will stabilise the perturbation.

In this way, MOND might be able to confer on disk galaxies just the right amount of stability $-$ neither too much to `smother' spiral structure nor too little to easily let the galaxy evolve into a pressure-supported system. The stabilising mechanism would not work for disk galaxies with a sufficiently high surface density as these would be in the Newtonian regime (if the central surface density $\pi G \Sigma_0 \gg a_0$). This may explain why such disks do not exist in nature \citep{Freeman_1970, McGaugh_1996}. If these disks formed at all, perhaps their central parts were unstable and formed into a bulge within a few dynamical times.

In $\Lambda$CDM, the large sizes of the DM halos around individual galaxies would lead to frequent collisions between them \citep{LCDM_1991}. These collisions would cause mergers through the process of dynamical friction, the tendency of a massive object moving through a field of lower mass particles to gravitationally focus those particles behind it \citep{Chandrasekhar_1943}. This wake then exerts a gravitational force on the object which created it, slowing it down.

As a result, soon after two similarly massive DM halos merge, their central galaxies should also merge \citep{Privon_2013}. The timescale for this would typically be the crossing time of the DM halo, much less than the age of the Universe. As a result, spiral galaxies ought to be quite rare. Even spiral galaxies that avoid major mergers should still have significant bulges due to star clusters losing angular momentum to the surrounding DM halo via dynamical friction \citep{Noguchi_1999, Combes_2014}. In theory, strong stellar feedback could disrupt the cluster quickly, preventing such an inspiral. However, recent observations suggest that star clusters can survive for hundreds of Myr \citep{Zanella_2015}. This leads to a major contradiction with recent observations indicating a high fraction of bulgeless disk galaxies \citep{Kormendy_2010}. To highlight the seriousness of the problem, \citet{Martig_2012} stated that ``no simulation has ever been able to produce a MW-mass bulgeless galaxy'' in a $\Lambda$CDM context.

The high frequency of not just disk galaxies but bulgeless disks poses problems to the idea that galaxies are surrounded by massive DM halos capable of exerting dynamical friction. Such friction arises regardless of the exact nature of the DM particles, as long as they exert gravity (as they must, if halos of such particles cause the acceleration discrepancies). Without such halos, dynamical friction would be reduced and galaxies would be much smaller, making them less likely to collide. This is almost certainly why bulges are much less prevalent and much smaller in MOND than in $\Lambda$CDM \citep{Combes_2014}.

\subsection{Satellite planes and tidal dwarf galaxies}
\label{Introduction_satellite_planes}

$\Lambda$CDM faces another problem with the detailed properties of galaxies in the Local Group (LG). Wide field surveys such as the Sloan Digital Sky Survey \citep{SDSS} and the Pan-Andromeda Archaeological Survey \citep{PANDAS} have shown that the satellite galaxies of the MW are preferentially located in a thin co-rotating planar structure \citep{Kroupa_2013}. The same is also true of Andromeda \citep{Ibata_2013}, though co-rotation can't be definitively confirmed without proper motions.

It appears very unlikely that these structures formed quiescently \citep{Pawlowski_2014, Ibata_2014}. For the MW, filamentary infall is considered unlikely because this would imply its satellites had very eccentric orbits, contrary to observations \citep{Angus_2011}. These require the accretion to have been long ago in order to give enough time to circularise the orbits via dynamical friction against the Galactic DM halo. However, interactions between satellites and numerous DM halos that are thought to surround the MW would cause the dispersal of any initially thin disk of satellites \citep{Klimentowski_2010}. A similar phenomenon would be expected around M31 \citep{Fernando_2018}. Even if the number of subhalos was smaller than predicted by $\Lambda$CDM, the triaxial nature of the potential would lead to any disk-like structure spreading out on a timescale of ${\ssim 5}$ Gyr unless it was fortuitously aligned with a symmetry axis of the potential \citep{Bowden_2013, Fernando_2016}. This issue is less serious in MOND as the matter distribution is much more concentrated, leading to a nearly spherical potential beyond ${\sim 40}$ kpc (Figure \ref{v_esc_profile}).

After careful consideration of several proposed explanations for why primordial satellites now lie in a thin plane, \citet{Pawlowski_2014} concluded that none of them agreed with observations for either the MW or M31. It was later shown that baryonic effects are unlikely to provide the necessary anisotropy if one sticks to a primordial origin for the satellites \citep{Pawlowski_2015}. This issue was revisited by performing a high-resolution $\Lambda$CDM hydrodynamical simulation of a MW analogue in a cosmological context \citep{Maji_2017}. Although this unpublished article claimed that the results were consistent with observations, it has recently been shown that this is not the case \citep{Pawlowski_2017}. Those authors showed that the satellite galaxy distribution of \citet{Maji_2017} was consistent with isotropy. However, the actual MW satellite system is inconsistent with isotropy at more than ${5 \sigma}$ once the survey footprint is taken into account \citep{Pawlowski_2016}.

More recent hydrodynamical $\Lambda$CDM simulations also fail to yield highly flattened satellite systems like those observed around the MW and M31 \citep{Ahmed_2017}. The mild flattening in these simulations might not even be related to baryonic effects as similar results arise in DM-only simulations \citep{Garaldi_2017}. In any case, it is difficult to see how baryonic effects like radiative cooling can explain a ${\sim 200}$ kpc-wide plane of primordial satellites composed mostly of DM.

This raises the possibility that most LG satellites are not primordial $-$ perhaps they formed as second-generation tidal dwarf galaxies (TDGs) during an ancient galactic interaction \citep{Kroupa_2005}. After all, we do see galaxies forming from material pulled out of interacting progenitor galaxies \citep[e.g. in the Antennae,][]{Mirabel_1992}. This naturally leads to anisotropy because the tidal debris tend to be confined within the common orbital plane of the interacting progenitor galaxies.


TDGs form by self-gravitating collapse, requiring a high density. This is easy to obtain by tidally perturbing baryons originally on near-circular orbits in a rotating disk, leading to a thin dense tidal tail. However, the DM halos hypothesised to surround galaxies need to be dynamically hot \citep{Hohl_1971}. Tidally perturbing this rather sparse (albeit massive) halo would yield only a very low density, insufficient to reach the threshold for Jeans instability. Consequently, TDGs should be free of DM \citep{Barnes_1992, Wetzstein_2007}. Their rather low escape velocity also precludes them from subsequently accreting significant amounts of DM out of their host galaxy's halo.

Thus, a surprising aspect of LG satellites is their high Newtonian dynamical masses compared to their low luminosities \citep[e.g.][]{McGaugh_2013}. These $M/L$ ratios are calculated assuming dynamical equilibrium, an assumption which could be invalidated by tides from the host galaxy. However, tides are likely not strong enough to do this \citep[][figure 6]{McGaugh_2010}. As DM is unlikely to be present in these systems, some other explanation must be found for their high inferred $M/L$ ratios.

This is true even in unconventional models of DM where it has significant non-gravitational interactions with baryons \citep{Famaey_2017}. This model is designed to explain the RAR and arguably can do so in the visible regions of both spiral and elliptical galaxies. However, their table 1 shows that TDGs are expected to be free of DM and thus follow standard Newtonian behaviour. Similarly, the MW and M31 satellite planes are rather extended \citep[e.g.][figure 2]{Kroupa_2013} and would very likely reach beyond the hypothetical superfluid DM halos of their host galaxies \citep{Berezhiani_2015, Khoury_2016}. This would cause the more distant MW and M31 satellite plane members to follow Newtonian dynamics.

Without DM, the strong self-gravity needed to maintain high internal velocity dispersions arises most naturally from a modification to gravity. In the context of the most widely investigated such model, the MW and M31 would have undergone an ancient close flyby ${\ssim 8}$ Gyr ago \citep[][figure 4]{BANIK_2017_ANISOTROPY}. Initial $N$-body simulations of this flyby in MOND suggest that this is a plausible scenario, though it is not yet clear if it can match LG properties in detail \citep{Bilek_2017}.

A past encounter with M31 might naturally account for the MW thick disk \citep{Gilmore_1983}, a structure which seems to have formed fairly rapidly from its thin disk \citep{Hayden_2015} ${9 \pm 1}$ Gyr ago \citep{Quillen_2001}. More recent investigations suggest that there was a burst of star formation at that time \citep[][figure 2]{Snaith_2014}. The star formation rate of M31 also appears to rise sharply for lookback times ${\ga 9}$ Gyr \citep[][figure 12]{Dolphin_2017}. The disk heating which likely formed the MW thick disk appears to have been stronger in its outer parts, characteristic of a tidal effect \citep{Banik_2014}. This may explain why the Galactic thick disk has a longer scale length than its thin disk \citep{Juric_2008, Jayaraman_2013}.

One possible objection to this theory is that the heavy element abundances of the planar M31 satellites seem rather similar to those outside its satellite plane \citep{Collins_2015}. One might expect there to be a difference if some M31 satellites formed from material already enriched by virtue of being within the disk of a massive galaxy (M31) while others formed primordially. However, this difference becomes very small if the MW-M31 interaction was a very long time ago. This is because there would have been little time to enrich the gas in the M31 disk. M31 would very likely have been much more gas-rich than at present, diluting any heavy elements formed by stars. Moreover, the material that formed into M31 satellites would necessarily have come from the outer parts of the M31 disk, which is generally less enriched \citep[e.g.][figure 9]{Gregersen_2015}. For all these reasons, it is quite feasible that there would be no observable difference between the chemical abundances of M31 satellites even if they had very different formation scenarios \citep{Kroupa_2015}. It will be interesting to see if some difference is eventually found, although this might be much easier around our Galaxy than around M31. This could take advantage of Sextans not being part of the MW satellite plane \citep{Dinescu_2018}.

Similarly to the MW and M31, the satellite system of Centaurus A (Cen A) is also highly flattened and co-rotating, as evidenced by a radial velocity gradient across it \citep{Muller_2018}. Such structures thus appear to be common, a claim also made by \citet{Ibata_2014_Nature} based on their finding that satellites on either side of a host galaxy have radial velocities of opposite signs, once the host systemic motion is accounted for. This is too recent for the debate to have settled \citep{Cautun_2015, Ibata_2015}. Nonetheless, it does seem like the Universe may well be full of TDGs if even just a few form in each galactic interaction \citep{Okazaki_2000}. A high frequency of TDGs is also suggested by the correlation between cases where their existence is confirmed and the bulge mass fraction of the central galaxies \citep{Lopez_Corredoira_2016}. If TDGs are more common, it would be easier to test whether the acceleration discrepancy persists in such systems, potentially resolving the question of how it arises in general.

\section{Overview of the portfolio}


To investigate the cause of the acceleration discrepancy, this portfolio considers several tests of the $\Lambda$CDM paradigm and some tests of MOND. The first of these \citep{BANIK_2015_MCE} is described in Section \ref{Bullet_Cluster} and relates to the rather high collision velocity of the components of the Bullet Cluster, two interacting galaxy clusters \citep{Tucker_1995}. Such a high velocity appears difficult to reconcile with $\Lambda$CDM \citep{Thompson_Nagamine_2012, Kraljic_2015}. However, the relative velocity between the clusters is mostly in the plane of the sky. Thus, it has not been directly measured but only estimated based on hydrodynamic simulations attempting to reproduce observed properties of the Bullet Cluster \citep{Lage_Farrar_2014}.

Fortunately, \citet{Molnar_2013} showed that it should soon become possible to measure the proper motion of the components of this cluster using the Moving Cluster Effect \citep[MCE,][]{Birkinshaw_Gull_1983}. The MCE involves measuring redshifts of a background object multiply imaged by a foreground lens. Motion of the lens makes its potential time-dependent, thus giving the images different redshifts. However, the images could also have different redshifts because they have different magnification patterns across the source, provided this has a redshift gradient e.g. due to rotation. Although the issue could be resolved by taking integral field unit spectra at the appropriate velocity resolution, this is extremely challenging $-$ only a spatially unresolved spectrum of each image is likely to be available for the foreseeable future.

Thus, I considered how these different effects could be disentangled using spectral line profiles of the individually unresolved images \citep{BANIK_2015_MCE}. I also considered observational strategies to minimise the effects of such systematic errors, thus clarifying the kinematics of the Bullet Cluster. The same techniques could be applied to other interacting galaxy clusters like El Gordo (ACT-CL J0102-4915), which may be particularly problematic for $\Lambda$CDM due to its combination of high redshift \citep[$z = 0.87$,][]{Menanteau_2012}, high mass \citep{Jee_2014} and high inferred collision speed \citep{Molnar_2014}.

Although relative proper motions may eventually be obtained in such systems, full 3D position and velocity information is only available within the LG out to about the distance of M31 \citep{Van_der_Marel_2012} and M33 \citep{Brunthaler_2005}. Thus, the remainder of this portfolio focuses on the LG. The second work in the portfolio \citep{BANIK_ZHAO_2016} $-$ described in Section \ref{Local_Group_2D} $-$ describes the construction of an axisymmetric dynamical model of the LG in $\Lambda$CDM, building on earlier spherically symmetric models \citep{Kahn_Woltjer_1959, Sandage_1986, Jorge_2014}. An axisymmetric model is expected to be rather accurate due to the very small MW-M31 tangential velocity \citep{Van_der_Marel_2012} and the close alignment of Cen A with the MW-M31 line \citep{Ma_1998}. This model is used to perform a timing argument analysis i.e. see if some combination of model parameters can match the observed positions and radial velocities of M31 and LG dwarf galaxies using cosmological initial conditions ($\bm{v}_{pec} = \bm{0}$ at early times). Despite a reasonable allowance for observational uncertainties and inaccuracies in my model as a representation of $\Lambda$CDM, a full grid search through the model parameters did not yield a model matching the observed kinematics of the LG. This is because some galaxies have very high radial velocities.

To investigate this issue further, I used a 3D model of the LG in $\Lambda$CDM \citep{BANIK_ZHAO_2017}, the third work in this portfolio (Section \ref{Local_Group_3D}). It is based on a \textsc{fortran} algorithm borrowed from P. J. E. Peebles \citep{Shaya_2011}. Despite using a different code written by a different author in a different programming language, my results still indicated that several LG galaxies have much higher radial velocities than expected in $\Lambda$CDM. The typical discrepancy between observations and the best-fitting 3D model was actually slightly higher than in the 2D case, even though the 3D model includes the major mass concentrations within ${\ssim 10}$ Mpc.

The fourth work in this portfolio \citep{BANIK_2017_ANISOTROPY} is described in Section \ref{HVG_plane} and looks at these high-velocity galaxies (HVGs) in more detail. As part of this, I visited Peebles at Princeton in order to revisit the work of Section \ref{Local_Group_3D} by performing a more thorough search for the best-fitting 3D model. This only slightly improved the fit to observations, still leaving several HVGs. A similar conclusion was also reached by \citet{Peebles_2017}. Confident that the HVGs are real, \citet{BANIK_2017_ANISOTROPY} shows that they preferentially lie very close to a well-defined plane which passes close to both the MW and M31. In this work, I use a restricted $N$-body model of the LG in MOND to argue that such a HVG plane is a natural consequence of a past MW-M31 flyby. Several $\Lambda$CDM-based explanations for the observations are also considered, but none of them seem plausible.

As a result, the fifth and final work in this portfolio (Section \ref{v_esc_article}) considers MOND in more detail, in particular how it works in our own Galaxy \citep{BANIK_2017_ESCAPE}. Although MOND has often been tested using rotation curves of galaxies \citep[e.g.][]{Famaey_McGaugh_2012}, I focus on comparing it to the recently measured Galactic escape velocity curve over distances of 8$-$50 kpc \citep{Williams_2017}. Both its amplitude and radial gradient are well matched in a MOND Galactic model that also accounts for its rotation curve. In future, the constraints should tighten considerably with GAIA data \citep{Perryman_2001}, much of which is expected to be released in April 2018.

In Section \ref{Future_prospects}, I suggest future avenues of investigation and give my conclusions in Section \ref{Conclusions}. Despite MOND not being a complete theory, it is well-defined and highly predictive in a wide range of circumstances. Therefore, it should be directly testable in the near future.


\section{Effects of Lens Motion \& Uneven Magnification On Image Spectra \citep{BANIK_2015_MCE}}
\label{Bullet_Cluster}

On a large scale, the collision speed distribution of interacting galaxy clusters can be quite sensitive to the underlying law of gravitation \citep{Cai_2014}. Indeed, the high collision speed of the components of the Bullet Cluster \citep[1E0657-56,][]{Tucker_1995} has been argued in favour of modified gravity \citep{Katz_2013}. However, this speed is not directly measured as the collision is mostly in the plane of the sky. Instead, the speed is estimated using simulations of the shock generated in the gas by the collision \citep{Lage_Farrar_2014}. The separation of the DM and gas \citep{Dark_Matter_Proof} also plays an important role $-$ there is less gas drag at lower speeds, reducing the separation. 

A collision speed close to 3000 km/s is considered necessary to explain the observed properties of the Bullet Cluster \citep{Mastropietro_Burkert_2008}. For the inferred masses of its components \citep{Clowe_2004}, this appears difficult to reconcile with $\Lambda$CDM \citep{Thompson_Nagamine_2012}. This work suggested that a cosmological simulation requires a co-moving volume of $(4.48 h^{-1} \text{Gpc})^3$ to see an analogue to the Bullet Cluster. A subsequent analysis also found that systems analogous to the Bullet Cluster are expected to be rare in $\Lambda$CDM \citep{Kraljic_2015}.

Moreover, a few other massive colliding clusters with high infall velocities have been discovered in the last few years \citep{Gomez_2012, Menanteau_2012, Molnar_2013_Abell}. The El Gordo Cluster (ACT-CL J0102-4915) may be particularly problematic due to its combination of high redshift \citep[$z = 0.87$,][]{Menanteau_2012}, high mass \citep{Jee_2014} and high inferred collision speed \citep{Molnar_2014}.

\citet{Molnar_2013_Abell} argue that inferring collision speeds from observations of the shock can be non-trivial just due to projection effects, let alone other complexities of baryonic physics. To see if there is any tension with $\Lambda$CDM, collision speeds should be determined in a more direct way. This is normally achieved using proper motions, but obtaining them is not feasible over cosmological distances.

Fortunately, the tangential motion of a massive object can be constrained using the Moving Cluster Effect \citep{Birkinshaw_Gull_1983}. The MCE relies on the gravitational potential of an object being time-dependent due to its motion. Consequently, if a source behind the object were multiply imaged, the images would have slightly different redshifts. Moreover, as DM generally outweighs gas on cluster scales \citep{Blaksley_Bonamente_2009}, the MCE is mostly sensitive to motion of the DM. This is simpler to model than gas, making the results easier to compare with cosmological simulations.

\subsection{Method}

\begin{figure}
	\centering
		\includegraphics[width = 8.5cm]{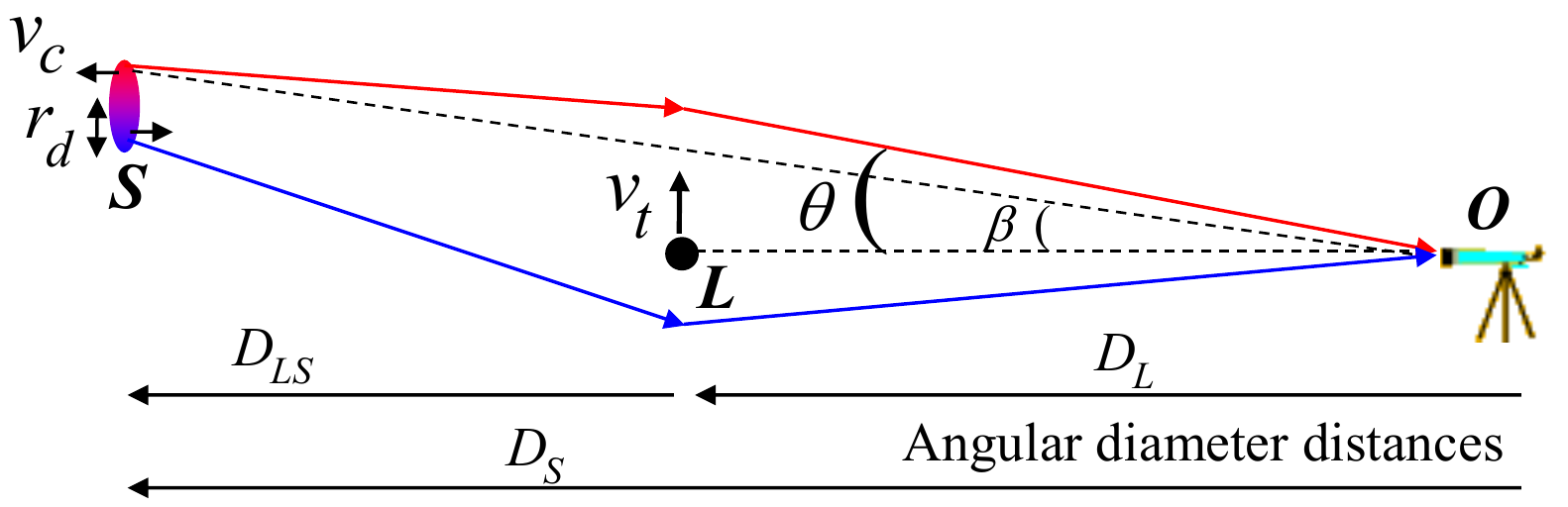}
	\caption{The lensing geometry is depicted here. Upper photon trajectory $=$ primary image (same side as unlensed source), lower trajectory $=$ secondary image. Relevant angular diameter distances are indicated at bottom. The lens $L$ is treated as a point mass moving transversely to the viewing direction at speed $v_t$. The source $S$ is an extended disk galaxy with scale length $r_d$. There is a redshift gradient across it due to rotation at speed $v_c \left( r \right)$, where $r$ is distance from the source galaxy's centre.}
	\label{Lensing_Geometry}	
\end{figure}

The basic geometry is shown in Figure \ref{Lensing_Geometry}. To understand the MCE, consider a static universe in which the observer and source have negligible peculiar motions compared to the lens, which has a transverse velocity $\bm{v}_t$ in addition to any line of sight velocity that is not relevant for this analysis. It helps greatly to transform reference frame to the one in which the lens is static but the observer and source are moving at $-\bm{v}_t$. The lensing potential is now static. Emitted and received photons have different frequencies because the photon trajectories are not orthogonal to the velocities of the observer or the source. The difference in image redshifts can be expressed as a velocity shift
\begin{eqnarray}
	{\left. \Delta {v_r} \right|}_{MCE} ~=~ -\bm{v}_t \cdot \left( \bm{\alpha}_1 - \bm{\alpha}_2 \right)
\end{eqnarray}

Here, the light deflection angle for each image $i$ is $\bm{\alpha}_i$. Using the thin-lens approximation\footnote{i.e. the deflection occurs over a very small fraction of the entire photon path}, this can be expressed in terms of the observed image positions $\bm{\theta}_i$ as
\begin{eqnarray}
	{\left. \Delta {v_r} \right|}_{MCE} ~=~ \frac{D_{s}}{D_{ls}} \bm{v}_t \cdot \left( \bm{\theta}_1  - \bm{\theta}_2 \right)
\end{eqnarray}

The angular diameter distances relevant to this problem are illustrated in Figure \ref{Lensing_Geometry}, with ${D_{ls}}$ representing the angular diameter distance to the lens as perceived by an observer at the source when the photons we detect now most closely approached the lens. A source perfectly aligned with a point-like lens of mass $M$ would appear as an Einstein ring of angular radius
\begin{eqnarray}
	{\theta }_E ~\equiv~ \sqrt{\frac{4GM}{c^2}\frac{D_{ls}}{D_l D_s}}
\end{eqnarray}

Combined with the source and lens positions, the lens mass $M$ thus sets a typical angular scale for the problem. I use it to define
\begin{eqnarray}
	u \equiv \frac{\beta}{\theta_E}   ~~~~~   \text{and}   ~~~~~  y \equiv \frac{\theta}{\theta_E}
\end{eqnarray}

In terms of the lens and source physical properties and their true (unlensed) positions,
\begin{eqnarray}
	{\left. \Delta {v_r} \right|}_{MCE} ~=~ \frac{2 v_t \sqrt{GM \left( u^2 + 4 \right) }}{c} \sqrt{\frac{D_s}{D_{ls}D_l}}
	\label{MCE}
\end{eqnarray}

$v_t$ is the component of the lens transverse velocity $\bm{v}_t$ directed along the separation between the observed images.

I estimated the effects of source \& observer peculiar velocities $-$ they should not affect $\Delta {v_r}$ much. There is also a time delay between the images, causing us to observe the source at an earlier epoch in one image than in the other. Due to cosmic expansion, this creates a redshift difference, but only a very small one (${<1}$ m/s).


\begin{figure}
	\centering
		\includegraphics [width = 8cm] {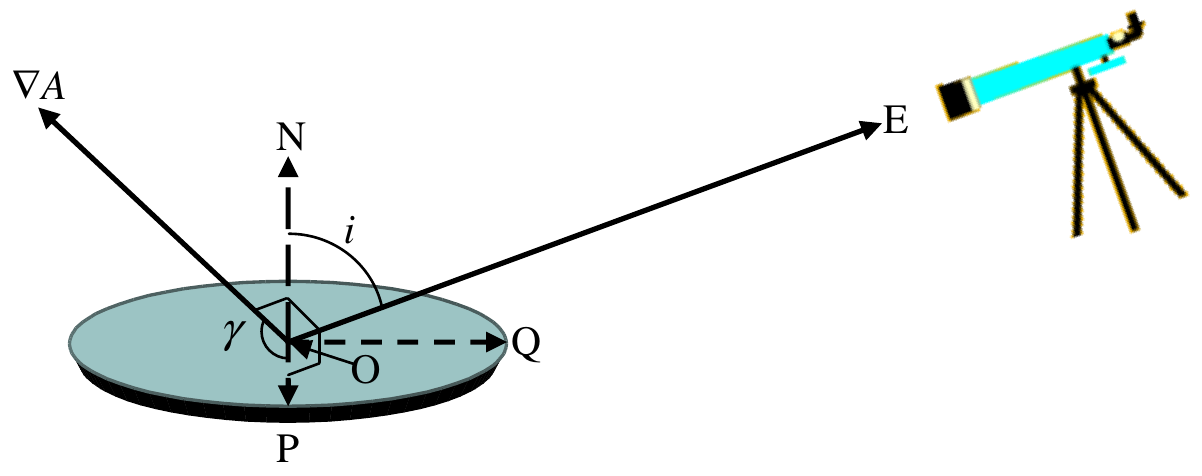}
	\caption{The observing geometry is shown here. The source galaxy has centre O and normal to its plane $\bm{ON}$. Earth is towards $\bm{OE}$, so the galaxy's inclination to the sky plane is $i$. $\bm{OQ}$ and $\bm{OP}$ are in the galaxy's plane and orthogonal to each other, with $\bm{OQ}$ as closely aligned with $\bm{OE}$ as possible. Thus, $\bm{OP}$ and $\bm{OE}$ are orthogonal. $\bm{\nabla} A$ is directed within the source plane, so must also be orthogonal to $\bm{OE}$. $\bm{\nabla} A$ is at an angle $\gamma$ to $\bm{OP}$. The source is parametrised using cylindrical polar co-ordinates ($r$, $\phi$), with centre $O$ and initial direction ($\phi = 0$) along $\bm{OQ}$.}
	\label{Source_geometry}
\end{figure}

The main systematic uncertainty in MCE measurements is likely to be the Differential Magnification Effect (DME), an observational artefact due to our inability to take highly accurate spectra of the images while also spatially resolving them. This causes parts of the source with different redshifts to get blended together in spectra. The precise way in which this blending occurs differs between the images.

To model how such single-pixel spectra might look, I modelled the source as a typical spiral galaxy with exponential surface density profile \citep{Freeman_1970} and a realistic rotation curve based on Equation \ref{Simple_mu}. The lens is treated as a point mass. The parameters considered (Table \ref{Inputs}) are designed for the Bullet Cluster \citep{Tucker_1995}.

The basic idea is that spatially unresolved spectra can determine the intensity-weighted mean redshift $\overline{v_r}$ of each image. This may be affected by rotation of the source galaxy. The effect isn't reliant on an expanding Universe. Neglecting cosmic expansion for the moment, the mean redshift velocity of each image is
\begin{eqnarray}
	\overline{v_r} ~\equiv~ \frac{\int_{\text{Image}}{A\Sigma {{v}_{r}}}~dS}{\int_{\text{Image}}{A\Sigma }~dS}
	\label{Governing_Equation}
\end{eqnarray}

The integrals are over area elements of the source $S$, which I treat as an exponential disk with surface density profile
\begin{eqnarray}
	\Sigma ~=~ \Sigma_0 ~ {e}^{-\tilde{r}} ~~\text{ where } \tilde{r} \equiv \frac{r}{r_d}
	\label{Surface_density}
\end{eqnarray}

The magnification $A$ varies little over the source galaxy as $\frac{r_d}{D_s} \ll \theta_E$ (Table \ref{Inputs}). Thus, a linear approximation to $A$ is sufficient.
\begin{eqnarray}
	A ~\approx~ A_0 + \frac{\partial A}{\partial u} du ~~~(A_0 \equiv A ~\text{at centre of source})
\end{eqnarray}

\begin{samepage}
The mean redshift of each image due to the DME is
\begin{equation}
	\label{Intermediate_Equation}
	\left| \overline{v_r} \right| =
\end{equation}
\begin{eqnarray}		
	\frac{v_{_f} r_d \sin i\cos \gamma }{D_s \theta _E} \frac{\int\limits_{0}^{\infty }{\int\limits_{0}^{2\pi }{\overbrace{{{e}^{-\tilde{r}}}}^{\propto \Sigma }\widetilde{v_c}(\tilde{r}){{\tilde{r}}^{2}}}\overbrace{\frac{4}{{{u}^{2}}{{({{u}^{2}}+4)}^{\frac{3}{2}}}}}^{-\frac{\partial A}{\partial u}}{{\sin }^{2}}\phi ~ d\phi ~ d\tilde{r}}}{\pi \underbrace{\left( \frac{{{u}^{2}}+2}{u\sqrt{{{u}^{2}}+4}}\pm 1 \right)}_{\propto A}} \nonumber
\end{eqnarray}
\end{samepage}

I obtained a family of rotation curves using the `simple $\mu$-function' in MOND \citep{Famaey_Binney_2005}, as discussed just before Equation \ref{Simple_mu}. The rotation curve shape is determined by the central surface density, which I parametrise using
\begin{eqnarray}
	k ~\equiv~ \frac{\Sigma_0 G}{a_0}
	\label{k}
\end{eqnarray}

Thus, my rotation curves flatline at
\begin{eqnarray}
	v_{_f} ~&=&~ \sqrt[4]{GMa_0} \\
	    ~&=&~ \sqrt[4]{2 \pi k} \sqrt{r_d a_0}
\end{eqnarray}

After making a few approximations to estimate $\bm{g}_{_N}$, I obtained the overall rotation curve shape
\begin{samepage}
\begin{eqnarray}
	\label{Rotation_Curve}
	\widetilde{v_c}(\tilde{r}) \equiv \frac{v_c(\tilde{r})}{v_{_f}} = \frac{\sqrt{\pi k \tilde{r} f(\tilde{r}) + \tilde{r} \sqrt{\left( {\pi k f(\tilde{r} ) + 1} \right)^{2} - 1}}}{\sqrt[4]{2 \pi k}}
\end{eqnarray}
\begin{eqnarray}
		f(\tilde{r}) = \frac{1 - \frac{13}{4} {e}^{-\tilde{r}}}{\tilde{r}^2} - \frac{7 {e}^{-\tilde{r}}}{4\tilde{r}} + \frac{9 \left( 1 - {e}^{-\tilde{r}} - \tilde{r}{e}^{-\tilde{r}}\right)}{2 \tilde{r}^4}
\end{eqnarray}
\end{samepage}

\begin{figure}
	\centering
		\includegraphics [width = 8.5cm] {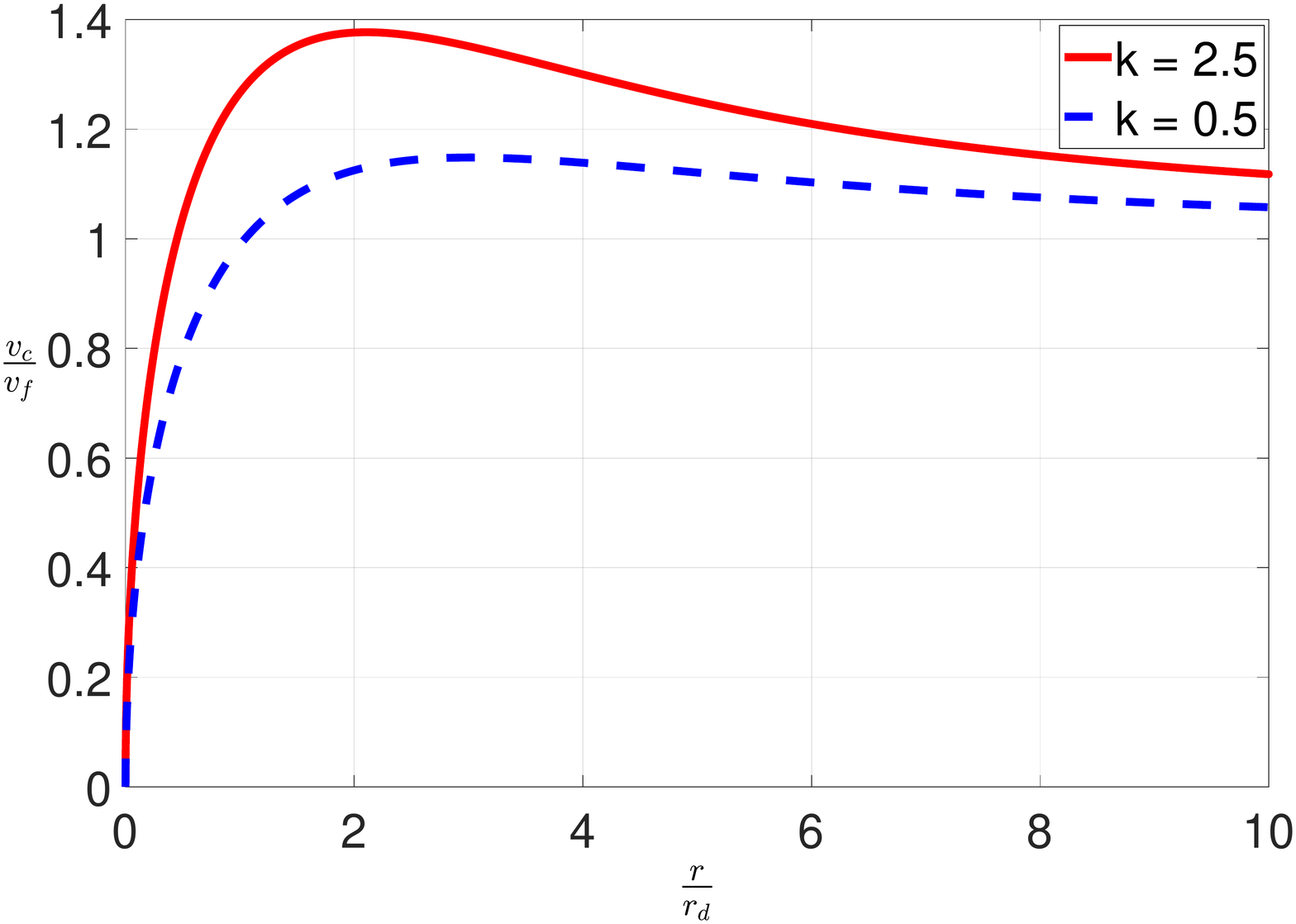}
		\includegraphics [width = 8.5cm] {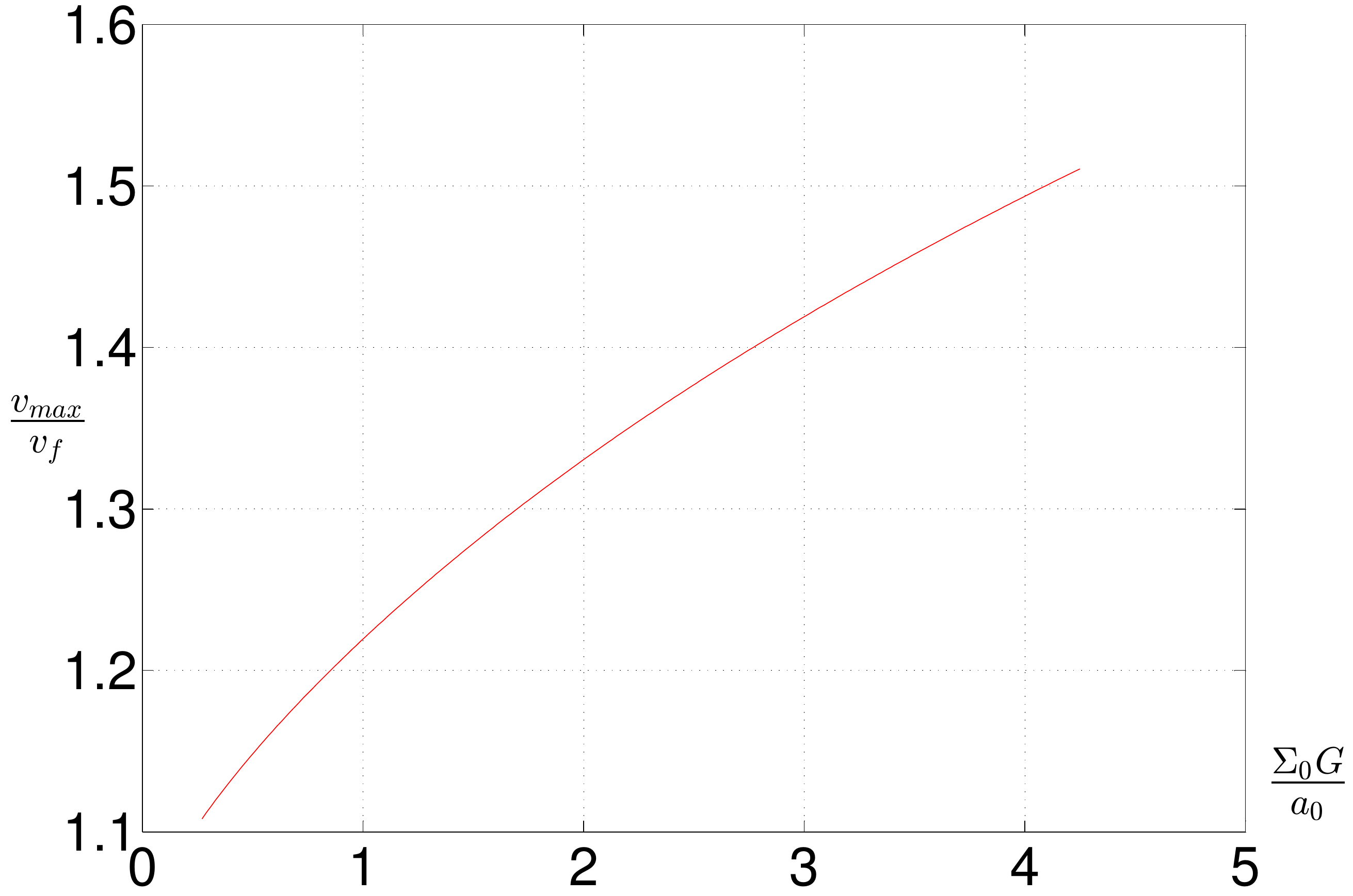}
	\caption{\emph{Top}: Rotation curves resulting from Equation \ref{Rotation_Curve}. $v_c \left( r \right)$ flatlines at $v_{_f}$. The surface density $\Sigma = \Sigma_0 e^{-\frac{r}{r_d}}$. The parameter $k \equiv \frac{G \Sigma_0}{a_0}$ controls the shape of the rotation curve. \emph{Bottom}: The ratio of maximum to flatline rotation speed as a function of $k$.}
	\label{Maximum_Rotation_Speeds}
\end{figure}

The normalised rotation speed at radius $\tilde{r} \equiv \frac{r}{r_d}$ is $\widetilde{v_c} \equiv \frac{v_c}{v_{_f}}$, with $v_{_f}$ the flatline level of the source galaxy rotation curve. Two example rotation curves are shown in the top panel of Figure \ref{Maximum_Rotation_Speeds}. Its bottom panel shows the ratio between $v_{max}$ and $v_{_f}$ for disks with different central surface densities.

\subsection{Results}

Combining Equation \ref{Intermediate_Equation} with the rotation curve shape from Equation \ref{Rotation_Curve}, I get that
\begin{eqnarray}
	\label{DME}
	{{\left. \Delta \overline{v_r} \right|}_{DME}} &=& \frac{v_{_f} ~ r_d ~ \sin i ~ \cos \gamma ~ I ~ c ~ \sqrt{D_l}}{\sqrt{u^2 + 4} ~ \sqrt{GM{D}_{ls}D_s}} \\
	I &\equiv& \int_0^\infty{{\rm{e}}^{-\tilde{r}} \widetilde{v_c}(\tilde{r})~\tilde{r}^2} d\tilde{r}
\end{eqnarray}

The integral $I$ depends on the central surface density $k$. However, the ratio $\frac{I}{\widetilde{v}_{max}} = 1.89 \pm 0.02$ over the range $k = 0.1 - 5$. Thus, $k$ is not needed very precisely if $v_{max}$ were available rather than $v_{_f}$. In practice, it is much easier to obtain $v_{max}$.

\begin{figure}
	\centering
		\includegraphics [width = 8.5cm] {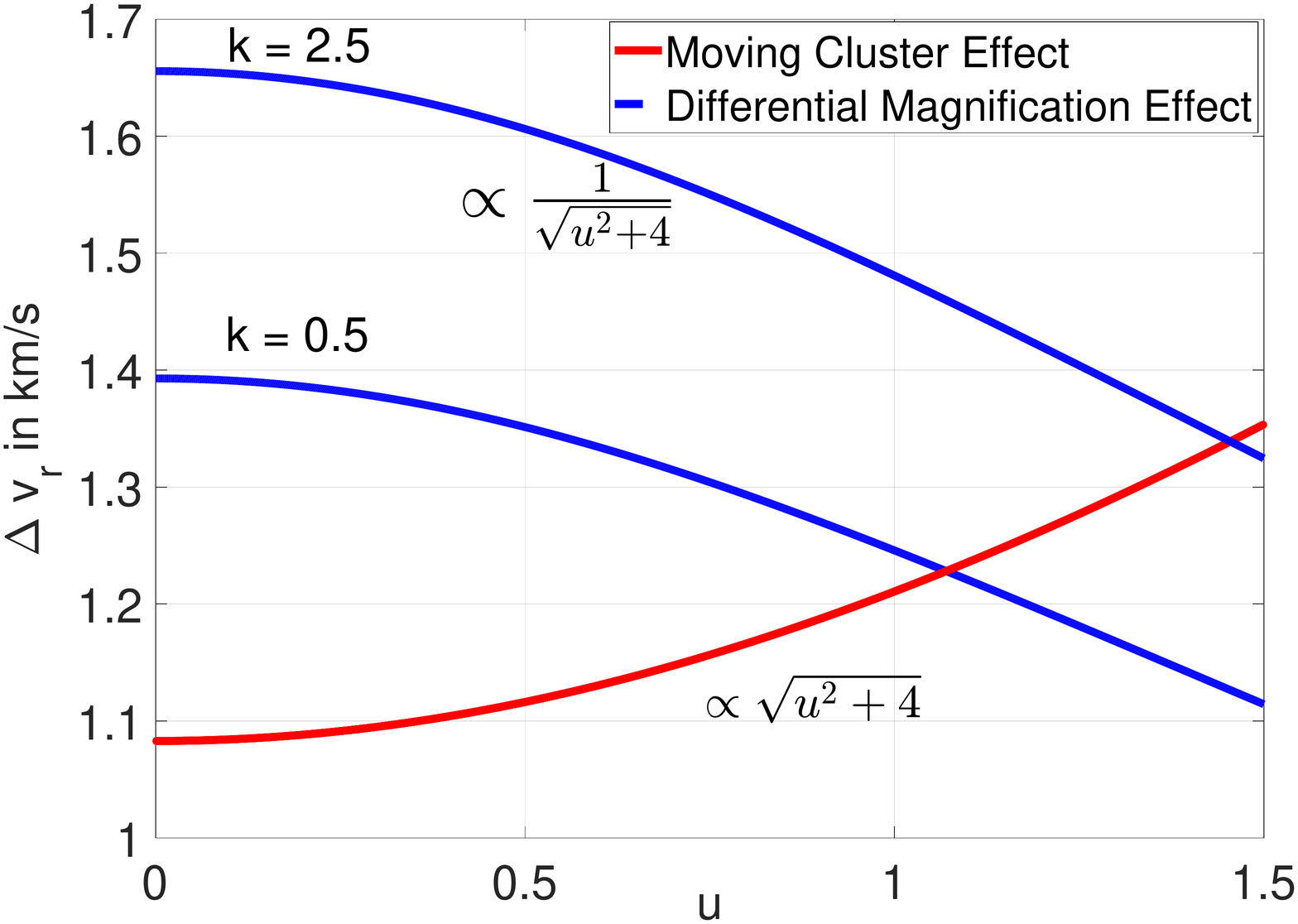}
	\caption{The difference in redshift between double images of a typical background galaxy as a function of its position, due to the effects described in the text (Equations \ref{MCE} and \ref{DME}). Parameter values used here are listed in Table \ref{Inputs}. The DME is affected by the shape of the rotation curve (governed by the central surface density $k$) once its flatline level $v_{_f}$ is fixed. If instead $v_{max}$ is known, then $k$ has only a very small (${\sim 1\%}$) impact on the DME.}
	\label{Overall_Effect}
\end{figure}

\begin{table} 
	\caption{Parameters used for Figure \ref{Overall_Effect}. The source galaxy is assumed positioned so as to maximise the MCE (i.e. it is separated from the lens on the sky along the direction of motion of the lens, which is clear from images). The lens mass should roughly correspond to the sub-cluster in the Bullet. A flat $\Lambda$CDM cosmology is adopted \citep{Planck_Cosmological_Parameters}.}
	\label{Inputs}	
		\begin{tabular}{lll}
			\hline
			Parameter & Meaning & Value\\
			\hline
			$H_0$ & Present Hubble constant & 67.3 km/s/Mpc \\
			$\Omega_m$ & Present matter density & 0.315 \\
			$D_l$ & (Angular diameter)  & 0.945 Gpc \\
			& distance to lens at $z_l = 0.296$ & \\
			$D_s$ & Distance to source at $z_s = 1.7$ & 1.795 Gpc\\
			${D}_{ls}$ & Distance to source from lens & 1.341 Gpc\\
			& position in spacetime & \\
			$M$ & Mass of lens & $1.2 \times {10}^{14} M_\odot$\\
			$r_d$ & Scale length of source galaxy & 3.068 kpc\\
			$v_t$ & Tangential velocity of lens & 3000 km/s\\
			$v_{_f}$ & Flatline level of source galaxy& 100 km/s\\
			& rotation curve & \\
			$\sin i \cos \gamma$ & See Figure \ref{Source_geometry}. Isotropic average. &$\frac{1}{2}$\\ [2pt]
			\hline
		\end{tabular}
\end{table}

Using the parameters in Table \ref{Inputs}, I obtained the results shown in Figure \ref{Overall_Effect}. The MCE and DME are comparable if the source is a typical spiral galaxy. This suggests that the DME might well confuse measurements of the MCE without certain precautions. Unfortunately, it can be difficult to calculate the DME and adjust for it because it relies on quantities that may be difficult to determine e.g. the variation of $A$ with position and which side of the galaxy is the approaching side. Thus, I considered whether the detailed profiles of individual spectral lines could be used to distinguish the DME from the MCE. A detailed line profile would contain much more information than just the centroid location $\overline{v_r}$.

I began by determining if existing observatories could attain the required spectral resolution within a reasonable timeframe. To this end, I considered the Atacama Large Millimetre Array (ALMA). Using the online calculator, I found that ALMA probably can resolve individual spectral lines well enough to distinguish the MCE from the DME (Table \ref{ALMA_inputs}).

\begin{table}	 
	\caption{Input parameters used for the ALMA exposure time calculator, available at:
	\newline
	\newline
	\href{https://almascience.eso.org/proposing/sensitivity-calculator}{https://almascience.eso.org/proposing/sensitivity-calculator}
	\newline
	\newline
	The dual polarisation mode should be used as polarisation is unimportant here. The angular resolution does not affect the result, which was 6.17 hours. \newline}
	\label{ALMA_inputs}	
		\begin{tabular}{ll}
			\hline
			Parameter & Value \\
			\hline
			Declination & $-56^\circ$ \\			
			Frequency & 150 GHz \\
			Bandwidth per polarisation & 100 m/s \\
			Water vapour column density & 5$^\text{th}$ octile (1.796 mm) \\ 
			Number of antennas & 50 $\times$ 12 metre \\
			Root mean square (rms) sensitivity & 1.5 mJy \\			
			\hline
		\end{tabular}
\end{table}

To take advantage of this, I performed calculations to see how the DME and MCE affect individual line profiles. I began by mapping the radial velocity of an edge-on disk galaxy.
\begin{eqnarray}
	v_r \left( r, \phi \right) = v_c \left( r \right) \sin \phi
	\label{v_r}
\end{eqnarray}

\begin{figure}
	\centering
		\includegraphics [width = 7cm] {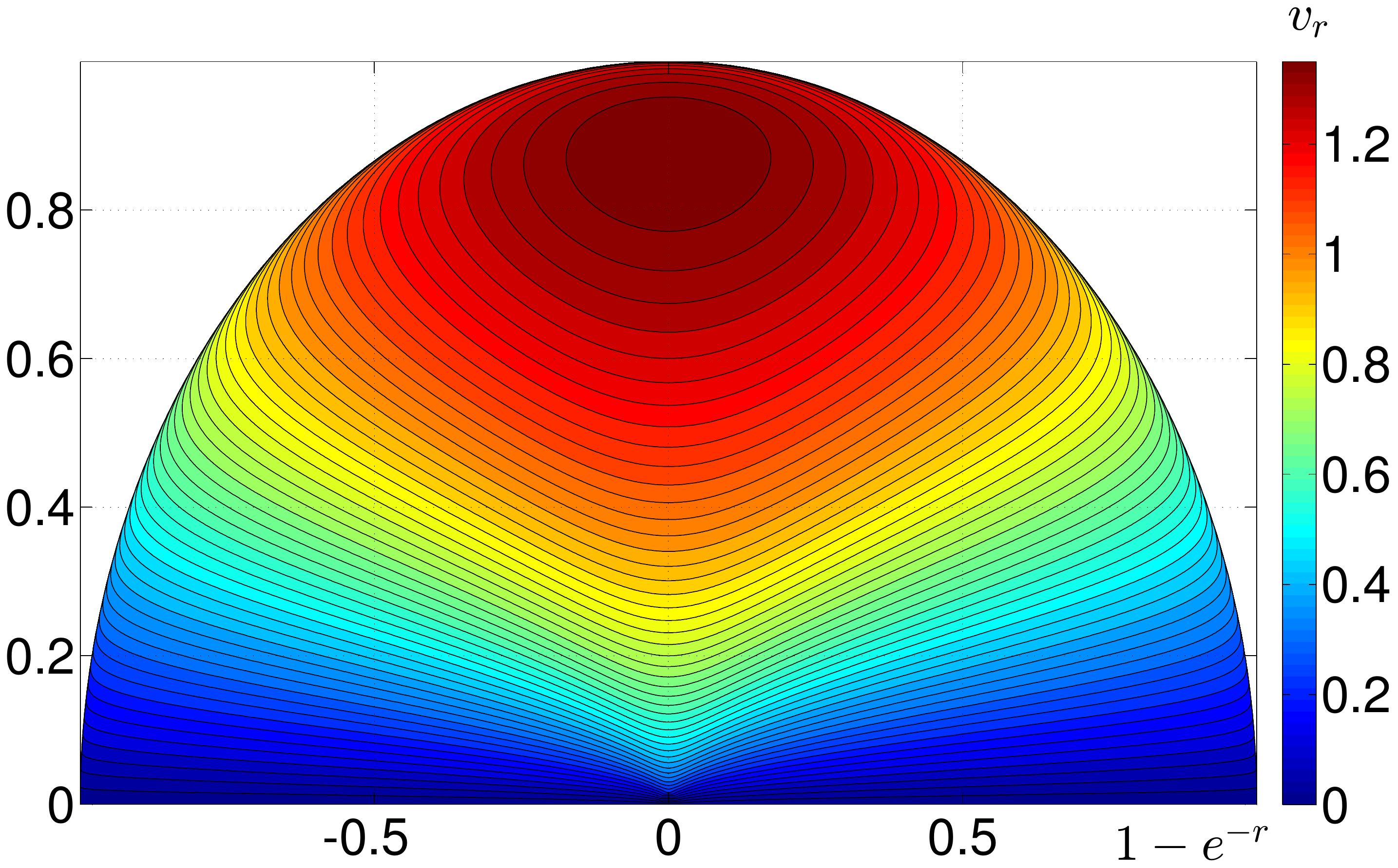}
	\caption{Radial velocity map of a disk galaxy viewed by an observer within its plane at large $x$ (far off to the right), for the case $k = 2.5$ (similar to the MW). Radial velocities are antisymmetric about the $x$-axis. The radial co-ordinate has been rescaled so the displayed size of each region is proportional to its brightness. The units are such that $r_d = 1$ and $v_{_f} = 1$. Note the large region with $v_r$ close to its maximum value. The result for $k = 0.5$ is very similar, although $v_{max}$ is much closer to $v_{_f}$ (Figure \ref{Maximum_Rotation_Speeds}).}
	\label{Radial_Velocity_Contour_k_2_5}
\end{figure}

\begin{figure}
	\centering
		\includegraphics [width = 8.5cm] {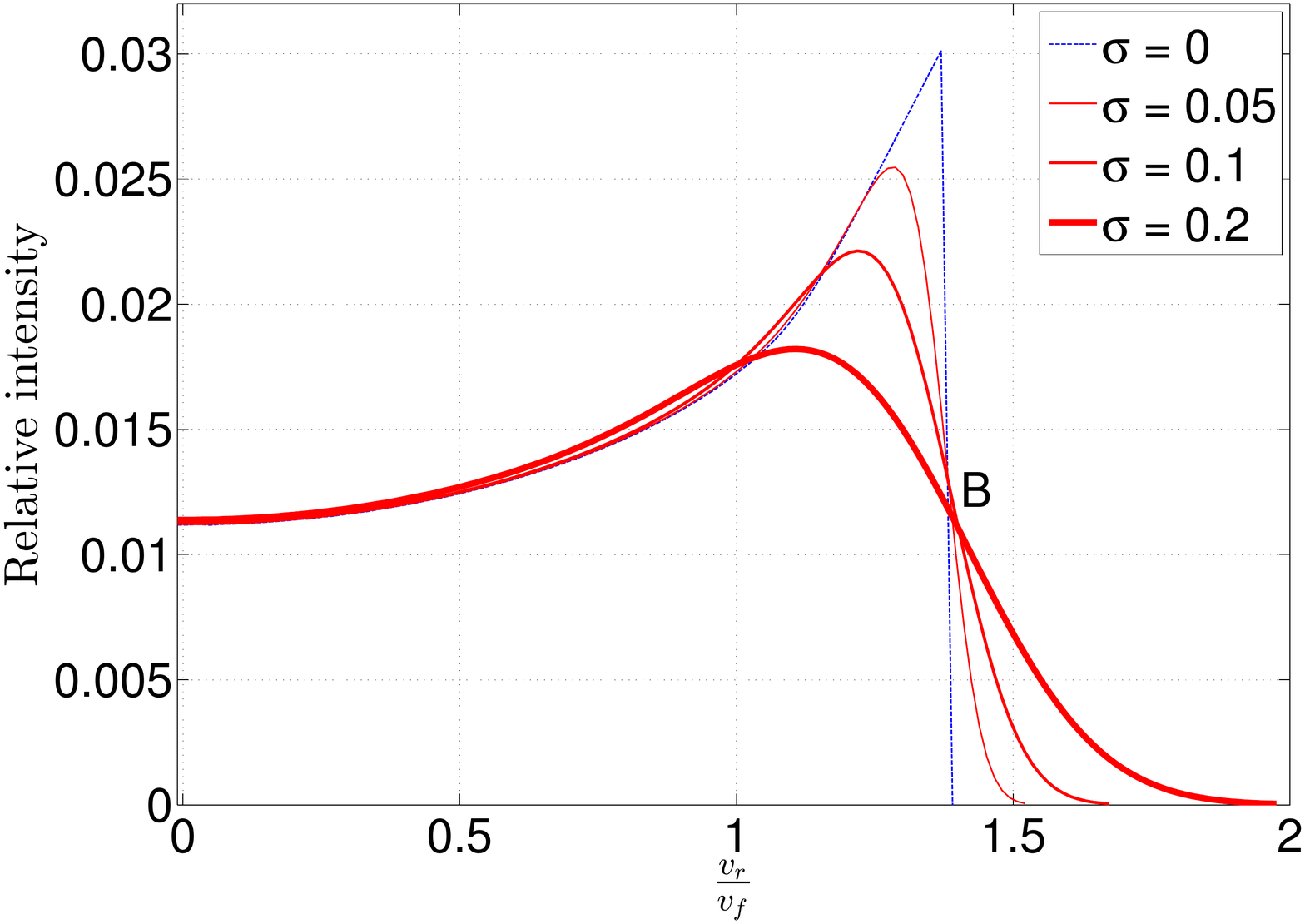}
	\caption{The synthetic line profile of an intrinsically narrow spectral line in an unlensed galaxy with $k = 2.5$, viewed edge-on. The profile is symmetric about $v_r = 0$. The sharp drop in the line profile (blue) would probably get blurred (e.g. by random motions), so I convolved the profile with Gaussians of width $\sigma$ (given in units of the flatline rotation curve level $v_{_f}$). The results are shown as red lines with thickness $\propto \sigma$. Notice how all 4 profiles pass close to the point marked B. The result for $k = 0.5$ is similar, if velocities are scaled to $v_{max}$ rather than $v_{_f}$.}
	\label{Spectral_profiles}
 \end{figure}

\begin{figure}
	\centering
		\includegraphics [width = 8.5cm] {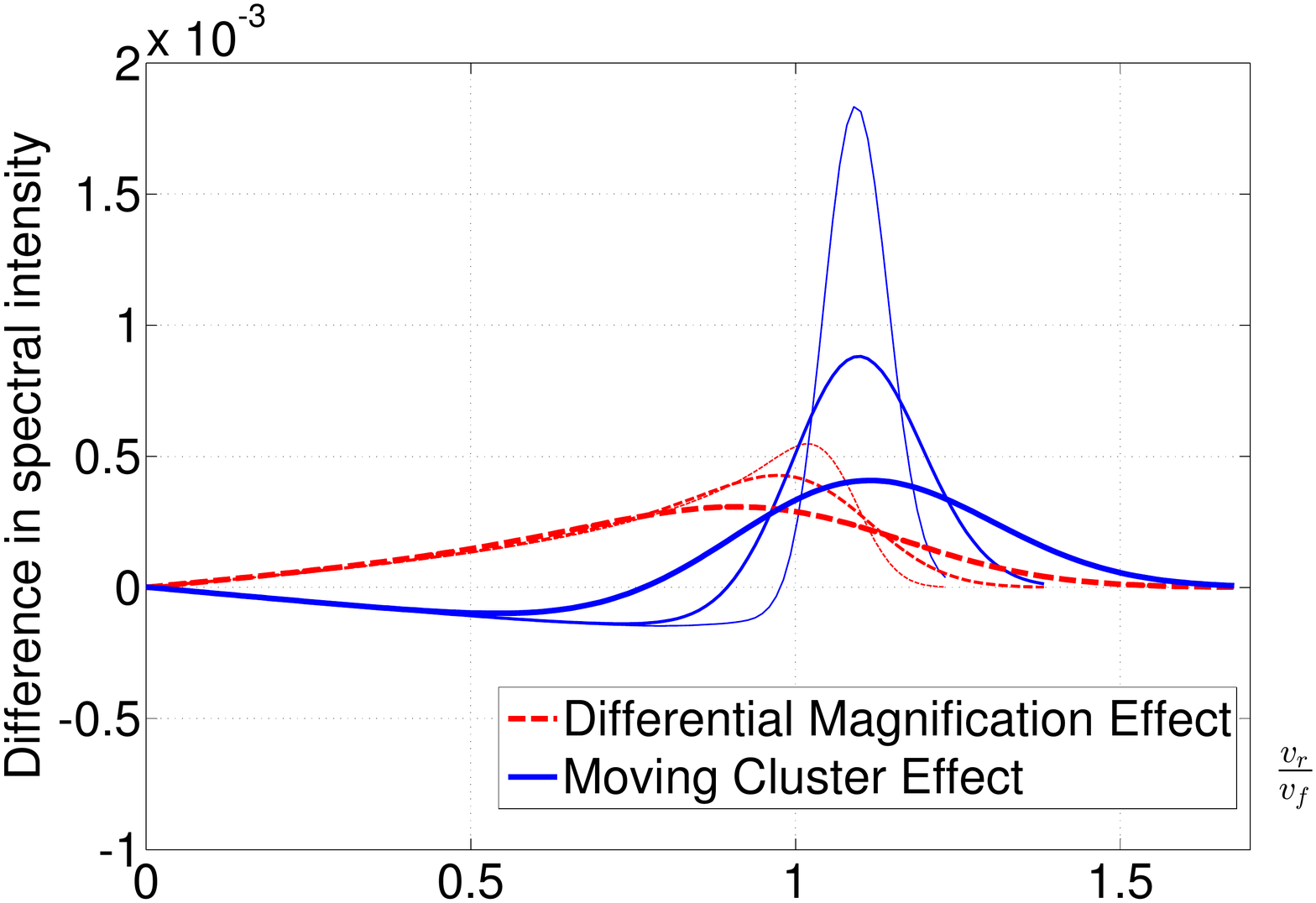}
		\includegraphics [width = 8.5cm] {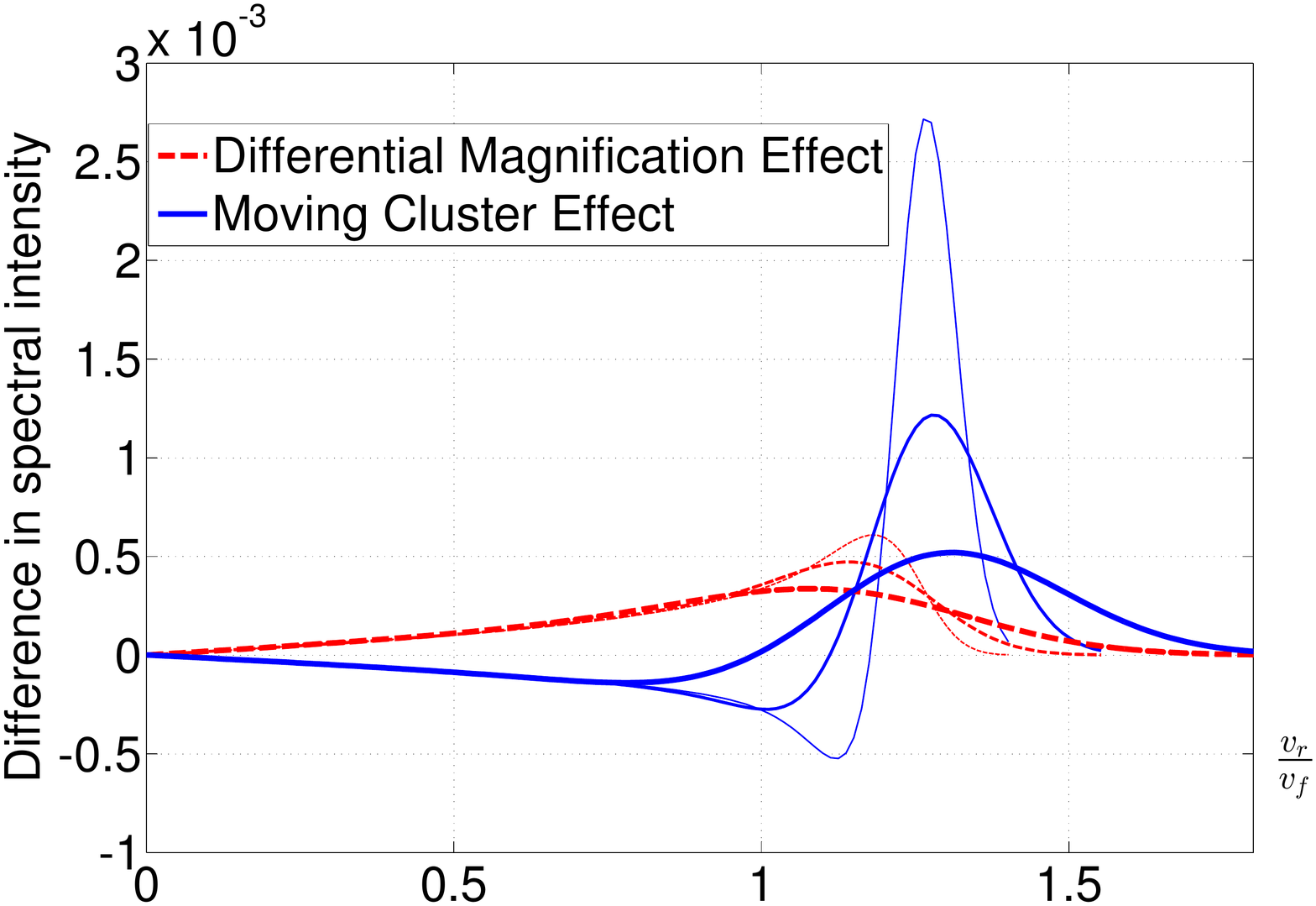}		
	\caption{The residuals in the spectral profile due to the DME (Equation \ref{Equation_39}) and the MCE (horizontal shift of profile), obtained by subtracting a control line profile (Equation \ref{Control}). The patterns are antisymmetric about $v_r = 0$. Results are shown for an edge-on galaxy with $k = 0.5$ (\emph{top}) and $k = 2.5$ (\emph{bottom}). Both effects change the mean redshift by 1\% of the maximum rotation speed, representing 1.08\% of $v_{_f}$ for $k = 0.5$ and 1.26\% for $k = 2.5$. The spectra were convolved with Gaussians of widths 0.05, 0.1 and $0.2~v_{_f}$ (higher $\sigma$ indicated by thicker line). The MCE can't change the amplitudes of the horns while the DME can $-$ it makes one more pronounced and the other less.}
	\label{Residuals_k_0_5}
\end{figure}

The results are shown in Figure \ref{Radial_Velocity_Contour_k_2_5} for the case ${k = 2.5}$, similar to the MW value. I then binned the galaxy in $r$ and $\phi$ in order to take advantage of Equation \ref{v_r} being separable in $\left( r, \phi \right)$. The key trick is to calculate $v_c$ only once at each $r$, for all $\phi$. In this way, I determined $v_r$ and thereby classified the luminosity of the galaxy according to the radial velocity of the region emitting the light. This allows a synthetic line profile to be constructed. Four examples are shown in Figure \ref{Spectral_profiles}, three of which allow for some random motion as well as ordered circular motion. The line profiles show a distinct horn corresponding to the `bull's-eye' towards the top of Figure \ref{Radial_Velocity_Contour_k_2_5}. This feature is due to a turning point in the rotation curve, causing a large part of the galaxy to have a similar ${v_r}$.

To allow for the MCE, I simply translated the line profile. For the DME, I let
\begin{eqnarray}
	A ~=~ 1 ~+~ n \tilde{r} \sin \phi
	\label{Equation_39}
\end{eqnarray}

For reference, I constructed a control line profile with
\begin{eqnarray}
	A ~=~ 1 ~ {\forall}_{r,\phi}
	\label{Control}
\end{eqnarray}

This control line profile was subtracted from the profiles modified by the DME and the MCE. The residuals are shown in Figure \ref{Residuals_k_0_5} for $n \ll 1$ but large enough to avoid numerical issues. The pattern of residuals is quite different in the two cases, even though both cause the same ${\Delta \overline{v_r}}$. This might well allow the MCE and DME to be distinguished. In particular, the MCE can lead to very large residuals close to the peaks in the line profile, depending on how sharp they are. Even the less dramatic features near $v_r = 0$ might be useful $-$ the MCE and DME give opposite signs for the residuals in this spectral region despite causing the same overall redshift difference between the images.

\subsection{Observational strategies}

The MCE can be enhanced relative to the DME by a number of strategies, especially if there is a choice of which multiple images to target for detailed spectroscopic follow-up. Avoiding a spiral galaxy as a target reduces the DME, although an elliptical can still rotate. Even a low-resolution spectrum should be able to distinguish a fast-rotating spiral from an elliptical with mild rotation in the region emitting most of the light. This is because spirals ought to have a characteristic double-horned spectral profile whereas ellipticals would have a roughly Gaussian profile. However, the latter lacks sharp features, making the MCE itself harder to detect by raising random errors.

Spiral galaxies are acceptable targets if they are viewed face-on as such objects have little gradient in $v_r$ across them. However, even an edge-on spiral can make a good target if it is oriented so its major axis is a direction along which $A$ hardly changes. For a point mass lens, this would mean the source galaxy's major axis was orthogonal to the apparent lens-source line. In a more complicated lens, it might be possible to estimate how $A$ varies with sky position and use this magnification map to guide the selection of targets.

Some interesting possibilities arise if the source is very inhomogeneous. A small region might be forming stars rapidly and emit strongly in the far-infrared due to dust. Targeting only spectral lines at these wavelengths then reduces the DME because the emitting region is small and $A$ varies only a little over it. However, the MCE is unaffected by the size of the emitting region (as long as it is much smaller than the Einstein radius).

One can target fainter spiral galaxies so the source is likely to be smaller and slower-rotating. This strategy may be difficult to implement with current technology. It is promising in the long run because there are many more fainter galaxies than brighter ones \citep{Schechter_1976}.

I expect that the MCE can be measured in the near future if careful consideration is put into reducing the impact of the DME. Important insights may be gained by comparing the detailed line profiles between different images of the same object. If a target was used for which the DME should be negligible, then the validity of any claimed MCE detection could be checked by comparison of the observed pattern of residuals between appropriately scaled spectra of the multiple images. If instead the DME is not negligible, its magnitude could be estimated from this pattern of residuals because of the very different ways in which differential magnification and lens motion affect spectral line profiles, even if they cause an equal difference between the mean image redshifts (Figure \ref{Residuals_k_0_5}). A promising target for detecting the MCE might be the triply imaged galaxy discovered by \citet{Gonzalez_2009}.

\section{Dynamics of the Local Group in $\Lambda$CDM \citep{BANIK_ZHAO_2016}}
\label{Local_Group_2D}

Section \ref{Bullet_Cluster} discussed possible tests of gravity based on galaxy clusters, where the collision velocity may be several hundred km/s faster in MOND than would be feasible in $\Lambda$CDM \citep[][figure 8]{Katz_2013}. Even so, it is difficult to actually perform this test because of the cosmological distances to these interesting systems \citep[e.g. the Bullet Cluster has a redshift of $z = 0.296$,][]{Tucker_1998}. This makes it difficult to know the 3D position and velocity structure of the system.

Although gravity could be tested without all 6 phase space co-ordinates, it is at least necessary to know the peculiar velocities. As velocity errors are relatively small nowadays, the dominant uncertainty arises from redshift-independent distances. Assuming a 10\% distance uncertainty and that peculiar velocities might differ by $\Delta v = 100$ km/s between the different gravity theories, this suggests that it is difficult to properly test MOND beyond a distance
\begin{eqnarray}
		d_{max} ~&=&~ \frac{\Delta v}{0.1H_{_0}} \\
		&=&~15~\text{Mpc}
\end{eqnarray}

A good test of MOND should involve a system where its prediction differs by ${\ssim 5 \sigma}$ from that of $\Lambda$CDM. This would reduce $d_{max}$ to only 3 Mpc. Therefore, the remainder of this portfolio will focus on the LG.

A major alteration to the gravitational field would undoubtedly have far-reaching implications for the motions of LG galaxies. In particular, MOND implies there was a past close MW-M31 flyby \citep{Zhao_2013} at a much faster relative velocity than expected in $\Lambda$CDM, a model in which such a flyby is precluded due to dynamical friction between DM halos \citep{Privon_2013}. These high velocities would allow the MW and M31 to gravitationally slingshot any passing LG dwarf galaxies out at high speeds in 3-body interactions (MW, M31 and dwarf). In this section, I summarise the work of \citet{BANIK_ZHAO_2016} where I investigated whether there is any evidence for such a scenario using an axisymmetric dynamical model of the LG in $\Lambda$CDM.

The basic idea behind this is called the timing argument \citep{Kahn_Woltjer_1959}. It involves using observed peculiar velocities $\bm{v}_{pec}$ to estimate the gravitational field $\bm{g} \left( t \right)$ within the LG throughout cosmic history, exploiting the fact that $\bm{v}_{pec}$ was rather small early in the history of the Universe \citep{Planck_2013}. Similar analyses were done previously by \citet{Sandage_1986} and by \citet{Jorge_2014}, the latter using a spherically symmetric model of the LG to estimate how the MW and M31 should slow the outward recession of LG dwarfs in $\Lambda$CDM. Consideration of 12 LG analogues in cosmological $\Lambda$CDM simulations \citep{Fattahi_2016} shows that such timing argument calculations are fairly accurate \citep{Fattahi_2017}.

To improve the accuracy further, I developed a more advanced axisymmetric model of the LG. Like \citet{Jorge_2014}, I also assumed that the mass in the LG is entirely contained within the MW and M31, which I took to be on a radial orbit. Recent proper motion measurements of M31 indicate only a small tangential velocity relative to the MW, making the true orbit almost radial \citep{M31_motion}. I also included Cen A in my models because it lies rather close to the MW-M31 line \citep{Ma_1998}.

\subsection{Method}
\label{Method_2D}

I adopt a standard flat $\Lambda$CDM cosmology. Ignoring components other than matter and dark energy, the evolution of the cosmic scale factor $a \left( t \right)$ can be determined analytically and is fully specified by the present Hubble constant $H_{_0}$ and matter density parameter $\Omega_{m,0}$.
\begin{eqnarray}
	\frac{\overset{..}{a}}{a} &=& -\frac{4 \pi G}{3} \left( \rho_m - 2 \rho_\Lambda \right) \\
	&=& {H_{_0}}^2 \left(- \frac{1}{2} \Omega_{m,0}~a^{-3} + \Omega_{\Lambda, 0}  \right) \text{. Therefore, } \nonumber \\
 a(t) &=& {{\left( \frac{{{\Omega }_{m,0}}}{{{\Omega }_{\Lambda ,0}}} \right)}^{\frac{1}{3}}}{{\sinh }^{\frac{2}{3}}}\left( \frac{3}{2}\sqrt{{{\Omega }_{\Lambda ,0}}}{{H}_{0}}t \right)
	\label{Expansion_history}
\end{eqnarray}

To integrate test particle trajectories, it is necessary to first have trajectories for the MW, M31 and Cen A that match the presently observed distances of M31 and Cen A to within 1 kpc. I do this using a 2D Newton-Raphson algorithm on the initial relative positions of all three galaxies along a line.\footnote{For stability, I under-relaxed the algorithm i.e. in each iteration, I altered the parameters by 80\% of what the algorithm would normally have altered them by} Initial velocities were found using
\begin{eqnarray}
	\bm{v}_{_i} = H_{_i} \bm{r}_{_i} ~~\text{ where  } H_{_i} \equiv \frac{\dot{a}}{a} ~\text{ when } t = t_i
	\label{Initial_conditions}
\end{eqnarray}

\begin{figure}
	\centering 
		\includegraphics [width = 8.5cm] {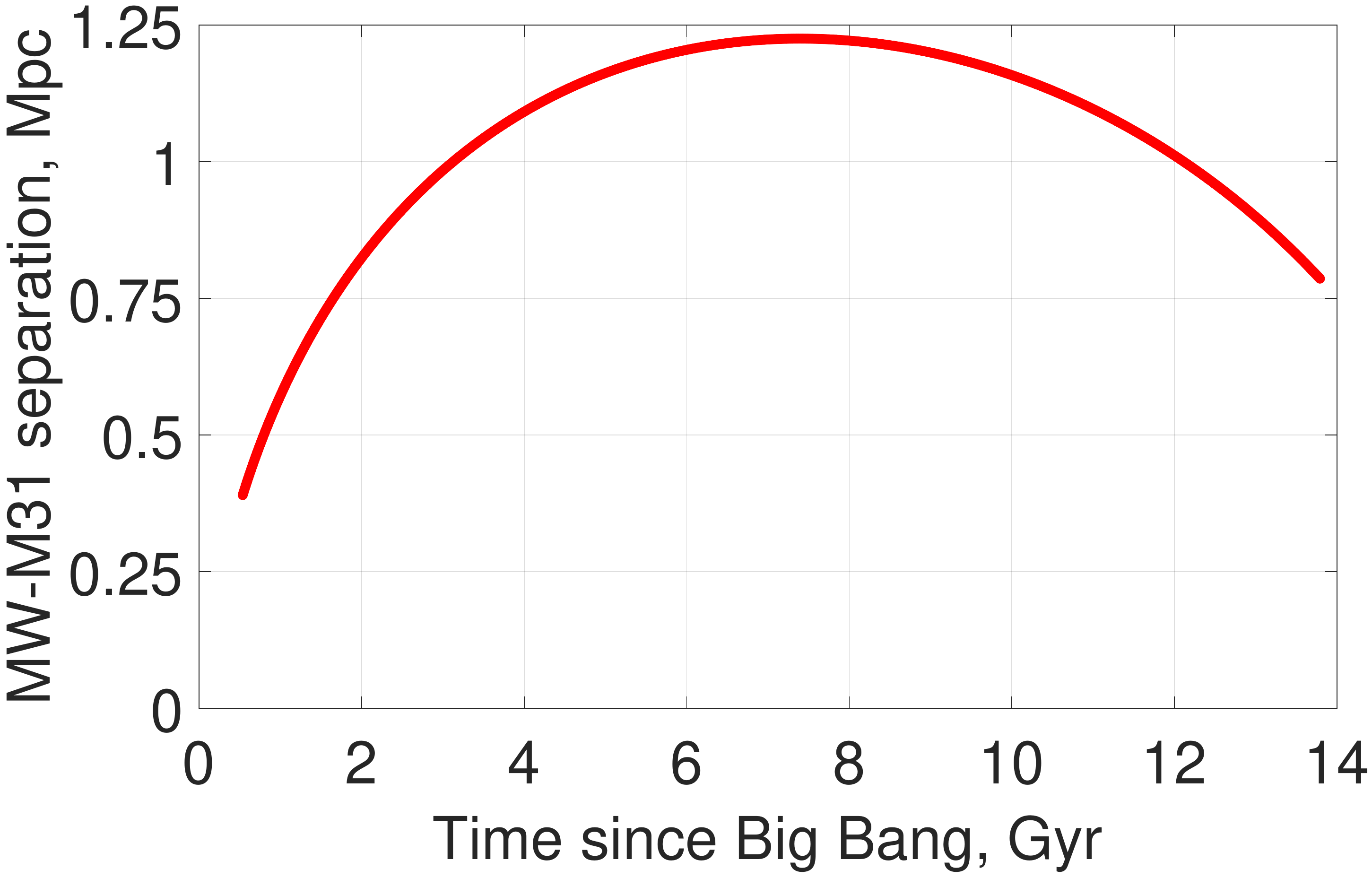}
	\caption{MW$-$M31 separation $d(t)$ for a typical model where $q_{_1} = 0.2$ and $M_i = 3.4 \times {10}^{12} M_\odot$ (parameters defined in Table \ref{Priors}). $d(t)$ always looks broadly similar $-$ in $\Lambda$CDM, the MW and M31 have never approached each other closely for any plausible model parameters.}
	\label{MW_M31_separation_history}
\end{figure}

To minimise convergence issues, I used each solution to the problem for the next set of model parameters, keeping the parameter changes small. Figure \ref{MW_M31_separation_history} shows the MW-M31 separation in the massive object solution for one of my models.

With the trajectories of the massive galaxies in hand, I solved test particle trajectories starting at some early initial time $t_i$. I took the barycentre of the LG then as the centre of expansion. The initial velocities followed a pure Hubble flow (Equation \ref{Initial_conditions}) because the Universe was nearly homogeneous at early times $-$ peculiar velocities on the last scattering surface are only ${\ssim 3}$ m/s \citep{Planck_2015} whereas the present value is over 100 km/s for M31 \citep{McConnachie_2012, Van_der_Marel_2012}.


Test particle trajectories were advanced using a fourth-order Runge-Kutta scheme with an adaptive but quantised timestep that was varied in powers of 2. Any close approaches to the MW or M31 were treated as an accretion event, increasing the mass of the accreting galaxy by an amount proportional to the volume represented by each test particle. This required the test particle trajectories to be recalculated. I found that the MW and M31 masses converged very well after just a single iteration. The resulting velocity field for the LG is shown in the top panel of Figure \ref{LG_Hubble_Diagram}. Its bottom panel shows the resulting LG Hubble diagram, comparing velocity directly away from the LG barycentre with distance from there.

To compare with observations, I obtained a test particle trajectory ending at the same position as a `target' LG galaxy. The Galactocentric radial velocity (GRV) of the observed target and the simulated test particle were then compared to judge how well the model fits observations. My list of targets was almost the same as that used by \citet{Jorge_2014}, with a few minor alterations. A wide range of plausible model parameters (Table \ref{Priors}) were investigated within the context of $\Lambda$CDM.

In a homogeneous Universe, $\bm{r} \propto a \left( t \right)$ so that $\overset{..}{\bm{r}} = \frac{\overset{..}{a}}{a}\bm{r}$. Including additional forces arising from the gravity of the MW, M31 and Cen A, this makes the equation of motion for each test particle
\begin{eqnarray}
	{\overset{..}{\bm r} } ~=~ {\frac{\overset{..}{a}}{a}}  {\bm r} - \sum_{\begin{array}{c} \text{j = MW,}\\ \text{M31, Cen A}\end{array}}{ \frac{G M_j \left( \bm r - \bm r_{_j} \right)} {{\left( |\bm r - \bm r_{_j} |^2 + {r_{_{S,j}}}^2 \right)}^{\frac{1}{2}} |{\bm r}- \bm r_{_j} |^2}}
	\label{Equation_of_motion_2D}
\end{eqnarray}

The force towards each massive galaxy is not $\propto \frac{1}{r^2}$ at low distance $r$ from it because this would contradict observed flat rotation curves of major LG galaxies \citep[e.g.][]{Carignan_2006}. To be consistent with the observed values of $v_{_f}$ for the MW and M31 (180 and 225 km/s, respectively); I had to soften the force below a radius $r_{_S} = \frac{GM}{{v_{_f}}^2}$. This radius is different for the MW and M31. For Cen A, I used $r_{_S} = 100$ kpc because this analysis does not consider LG particles ending up near Cen A, making the precise force law used there irrelevant.

Test particles were started on a grid of plane polar co-ordinates. I assumed that the initial masses of the MW and M31 arose by completely depleting all the mass in some region, so I did not start any test particles within it. Assuming the `feeding zone' to be delimited by an equipotential, I first determined the potential resulting from the MW and M31.
\begin{eqnarray}
	\label{U}
	U ~&=&~ \sum_{j = MW,M31}{{-\frac{GM}{r_{_{S,j}}} Ln\left( \frac{\sqrt{1 + {b_j}^2} - 1}{b_j} \right)}} \\
	b_j &\equiv& \frac{\left| \bm{r} - \bm{r_{_j}} \right|}{r_{_{S,j}}}
\end{eqnarray}

Next, I determined the volume $V$ of the accretion region such that $\rho_{_M} V = M_i$, where $M_i$ is the initial combined mass of the MW \& M31 in my model while $\rho_{_M}$ is the cosmic mean matter density at $t_i$.\footnote{$\rho_{_M}$ includes both baryons and DM.} Finally, I determined the equipotential $U_{exc}$ such that the region with $U < U_{exc}$ has volume $V$. Test particle trajectories were not started in the region where $U < U_{exc}$ at $t = t_i$.

To get target galaxies in the same co-ordinate system as used in my simulation (which has its $y-$axis aligned with the MW-M31 separation), I used
\begin{eqnarray}
	x ~&=&~ d_{_{MW}} ~|\hat{\bm d}_{_{MW}} \times \hat{\bm r}_{_{MW}}| \\
	y_{rel} ~\equiv ~ y - y_{_{MW}} ~&=&~  d_{_{MW}} \left( \hat{\bm d}_{_{MW}} \cdot \hat{\bm r}_{_{MW}} \right)
	\label{y_rel}
\end{eqnarray}

The vector from the MW to a target galaxy is denoted ${\bm{d}}_{_{MW}}$ while the direction from M31 towards the MW is denoted $\hat{\bm r}_{_{MW}}$. I use the convention that $\hat{\bm{v}} \equiv \frac{\bm{v}}{v}$ for any vector $\bm{v}$ with length $v \equiv \left| \bm{v} \right|$.

\subsection{Comparison with observations}

\begin{figure}
	\centering 
		\includegraphics [width = 8.5cm] {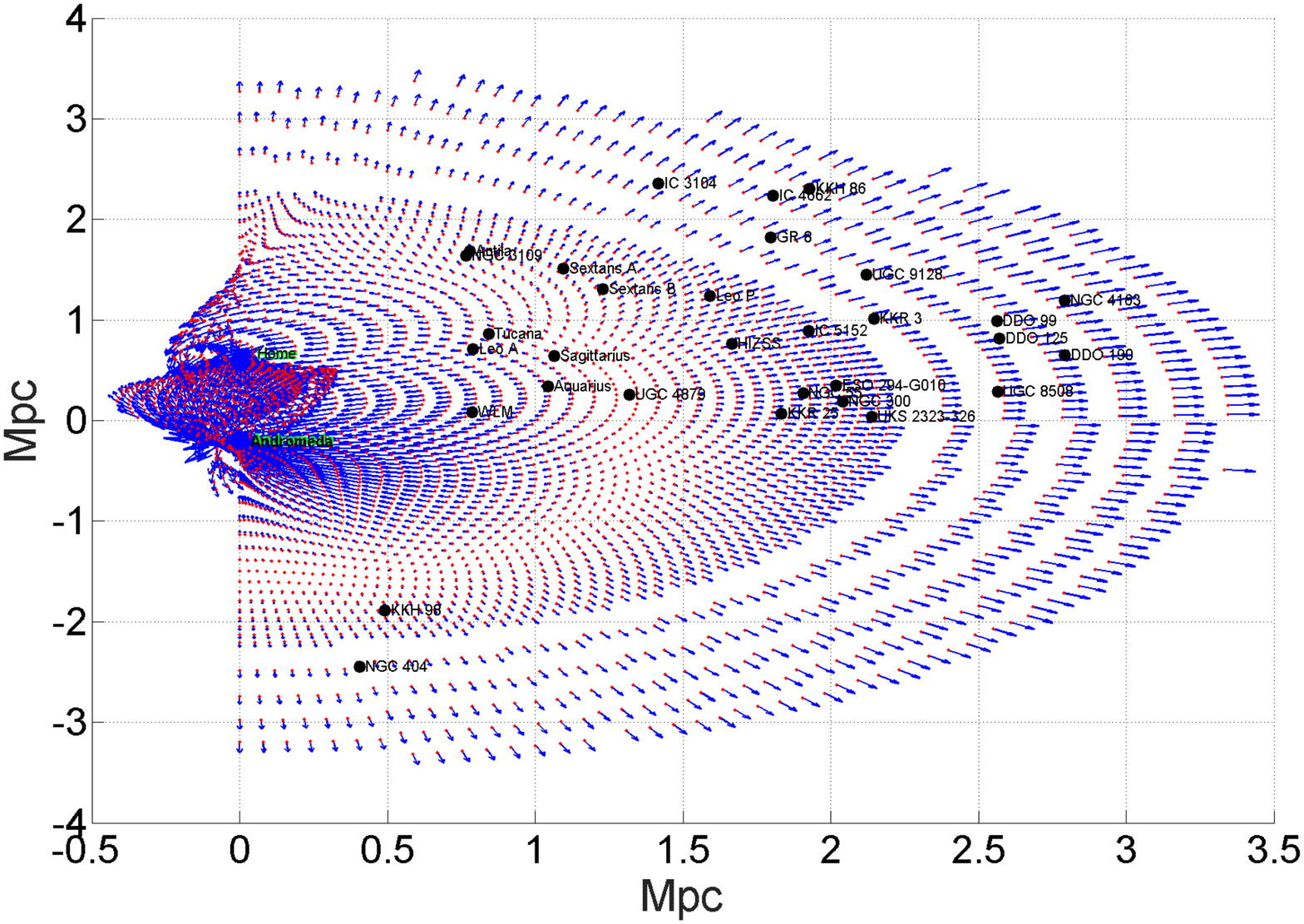}
		\includegraphics [width = 8.5cm] {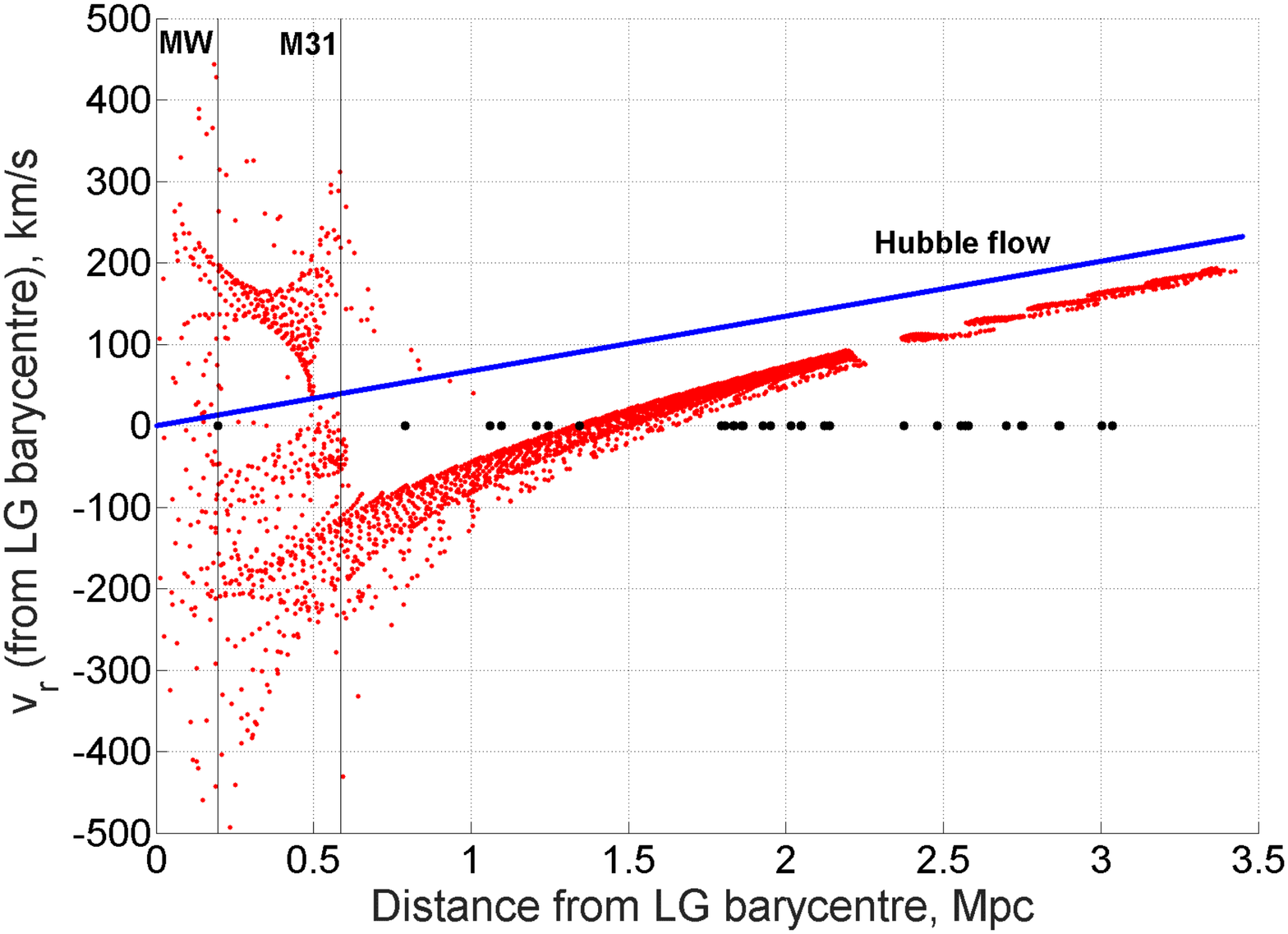}
	\caption{\emph{Top:} Local Group velocity field for the case $q_{_1} = 0.3$, $M_i = 4 \times {10}^{12} M_\odot$. Locations of target galaxies are overplotted as black dots with galaxy names given beside them. The MW is just above the centre. Only particles starting at $x>0$ (and thus $v_x > 0$) were considered. Thus, the presence of particles at $x < 0$ indicates intersecting trajectories and a disturbed velocity field. \emph{Bottom:} Radial velocities of test particles with respect to the LG barycentre. Vertical lines represent distances of the MW and M31 from there. Black dots indicate positions of target galaxies. Without proper motions, they can't be put on such a Hubble diagram at the correct velocity, so I show this as 0.}
	\label{LG_Hubble_Diagram}
\end{figure}

Close to the MW and M31, the velocity field is complicated because there are intersecting trajectories (Figure \ref{LG_Hubble_Diagram}). This makes it impossible to uniquely predict the velocity based on position. As a result, I had to exclude any target galaxies which fell in such regions. Fortunately, this was extremely rare, mainly because of pre-selection of targets by \citet{Jorge_2014}.

The key aspect of my algorithm was carefully determining the model-predicted GRV of each target galaxy. To do this, I obtained a test particle trajectory landing at the same position as each target. I started with the test particle that landed nearest to its observed position. I then applied the 2D Newton-Raphson algorithm to the initial position of this test particle, updating its initial velocity according to Equation \ref{Initial_conditions}. I considered this process to have converged once the final position error fell below 0.001\% of the target's distance from the LG barycentre. This trajectory was then used to determine the model-predicted GRV of the target galaxy.
\begin{eqnarray}
	GRV_{model} ~~=~~ \frac{v_x x ~+~ \left( v_y - \dot{y}_{_{MW}} \right) \left( y_{rel} \right)}{\sqrt{x^2 ~+~ {y_{rel}}^2}}
	\label{Model_GRV}
\end{eqnarray}

To account for distance uncertainties, this procedure was repeated for each target with its heliocentric distance raised to the 1$\sigma$ upper limit of the observed value. The difference between these GRV estimates is $\sigma_{pos}$, the uncertainty in the model-predicted GRV of a target due to uncertainty in its position along the line of sight.

To compare with observations, the observed GRV of each target is obtained by adjusting its observed heliocentric radial velocity (HRV) for the motion of the Sun within the MW. The solar motion $\bm{v_\odot}$ is mostly circular motion within the MW disk at speed $v_{c, \odot}$, the speed of a test particle on a circular Galactic orbit at the position of the Sun. It is useful to define a Local Standard of Rest (LSR), a reference frame rotating at this speed. Naturally, the Sun has a small amount of motion with respect to the LSR (magnitude and direction given in Table \ref{Priors}).
\begin{eqnarray}
	GRV_{obs} ~=~ HRV + \mathbf{v_\odot} \cdot \hat{\mathbf d}_{_{MW}}
	\label{GRV_obs}
\end{eqnarray}

As well as the uncertainty on each GRV due to position $\sigma_{pos}$ and error on the observed HRV $\sigma_{v_h}$, I added another term $\sigma_{extra}$ to account for large scale structure, interactions between LG dwarf galaxies and other effects not included in my model.
\begin{eqnarray}
	\sigma = \sqrt{{\sigma_{pos}}^2 + {\sigma_{v_h}}^2 + {\sigma_{extra}}^2}
	\label{sigma}
\end{eqnarray}

$\sigma_{extra}$ should not exceed typical velocity dispersions/rotation speeds of LG dwarfs, which I estimate as ${\ssim 15}$ km/s \citep[e.g.][]{Kirby_2014}. As the motion of M31 would be much harder to alter than the motion of a less massive LG dwarf, I kept $\sigma_{extra}$ for M31 at $\frac{1}{10}$ of the value for other LG galaxies. However, the results are not much altered if M31 is treated in the same way as other LG dwarfs.

\subsection{Results and discussion}

\begin{figure}
	\centering
		\includegraphics [width = 8.5cm] {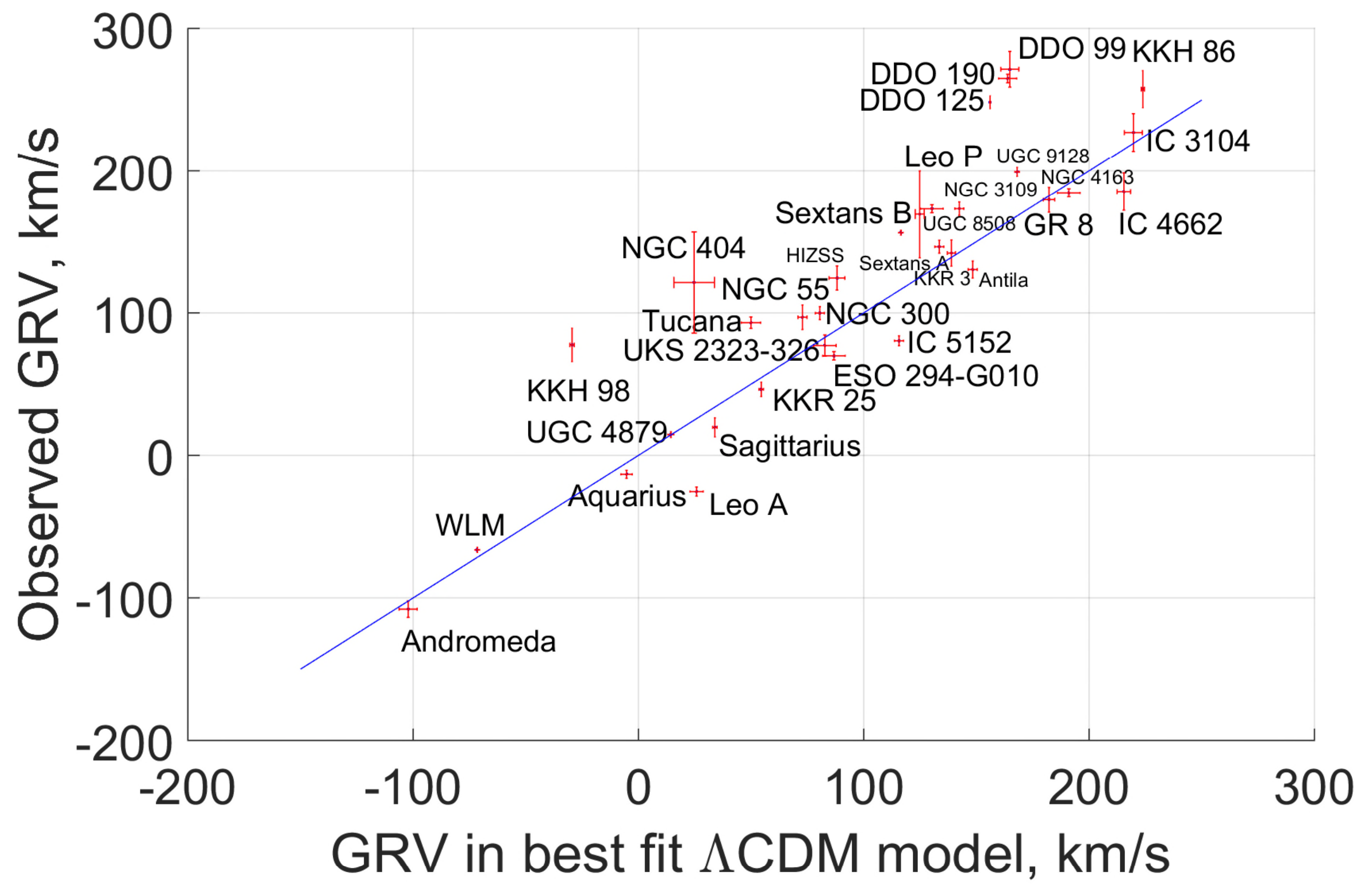}		
	\caption{Comparison between model-predicted and observed Galactocentric radial velocities based on the most likely model parameters ($q_{_1} = 0.2$, $M_i = 3.4 \times 10^{12} M_\odot$). The line of equality is also shown in blue.}
	\label{GRV_LCDM_Comparison}
\end{figure}

The resulting probability distributions of the model parameters are shown in Figure \ref{Triangle_plot} and summarised in Table \ref{Priors}. The posterior on $\sigma_{extra}$ only allows values ${\ga 40}$ km/s, which is rather high. I considered explanations such as a higher Hubble constant, altered start time, interactions with massive satellites like M33 and tides from objects outside the LG. None of these seem to work, though the last two possibilities are considered more thoroughly in Sections \ref{Local_Group_3D} and \ref{HVG_plane}.

Additional inaccuracies in the model may arise from the effects of large scale structure and distant encounters between LG dwarfs. The likely magnitude of such effects can be estimated based on more detailed $\Lambda$CDM cosmological simulations. Considering LG analogues in such simulations, it has been found that the dispersion in radial velocity with respect to the LG barycentre at fixed distance from there should be $\sigma_{_H} \sim 30$ km/s \citep{Aragon_Calvo_2011}, insufficient to explain my high inferred $\sigma_{extra}$.

\begin{figure}
	\centering
		\includegraphics [width = 8cm] {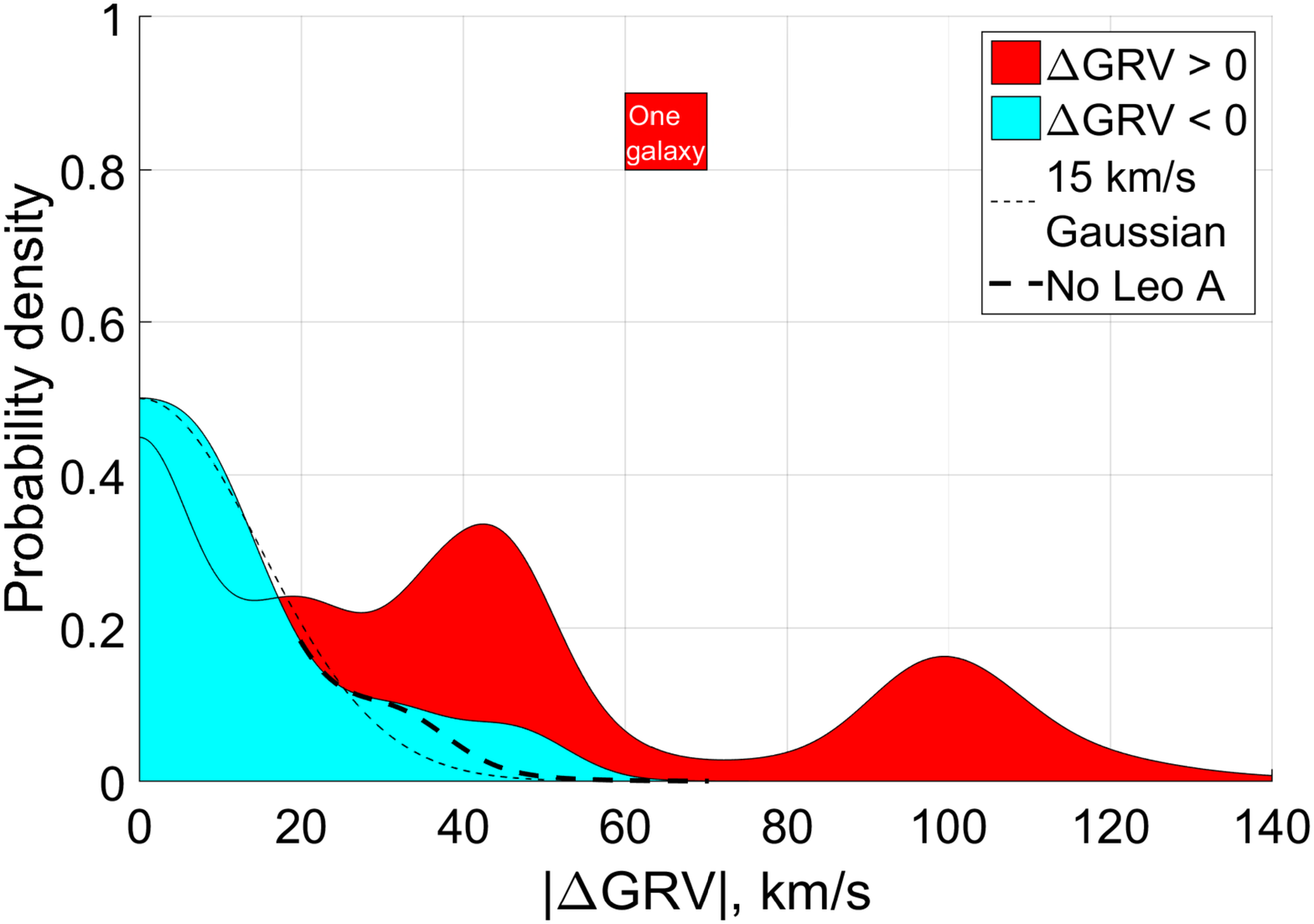} 
	\caption{Histogram showing observed $-$ predicted GRVs (i.e. $\Delta GRV$s) of target galaxies using the most plausible model ($q_{_1} = 0.2$ instead of 0.14, other parameters as in Figure \ref{GRV_LCDM_Comparison}). Each data point was convolved with a Gaussian of width $\sigma = \sqrt{{\sigma_{pos}}^2 + {\sigma_{v_h}}^2 + {\sigma_{v_{c, \odot}}}^2}$. I divided the sample into targets with $\Delta GRV < 0$ (blue) and those with $\Delta GRV > 0$ (red). The area corresponding to one galaxy is shown as a red square. A Gaussian of width 15 km/s is overplotted as a short-dashed line. This matches the $\Delta GRV < 0$ subsample quite well, especially when Leo A is excluded (long-dashed line).}
	\label{Delta_GRV_histogram_detailed}
\end{figure}

\begin{figure}
	\centering
		\includegraphics [width = 16cm] {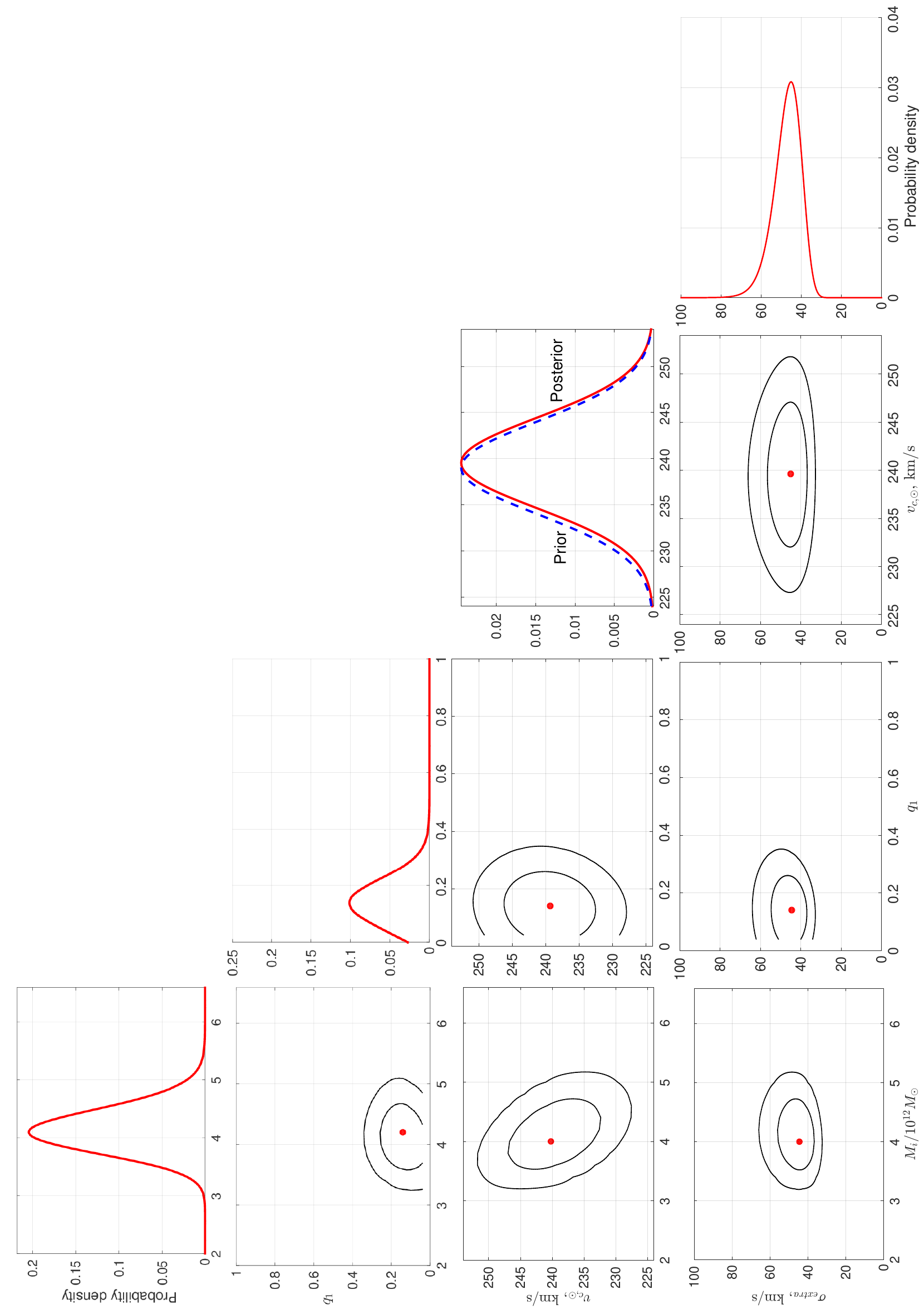}
	\caption{Marginalised posterior probability distributions of the free parameters defined in Table \ref{Priors}, with other parameters marginalised over. For variables plotted against other variables, I show the contours of the probability density which correspond to the usual ${1 \sigma}$ and ${2 \sigma}$ confidence levels, as well as the most likely pair of values. Rotate figure ${90^\circ}$ clockwise for viewing.}
	\label{Triangle_plot}
\end{figure}

To better understand the discrepancy, I compare model-predicted and observed GRVs for the best-fitting model (Figures \ref{GRV_LCDM_Comparison} and \ref{Delta_GRV_histogram_detailed}). These show that most velocities are more outwards than in the model, making it difficult to argue that the discrepancy arises from unmodelled interactions between LG dwarfs. Instead, I suggested that gravitational slingshot encounters with the MW or M31 flung out LG dwarfs at high speed. As this process is already accounted for, these massive galaxies must have been moving much faster than in my model. This would occur in MOND $-$ their relative speed at closest approach would have been ${\sim 600}$ km/s \citep{Bilek_2017}, very different to $\Lambda$CDM where it would rarely have been faster than the 110 km/s it is today (Figure \ref{MW_M31_separation_history}).

\begin{table}
  \centering
	\begin{tabular}{llll}
		\hline
		Name & Meaning and units & Prior & Result\\
		\hline
		$\sigma_{extra}$ & Extra velocity dispersion & 0 $-$ 100&${45.1}_{-5.7}^{+7.0}$\\
		& along line of sight, km/s & & \\ [5pt]
		$M_i$ & Initial MW $+$ M31 mass, & 2 $-$ 6.6&4.1$\pm$0.3\\
		& trillions of solar masses & & \\ [5pt]
		$q_{_1}$ & Fraction of MW $+$ M31 &0.04$-$0.96&0.14$\pm$0.07\\
		& mass initially in the MW & & \\ [5pt]
		$v_{c, \odot}$ & Circular speed of MW at & $239 \pm 5$&239.5$\pm$4.8\\
		& position of Sun, km/s & & \\ [5pt]
		\hline
		\multicolumn{4}{c}{Fixed parameters} \\
		\hline
		$d_0$ & Distance to M31, kpc & $783 \pm 25$ &\\ [5pt]
		$H_{_0}$ & Hubble constant at the & 67.3&\\
		& present time, km/s/Mpc & & \\ [6pt]
		$\Omega_{m,0}$ & Present matter density in & 0.315&\\
		& the Universe $\div \frac{3{H_{_0}}^2}{8 \rm{\pi} G}$ & & \\ [5pt]
		$a_{_i}$ & Scale factor of Universe & 0.1 &  \\
		& at start of simulation & & \\ [5pt]
		$r_{_{acc, MW}}$ & Accretion radius of MW & \multicolumn{2}{l}{15,337 parsecs} \\
		$r_{_{acc, M31}}$ & Accretion radius of M31 & \multicolumn{2}{l}{21,472 parsecs} \\ [5pt]
		$U_\odot$ & Components of & \multicolumn{2}{l}{14.1 km/s}  \\
		$V_\odot$ & non-circular motion of & \multicolumn{2}{l}{14.6 km/s}  \\
		$W_\odot$ & Sun within Milky Way & \multicolumn{2}{l}{6.9 km/s}  \\
		\hline
	\end{tabular}
  \caption{Priors and $1 \sigma$ confidence levels on model parameters. The latter are far from the boundaries imposed by the former, showing that the results are not strongly affected by prior assumptions. I used the measurement of $d_0$ by \citet{McConnachie_2012}, cosmological parameters from \citet{Planck_2015}, $v_{c, \odot}$ from \citet{McMillan_2011} and the Sun's non-circular velocity from \citet{Francis_2014}, given here in standard notation. Uncertainty in the latter is much less than in $v_{c, \odot}$, which was assumed to be within 3$\sigma$ of its observed value.}
  \label{Priors}
\end{table}

\onecolumn
\twocolumn

\section{Dynamical History of the Local Group in $\Lambda$CDM $-$ II. Including External Perturbers in 3D \citep{BANIK_ZHAO_2017}}
\label{Local_Group_3D}

To better understand if $\Lambda$CDM really faces a problem with the dynamics of LG galaxies, I followed up the work of Section \ref{Local_Group_2D} using a 3D model kindly lent by P. J. E. Peebles, who had previously used it to understand LG dynamics \citep{Peebles_2013}. The algorithm works by applying the numerical action method to solve the governing equations \citep{Phelps_2013}. They tested the method by applying it to results of cosmological simulations, recovering the galaxy masses fairly well (see their figure 2).

\subsection{Method}
\label{3D_method}

The 3D model lent by Peebles already included the LG galaxies and surrounding perturbers listed in table 1 of \citet{Shaya_2011}. This brightness-based catalogue misses the vast majority of the galaxies analysed in Section \ref{Local_Group_2D}, a major shortcoming because LG dwarfs $\ssim 1-3$ Mpc away turned out to be crucial to its conclusions. Thus, if not already present in the 3D model, I added the galaxies analysed in that work to the catalogue. LG dwarfs have very low masses, allowing me to add them as test particles satisfying the equation of motion
\begin{eqnarray}
	{\overset{..}{\bm r} } ~&=&~ {H_{_0}}^2\Omega_{_{\Lambda, 0}}{\bm r} ~- \sum_{\begin{array}{r} \text{j = Distant}\\ \text{massive}\\ \text{particles}\end{array}} \frac{G M_j \left( \bm r - \bm r_{_j} \right)} {|\bm r - \bm r_{_j} |^3} \nonumber \\
	&~&- \sum_{\begin{array}{r} \text{j = Nearby}\\ \text{massive}\\ \text{particles}\end{array}} \frac{G M_j \left( \bm r - \bm r_{_j} \right)  \left( {r_{_c}}^2 + {r_{_{S,j}}}^2 \right)} {\left( |\bm r - \bm r_{_j} |^2 + {r_{_c}}^2 \right) {r_{_{S,j}}}^3 }\nonumber \\
	\label{Equation_of_motion_3D}
\end{eqnarray}

When determining the force between any pair of massive galaxies, the same equation applied but I used the value of $r_{_{S}}$ corresponding to the galaxy with the larger $r_{_{S}}$. The massive galaxies in this analysis are given in Table \ref{Massive_galaxy_list_3D}. The distances and HRVs shown are best-fitting values considering all observational constraints within their uncertainties (Section \ref{Chi_sq_analysis_3D}).

The gravitational field near massive particles is handled slightly differently than in my 2D model (Section \ref{Local_Group_2D}). For any given test particle $A$, an explicit distinction is now drawn between massive particles whose $r_{_{S}}$ is below the distance to $A$ and masses for which this is not the case, forces from which are handled using a pure inverse square law. Forces from nearby masses at first rise linearly with separation before falling as $g \propto \frac{1}{r}$, recovering the observed flat rotation curves of galaxies. The transition occurs near $r_{_c} = 10$ kpc.

For the MW and M31, $r_{_{S}}$ is defined in the same way as in Section \ref{Local_Group_2D}, though with the added assumption that the LSR speed is the same as $v_{_f}$ for the MW. Its value is allowed to float, with a prior assumption of ${240 \pm 10}$ km/s. The same value is used for M31. For other massive galaxies, I assume $r_{_{S}} = 100$ kpc to avoid an adjustment each time their masses are altered.

Equation \ref{Equation_of_motion_3D} is slightly different to that used in my 2D analysis of the LG (Equation \ref{Equation_of_motion_2D}). The matter portion of the cosmological acceleration $\ddot{a}$ is handled in a different way, though the dark energy is handled similarly. Previously, the Universe was treated as homogeneous except for a few massive particles. The cosmic expansion $\bm r \propto a$ could then be recovered at long range because $\ddot{\bm{r}} = \frac{\ddot{a}}{a}\bm r$. Here, I treated the Universe as empty except for the massive particles that I explicitly include. As the Universe is homogeneous on large scales, an accurate understanding of all the matter interior to a sufficiently distant test particle also leads to its separation from us changing with time as $\bm r \propto a$.

To see if this applies to my model, I determined how much mass was in the simulation out to the distance of M101, the most distant galaxy in my sample. The result of $4.9 \times 10^{13} M_\odot$ corresponds to a sphere of radius 7.01 Mpc filled with matter at a density equal to the present cosmic mean value. This is similar to the observed distance of M101 \citep{Shappee_2011}, suggesting that the massive galaxies in my model mimic a smooth distribution on large scales with the correct density. A similar conclusion would be reached if I only considered the galaxies in my catalogue that lie within 3 Mpc of the MW. Thus, the matter distribution used should be accurate enough out to a sufficiently large distance that it enables the construction of an accurate dynamical model for motions within the LG.

\begin{table}
	\begin{tabular}{lcrc}
	\hline
	Galaxy & Distance, & HRV, & Mass, \\
	& Mpc & km/s & $10^{12} M_\odot$ \\
	\hline
	Milky Way & 0.008 & $-$11.10 & 1.8302 \\
	Andromeda (Messier 31) & 0.707 & $-$309.18 & 2.0567 \\
	Centaurus A & 3.736 & 504.52 & 5.8831 \\ \hline
	Messier 101 & 7.391 & 439.62 & 9.3108 \\
	Messier 94 & 4.366 & 324.31 & 8.8144 \\
	Sculptor & 4.095 & 246.97 & 6.9296 \\
	NGC 6946 & 5.859 & 107.38 & 4.6142 \\
	Messier 81 & 3.625 & 73.48 & 4.0625 \\
	Maffei & 3.988 & $-$28.75 & 3.4924 \\
	IC 342 & 3.350 & $-$12.98 & 1.2994 \\
	Triangulum (Messier 33) & 0.948 & $-$192.72 & 0.2214 \\
	Large Magellanic Cloud & 0.065 & 235.97 & 0.2007 \\
	NGC 55 & 2.035 & 163.16 & 0.1323 \\
	NGC 300 & 1.963 & 158.70 & 0.1073 \\
	IC 10 & 0.781 & $-$338.02 & 0.0437 \\
	NGC 185 & 0.706 & $-$213.37 & 0.0129 \\
	IC 5152 & 1.878 & 138.56 & 0.0094 \\
	NGC 147 & 0.679 & $-$201.04 & 0.0064 \\
	NGC 6822 & 0.510 & $-$69.93 & 0.0059 \\
	\hline
	\end{tabular}
	\caption{Data on the massive galaxies in my 3D model using a similar catalogue to \citet[][table 1]{Shaya_2011}. Distances and masses are allowed to vary to best match observations, though their prior distributions are not uniform (see text). The masses derived in my model correspond to the total halo mass of each system, some of which is located beyond its virial radius \citep{Fattahi_2017}. The top section of this table contains galaxies which are also directly included as massive extended objects in the 2D model (Section \ref{Local_Group_2D}). The remaining galaxies are sorted in descending order of simulated mass. For clarity, I abbreviated the names of galaxies from the New General Catalogue (NGC) and Index Catalogue (IC).}
\label{Massive_galaxy_list_3D}
\end{table}

The heart of this model involves solving the equations of motion by adjusting a trial trajectory towards the true one. An incorrect trajectory will have a mismatch between the acceleration along it and that expected due to the gravity of other particles. Thus, at each timestep, the positions of all the particles are adjusted to try and equalise the gravitational field acting on each one with the acceleration $\overset{..}{\bm r}$ it experiences along its trajectory. This is done assuming both respond linearly to a position adjustment, although only the latter does. Thus, a solution can only be obtained after several iterations, each of which is reliant on a matrix inversion to handle the highly inter-connected nature of the problem. Some shortcuts are taken for test particles because their positions do not affect forces felt by other particles.

This method of solution is second-order accurate because of the standard finite differencing scheme used to obtain accelerations from a series of discrete positions valid at known times. Due to the large number of particle pairs, an adaptive timestep is impractical. Instead, I adapt the temporal resolution to the problem in a fixed way based on physical considerations. Each timestep corresponds to an equal increment in the cosmic scale-factor $a$, with 500 steps used between when $a = 0.1$ and the present time ($a \equiv 1$).

To check if I had adequate resolution, I solved the problem by forward Runge-Kutta integration instead, using 5000 timesteps equally spaced in $a$. The maximum error in the present position was 0.23 kpc while that in the velocity was 0.84 km/s. Both errors are very small, suggesting that there was enough resolution. Some other checks were also done to verify the numerical accuracy of the solution \citep[][section 2.4]{Shaya_2011}.

\subsection{Determining $\chi^2$ and finding the best model}
\label{Chi_sq_analysis_3D}

Like my axisymmetric model (Section \ref{Local_Group_2D}), the 3D model accurately matches the observed sky positions of target galaxies. However, this is achieved rather differently. Instead of integrating the equations of motion forwards in time and using the Newton-Raphson method to very precisely match present positions, the 3D model integrates backwards in time starting from a position along the line of sight towards a target galaxy. I no longer require agreement between simulated and observed heliocentric distances. Instead, I add a contribution to the total $\chi^2$ of the model if there is a mismatch. Handling distance uncertainties in this way makes error budgets model-independent, allowing relative model likelihoods to be determined simply by comparing their $\chi^2$.

The distance errors $\sigma_{d}$ come from observations. For M31, I use a slightly closer and more uncertain estimate \citep[${770 \pm 40}$ kpc,][]{Ma_2010}. Galaxies outside the LG might be affected by objects beyond the region covered by the analysis. It can also be difficult to determine the mass ratios between galaxies in an extended group and thus the location of its barycentre. For these reasons, I allow a fairly large distance uncertainty $\sigma_{d}$ for such objects.
\begin{eqnarray}
	\frac{\sigma_{d}}{d_{_{MW}}} ~=~ \frac{1}{10}~~\text{ if } d_{_{MW}} > 3.2~\text{Mpc}
	\label{Distance_uncertainty_perturbers}
\end{eqnarray}

Mismatches between observed and simulated GRVs are handled similarly, based on a tolerance of 20 km/s rather than the actual HRV measurement uncertainty. This is because I do not expect the model to be much more accurate as a representation of $\Lambda$CDM considering the level of scatter about the Hubble flow in more detailed cosmological simulations of the paradigm \citep{Aragon_Calvo_2011}. As the observational uncertainty $\sigma_{v_h}$ is always much smaller than this, the effect of raising it to 20 km/s is similar to setting ${\sigma_{extra} = 20}$ km/s in Equation \ref{sigma}. Either method handles modelling uncertainties by preventing the analysis from placing undue statistical weight on a very precisely observed galaxy.


Some LG galaxies have proper motion measurements. I made use of such data for M31, M33, the Large Magellanic Cloud (LMC), IC 10 and Leo I by adding a penalty to $\chi^2$ when simulated and observed values disagree. Observational proper motion error estimates are taken at face value.

Unlike in my 2D model, Equation \ref{Initial_conditions} is no longer strictly enforced at the start of the simulation because this is difficult to achieve when integrating backwards. Instead, I penalise models which fail to enforce it.
\begin{eqnarray}
	\Delta \chi^2 ~=~ \frac{| \overbrace{\bm{v}_{_i} - H_{_i} \bm{r}_{_i}}^{\bm{v}_{pec} \left( t = t_i \right)} |^2}{{\sigma_{_v}}^2}
	\label{chi_sq_contribution_v_pec}
\end{eqnarray}

\begin{figure}
	\centering 
		\includegraphics [width = 8.5cm] {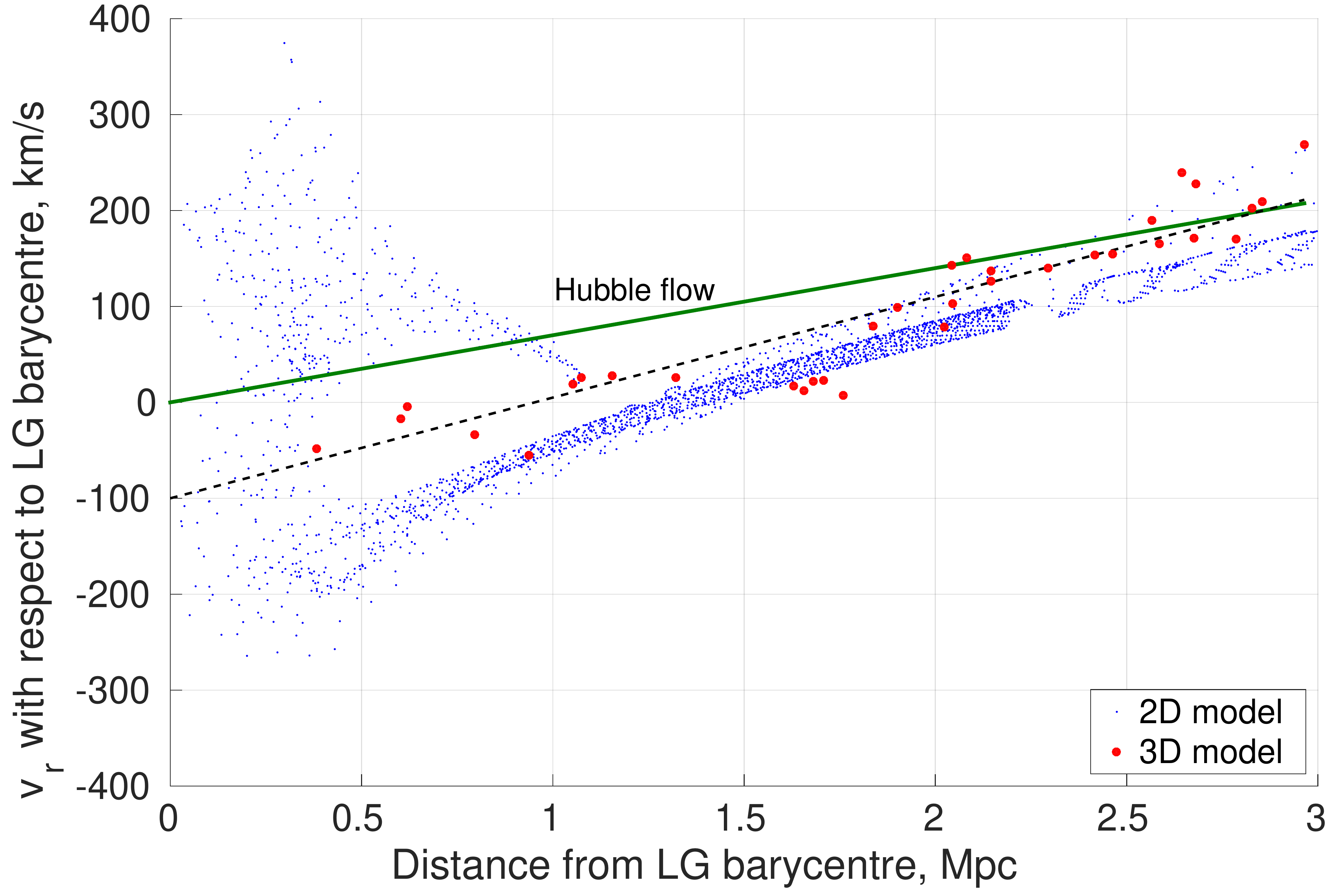}
	\caption{Radial velocities of test particles with respect to the LG barycentre are shown in blue for my 2D model with parameters matched to my best-fitting 3D model (Table \ref{Best_fit_parameters}), results of which are shown as large red dots. The solid green line is the Hubble flow relation for $H_{_0} = 70$ km/s/Mpc, the adopted value. The dashed black line has a gradient ${1.5\times}$ larger. Due to the effect of gravity, it provides a better fit to the 3D model within the LG (Section \ref{Results_3D}).}
	\label{LG_Hubble_diagram_3D}
\end{figure}

Based on present-day deviations from the Hubble flow (Figure \ref{LG_Hubble_diagram_3D}), I assume that the typical peculiar velocity $\bm{v}_{pec}$ was $\sigma_{_v} = 50$ km/s when $a = 0.1$. This is a 1D measure which underestimates typical values of ${v}_{pec}$ today. However, the nearly homogeneous state of the Universe at recombination \citep{Planck_2015} implies that ${v}_{pec}$ was typically smaller than today when my simulations started.

I do not fix the masses of any simulated galaxies which are treated as massive (Table \ref{Massive_galaxy_list_3D}). The prior used prefers a particular mass based on assuming $M/L$ is ${50 \times}$ the Solar value $\left( \frac{M}{L_K} \right)_\odot$ in the near-infrared $K$-band \citep{Tully_2013}. Observational estimates of the luminosity $L_K$ in this band are based on a particular distance to each target. If the simulated distance is lower, then the model implies that the target is likely closer to us and thus intrinsically fainter for the same apparent magnitude. This makes it likely to be less massive. Accounting for this, I define a preferred mass estimate
\begin{eqnarray}
	M_c ~\equiv~ 50 L_K \left( \frac{M}{L_K} \right)_\odot \left( \frac{d_{model}}{d_{obs}} \right)^2
	\label{M_c}
\end{eqnarray}

Using a different mass $M$ incurs a $\chi^2$ penalty of
\begin{eqnarray}
	\Delta \chi^2 ~=~ \left[ \frac{Ln \left( \frac{M}{M_c}\right)}{Ln ~1.5} \right]^2
	\label{chi_sq_contribution_M}
\end{eqnarray}

For the MW and M31, there is no a priori preference towards any particular mass for either galaxy. However, a particular \emph{ratio} between their masses is preferred. This is the ratio of their $M_c$ values.
\begin{eqnarray}
	\Delta \chi^2 ~= \left( \frac{Ln~\frac{M_{MW}}{M_{M31}} ~-~ Ln~\frac{M_{MW,c}}{M_{M31,c}}}{Ln~1.25}\right)^2
	\label{chi_sq_contribution_MW_M31_ratio}
\end{eqnarray}

The model now has too many parameters to permit a grid search through them. Thus, I focus on results from the best-fitting 3D model, which I obtain by minimising $\chi^2$ using a downhill-seeking walk through parameter space \citep[][section 2.2]{Shaya_2011}. Each parameter $A$ is varied by a small amount $\Delta A$ in an attempt to reduce $\chi^2$. If this does not happen, then the algorithm restores the previous solution and reverses $\Delta A$ while also reducing its magnitude. This is necessary because the increase in $\chi^2$ is often due to overshooting the minimum.
\begin{eqnarray}
	\Delta A ~\to ~-\frac{1}{2} \Delta A ~~ \left( \chi^2 \text{ increased} \right)
\end{eqnarray}

When a parameter adjustment reduces $\chi^2$, I accelerate the convergence by setting
\begin{eqnarray}  
	\Delta A ~\to~ \frac{5}{4} \Delta A ~~~~~ \left( \chi^2 \text{ decreased} \right)
\end{eqnarray}

To avoid the parameter adjustments being too large or too small, a cap is imposed on $\left| \Delta A \right|$ such that
\begin{eqnarray}
	\left| \frac{\Delta A}{A} \right| ~<~ 10^{-1}
\end{eqnarray}

I assume the process has converged once $\left| \frac{\Delta A}{A} \right| < 10^{-5}$.

\subsection{Results and discussion}
\label{Results_3D}

A comparison between my best-fitting 2D and 3D models is complicated somewhat by the latter having many more degrees of freedom. In particular, it is not required to match the observed distances of LG galaxies. This allows it to place a galaxy further away than observed, increasing its predicted GRV and better explaining a very high observed GRV. I handle this by applying a correction to the predicted GRV of each galaxy based on how its simulated distance differs from the observed value. Thus, I set
\begin{eqnarray}
	\label{GRV_adjustment_distance_error}
	GRV_{model} &\to&  GRV_{model} + \left( d_{obs} - d_{model}\right) \alpha H_{_0} \\
	\alpha &\equiv& \frac{1}{H_{_0}}\frac{dv_{_r}}{dr}
\end{eqnarray}

I use $\alpha = 1.5$ because this provides a reasonable description of how radial velocities $v_{_r}$ depend on distances within the LG (Figure \ref{LG_Hubble_diagram_3D}). At long range, $\alpha = 1$ but within the LG, gravity from the MW and M31 becomes important. Thus, an object further from them has been decelerated less by their gravity. Consequently, its radial velocity will be higher than for the more nearby object by a greater amount compared with a homogeneously expanding Universe. Neglecting projection effects (which become small a few Mpc from the LG), it is clear that $\alpha$ should slightly exceed 1.

\begin{table}
	\begin{tabular}{lcc}
	\hline
	Galaxy & $~~\Delta GRV$ & Distance from LG \\
	& $~~$(km/s) & barycentre (Mpc) \\
	\hline
	HIZSS 3 & $~~~123.2 \pm 10.6$ & $1.76 \pm 0.11$ \\
	NGC 3109 & $~~~110.7 \pm 7.3~$ & $1.63 \pm 0.05$ \\
	Sextans A & $~~~~95.1 \pm 7.2~$ & $1.66 \pm 0.02$ \\
	Sextans B & $~~~~75.4 \pm 5.4~$ & $1.71 \pm 0.05$ \\
	Antlia & $~~~~61.6 \pm 8.3~$ & $1.68 \pm 0.06$ \\
	\hline
	UGC 4879 & $~-31.1 \pm 5.5~$ & $1.32 \pm 0.02$ \\
	KKR 3 & $~-33.6 \pm 10.9$ & $2.30 \pm 0.12$ \\
	GR 8 & $~-40.0 \pm 10.5$ & $2.42 \pm 0.12$ \\
	NGC 55 & $~-42.0 \pm 10.4$ & $2.08 \pm 0.11$ \\
	NGC 4163 & $-130.6 \pm 7.7~$ & $2.96 \pm 0.04$ \\
	\hline
	\end{tabular}
	\caption{$\Delta GRV$s with respect to my 3D model for the 5 LG galaxies with the most positive and negative $\Delta GRV$s (excluding NGC 404 and Leo P due to large distance uncertainties). Errors are estimated using Equation \ref{sigma}. The LG barycentre is put almost exactly at the MW-M31 mid-point (Table \ref{Best_fit_parameters}). Errors in the distance from there are obtained from those on heliocentric distances in the usual way. Notice how the tails of the $\Delta$GRV distribution are rather asymmetric, as shown in Figure \ref{Delta_GRV_histogram_3D}.}
\label{Delta_GRV_3D_list}
\end{table}

\begin{table}
	\begin{tabular}{  m{1.15cm}  m{2.95cm} m{1.1cm}  m{1.32cm}}
	\hline
	\multicolumn{1}{l}{Parameter} & Meaning \& units & Best- & Best- \\
	 &   & fitting & fitting \\
	& & value & value \\	 
	& & in 2D &  in 3D \\ \hline
	$M$ & LG mass, ${10}^{12} M_\odot$ & 2.756 & 4.088 \\
	$q_{_{MW}}$ & ${M_{_{MW}}} \div {M}$ & 0.356 & 0.497 \\
	$q_{_{LMC}}$ & ${M_{_{LMC}}} \div {M_{_{MW}}}$ & 0.157 & 0.099 \\
	$v_{c, \odot}$ & LSR speed, km/s & 239 & 223.0 \\ \cline{1-3}
	$v_{f, M31}$ & $v_{_f}$ of M31, km/s & \multicolumn{1}{l|}{225} & 240.3 \\	
	$d_{_{M31}}$ & Distance to M31, kpc & \multicolumn{1}{l|}{783} & 707 \\
	$M_{\text{Cen A}}$ & Cen A mass, ${10}^{12} M_\odot$ & \multicolumn{1}{l|}{4} & 5.883 \\ [5pt] \cline{4-4}
	$U_\odot$ & Components of the & 14.1 & 11.1\\
	$V_\odot$ & non-circular motion of & 14.6 & 12.2\\
	$W_\odot$ & Sun in the MW, km/s & 6.9 & 7.2\\
	\hline
	$H_{_0}$ & Hubble constant & 67.3 & 70 \\ 
	$\Omega_{m,0}$ & Present matter density & 0.315 & 0.27\\
	& in the Universe $\div \frac{3{H_{_0}}^2}{8 \rm{\pi} G}$ & & \\ [5pt]
	\hline
	\end{tabular}
	\caption{The parameters of my best-fitting axisymmetric (2D) and 3D models. $q_{_{LMC}}$ is the LMC mass as a fraction of the MW mass, which I take to include the LMC. The top section of this table contains the parameters I varied using a grid search in my 2D model (Section \ref{Local_Group_2D}) or using gradient descent in 3D (Section \ref{Local_Group_3D}). The central section contains the parameters associated with the non-circular motion of the Sun in the Milky Way, which I obtain from \citet{Francis_2014} for the 2D model and from \citet{Schonrich_2012} for the 3D model. This section also contains two parameters related to M31. In the 2D model, its distance estimate is from \citet{McConnachie_2012} while the 3D model uses a prior of ${770 \pm 40}$ kpc \citep{Ma_2010}. Its rotation curve flatlines at a level $v_{f, M31}$ which is fixed in the 2D model but has a prior of ${240 \pm 10}$ km/s in the 3D model \citep{Carignan_2006}. This model assumes $v_{_f} = v_{c,\odot}$ for the MW whereas the 2D model fixes the former at 180 km/s \citep{Kafle_2012} and uses a prior on the latter of ${239 \pm 5}$ km/s \citep{McMillan_2011}. I adopt a flat dark energy-dominated cosmology whose parameters are fixed at the values given in the bottom section, with the 2D results based on those of \citet{Planck_2015} while the 3D results are based on \citet{Komatsu_2011}. Both models start when the cosmic scale-factor ${a = 0.1}$.}
\label{Best_fit_parameters}
\end{table}

In Section \ref{Local_Group_2D}, I added an extra dispersion term $\sigma_{extra}$ to Equation \ref{sigma} and then marginalised $\sigma_{extra}$ over other variables to obtain its probability distribution. The most likely value (using the optimal LMC mass) was 40.43 km/s. Using the same target galaxies, the rms dispersion in $\Delta GRV$ with respect to the best-fitting 2D model is 40.65 km/s, almost exactly the same. This suggests that the two statistics are very similar, even though the former uses integration over model parameter space while the latter is based on just one model. Thus, the rms $\Delta GRV$ of the best-fitting model should provide a very good guide to the results of a more thorough statistical analysis attempting to pin down how inaccurate each model is as a representation of the data. To draw conclusions about the validity of $\Lambda$CDM, this would then have to be compared with how accurately the models can be expected to represent $\Lambda$CDM.

\begin{figure}
	\centering 
		\includegraphics [width = 8.5cm] {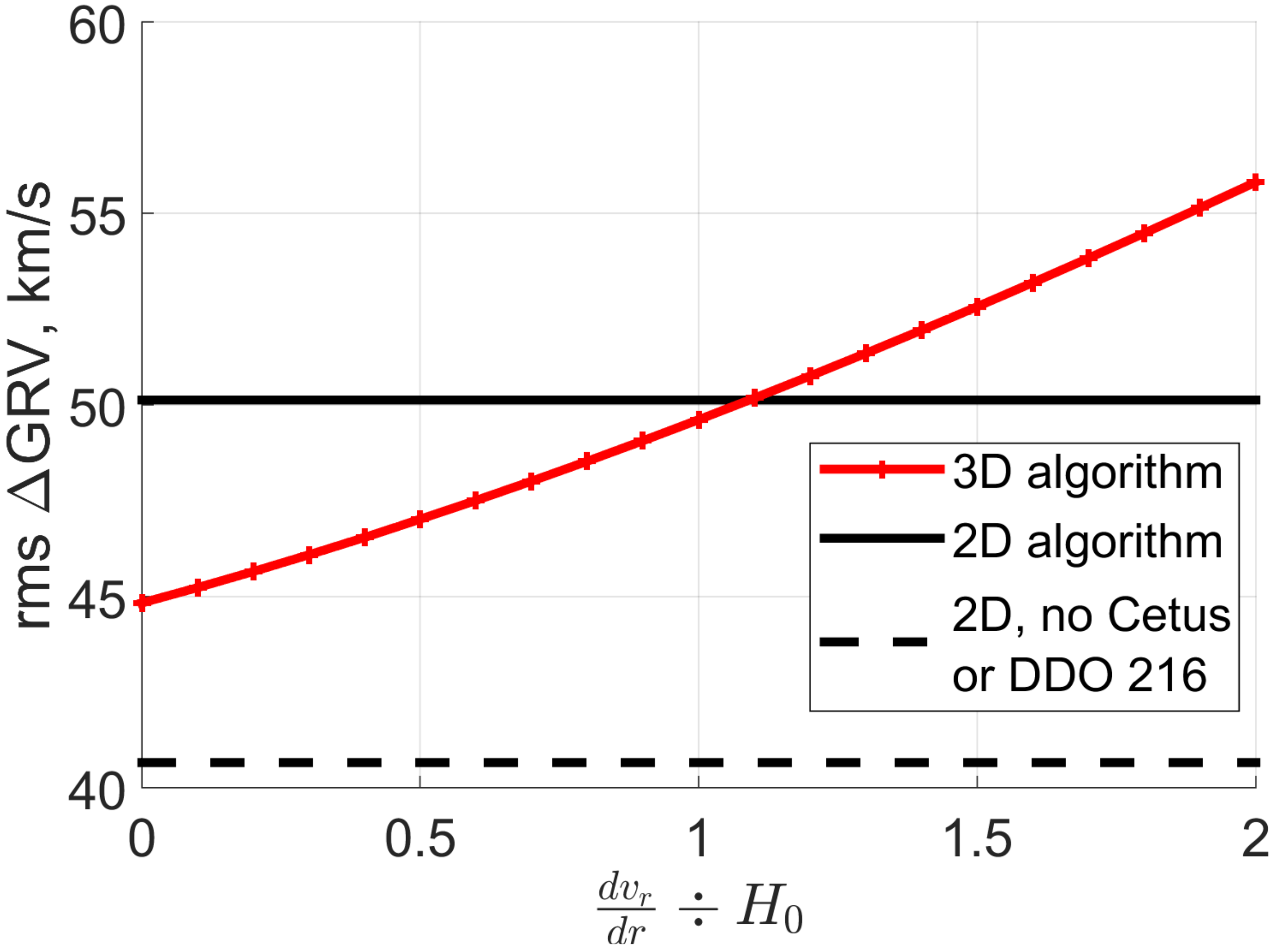}
	\caption{The root mean square value of $\Delta GRV$ for my best-fitting 2D (black) and 3D (red) models as a function of $\alpha$, which governs how 3D model predictions are adjusted to put them on an equal footing with my 2D model (Equation \ref{GRV_adjustment_distance_error}). The adjustment is unnecessary for the latter. This model is likely unreliable close to M31 as it lacks M33, making its predictions for Cetus and DDO 216 unreliable. Results including these galaxies (solid black) and without them (dashed black) are shown. In the 3D model, removing them increases the results by only $\ssim 0.7$ km/s, thus leaving them almost unchanged. This model treats the LG as empty apart from a few point masses. Using a similar assumption in my 2D models would reduce the rms value of $\Delta GRV$ by ${\ssim 6}$ km/s (not shown).}
	\label{rms_Delta_GRV}
\end{figure}

After obtaining corrected GRV predictions for my 3D model using Equation \ref{GRV_adjustment_distance_error}, I subtracted them from observed GRVs (Equation \ref{GRV_obs}) to obtain a list of $\Delta GRV$s. The rms of these $\Delta GRV$s is then found for a range of plausible assumptions regarding $\alpha$ (Figure \ref{rms_Delta_GRV}). For comparison, I also show the result of the same calculation for my best-fitting 2D model using the same target galaxies. This model requires an extremely precise match between their simulated and observed distances, making the result independent of $\alpha$. Although it was technically difficult to operate the 3D model in this way, one can gain a conservative lower bound on the rms value of $\Delta GRV$ had this been done by setting $\alpha = 0$ in Equation \ref{GRV_adjustment_distance_error}. This corresponds to taking the GRV predictions of the 3D algorithm at face value, even though it has some flexibility with distances. Removing this flexibility can only worsen the agreement between predicted and observed GRVs.

\begin{figure}
	\centering 
		\includegraphics [width = 8.5cm] {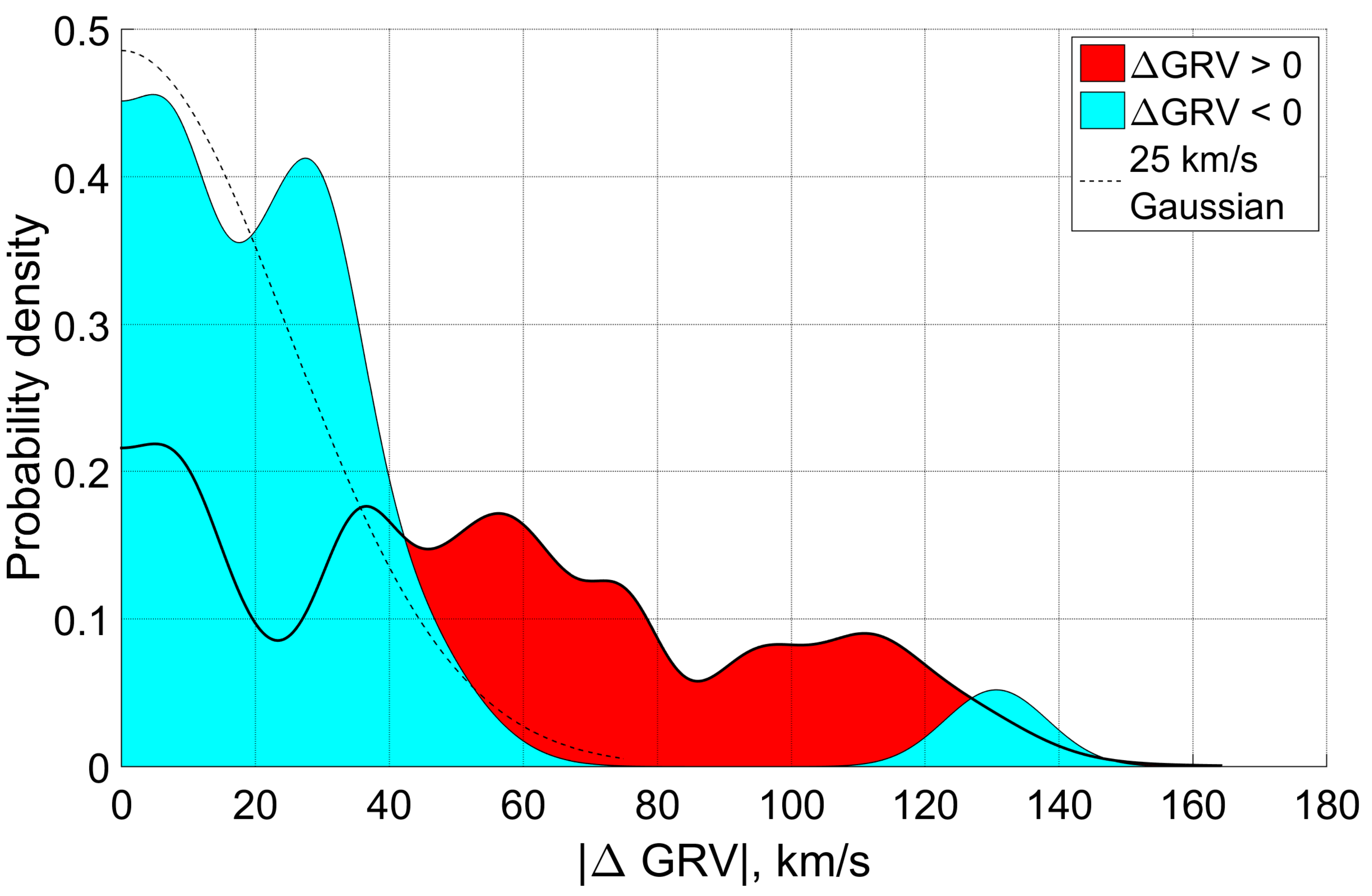}
	\caption{Histogram of $\Delta GRV$s with respect to my 3D model, shown separately according to the sign of $\Delta GRV$. The area of each square corresponds to 2 galaxies. A similar pattern emerges to my 2D model of the LG (Figure \ref{Delta_GRV_histogram_detailed}), with the blue bump near 130 km/s caused by NGC 4163. Otherwise, the galaxies with $\Delta GRV < 0$ (solid blue) are well described by a 25 km/s Gaussian (dashed line). This is not the case for galaxies with $\Delta GRV > 0$ (solid red).}
	\label{Delta_GRV_histogram_3D}
\end{figure}

With $\alpha$ irrelevant for the 2D model, its main uncertainty is whether Cetus and DDO 216 should be included in the analysis as they are very discrepant with this model. I suggest that they should not be included as they are quite close to M31. Unlike in the 2D case, excluding them from the 3D analysis hardly affects its rms value of $\Delta GRV$ (this rises $\ssim 0.7$ km/s), which then greatly exceeds the 2D result for the same sample. Even if these galaxies are included in both models, any value of ${\alpha > 1.1}$ implies that the rms $\Delta GRV$ is larger in the 3D analysis. Thus, modelling the LG in 3D does not alleviate the discrepancy with $\Lambda$CDM first highlighted using my 2D model (Section \ref{Local_Group_2D}). In fact, the discrepancy is slightly worse in the 3D case.

To better characterise this discrepancy, the residuals between model predictions and observations ($\Delta GRV$s) are shown as a histogram in Figure \ref{rms_Delta_GRV}. These results show a similar pattern to my 2D results (Figure \ref{Delta_GRV_histogram_detailed}) in that the $\Delta GRV$ distribution for galaxies with $\Delta GRV < 0$ (shown in blue) can broadly be understood using a Gaussian of width 25 km/s, a reasonable estimate of the modelling uncertainty (e.g. due to neglecting interactions between LG dwarfs). The only exception is NGC 4163, a very distant LG galaxy whose observed GRV is much lower than nearby galaxies at similar heliocentric distances in the Canes Venatici I cloud \citep[][table 2]{Makarov_2013}. NGC 4163 may have been
flung towards the LG by a close interaction outside it that
was not captured by my model.

Unlike galaxies with $\Delta GRV < 0$, those with $\Delta GRV > 0$ (shown in red) have a $\Delta GRV$ distribution completely different from a 25 km/s Gaussian. This discrepancy is not due to just one galaxy. To emphasise this, I use Table \ref{Delta_GRV_3D_list} to list the five galaxies with the highest and lowest (most negative) $\Delta GRV$s compared to my best-fitting 3D model. The inferred LG parameters in this model are given in Table \ref{Best_fit_parameters}, which compares the results to those of my best-fitting 2D model. Neither model matches LG observations particularly well.

\begin{figure}
	\centering 
		\includegraphics [width = 8.5cm] {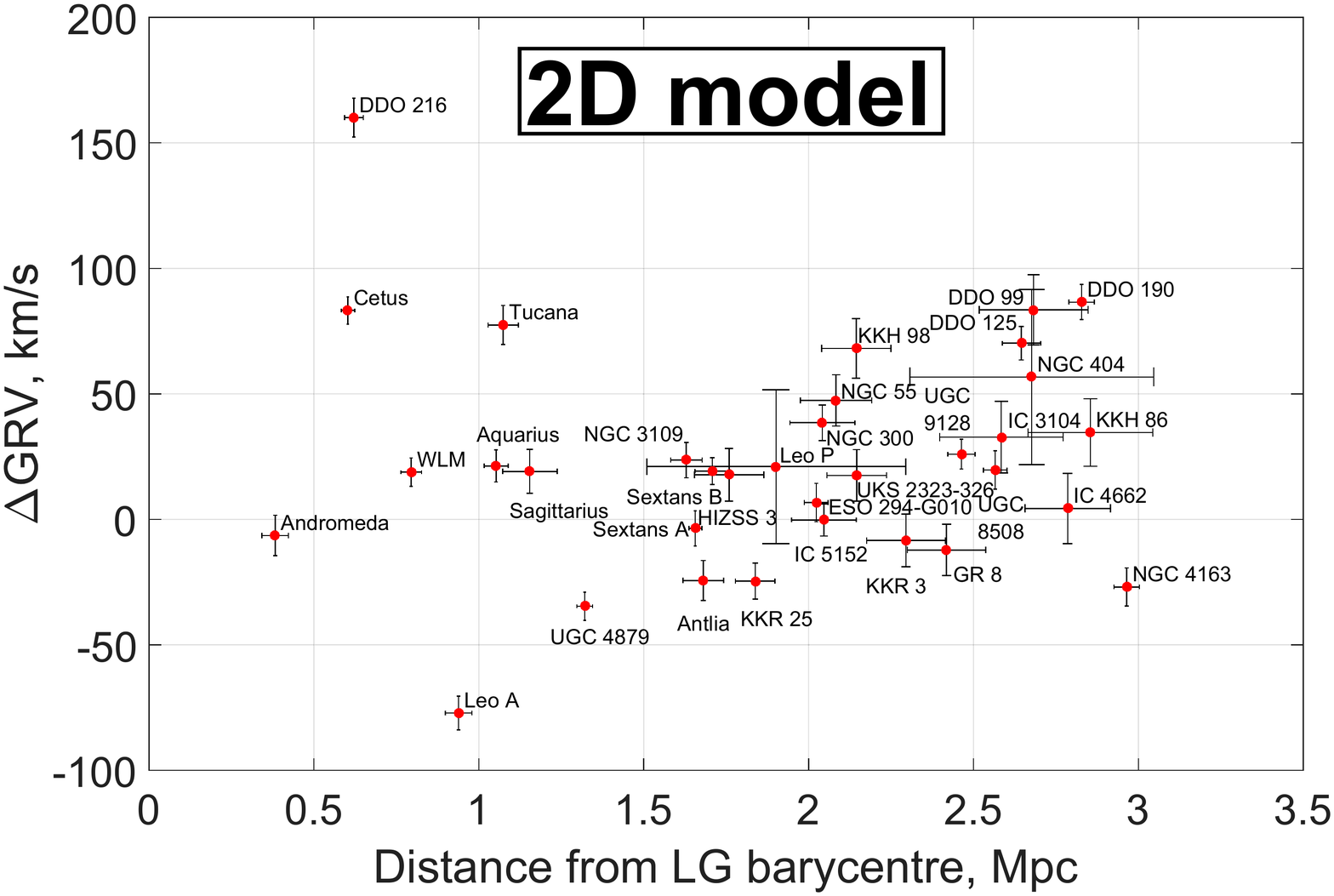}
		\includegraphics [width = 8.5cm] {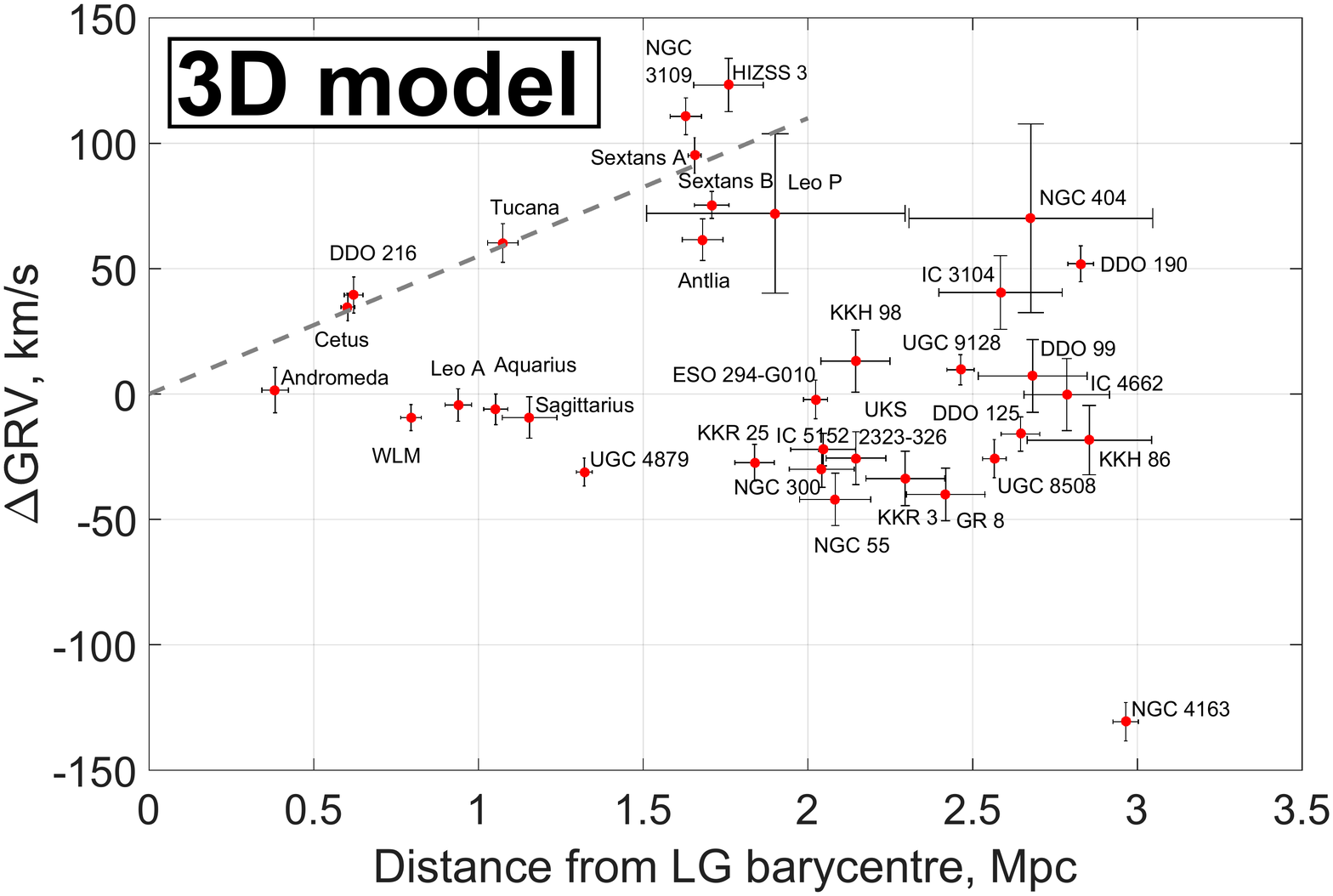}
	\caption{The $\Delta GRV$ of each target galaxy is shown against its distance from the LG barycentre. Parameters of the models used are given in Table \ref{Best_fit_parameters}, with the best-fitting ones used for the relevant number of dimensions in each model. Errors shown tend to be anti-correlated because a larger distance to a target increases its predicted GRV, reducing its $\Delta GRV$.}
	\label{Distance_GRV_correlation}
\end{figure}

As discussed in Section \ref{Local_Group_2D}, the kinematics of the HVGs might be due to enhanced gravitational slingshot interactions with the MW and M31 around the time of their flyby. If this is correct, $\Delta GRV$ should be larger for galaxies further from the LG. A trend of ${u \ssim 50}$ km/s/Mpc is apparent in Figure \ref{Distance_GRV_correlation} (dashed line), suggesting a MW-M31 flyby $\sim \left( H_{_0} + u \right)^{-1} = 8$ Gyr ago. This is roughly when the MW-M31 flyby is expected to have taken place in MOND \citep{Zhao_2013} and when the Galactic thick disk formed \citep{Quillen_2001}. In Section \ref{HVG_plane}, I consider other patterns that would be expected in this scenario.

\newpage
\section{A Plane of High-Velocity Galaxies across the Local Group \citep{BANIK_2017_ANISOTROPY}}
\label{HVG_plane}

Section \ref{Local_Group_3D} confirmed my earlier result (Section \ref{Local_Group_2D}) that the LG has several galaxies with unusually high radial velocities compared to $\Lambda$CDM expectations. In this section, I revisit that analysis with a more thorough search for model parameters and dwarf trajectories that agree better with observations (Section \ref{Model_refinements}). As this still leaves several HVGs, I try and understand them using a MOND model of the LG (Section \ref{MOND_simulation_LG}). A past MW-M31 flyby could yield HVGs at roughly the observed radial velocities, but it would tend to concentrate the HVGs within a particular plane. I consider how to test this statistically in Section \ref{Statistical_analysis_method} and conduct this test in Section \ref{Monte_Carlo_trials}, where I also consider other properties expected of the HVGs in this scenario (Table \ref{Criteria_definitions}).

\subsection{Revisiting the Local Group in $\Lambda$CDM}
\label{Model_refinements}

To better identify which galaxies may have been flung out by a fast-moving MW/M31, I refine the LG timing argument analysis of Section \ref{Local_Group_2D}. The input catalogue is updated, with the main changes being a more accurate distance to NGC 404 \citep{Dalcanton_2009} and Leo P \citep{McQuinn_2015}. For NGC 4163, I use a less accurate distance of ${2.95 \pm 0.07}$ Mpc to bracket the range between different Hubble-based measurements \citep{Dalcanton_2009, Jacobs_2009}.

The model is improved by relaxing the assumption that the flatline level of the MW rotation curve $v_{f, MW}$ is equal to its amplitude $v_{c, \odot}$ at the position of the Sun. Instead, I let $v_{f, MW}$ vary with a prior of $205 \pm 10$ km/s \citep{McGaugh_2016_MW} while $v_{c, \odot}$ is fixed at 232.8 km/s \citep{McMillan_2017}. The time resolution is improved ${10 \times}$ so that the history of the Universe since redshift 9 (${a = 0.1}$) is covered with 5000 steps, allowing for a much better handling of close encounters.

I also improved the $\chi^2$ minimisation procedure, which is now done by applying gradient descent to all model parameters, as described in Section \ref{Best_fitting_plane_finding}. To maximise the chance of matching observations, I ran a grid search through the trajectories of all the dwarf galaxies (treated as test particles). As explained in Section \ref{3D_method}, trajectories were solved by relaxing an initial guess towards a solution that satisfies the equations of motion. The initial guess has the co-moving position varying linearly with $a$. Each dwarf's current $\bm{v}_{pec}$ was varied over a 3D grid of possibilities, giving the algorithm a much better chance of finding close encounters that might otherwise get missed if the initial trajectory went nowhere near the spacetime location of the encounter. Thus, the grid search is complementary to gradient descent, which can find the minimum more precisely but is more prone to finding a local minimum rather than the true global minimum.

As some improvements were indeed found in this way, I repeated the gradient descent stage and the grid search in an alternating manner until the algorithm converged (i.e. the grid search did not further reduce $\chi^2$). This process took a few days and yielded reliable trajectories for all simulated galaxies $-$ their present positions and velocities were almost perfectly recovered compared to a forward integration using the fourth-order Runge-Kutta method with $10\times$ finer resolution (maximum errors of 9 pc and 16 m/s, respectively).

\begin{figure}
	\centering 
		\includegraphics [width = 8.5cm] {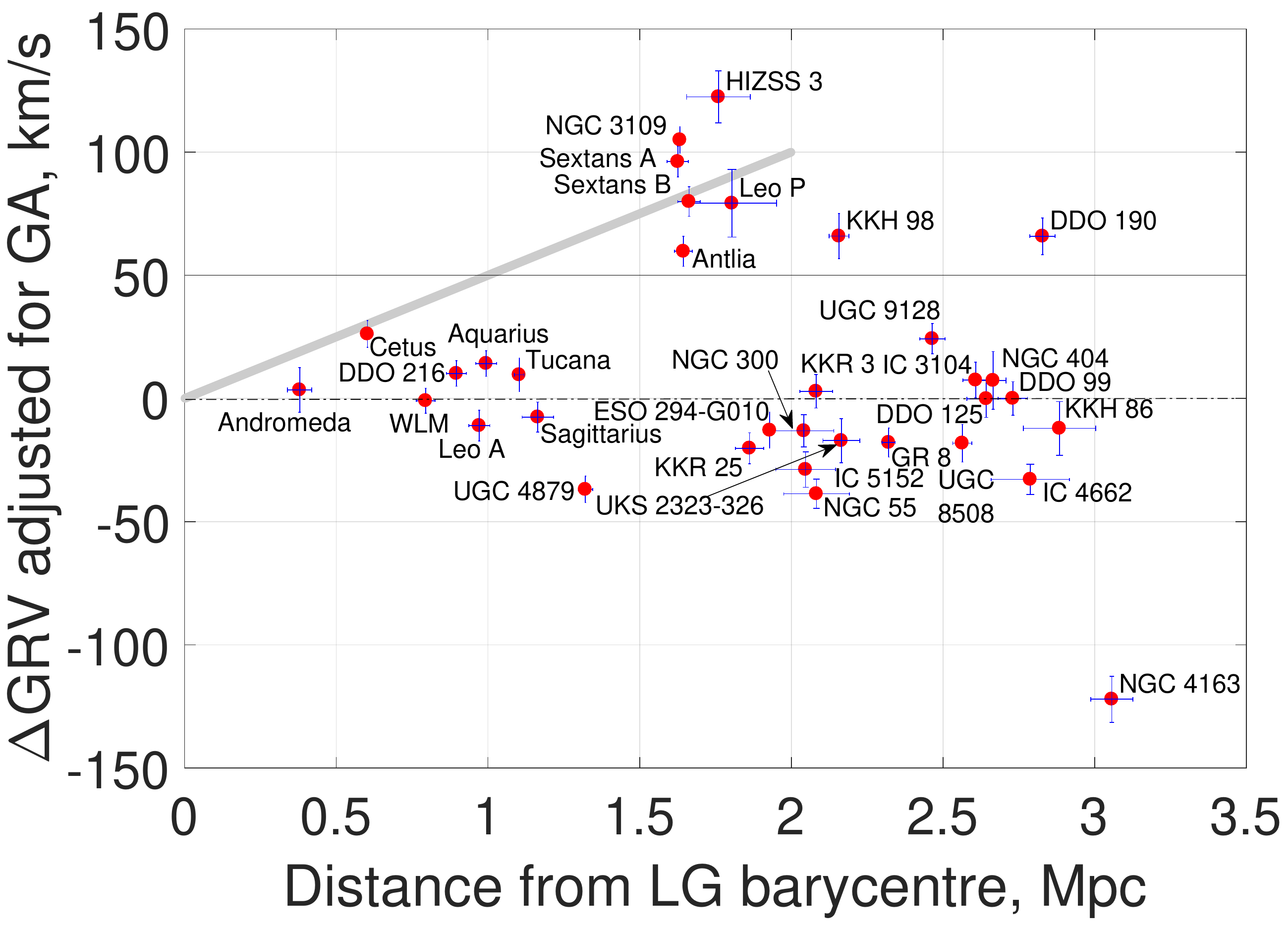}
		\caption{The deviation $\Delta GRV$ of each target galaxy from my best-fitting $\Lambda$CDM model, shown against its distance from the LG barycentre. An adjustment is applied to account for tides raised by the Great Attractor \citep[][equation 30]{BANIK_ZHAO_2017}. If the model worked perfectly, then all galaxies would have ${\Delta GRV \equiv 0}$ as model predictions are subtracted. Given likely model uncertainties of ${\ssim 25}$ km/s \citep{Aragon_Calvo_2011}, $\Lambda$CDM would thus find it difficult to explain galaxies with ${\Delta GRV > 50}$ km/s. In the MOND scenario of a past MW-M31 flyby, the HVGs should broadly follow a trend of 50 km/s/Mpc (diagonal grey line) and reach distances up to ${\ssim 2}$ Mpc (Figure \ref{Hubble_diagram_coloured_FL}).}
	\label{Distance_GRV_correlation_3D_GA}
\end{figure}

Using equation 30 from \citet{BANIK_ZHAO_2017}, I adjusted the predictions of this best-fitting model for the effect of tides raised on the LG by the Great Attractor (GA). This only slightly affects the results, which are shown in Figure \ref{Distance_GRV_correlation_3D_GA}. Compared to Section \ref{Local_Group_2D}, the main difference is that Tucana is now consistent with $\Lambda$CDM expectations. Given modelling uncertainties, this applies to any galaxy with $\left| \Delta GRV \right| < 50$ km/s. However, several galaxies still have a larger $\Delta GRV$.

In the MOND flyby scenario, the HVGs passed close to the spacetime location of the MW-M31 flyby. Thus, the HVGs should follow a $\Delta GRV \appropto d$ relation of the sort marked on Figure \ref{Distance_GRV_correlation_3D_GA}. DDO 190 does not fit neatly into this picture. However, given 34 target galaxies, it would not be particularly unusual to find one with a $\Delta GRV$ as large as the $66 \pm 7$ km/s value for DDO 190 if the model uncertainty is taken to be 25 km/s (probability ${P = 12\%}$). A second such instance would be unexpected (${P = 0.8\%}$). Bearing this in mind, I obtain the HVG sample listed in Table \ref{Planar_backsplash_galaxies}. This allows me to quantify the statistical properties of the HVGs and thereby better understand how their unusual kinematics arose (Section \ref{Statistical_analysis_method}).


\begin{table}
 \begin{tabular}{ccc}
		\hline
		Galaxies included & Distance from MW-M31 & $\Delta GRV$, km/s \\
    in the plane fit & mid-point, Mpc & \\
		\hline
		Milky Way & $0.382 \pm 0.04$ & NA \\
		Andromeda & $0.382 \pm 0.04$ & $3.5 \pm 9.1$ \\ [5pt]
		Sextans A & $1.624 \pm 0.036$ & $96.1 \pm 6.3$ \\
		Sextans B & $1.661 \pm 0.037$ & $79.9 \pm 6.0$ \\
		NGC 3109 & $1.631 \pm 0.014$ & $105.0 \pm 5.3$ \\
		Antlia & $1.642 \pm 0.030$ & $59.7 \pm 6.1$ \\
		Leo P & $~1.80 \pm 0.15$ & $79 \pm 14$ \\
		KKH 98 & $2.160 \pm 0.033$ & $65.5 \pm 9.1$ \\
		\hline
 \end{tabular}
 \caption{Galaxies considered HVGs based on Figure \ref{Distance_GRV_correlation_3D_GA}. The MW and M31 are shown here for reference.}
 \label{Planar_backsplash_galaxies}
\end{table}

\subsection{Simulating the Local Group in MOND}
\label{MOND_simulation_LG}

\subsubsection{Governing equations}
\label{MOND_governing_equations}

As in $\Lambda$CDM models of the LG, it is first necessary to obtain the MW-M31 trajectory. This is done by advancing them according to their mutual gravity supplemented by the cosmological acceleration term (e.g. Equation \ref{Equation_of_motion_2D}).
\begin{eqnarray}
      \label{MW_M31_governing_equation}
	\ddot{\bm{r}}_{_{rel}} &=& \bm{g}_{_{M31}} - \bm{g}_{_{MW}} + \frac{\ddot{a}}{a}\bm{r}_{_{rel}} ~\text{  where} \\
	\bm{r}_{_{rel}} &\equiv & \bm{r}_{_{M31}} - \bm{r}_{_{MW}}
\end{eqnarray}

$\bm{r}_{_i}$ is the position vector of galaxy $i$ (MW or M31), at whose location the gravitational field (excluding self-gravity) is $\bm{g}_{_i}$. All position vectors are with respect to the LG barycentre, which I take to be 0.3 of the way from M31 towards the MW. This is based on the MW rotation curve asymptotically reaching a flat level of ${\ssim 180}$ km/s \citep{Kafle_2012} while the equivalent value for M31 is ${\ssim 225}$ km/s \citep{Carignan_2006}. In the context of MOND, this suggests that the mass of M31 is $\left( \frac{225}{180} \right)^4 \approx {2.3 \times}$ that of the MW (Equation \ref{BTFR}).

To find the gravitational field $\bm{g}$ due to the MW and M31 at some position $\bm{r}$, I treat them as point masses and use the quasilinear formulation of MOND \citep[QUMOND,][]{QUMOND}. In spherically symmetric situations, this yields identical forces to the more traditional aquadratic Lagrangian formulation \citep[AQUAL,][]{Bekenstein_Milgrom_1984}. Even in less symmetric systems, the forces differ by only a few percent \citep[e.g.][]{Banik_2015_analytic, Candlish_2016}. As discussed in Section \ref{MOND_introduction}, I use the `simple' interpolating function between the Newtonian and deep-MOND regimes.
\begin{eqnarray}
	\bm{g}_{_N} &\equiv & -\sum_{i=MW,M31} \frac{GM_{i} \left( \bm{r} - \bm{r}_{_{i}} \right)}{ \left|\bm{r} - \bm{r}_{_{i}} \right|^3}  \\
	\label{Simple_interpolating_function}
	\nabla \cdot \bm{g} &\equiv & \nabla \cdot \left[\nu \left( \frac{\left| \bm{g}_{_N} \right|}{a_{_0}}\right) \bm{g}_{_N} \right] ~~\text{ where} \\
	\nu \left( x \right) &=& \frac{1}{2} ~+~ \sqrt{\frac{1}{4} + \frac{1}{x}} \nonumber
\end{eqnarray}

For an isolated point mass, $\bm{g}$ is given by Equation \ref{Simple_mu}. In the more complicated axisymmetric situation relevant to this problem, I use direct summation to obtain $\bm{g}$ from its divergence.
\begin{eqnarray}
	\bm{g} \left( \bm{r} \right) ~=~ \int \nabla \cdot \bm{g} \left( \bm{r'}\right) \frac{\left( \bm{r} - \bm{r'} \right)}{4 \pi \left| \bm{r} - \bm{r'} \right|^3} ~d^3\bm{r'}
	\label{g_direct_sum}
\end{eqnarray}

$\nabla \cdot \bm{g}$ is calculated out to almost 150 $r_{_{rel}}$, beyond which it should be very nearly spherically symmetric. Due to the shell theorem, it is unnecessary to consider $\nabla \cdot \bm{g}$ (or `phantom dark matter') at even larger radii. As $\bm{g}$ is only determined out to 66.5 $r_{_{rel}}$, my results should be nearly free of edge effects. At larger distances, the MW and M31 are treated as a single point mass located at their barycentre, yielding $\bm{g} = \nu \bm{g}_{_N}$ (Equation \ref{Simple_mu}).

In general, the galaxies will not be on the Hubble flow at the start time of the simulations $t_{_i}$, when the cosmic scale-factor $a_{_i} = 0.05$ and $H \equiv H_{_i}$. However, deviations from the Hubble flow are observed to be very small at early times \citep{Planck_2013}. In order to satisfy this condition, I varied the total MW and M31 mass $M$ using a Newton-Raphson root-finding algorithm to ensure that
\begin{eqnarray}
	\dot{\bm{r}}_{_{rel}} = H_{_i} \bm{r}_{_{rel}} ~\text{ when } t = t_{_i}
	\label{Hubble_flow_initial}
\end{eqnarray}

The MW and M31 are not on a purely radial orbit. Their mutual orbital angular momentum prevents them from converging onto the Hubble flow at very early times. This is unrealistic as any non-radial motion must have arisen due to tidal\footnote{affecting the MW and M31 differently} torques well after the Big Bang. Thus, I take the MW-M31 orbit to be purely radial prior to their first turnaround at $t \approx 3$ Gyr. After this time, I assume their trajectory conserves angular momentum at its present value. This implies the MW-M31 angular momentum was gained near the time of their first turnaround, when their large separation would have strengthened tidal torques. At later times, the larger scale factor would weaken tidal torques, suggesting that these have a much smaller effect around the time of the second MW-M31 turnaround than the first.

\begin{figure}
	\centering 
		\includegraphics [width = 8.5cm] {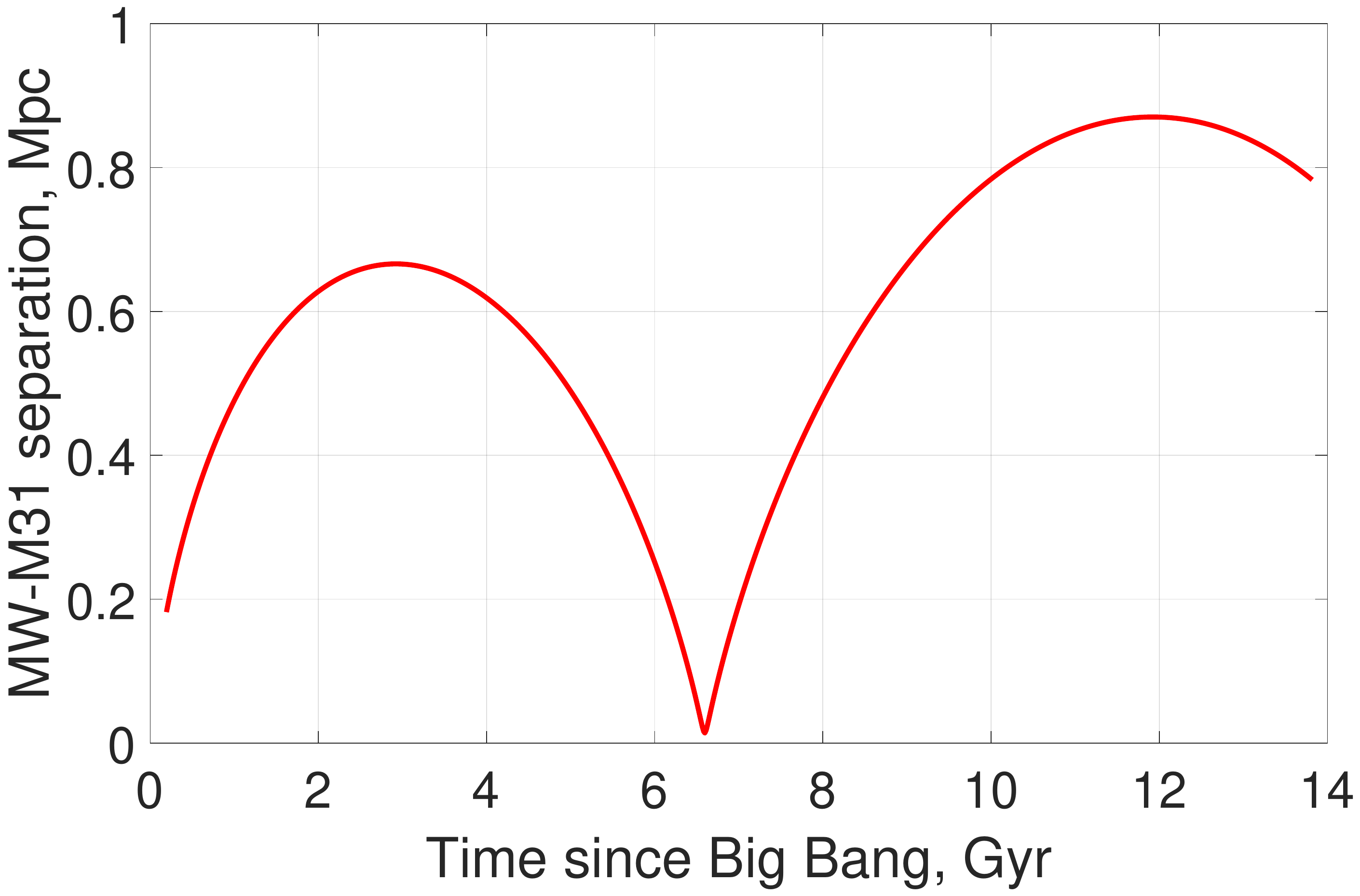}
		\caption{MW-M31 separation in my MOND simulation, showing a past close flyby 6.59 Gyr after the Big Bang at a closest approach distance of 14.17 kpc. At that time, their relative velocity was 716 km/s, of which 501 km/s was due to motion of the MW. The higher second apogalacticon is partly due to the effect of cosmology (Equation \ref{MW_M31_governing_equation}) and my assumption that the MW and M31 lose 5\% of their mass around the time of their encounter. Some other reasons for this are explained in section 5.1.2 of \citet{Banik_2018}.}
	\label{MW_M31_trajectory_FL}
\end{figure}

The resulting MW-M31 trajectory is shown in Figure \ref{MW_M31_trajectory_FL}. With this information, $\bm{g}$ can be found everywhere within the LG at all times using Equation \ref{g_direct_sum}, assuming only these point masses are present in an otherwise homogeneous Universe.

Test particle trajectories can now be advanced using
\begin{eqnarray}
	\label{Test_particle_acceleration}
	\ddot{\bm{r}} ~&=&~ \frac{\ddot{a}}{a}\bm{r} ~+~ \bm{g} \\
	\dot{\bm{r}} ~&=&~ H_{_i} \bm{r} ~\text{ when }t = t_{_i}	
\end{eqnarray}

Because the MW and M31 must have accreted matter from some region prior to the start of the simulation, I exclude all test particles starting within a distance $r_{_{exc,i}}$ of galaxy $i$. This distance is determined by requiring that the excluded volume has as much baryonic matter as galaxy $i$, taking the density of baryons to be the cosmic mean value. To obtain this, I assume baryons currently comprise a fraction $\Omega_{b,0} = 0.049$ of the cosmic critical density, which I found by taking $H_{_0} = $ 67.3 km/s/Mpc \citep[][table 4]{Planck_2015}. The cosmic baryon density can be estimated using Big Bang nucleosynthesis $-$ only a narrow range of values is consistent with the primordial abundances of light elements like deuterium \citep{Cyburt_2016}.
\begin{eqnarray}
	\frac{4 \pi}{3}{r_{{exc,i}}}^3 \times \overbrace{\frac{3{H_{_0}}^2}{8\pi G} \Omega_{_{b,0}} {a_{_i}}^{-3}}^{\text{Baryon density at }t_{_i}} ~\equiv ~ M_i ~~ \left( \text{for } r_{{exc,i}} \right)
	\label{r_exc}
\end{eqnarray}

To avoid the excluded regions overlapping, it is necessary that their sizes satisfy
\begin{eqnarray}
	r_{_{exc,MW}} ~+~ r_{{exc,M31}} ~\leq ~r_{_{rel}} ~\text{ when } t = t_{_i}
\end{eqnarray}

This inequality applies because $r_{_{exc}}$ is 77.7 kpc for the MW and 102.5 kpc for M31, leading to a total of 180.2 kpc $-$ interestingly, this is just smaller than $r_{_{rel}} \left( t_{_i} \right) = 182.1$ kpc, suggesting that the two galaxies accreted matter from regions which just touched. This remains the case with a slightly different start time as ${r_{{exc,i}}} \propto a_{_i}$, similarly to $r_{_{rel}} \left( t_{_i} \right) - $ at such early times, both galaxies would follow the Hubble flow rather closely. However, this coincidence does not occur in $\Lambda$CDM, a model in which the excluded regions would very likely overlap \citep[][section 2.2.1]{BANIK_ZHAO_2016}.

\subsubsection{Results}

\begin{figure}
	\centering 
		\includegraphics [width = 8.5cm] {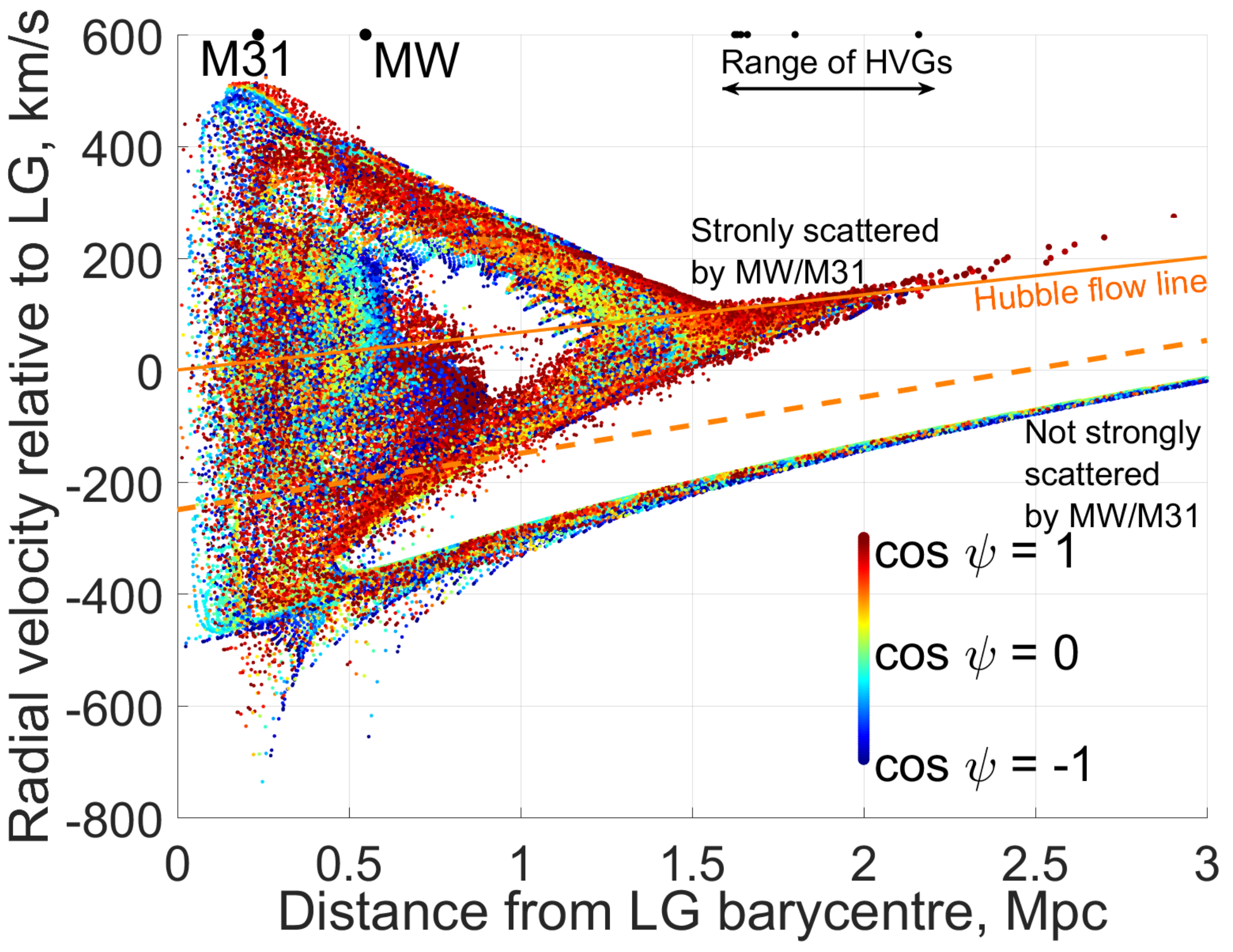}
		\caption{Hubble diagram of the test particles in my simulation coloured by their value of $\cos \psi$, which parametrises how well their orbital angular momenta align with that of the MW-M31 orbit (Equation \ref{cos_psi_definition}). The Hubble flow line is shown in solid orange. I also show a ${1.5 \times}$ steeper line (dashed orange) that is used to select analogues of HVGs in Figure \ref{z_distribution}. Particles below this line have generally never interacted closely with the MW or M31, unlike particles above the line. The black dots along the top edge of the figure indicate distances to the MW, M31 and the HVGs (Table \ref{Planar_backsplash_galaxies}).}
	\label{Hubble_diagram_coloured_FL}
\end{figure}

Figure \ref{Hubble_diagram_coloured_FL} shows the distances and radial velocities of test particles with respect to the LG barycentre, colour-coded according to the orientation of their orbital plane relative to that of the MW-M31 orbit. This is quantified based on each particle's specific angular momentum $\bm{h}$, whose direction can readily be compared with the MW-M31 orbital pole $\widehat{\bm{h}}_{_{LG}}$.
\begin{eqnarray}
	\bm{h} &\equiv & \bm{r} \times \dot{\bm{r}}	\\
	\cos \psi &\equiv & \widehat{\bm{h}} \cdot \widehat{\bm{h}}_{_{LG}}
	\label{cos_psi_definition}
\end{eqnarray}


For this section, the important feature is the upper branch of the Hubble diagram. Its upward slope arises because these particles must have passed close to the spacetime location of the MW-M31 encounter and gained a substantial amount of kinetic energy in what was essentially a 3-body interaction. Thus, for such particles to be further away from the LG now, they must have a larger outwards velocity.

$\Lambda$CDM also allows slingshot encounters with the MW and M31, but their fairly slow motion only allows them to fling galaxies out to ${\ssim 1}$ Mpc from the LG. At this point, the upper branch of the Hubble diagram simply stops (Figure \ref{LG_Hubble_Diagram}). Even in more detailed cosmological simulations of $\Lambda$CDM that include encounters with satellites of MW and M31 analogues, dwarf galaxies do not get flung out beyond this distance \citep[][figures 3 and 6]{Sales_2007}. For MOND, the corresponding limit is ${\ssim 2.5}$ Mpc due to the MW-M31 flyby, which therefore makes a dramatic difference to the Hubble diagram at distances of ${\ssim 1 - 2}$ Mpc (Figure \ref{Hubble_diagram_coloured_FL}). In this distance range, my simulation yields a bimodal distribution of radial velocities, with the HVGs corresponding to particles in the upper branch. A pattern of this sort is apparent in the kinematics of the observed LG (Figure \ref{Distance_GRV_correlation_3D_GA}).

\begin{figure}
	\centering 
		\includegraphics [width = 8.5cm] {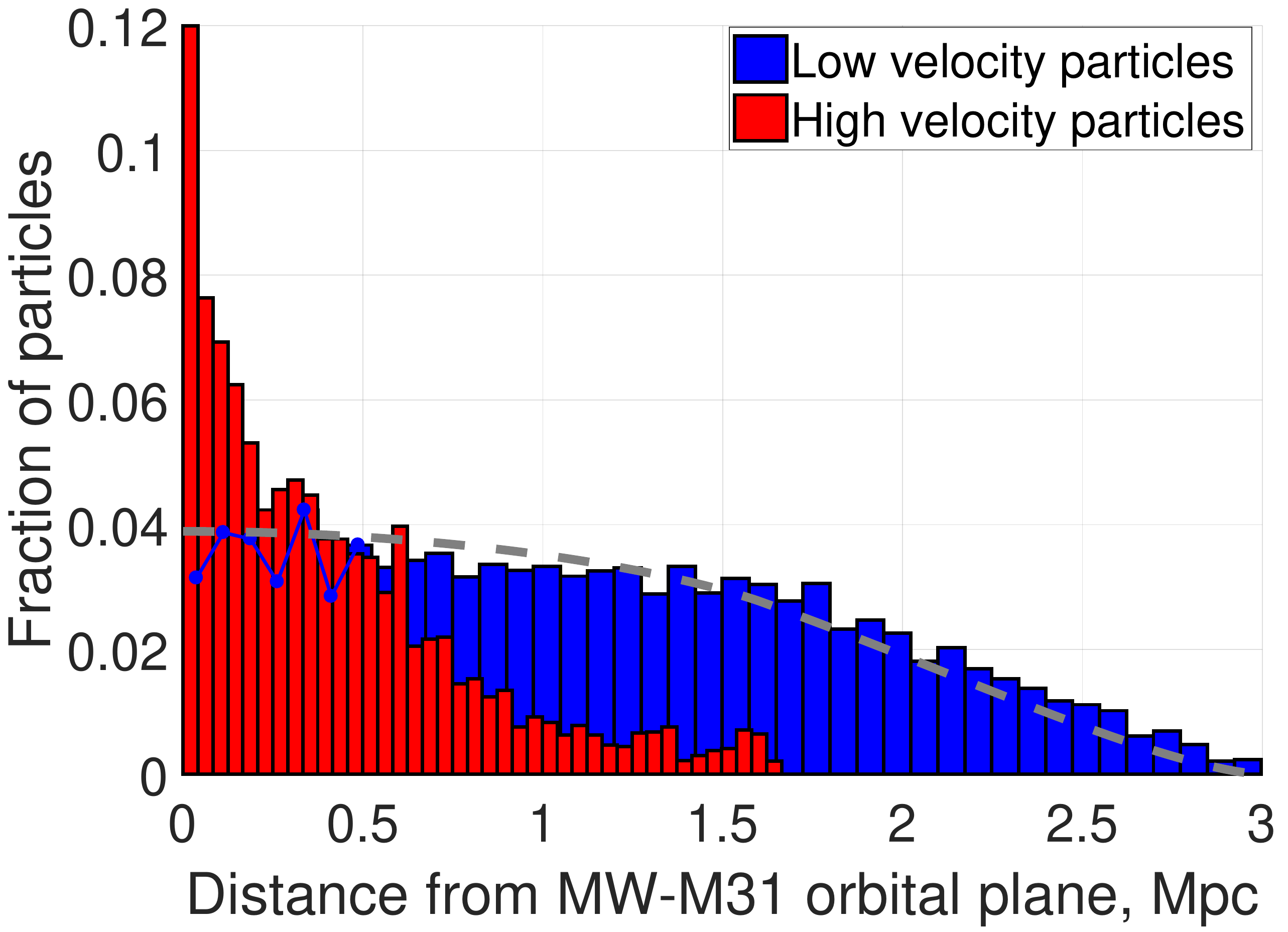}
		\caption{Histogram showing how far simulated particles are from the MW-M31 orbital plane. This only shows particles currently ${1.6-3}$ Mpc from the LG, sorted according to whether they are in the high-velocity branch of the Hubble diagram (above dashed orange line in Figure \ref{Hubble_diagram_coloured_FL}). If they are, I show them as red. The remaining particles (shown in blue) are well described by an isotropic distribution (dashed grey line).}
	\label{z_distribution}
\end{figure}

Gravitational slingshot interactions with the MW or M31 would be most efficient for particles flung out roughly parallel to the motion of the perturber. Considering that the MW-M31 flyby occurred a fixed time in the past, these particles should currently be furthest away from the LG. Thus, it is not very surprising that the spatial distribution of such distant HVG analogues is indeed highly flattened with respect to the MW-M31 orbital plane (Figure \ref{z_distribution}).

\subsection{Statistical analysis}
\label{Statistical_analysis_method}

\subsubsection{Finding the best-fitting plane}
\label{Best_fitting_plane_finding}

To quantify whether a set of galaxies is distributed anisotropically, it is necessary to define a measure of anisotropy and determine how unusual its value is. The statistic I used was $z_{_{rms}}$, the rms of the minimum distances between the galaxies I consider and the best-fitting plane through them (i.e. the one that minimises $z_{_{rms}}$). With respect to a plane having normal $\widehat{\bm{n}}$ and containing the vector $\bm{r}_{_0}$, the vertical dispersion is
\begin{eqnarray}
	\label{z_rms}
	{z_{_{rms}}}^2 ~&=&~ \frac{1}{N}\sum_{i = 1}^{N} \left[ \left( \bm{r}_{_i} - \bm{r}_{_0} \right) \cdot \widehat{\bm{n}} \right]^2 \\
	 ~&=&~ \widehat{\bm{n}} \cdot \left(\mathbf{I} \widehat{\bm{n}} \right) ~~\text{ where} \\
	\mathbf{I}_{jk} ~&\equiv &~ \frac{1}{N}\sum_{i = 1}^{N} \left( \bm{r}_{_i} - \bm{r}_{_0} \right)_j \left( \bm{r}_{_i} - \bm{r}_{_0} \right)_k
\end{eqnarray}

\begin{table}
 \begin{tabular}{llll}
\hline
  Quantity & Full sample & Without Antlia\\ 
  \hline
  Galaxies in plane & 8 & 7 \\ [5pt]
  Normal to plane of & \multirow{2}{*}{$\begin{bmatrix} 204.4^\circ \\ -30.1^\circ \end{bmatrix}$} & \multirow{2}{*}{$\begin{bmatrix} 206.6^\circ \\ -31.8^\circ \end{bmatrix}$} \\
  high $\Delta GRV$ galaxies  &  \\ [5pt]
  rms plane width, kpc & 101.1 & 101.9 \\
  Aspect ratio (Eq. \ref{Aspect_ratio}) & 0.0763 & 0.0750 \\ [5pt]
  MW offset from plane & 224.7 & 195.4 \\
  M31 offset from plane & -0.6 & -12.8 \\
	Angle of MW-M31  & \multirow{2}{*}{$16.2^\circ$} & \multirow{2}{*}{$14.9^\circ$} \\
	line with plane & & \\
  \hline
 \end{tabular} 
 \caption{Information about the plane best fitting the galaxies listed in Table \ref{Planar_backsplash_galaxies}, with distances in kpc and plane normal direction in Galactic co-ordinates (latitude last). The last column shows how the results change if Antlia is removed from the sample as it could be a satellite of NGC 3109 \citep{Van_den_Bergh_1999}.}
 \label{Plane_parameters} 
\end{table}

The galaxies are at heliocentric positions $\bm{r}_{_i}$. The minimum of $z_{_{rms}}$ is attained when $\bm{r}_{_0} = \frac{1}{N} \sum_{i = 1}^{N} \bm{r}_{_i}$, corresponding to the geometric centre of the $N$ galaxies to which a plane is being fit. I find its best-fitting orientation $\widehat{\bm{n}}$ using a gradient descent method \citep[e.g.][]{Fletcher_1963}. Issues of local minima are resolved by starting the gradient descent at whichever $\widehat{\bm{n}}$ yields the smallest $z_{_{rms}}$ in a low resolution grid of possible directions for $\widehat{\bm{n}}$. Once the angular step size is below ${0.006^\circ}$, further iterations are stopped and I assume the algorithm has converged to an acceptable precision.

\begin{figure}
	\centering 
		\includegraphics [width = 8.5cm] {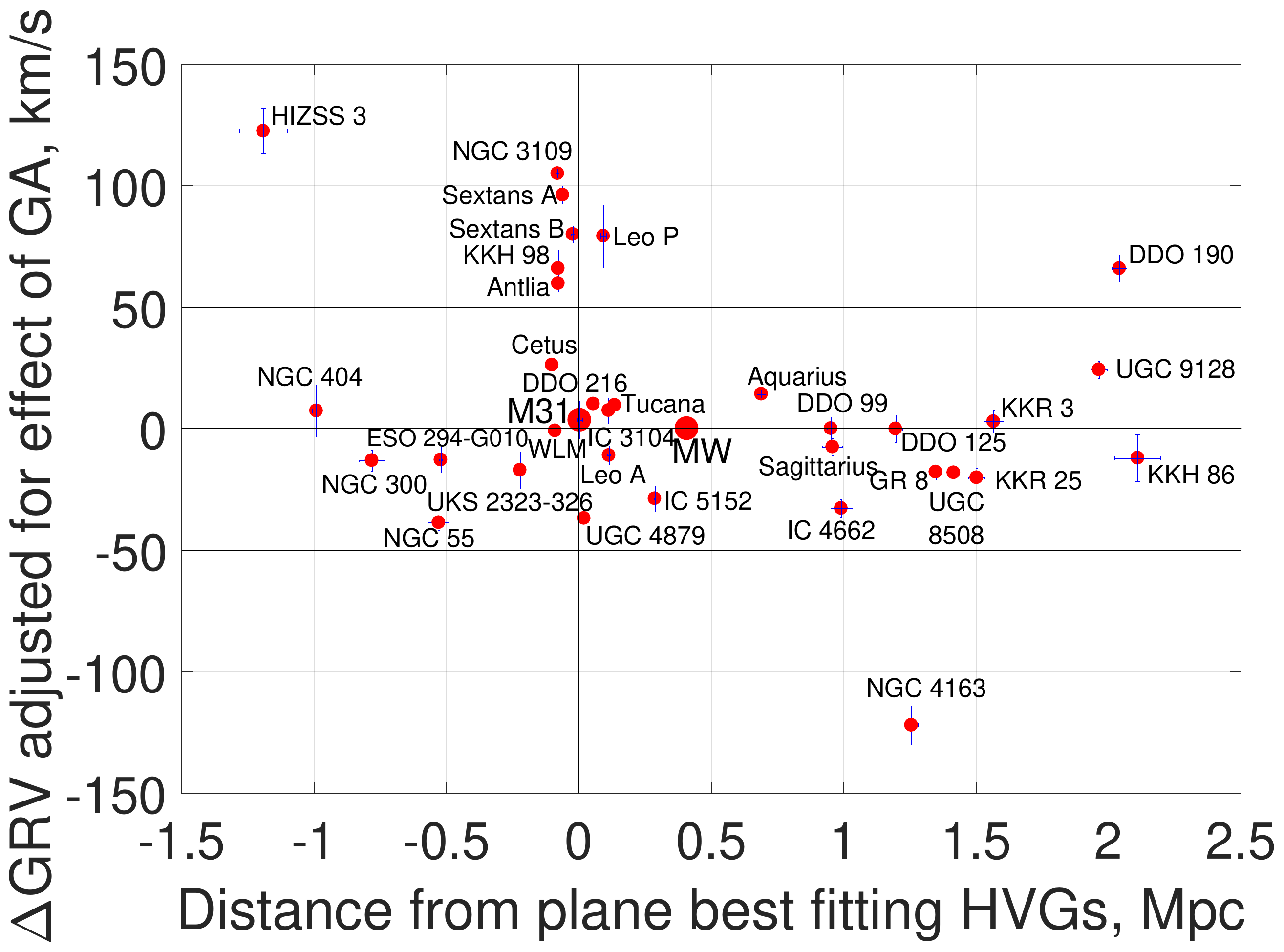}
	\caption{$\Delta GRV$s of target galaxies are shown against their offsets from the best-fitting plane through the ones with the largest $\Delta GRV$s except HIZSS 3 (parameters of this plane given in central column of Table \ref{Plane_parameters}). By definition, $\Lambda$CDM predicts $\Delta GRV = 0$ with an uncertainty of $\ssim 25$ km/s \citep{Aragon_Calvo_2011}. Thus, it can't easily explain galaxies with ${\Delta GRV > 50}$ km/s (above upper horizontal line). Most of these galaxies lie very close to a plane (near vertical gridline), unlike the rest of the sample. The concept of a $\Delta$GRV is meaningless for the MW, so I show this as 0.}
	\label{Plane_offset_Delta_GRV_GA}
\end{figure}

Applied to the major LG galaxies along with the HVGs except HIZSS 3 (full list in Table \ref{Planar_backsplash_galaxies}), this reveals that they define a rather thin plane whose parameters are given in Table \ref{Plane_parameters}. This allows a comparison between the $\Delta GRV$ of each galaxy\footnote{adjusted for the Great Attractor using equation 30 of \citet{BANIK_ZHAO_2017}} and its minimum distance from this plane. The galaxies in the full sample have a wide range of positions relative to it, with a similar number on either side (Figure \ref{Plane_offset_Delta_GRV_GA}). However, the HVGs tend to lie very close to it. The only exception is HIZSS 3, justifying my decision not to consider it when defining the HVG plane. In any case, the observations for HIZSS 3 are rather insecure due to its very low Galactic latitude \citep[${0.09^\circ}$,][]{Massey_2003}. Some of the issues caused by this are discussed in section 6.3 of \citet{BANIK_2017_ANISOTROPY}. Apart from HIZSS 3, my analysis has no target galaxies within $16^\circ$ of the Galactic plane.

\subsubsection{Monte Carlo trials}
\label{Monte_Carlo_trials}

To see if the situation shown in Figure \ref{Plane_offset_Delta_GRV_GA} is consistent with isotropy, I conduct a series of Monte Carlo (MC) trials in which I randomise the sky directions of the HVGs and recompute $z_{_{rms}}$. Formally, isotropy implies that the Galactic longitude $l$ has a uniform probability distribution while that of the Galactic latitude $b$ is
\begin{eqnarray}
	P(b)~db ~=~ \frac{1}{2}\cos b~db
	\label{Probability_distribution_b}
\end{eqnarray}

To mimic uncertainties in measured distances to LG galaxies, I randomly vary their heliocentric distances using Gaussian distributions of the corresponding widths. Although this happens very rarely, any negative distances are raised to 0.

To account for HIZSS 3 being excluded from the plane fit despite its high $\Delta GRV$, I used the procedure described in Section \ref{Best_fitting_plane_finding} to find the best-fitting plane through every combination of all HVGs but one as well as the MW and M31. The combination yielding the lowest $z_{_{rms}}$ is considered the analogue of the observed HVG system less HIZSS 3 for that particular randomly generated mock catalogue. The enforced inclusion of the MW and M31 is necessary because it would not make sense for them to lie far from a plane supposedly corresponding to their mutual orbital plane.

To check whether the particular choice of statistic affected my results, I also performed calculations where I selected the combination yielding the lowest aspect ratio $A$ rather than $z_{_{rms}}$.
\begin{eqnarray}
	\label{Aspect_ratio}
	A ~&\equiv&~ \frac{z_{_{rms}}}{\sqrt{{{r_{_{rms}}}^2 - {z_{_{rms}}}^2}}} ~~~\text{  where} \\
	{r_{_{rms}}}^2 ~&\equiv&~ \frac{1}{N} \sum_{i = 1}^{N} \left| \bm{r}_{_i} - \bm{r}_{_0} \right|^2 ~=~ Trace\left( \mathbf{I} \right)
\end{eqnarray}

$r_{_{rms}}$ is the rms distance of the galaxies from their geometric centre $\bm{r}_{_0}$. To get the rms extent of the system after projection into the best-fitting plane, I need to subtract $z_{_{rms}}$ in quadrature. Dividing $z_{_{rms}}$ by the result then gives a measure of the typical `vertical' extent of galaxies out of this plane relative to their `horizontal' extent within it. Identical probabilities would be obtained had I defined $A \equiv \frac{z_{_{rms}}}{r_{_{rms}}}$ instead, as long as it is defined in the same way for the actual HVGs and the mock sample in each MC trial. This is because $A$ is a monotonic function of $\frac{z_{_{rms}}}{r_{_{rms}}}$ with either definition. Moreover, both definitions yield very similar $A$ for a highly flattened structure.

In Table \ref{Criteria_definitions}, I give the criteria which I used to determine whether the HVGs in each MC trial are distributed in an analogous way to observations. I choose these criteria so that they should be satisfied if the LG behaves similarly to my MOND simulation of it (Section \ref{MOND_simulation_LG}). I consider one of the first two anisotropy-related criteria alongside both of the others. I used a toy model to crudely estimate the MW-M31 orbital pole required to explain the observed orientations of their satellite planes \citep[][section 2.2]{BANIK_2017_ANISOTROPY}. The MW-M31 orbital pole preferred by this toy model has only a small impact on my final results, which would be similar if the constraint it yields was not considered (Table \ref{Probability_table_criteria_combinations}).

If a past MW-M31 flyby is responsible for the unusual kinematics of the HVGs, then the plane they define should intersect the MW-M31 barycentre. I take this to be $0.3$ of the way from M31 towards the MW for reasons discussed at the start of Section \ref{MOND_governing_equations}. This puts the MW-M31 barycentre 67 kpc from the best-fitting plane, a rather small offset from a plane with a radial extent of ${\ssim 1.3}$ Mpc.

\begin{table}
 \begin{tabular}{ll}
		\hline
		Criterion & Meaning\\
		\hline
		Plane & There must be a plane of HVGs with rms\\
		thickness & thickness (Equation \ref{z_rms}) below that observed\\ [5pt]
		Aspect & There must be a plane of HVGs with aspect \\
		ratio & ratio (Equation \ref{Aspect_ratio}) below that observed\\ [5pt]
		Barycentre & Barycentre of MW and M31 (assuming\\
		offset &30\% of total mass in MW) closer to\\
		& plane than observed situation\\ [5pt]
		Direction & Normal to HVG plane closer than \\
		& observed to expectation of toy model \\
		\hline
 \end{tabular}
 \caption{Criteria used to judge whether a randomly generated population of galaxies is analogous to the observed HVG system. Only one of the first two criteria is used at a time. Note that the criteria are not all independent. For example, as the MW and M31 positions are fixed and my plane fitting procedure (Section \ref{Best_fitting_plane_finding}) always considers them, the thinnest planes are likely to be obtained when these galaxies are close to the plane best fitting the HVGs. This makes it more likely that the `barycentre offset' criterion is satisfied (compare corners of Table \ref{Probability_table_criteria_combinations}).}
 \label{Criteria_definitions}
\end{table}

\subsection{Results}
\label{Results}

Applying the criteria defined in Table \ref{Criteria_definitions} to 20 million MC trials based on my nominal sample of HVGs (Table \ref{Planar_backsplash_galaxies}), I obtained the results shown in Table \ref{Probability_table_criteria_combinations}. The uncertainties are found by repeating the MC trial using 4 different seeds for the random number generator, with each seed used for ${5 \times 10^6}$ trials. The variance between the results gave an indication of the uncertainty in the final result, which is a simple mean. I also estimated the error using binomial statistics. My final error estimate was based on whichever method gave a higher uncertainty (this was usually based on comparing different runs). In all cases, I determined the proportion of `successful' MC trials to within a few percent.

\begin{table}
 \begin{tabular}{cccc}
\hline
 & Thickness & Direction & Barycentre \\
 & & & offset \\ \hline
Thickness & $4.6 \pm 0.3$ &  & \\
Direction & $2.3 \pm 0.1$ & $417.4 \pm 0.5$ & \\
Barycentre offset & $2.4 \pm 0.2$ & $81.1 \pm 1.3$ & $181.8 \pm 2.5$ \\
\hline
 \end{tabular} 
 \caption{Monte Carlo trial-based probabilities in parts per thousand (\perthousand) of the HVG system (all galaxies in Table \ref{Planar_backsplash_galaxies}) matching various combinations of the criteria defined in Table \ref{Criteria_definitions}. These criteria are used to determine if a mock HVG system is analogous to the observed system, using the method outlined in Section \ref{Statistical_analysis_method}. When the same criterion appears in both the row and column headings, the result is the probability of matching that criterion alone, regardless of the others. The probability of all three criteria being met simultaneously is $1.48 \pm 0.10\perthousand$, which corresponds to the first row of Table \ref{Probability_table}.}
 \label{Probability_table_criteria_combinations}
\end{table}

The direction criterion was met in ${\ssim 417\perthousand}$ of the trials and was the least problematic criterion. This is due to the rather wide range of orientations allowed for the plane best fitting the mock galaxies.

The plane of HVGs is offset from the MW-M31 barycentre by 67 kpc, which is rather small considering the extent of the HVG plane ($\ssim 1$ Mpc). Thus, the `barycentre offset' criterion in Table \ref{Criteria_definitions} is only met around ${182\perthousand}$ of the time.

\begin{table}
 \begin{tabular}{lll}
	\hline
  Investigation & Sample & Probability (\perthousand)\\ 
   &  &   \\ 
  \hline
  Nominal (physical thickness) & All & $1.48 \pm 0.10$ \\
  $\widehat{\bm h}$ rotated $5^\circ$ south ($\theta = 75^\circ$) & All & $1.51 \pm 0.10$ \\
  Distances fixed & All & $1.45 \pm 0.01$ \\
  Nominal & \st{HIZSS 3} & $0.41 \pm 0.02$ \\
  Nominal & \st{Antlia} & $5.17 \pm 0.36$ \\
	Aspect ratio & All & $1.62 \pm 0.01$ \\
  Aspect ratio & \st{Antlia} & $5.35 \pm 0.02$ \\
  \hline
 \end{tabular} 
 \caption{How my results depend on various model assumptions. The final column shows the probability of a MC trial satisfying the criteria given in Table \ref{Criteria_definitions} based on randomising the directions towards the HVGs in Table \ref{Planar_backsplash_galaxies} but with a fixed M31 direction. Galaxies whose names have been crossed out are excluded from the sample in that particular investigation, with the nominal sample corresponding to the central column of Table \ref{Plane_parameters}. The exclusion of HIZSS 3 is achieved by altering Equation \ref{Probability_distribution_b} to impose the requirement that any observable galaxy be ${\geq 15^\circ}$ from the Galactic plane.}
 \label{Probability_table}
\end{table}

By far the most important criterion is the requirement that all but one mock HVG define a plane with rms thickness smaller than observed. This criterion is met in only ${5.2 \pm 0.2\perthousand}$ of the MC trials. Consequently, it is very unlikely (probability ${1.48 \pm 0.10\perthousand}$) that all three criteria are satisfied simultaneously. This remains a very unlikely situation despite various changes to my modelling assumptions and choice of sample, for example if I change the anisotropy statistic (Table \ref{Probability_table}). Neglecting distance uncertainties altogether has little impact on the results, suggesting that they should be robust to future improvements in distance measurements. The most significant change occurs if Antlia is removed from my HVG sample as it could be a satellite of NGC 3109 \citep{Van_den_Bergh_1999}. However, even this case yields a very low probability of 5.4\perthousand.

\begin{figure}
	\centering 
		\includegraphics [width = 8.5cm] {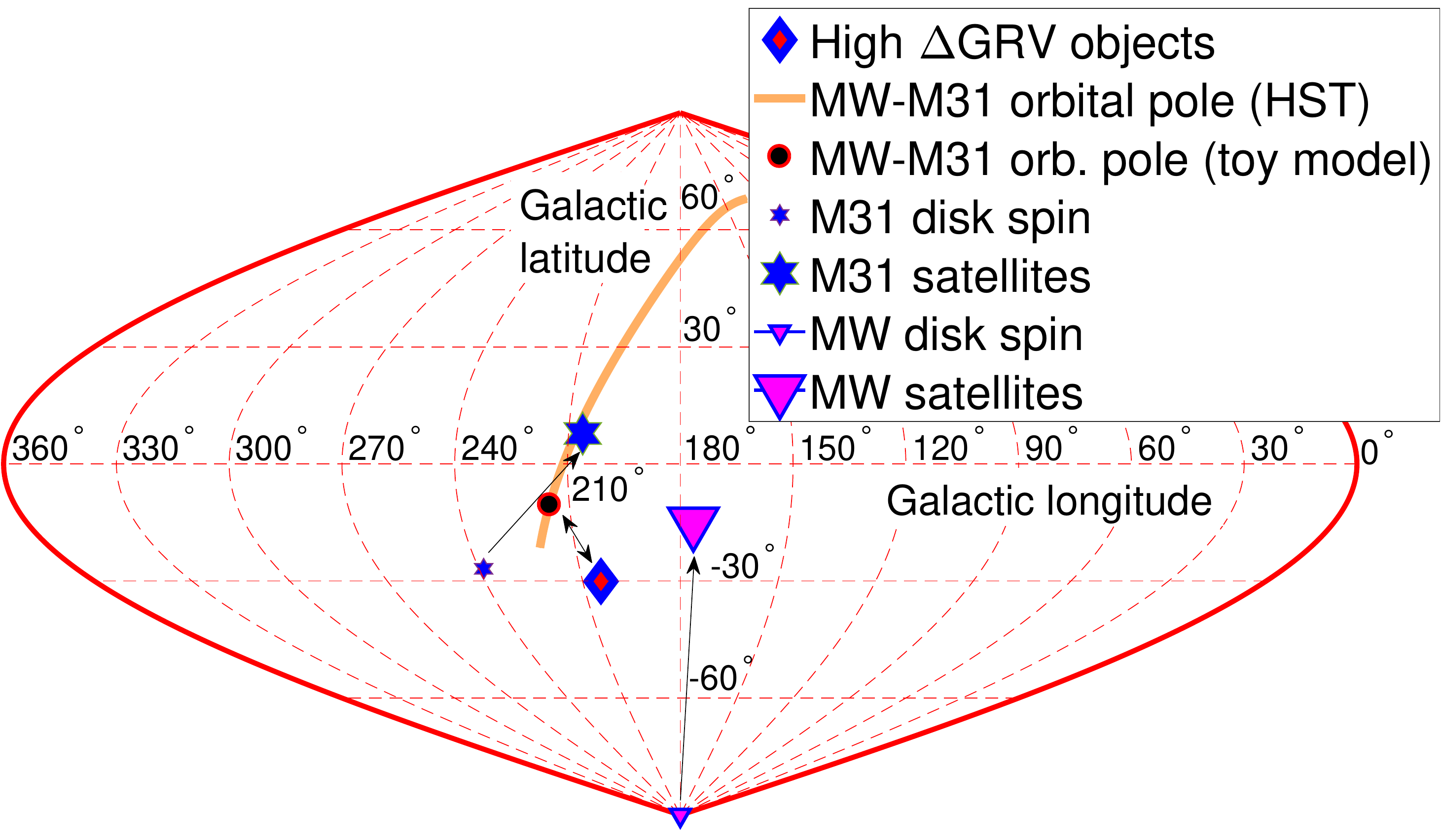}
	\caption{Normal directions to the important LG planes, shown in Galactic co-ordinates. Assuming a past close MW-M31 flyby, I expect tidal torque from M31 to explain the misalignment between the orientation of the MW disk (small triangle) and its plane of satellites (large triangle). The effect of such torques is illustrated with an upward arrow. Tidal torque from the MW explains a similar misalignment for M31 (hexagrams used instead of triangles). My MOND-based toy model is able to reproduce these orientations fairly well if the MW-M31 orbital pole lies in the direction of the black dot with red rim \citep[][section 2.2]{BANIK_2017_ANISOTROPY}. This is reasonably consistent with the normal to the plane defined by the HVGs (diamond), though its sense of rotation is unknown. The proper motion of M31 has recently been measured \citep{Van_der_Marel_2012}, suggesting a particular MW-M31 orbital pole (1$\sigma$ allowed region shown as orange line). This must be orthogonal to the present direction towards M31. Unfortunately, at 2$\sigma$, any direction consistent with this requirement is allowed.}
	\label{Directions_plotting_3D}
\end{figure}

Figure \ref{Directions_plotting_3D} shows how the HVG plane I found fits into other LG structures. If the HVG plane can be identified with the MW-M31 orbital plane, then this constrains models of a past MW-M31 flyby attempting to match the observed orientations of their satellite planes \citep[e.g.][]{Bilek_2017}.

To conclude this section, the LG has several galaxies with unusually high radial velocities in the context of $\Lambda$CDM. The spatial distribution of these HVGs is inconsistent with isotropy because they lie rather close to a well-defined plane. Thus, the spatial arrangement of the HVGs is similar to what would be expected in the MOND scenario where their unusual kinematics arose due to past gravitational slingshot interactions with the MW/M31 around the time of the MW-M31 flyby. In this scenario, the anisotropy arose because the LG dwarfs flung out at the highest speeds were those flung out nearly parallel to the motion of the perturber. If this is correct, the MW-M31 line should lie within the HVG plane. In reality, it is only ${16^\circ}$ off the plane, a possible consequence of tides raised by large scale structure and other effects not included in my simplified MOND model of the LG.

\section{The escape velocity curve of the Milky Way in Modified Newtonian Dynamics \citep{BANIK_2017_ESCAPE}}
\label{v_esc_article}

So far, I have used galaxies merely as tracers for a timing argument analysis or as light sources which are gravitationally lensed. This is because there is already an extensive literature on testing MOND with forces internal to galaxies using their rotation curves \citep[e.g.][]{Swaters_2009, Famaey_McGaugh_2012, Papastergis_2016}. It is difficult to analyse the forces within external galaxies much beyond measuring their radial acceleration profile in this way. However, additional measurements are possible for the MW because individual stars are resolved more easily, meaning their proper motions are often available \citep[e.g.][]{Zacharias_2017} in addition to their radial velocities \citep[e.g.][]{Kunder_2017}.

In this section, I focus on the Galactic escape velocity curve. Although MOND is often assumed to imply a $r^{-1}$ force law towards a point mass (Equation \ref{Deep_MOND_limit}), this is only true if the mass is isolated. Even if more distant masses impose a constant external gravitational field on a system, the non-linearity of MOND implies that this affects the internal forces within the system. This external field effect (EFE) arises because the theory is acceleration-dependent \citep[][section 2g]{Milgrom_1986}. Ultimately, the EFE and the inherent non-linearity of MOND are required by data indicating that the force towards a galaxy of mass $M$ scales more nearly as $\sqrt{M}$ rather than linearly with $M$ (Figure \ref{BTFR_diagram}).

To understand the EFE, consider a dwarf galaxy governed by MOND which has very low internal accelerations (${\ll a_{_0}}$) but is freely falling in the strong gravity (${\gg a_{_0}}$) of a distant massive galaxy. The overall acceleration at any point in the dwarf is rather high due to the dominant external field (EF) of the massive galaxy. Thus, the dwarf would obey Newtonian dynamics and forces in its vicinity would follow the usual inverse square law rather than Equation \ref{Deep_MOND_limit}. This is true even if the EF is uniform across the dwarf i.e. there are no tidal effects. However, without the massive galaxy, the dwarf's internal dynamics would be very non-Newtonian.

Realising that MOND with the EFE predicts potential wells of finite depth, \citet{Famaey_2007} used an analytic method to estimate the Galactic escape velocity $v_{esc}$ from the Solar neighbourhood. Similar results were later obtained by \citet{Wu_2008} using a numerical solution to AQUAL. Their estimated $v_{esc}$ agrees reasonably well with later measurements based on high-velocity MW stars \citep{Piffl_2014}. Recently, a similar technique was used to measure $v_{esc}$ over a wide range of Galactocentric radii \citep[8$-$50 kpc,][]{Williams_2017}. This work applied the method of \citet{Leonard_1990} to a variety of tracers detected in the ninth data release of the Sloan Digital Sky Survey \citep[SDSS,][]{Ahn_2012}. This section focuses on calculating the expected $v_{esc}$ in MOND at these positions for a range of plausible Galactic mass models, bearing in mind constraints from the Galactic rotation curve \citep{McGaugh_2016_MW}.

\subsection{Method}

As in Section \ref{MOND_governing_equations}, the gravitational field of the MW is determined using Equation \ref{g_direct_sum}. The main complication is in calculating $\bm{g}_{_N}$ accurately because the MW can no longer be treated as a point mass. The EF also has to be included. To avoid a total breakdown of symmetry in the problem, I assume the EF is aligned with the symmetry axis of the MW disk. The true EF direction may well be different, but this is expected to have only a very small effect on the results \citep[][section 4.3]{BANIK_2017_ESCAPE}.

The MW is assumed to consist of a hot gas corona surrounding two aligned and concentric infinitely thin exponential disks representing its gas and stellar components. Taking advantage of the fact that potentials superpose in Newtonian gravity, I simply add the potential of the corona to that of the other components. The corona is treated as a Plummer model \citep{Plummer_1911} with mass $M_{cor}$ and core radius $r_{_{cor}}$, yielding a corona potential at a Galactocentric distance $r$ of
\begin{eqnarray}
	\Phi_{cor} ~=~ -~\frac{GM_{cor}}{\sqrt{r^2 + {r_{_{cor}}}^2}}
	\label{Corona_potential}
\end{eqnarray}

For the disk components, the superposition principle means that it is only necessary to solve for a single exponential disk. I take this to have unit scale length and $GM$, scaling it up to the required values later. To determine the Newtonian potential $\Phi_N$ of this mass distribution $\rho_{_b} \left( \bm{r} \right)$, I numerically solve the Poisson equation
\begin{eqnarray}
	\label{Poisson_equation}
	\nabla^2 \Phi_N ~=~ 4\pi G \rho_{_b}
\end{eqnarray}

This is done using successive over-relaxation in spherical polar co-ordinates (polar angle $\theta$). Further details are provided in appendix A of \citet{BANIK_2017_ESCAPE}, which explains the discretisation scheme and convergence criteria.

To include the EF, I add the contribution from the Newtonian EF $\bm{g}_{_{N,ext}}$. This is what the EF would have been in Newtonian gravity. I assume the spherically symmetric MOND relation between it and the actual EF $\bm{g}_{ext}$.
\begin{eqnarray}
	\label{External_field_increment}
	\Phi_N ~\to~ \Phi_N ~-~ \bm{r} \cdot \bm{g}_{_{N,ext}} &~&\text{where} \\
	\overbrace{\nu \left(\frac{{g}_{_{N,ext}}}{a_{_0}} \right)}^{\nu_{_{ext}}} \bm{g}_{_{N,ext}} ~&=&~ \bm{g}_{ext}
\end{eqnarray}

\begin{table}
 \begin{tabular}{lll}
	\hline
  Variable & Meaning & Value \\ 
  \hline
  $R_\odot$ & Galactocentric distance of Sun & 8.2 kpc\\
  $r_*$ & Stellar disk scale length & 2.15 kpc\\
  $M_{*,0}$ & Nominal stellar disk mass & $5.51 \times 10^{10} M_\odot$\\
  $r_{_g}$ & Gas disk scale length & 7 kpc\\
  $M_{g,0}$ & Nominal gas disk mass & $1.18 \times 10^{10} M_\odot$\\ \hline
	$\frac{M_*}{M_{*,0}}$ & Disk mass scaling factor & $0.8-1.4$\\ 
  $r_{_{cor}}$ & Plummer radius of corona & $\left(20-60 \right)$ kpc\\
  $M_{_{cor}}$ & Corona mass & $\left(2 - 8 \right)\times 10^{10} M_\odot$\\
  $g_{_{ext}}$ & External field on MW & $\left(0.01 - 0.03\right)a_{_0}$\\
  \hline
 \end{tabular} 
 \caption{Parameters of the MW mass distribution, with $_0$ subscripts indicating nominal values while $_*$ and $_g$ subscripts refer to its stellar and gas components, respectively. I always use the same value of $\frac{M_g}{M_*}$ and the same disk scale lengths, but vary the other parameters using a grid search. The first part of the table contains the fixed parameters $R_\odot$ \citep{McMillan_2017}, $M_{*,0}$ \citep{McGaugh_2016_MW}, $r_*$ \citep{Bovy_2013}, $r_{_g}$ \citep{McMillan_2017} and $M_{g,0}$, which is based on applying the method described in \citet[][section 3.3]{McGaugh_2008} to the observations of \citet[][table D1]{Olling_Merrifield_2001}. The Galactic hot gas corona is modelled using Equation \ref{Corona_potential} \citep{Plummer_1911}.}
 \label{Parameters}
\end{table}

Once all component potentials have been appropriately scaled according to the parameters in Table \ref{Parameters}, it is easy to add them and thereby determine $\bm{g}_{_N} = -\nabla \Phi_N$. As in Section \ref{MOND_governing_equations}, I then determine $\nabla \cdot \bm{g}$ using Equation \ref{Simple_interpolating_function} and apply direct summation (Equation \ref{g_direct_sum}) to obtain $\bm{g}$ at positions of interest. Some small corrections are then applied for edge effects \citep[][section 2.2]{BANIK_2017_ANISOTROPY} using analytic approximations that are asymptotically correct \citep{Banik_2015_analytic}. I found the escape velocity $v_{esc}$ by integrating the radial component of $\bm{g}$ along a radial transect out to infinity. The nominal MW model used is designed to be consistent with its observed rotation curve in a MOND context \citep[][table 1 model Q4ZB]{McGaugh_2016_MW}, though I also consider several hundred other models.

\subsection{Results}

\begin{figure}
	\centering 
	\includegraphics [width = 8.5cm] {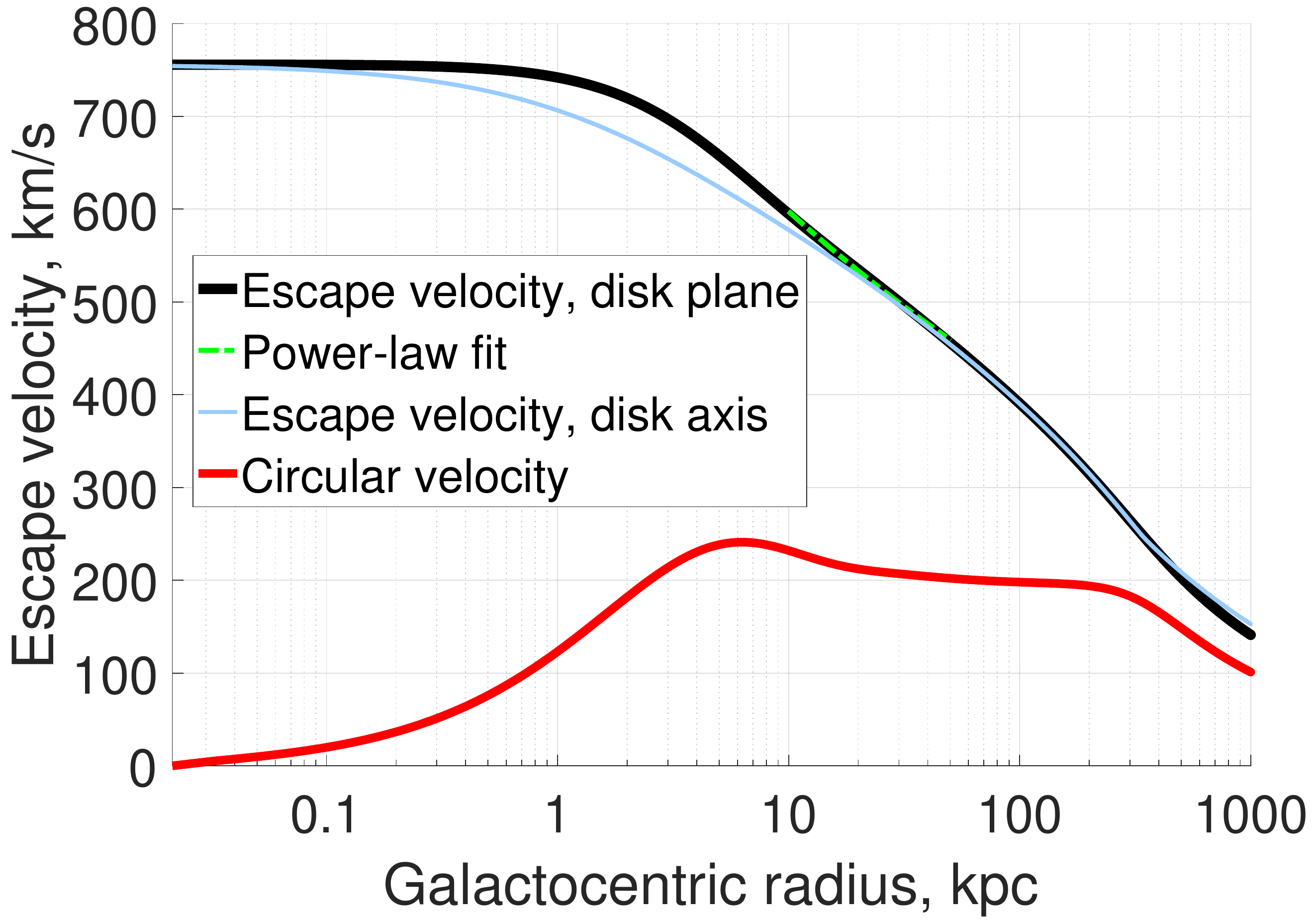}
	\caption{How the circular velocity of the MW (lower red curve) and its escape velocity (upper black curve) depend on position within its disk plane. The latter can be parametrised rather well as a power law (Equation \ref{Power_law_fit}) over the radial range $10-50$ kpc (dashed green curve). At the same distance from the MW, its escape velocity is lower along its disk axis (thin blue curve) for points close to the MW due to the effect of its disk. However, this pattern is reversed at long range because the EF on the MW is assumed to align with its disk axis, deepening the potential in this direction \citep[][equation 37]{Banik_2015_analytic}. The model shown here uses the nominal disk masses in Table \ref{Parameters} and $g_{_{ext}} = 0.03a_{_0}$, with the corona being as small and low-mass as possible.}
	\label{v_esc_profile}
\end{figure}

I begin by showing the circular and escape velocity curves of the MW with my nominal values for the MW stellar and gas disk masses (Figure \ref{v_esc_profile}). For comparison with observations, I fit a power-law model to $v_{esc}$ over the radial range 10$-$50 kpc. This assumes that
\begin{eqnarray}
	v_{esc} \left( r \right) ~\propto~ r^{-\alpha}
	\label{Power_law_fit}
\end{eqnarray}

Power-law fits become linear when considering the logarithms of both variables. Thus, if $y \equiv Ln~v_{esc}$ and $x \equiv Ln~r$ are lists of size $N$, then
\begin{eqnarray}
	\alpha ~&=&~ -\frac{\sum_{i = 1}^N \tilde{x}_{_i} \tilde{y}_{_i} }{\sum_{i = 1}^N \tilde{x}_{_i} \tilde{x}_{_i}} ~~\text{ where}\\
    \tilde{x} ~&\equiv&~ x - \frac{1}{N}\sum_{i = 1}^N x_{_i} ~~~\left(\tilde{y}~\text{defined analogously} \right)
\end{eqnarray}

\begin{figure}
	\includegraphics [width = 12.5cm] {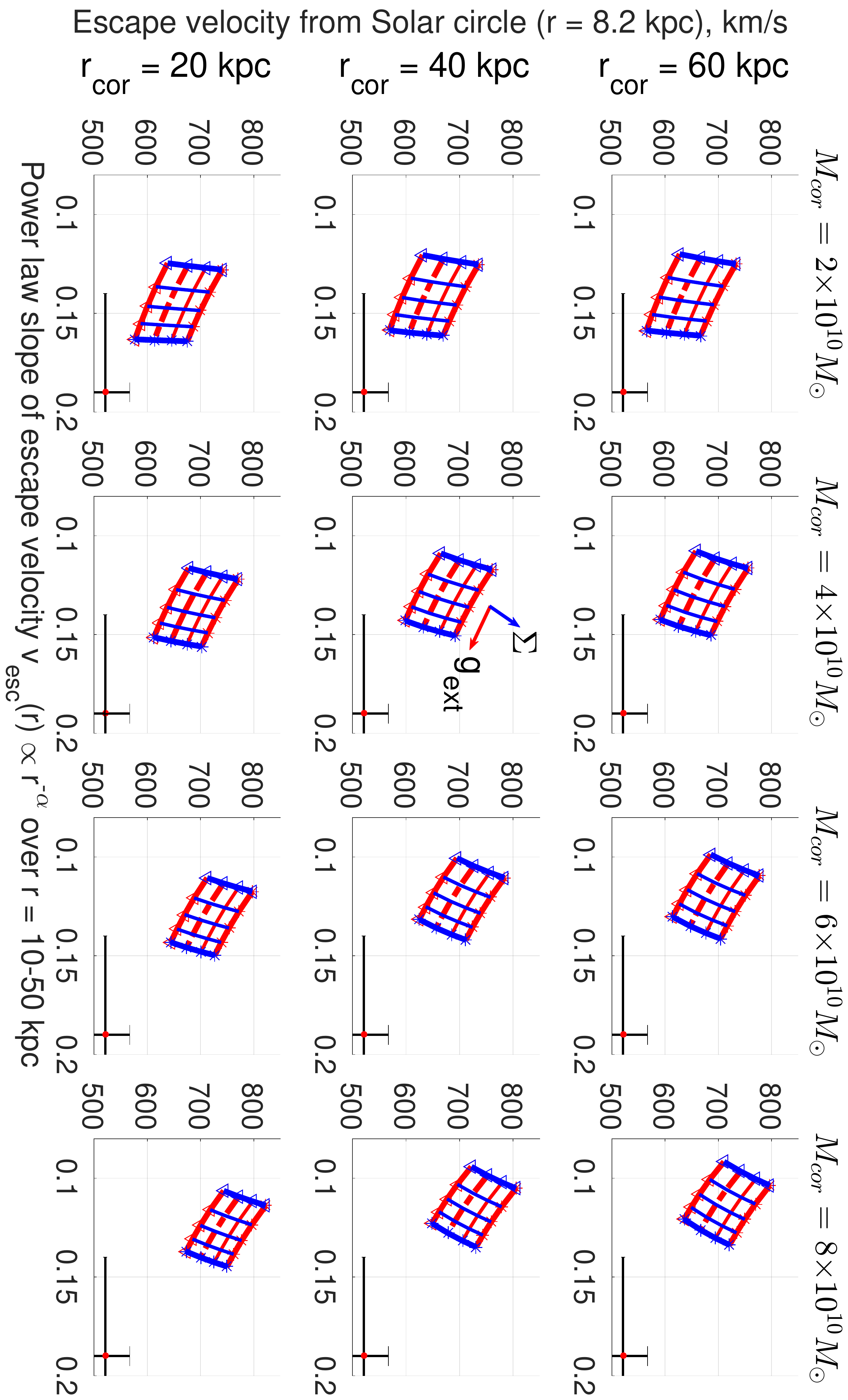}
	\caption{Escape velocity $v_{esc}$ from within the MW disk plane as a function of model parameters. The $x$-axis shows the value of $\alpha$ such that $v_{esc} \left( r \right) ~\propto~ r^{-\alpha}$ while the $y$-axis shows $v_{esc}$ near the Sun. The measured values of these quantities are shown as a red dot with black error bars towards the bottom right \citep{Williams_2017}. Each subplot has a fixed corona mass and scale length, with red tracks showing the effect of varying $g_{_{ext}}$ with constant disk mass (vice versa for blue tracks). In each case, an inverted triangle is used to show the result when the parameter being varied has the lowest value considered while a star is used for the largest value. This is also shown by the arrows in the central subplot, which point towards higher values of the indicated parameter. I consider disk masses scaled from the nominal value by factors given in Table \ref{Parameters}, where I also show the range in $g_{_{ext}}$ that I try (values of all parameters are spaced linearly). The dashed red lines show the results for the nominal stellar and gas disk masses, which is required to obtain the correct $v_{c, \odot}$ \citep[][table 1 model Q4ZB]{McGaugh_2016_MW}. I assume $\bm{g}_{_{ext}}$ is aligned with the disk symmetry axis. Rotate $90^\circ$ anti-clockwise for viewing.}
	\label{v_esc_disk_results}
\end{figure}

$v_{esc} \left( r \right)$ can be described rather well as a power law over the range $r = 10-50$ kpc (Figure \ref{v_esc_profile}). For a range of models, I determine the best-fit slope $\alpha$ and Solar circle normalisation for comparison with observations. It is unclear exactly which Galactic polar angles $\theta$ the observations of \citet{Williams_2017} correspond to, but most likely a range of angles is used in order to get enough of the relatively rare high-velocity stars that are necessary for an escape velocity determination. I show results within the disk plane ($\theta = \frac{\pi}{2}$) in Figure \ref{v_esc_disk_results}.

$v_{c, \odot}$ depends mainly on the disk surface density such that only the nominal value is able to correctly reproduce the observed LSR speed of $v_{c, \odot} \approx 235$ km/s \citep{McMillan_2017}. However, I also consider the effect of scaling the surface density by factors of 0.8$-$1.4. Raising this factor by 0.1 increases $v_{c, \odot}$ by ${\sim 8}$ km/s.

Within the range considered, adjusting $M_{_{cor}}$ affects $v_{c, \odot}$ by $\la 5$ km/s while adjusting $r_{_{cor}}$ has a smaller effect of ${\ssim 1}$ km/s. At the Solar circle, the MW is effectively isolated $-$ adjusting $g_{_{ext}}$ only affects $v_{c, \odot}$ by $\ssim 4$ m/s. These factors are more significant further from the MW, but the scarcity of tracers makes it difficult to directly measure $\bm{g}$ there. Fortunately, forces at large $r$ affect the escape velocity $v_{esc} = \sqrt{-2\Phi}$ near the Sun. A local $v_{esc}$ measurement could thus constrain the MW gravitational field at large distances, with the appropriate analysis. One of this section's objectives is to do just that, in a MOND context.

At the same $r$, escape velocities are slightly larger within the MW disk plane as the MW matter distribution is concentrated towards this plane. Within ${\ssim 100}$ kpc of the MW, this near-field effect is more important than the non-sphericity of the MW potential in the far-field EF-dominated region, where the MW exerts very little gravity in any case. However, beyond ${\ssim 100}$ kpc, the latter effect dominates because the MW can be considered as a point mass (compare black and blue curves in Figure \ref{v_esc_profile}). As shown by \citet{Banik_2015_analytic}, this leads to a deeper potential in the direction of $\bm{g}_{_{ext}}$ i.e. along the disk axis in my axisymmetric models.

An important constraint on the true Galactic $v_{esc} \left( r \right)$ curve arises because it is determined by the same potential $\Phi$ that governs the rotation curve $v_c \left( r \right)$.
\begin{eqnarray}
	\frac{\partial \left( \frac{1}{2} {v_{esc}}^2 \right)}{\partial r} ~&=& -\frac{\partial \Phi}{\partial r} ~=~ -\frac{{v_{c}}^2}{r} \\
	\alpha ~\equiv~ -\frac{\partial Ln~v_{esc}}{\partial Ln~r} ~&=&~ \left( \frac{v_{c}}{v_{esc}}\right)^2
\end{eqnarray}
\onecolumn
\twocolumn


If $v_{c, \odot} = 232.8$ km/s \citep{McMillan_2017}, then $\alpha = 0.200$ for a local escape velocity of 521 km/s. This is entirely consistent with the observed value of $0.19 \pm 0.05$. It is clear that my calculated escape velocities are towards the upper end of the range allowed by observations. Thus, my analysis disfavours a hot gas corona. I have included one because XMM-Newton \citep{Jansen_2001} observations at a range of Galactic latitudes indicate that one is present \citep{Nicastro_2016}. Their best-fitting model suggested that its mass is $2 \times 10^{10} M_\odot$ (see their table 2 model A) which is therefore the lowest value for $M_{_{cor}}$ that I consider. Similarly to my analysis, the best fit to their observations was obtained for the lowest mass corona model they tried out, though substantially more massive halos are far from ruled out.

A hot gas corona would also cause ram pressure stripping effects on MW satellites containing gas. This is thought to explain the asymmetry of the Magellanic Stream \citep{Hammer_2015} and perhaps also the truncation of the LMC gas disk at a much shorter distance than the extent of its stellar disk \citep{Salem_2015}. Those authors used this argument to estimate that $M_{_{cor}} = 2.7 \pm 1.4 \times 10^{10} M_\odot$, consistent with other estimates.

\begin{figure}
	\includegraphics [width = 8.5cm] {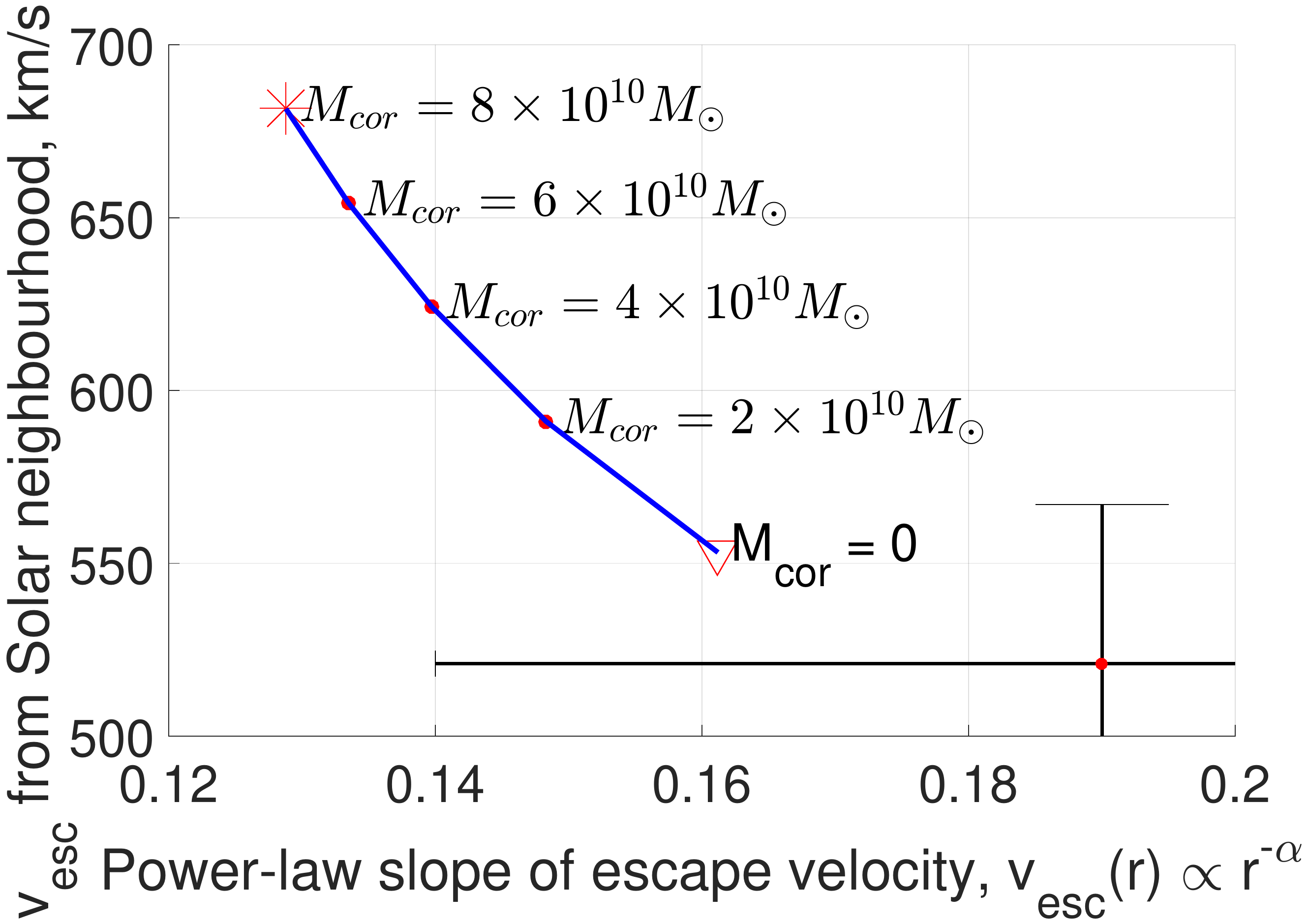}
	\caption{Effect of the MW corona mass $M_{_{cor}}$ on its escape velocity curve for points along its disk axis in the direction of the external field. Other model parameters are the same as in Figure \ref{v_esc_profile}. The $x$-axis shows the value of $\alpha$ such that $v_{esc} \left( r \right) ~\propto~ r^{-\alpha}$ while the $y$-axis shows $v_{esc}$ near the Sun. The measured values of these quantities are shown as a red dot with black error bars towards the bottom right \citep{Williams_2017}.}
	\label{v_esc_results_top_axis_no_corona}
\end{figure}

Although my analysis is consistent with this, it prefers an even lower $M_{_{cor}}$. I therefore considered lowering $M_{_{cor}}$ all the way down to 0. As expected, this makes the MW $v_{esc}$ curve slightly more consistent with observations in terms of both its amplitude and its radial gradient (Figure \ref{v_esc_results_top_axis_no_corona}).

The Galactic escape velocity curve \citep{Williams_2017} is consistent with expectations in MOND based on a MW mass model that also explains its rotation curve \citep[][table 1 model Q4ZB]{McGaugh_2016_MW}. A fairly low mass corona is preferred, consistent with independent measurements \citep[][table 2 model A]{Nicastro_2016}. It is presently difficult to use the Galactic $v_{esc}$ curve to meaningfully constrain how extended its corona is. Within the range considered, my analysis prefers a strong EF with $g_{_{ext}} = 0.03a_{_0}$, a value also assumed by \citet{Famaey_2007}.

\section{Future prospects}
\label{Future_prospects}

MOND represents a significant departure from Newtonian dynamics. This should allow definitive tests in the near future. In this section, I discuss some possible ways of distinguishing the theories at a variety of astrophysical scales.

\subsection{Beyond the Local Group}

In Section \ref{Bullet_Cluster}, I discussed how measurements of the Moving Cluster Effect could help determine the collision velocity of interacting galaxy clusters like the Bullet Cluster and El Gordo \citep{Menanteau_2012}. Much faster collision velocities are expected in MOND \citep{Candlish_2016_velocity}. Confirmation of the high estimated collision velocity in the Bullet Cluster and discovery of even a small number of other similar systems could severely challenge $\Lambda$CDM \citep{Kraljic_2015}.

In this paradigm, close interactions between galaxies rapidly end in a merger \citep{Privon_2013} due to dynamical friction \citep{Chandrasekhar_1943} between overlapping DM halos. Without these halos, merger rates are expected to be much lower. Unfortunately, it is difficult to test this directly because there is a degeneracy between the frequency and visible duration of galactic interactions. Merger rates could be constrained through gravitational waves (GWs) emitted from merging supermassive black holes (SMBHs). Null detections using pulsar timing arrays are in some tension with the expected frequency of such events \citep{Shannon_2015}.

The expected GW background is not expected to differ much depending on precisely how the progenitor black holes eventually merge, a question often called the `final parsec problem'. Either this occurs through dynamical friction against stars and gas, or this process becomes inefficient near the final parsec. In this case, SMBHs from several progenitor galaxies would orbit within a rather small region such that dynamical friction would effectively arise against this population of objects $-$ SMBHs would interact with each other \citep{Ryu_2018}. Moreover, those authors showed that a longer merging timescale would mean mergers typically occurred later and thus closer to the Earth, making for a stronger GW here. This nearly cancels the effect of fewer mergers occurring altogether (see their section 4.1.1). In any case, the GW background due to merging SMBHs ought to become detectable in the near future \citep{Wang_2017}, thereby constraining cosmological models. Indeed, existing measurements are already placing interesting constraints \citep{NANOGrav_2018}.

As well as detecting these GWs directly, the momentum they carry could be detected indirectly because it ought to cause the remnant SMBH to recoil. Given that the first GWs to be detected carried off $\ssim 5\%$ of the progenitor's rest mass as energy \citep{LIGO_2016}, the recoil could be significant if the GW emission is even slightly asymmetric. This would cause the SMBH to oscillate in the potential of its host galaxy, leading to an offset between the photometric centres of nearby elliptical galaxies and the positions of their central SMBH (often identifiable as an active galactic nucleus). \citet{Lena_2014} searched for these offsets but found only small ($\la 10$ pc) offsets. As frequent mergers are expected in $\Lambda$CDM, larger offsets should have been detected in some of the 14 cases considered. Moreover, even the small detected offsets were often aligned with the jet created by accretion onto the SMBH, suggesting that hemispherical asymmetries in its power are responsible for the observed offset (see their section 5.3).

Extending the analysis to more galaxies should give a better statistical understanding of whether significant SMBH-host galaxy offsets are common. More detailed modelling is required to understand what this implies about the merger rate of galaxies. If it is very difficult for the progenitor SMBHs to inspiral sufficiently for GW emission to become significant, then there should be galaxies with multiple `stalled' SMBHs near their centre that could ultimately be revealed through detailed multi-epoch kinematic measurements \citep[e.g.][]{Wang_2016}. If instead the binary SMBH orbit ought to rapidly decay, then it will be important to understand whether the eventual merger is likely to cause asymmetric emission of GWs and thus a detectable recoil. Without a convincing explanation for why the GWs ought to be symmetric, continued null detection of large SMBH-galaxy offsets would strongly suggest a low major merger rate (depending on how efficiently the oscillations are expected to be damped).

Reduced dynamical friction between galaxies would allow close interactions at a much higher relative velocity. This could be tested by searching for galaxies with a high pairwise relative velocity but a small separation, such as might be the case for NGC 1400 and NGC 1407 \citep{Tully_2013, Tully_2015} and perhaps also for NGC 6050 and IC 1179. A small physical separation might be discernible from tidal features connecting the galaxies. This could also help prove that they were observed past pericentre, when dynamical friction between their DM halos should have slowed them down.

As well as the dynamics of the interacting galaxies themselves, an important issue is the properties of any tidal dwarf galaxies (TDGs) that form out of the encounter. As discussed in Section \ref{Introduction_satellite_planes}, it is critical to understand whether the acceleration discrepancy persists in such systems as it should not in $\Lambda$CDM \citep{Wetzstein_2007}. This is precisely what was investigated by \citet{Gentile_2007} based on observations of the NGC 5291 system \citep{Bournaud_2007}. However, it was later realised that these TDGs formed rather recently, leaving them insufficient time to settle into dynamical equilibrium \citep{Flores_2016}. Even so, detailed observations of much older TDGs remain a promising way to understand how the acceleration discrepancies arise. One possible target is the ${\ssim 4}$ Gyr old TDG identified by \citet{Duc_2014}, which may well have settled into dynamical equilibrium by now. Another very promising set of targets are the members of the recently discovered satellite plane around Cen A \citep{Muller_2018}.

\subsection{Within the Local Group}

Although some systems outside the LG would behave very differently in MOND and $\Lambda$CDM, the large distance to these systems makes it difficult to tell whether this is actually the case. This is why most of my thesis has focused on the LG, even though any signatures of MOND are likely to be more subtle.

One obvious way to test MOND within the LG is to simulate the MW-M31 flyby in more detail using $N$-body and (eventually) hydrodynamic models that incorporate MOND gravity. This is feasible using the Phantom of RAMSES algorithm \citep{PoR}, an adaptation of the RAMSES algorithm widely used by astronomers \citep{Teyssier_2002}. Precisely this sort of simulation has recently been done \citep{Bilek_2017}. It is important to search the parameter space more thoroughly to see if some model can explain the observed orientations of the MW and M31 satellite planes as arising from tidal debris expelled during their interaction.

I recently conducted an investigation along these lines with the help of a summer student I hired \citep{Banik_2018}. Treating the MW and M31 as point masses, I considered a disk of test particles around each one and advanced their trajectories using a very similar method to that in Section \ref{MOND_simulation_LG}. I tried out a range of MW-M31 orbital poles consistent with the small observed proper motion of M31 \citep{Van_der_Marel_2012}. Unfortunately, this does not reliably constrain the orbital pole much beyond the fact that it must be orthogonal to the present direction in which we observe M31. Thus, I tried the full range of allowed directions and a range of MW-M31 tangential speeds, leading to a range of closest approach distances.

\begin{figure}
	\centering
		\includegraphics [width = 8.5cm] {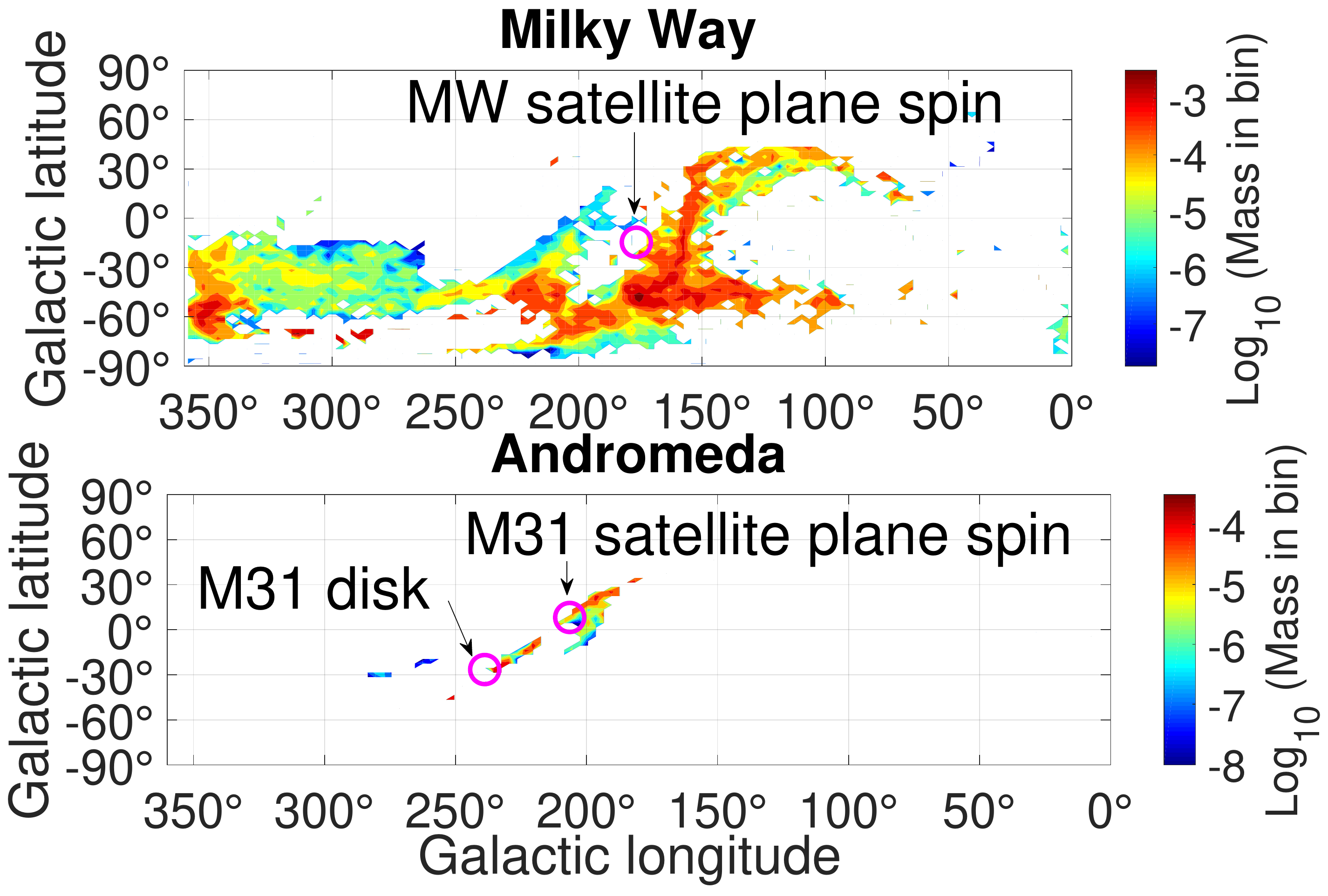}		
\caption{The distribution of orbital angular momentum directions (spin vectors) for tidal debris around the MW and M31 disks at the end of my best-fitting restricted $N$-body simulation of a past MW-M31 flyby \citep{Banik_2018}. The mass units are arbitrary. \emph{Top}: Results for the MW. Its disk spin vector points at the South Galactic Pole while the open pink circle shows the spin vector of its satellite plane. \emph{Bottom}: Results for M31. I use open pink circles to show the observed spin vectors of its disk (lower left) and satellite plane (upper right).}
	\label{Orbital_pole_histogram}
\end{figure}

In each model, I looked at the distribution of tidal debris outside the disk plane of each galaxy but within 250 kpc. The orbital poles of particles in this `satellite region' are shown in Figure \ref{Orbital_pole_histogram}, with each particle statistically weighted according to the mass it represents within the disk it originated in. Around each galaxy, the orbital poles show a clear clustering, as happens in most models. In the particular one shown, the preferred orbital poles align fairly well with the observed orientations of the MW and M31 satellite planes. Around the MW, the model also gets some material on orbits that are roughly counter-rotating with respect to the preferred rotation direction. This is interesting as the MW satellite Sculptor is counter-rotating within the MW satellite plane \citep{Sohn_2017}.

My model does not get counter-rotators around M31, a prediction that could be tested with proper motions of its satellites within its satellite plane. In particular, the radial velocities of And XIII and And XXVII suggest that they may be counter-rotators. However, they could just be interlopers whose orbits take them far from the satellite plane. This seems quite feasible given that only about half of the M31 satellites lie within its satellite plane structure \citep{Ibata_2013}.

As discussed further in \citet{Banik_2018}, the model also gets a reasonable radial distribution of tidal debris and thickens the MW disk by a similar amount to its observed thickness. The LG mass needed to satisfy the timing argument (Equation \ref{Hubble_flow_initial}) is similar to what the MW and M31 rotation curves imply, without assuming significant mass loss during their flyby. This occurs roughly when observations indicate the MW thick disk formed \citep{Quillen_2001}. However, the MW-M31 orbital pole in the model is nearly ${60^\circ}$ from the HVG plane identified in Section \ref{HVG_plane}. This may be because the HVGs identified there are mostly from a single bound association that has now disrupted after a close passage with the MW or M31. Such a common origin would naturally explain the filamentary nature of the NGC 3109 association \citep{Bellazzini_2013}. Thus, the orientation of the HVG plane might not be securely determined yet.

In addition to a past MW-M31 flyby, MOND also has more subtle consequences in the LG. One very interesting example is the EFE $-$ the internal dynamics of a system should be affected by the constant EF in which it is embedded, even in the absence of tidal effects \citep[e.g.][]{Banik_2015_analytic}. This violates the strong equivalence principle. Perhaps the most accurate current test of the EFE is the internal velocity dispersion $\sigma$ of the recently discovered MW satellite Crater 2 \citep{Torrealba_2016}. Without the EFE, $\sigma$ should have been ${\ssim 4}$ km/s in MOND but including the EFE (which is natural to MOND) reduces this to 2.1 km/s \citep{McGaugh_2016}, mainly because its internal accelerations are reduced by its rather large half-light radius of $1066 \pm 84$ pc \citep{Torrealba_2016}. $\sigma$ was later observed to be $2.7 \pm 0.3$ km/s \citep{Caldwell_2017}, a major topic of discussion at the Cleveland debate between MOND and $\Lambda$CDM.

In $\Lambda$CDM, large satellites like this ought to probe a significant part of their DM halo. This makes it difficult to argue that the visible extent of Crater 2 only probes the rising part of its rotation curve. As the baryon fraction needs to be very low in low mass DM halos to explain their internal dynamics, they must retain only a very small fraction of their baryons. However, if an object with such a low $\sigma$ can retain visible baryons at all, then there ought to be many more satellites with slightly higher $\sigma$. This would worsen the missing satellites problem whereby the MW satellite mass function does not match the distribution of DM subhalo masses expected in $\Lambda$CDM \citep{Klypin_1999}.

This issue could be resolved if the DM halo of Crater 2 was tidally stripped during close passage(s) with the MW \citep{Fattahi_2018}. This is possible if Crater 2 is on a very eccentric Galactic orbit. Otherwise, it might be difficult for $\Lambda$CDM to explain its very low internal velocity dispersion for its size while also remaining consistent with statistical properties of MW satellites. The orbital history of Crater 2 should become much clearer once its proper motion is known, making this an important test of the $\Lambda$CDM paradigm. Thus, it is fortunate that \citet{Sohn_2016} proposed taking this measurement.

\subsubsection{Within the Milky Way}

Several detailed tests of MOND should become possible with MW data collected by the GAIA mission \citep{Perryman_2001}. One of these is based on the vertical force towards the MW disk at a range of Galactocentric radii \citep{Bienayme_2009}. MOND predicts that the vertical force is boosted by the local factor of $\nu$, which is radius-dependent. Thus, the Newtonian dynamical disk surface density should decline outwards in a different way to that of the visible MW baryons. Another test is based on accurately measuring the shape of the stellar velocity dispersion tensor several kpc from the Galactic disk plane, possibly near the Solar Circle.

Due to an effect similar to the MCE (Section \ref{Bullet_Cluster}), some constraints can be placed based on the plethora of DM substructure that ought to arise in $\Lambda$CDM. The motion of a DM mini-halo between us and a pulsar would occasionally cause the observed period of the pulsar to decrease if such halos were sufficiently common. As pulsar periods generally increase, this has allowed interesting constraints to be placed on low-mass DM halos \citep{Clark_2016}. Future pulsar timing observations could improve these constraints further, regardless of assumptions concerning whether DM undergoes self-annihilation. The precise particle nature of the DM is still somewhat relevant because if it is of a sufficiently low mass, then it would not form very low mass halos \citep{Viel_2013}. However, such warm DM scenarios are strongly constrained by Lyman-$\alpha$ forest data which reveals plenty of very low mass gas clouds that likely delineate low mass DM halos \citep{Irsic_2017}.

MOND is an acceleration-dependent modification to gravity, so the transition from Newtonian to modified dynamics can arise at a much smaller length scale than the sizes of galaxies whose rotation curves originally motivated the theory. This is possible if one focuses on much lower mass systems. One such situation that has recently attracted some attention is wide binary stars \citep{Hernandez_2012}. This is based on the MOND radius of the Sun being only 7000 AU (Equation \ref{Deep_MOND_limit}), so two Sun-like stars separated by this distance or more should rotate around each other faster than expected in Newtonian gravity. In general, the actual rotation speed could exceed the Newtonian expectation by an arbitrarily large factor. However, in the Solar neighbourhood, the `external' gravitational field from the rest of the Galaxy limits the boost to gravity that could be provided by MOND. With the simple interpolating function \citep{Famaey_Binney_2005}, this boost is up to a factor of 1.56.

Such an enhancement to the self-gravity of wide binaries should become detectable in the GAIA era \citep{Scarpa_2017}. This issue was recently investigated in some detail by \citet{Pittordis_2017}, who showed that the prior distribution of orbital eccentricity and semi-major axis should not much affect the conclusions. Unbound wide binaries would disperse rather quickly $-$ at typical orbital velocities of $\ssim \sqrt{\frac{GM_\odot}{7000AU}} = 0.3$ km/s, the separation should rise to 1 pc in only a few Myr. But wide binaries in the MOND sense are still much closer together than field stars in the MW, making for only a very small chance that two unbound stars would randomly be so close together in 3D. Also requiring a similar 3D velocity would reduce the contamination further, making the wide binary test a compelling way of constraining what law of gravity governs gravitational systems with accelerations typical of galactic outskirts.

The wide binary systems necessary for this test are expected to be quite common \citep{Andrews_2017}. In fact, our nearest external star system consists of Proxima Centauri (Proxima Cen) orbiting the close binary $\alpha$ Cen A and B at a distance of 13000 AU \citep{Kervella_2017}. This puts the Proxima Cen orbit well within the regime where MOND would have a significant effect \citep{Beech_2009, Beech_2011}. As well as suggesting that wide binaries ought to be common, even this single system could allow a direct test of MOND with the proposed Theia mission \citep{Theia_2017}.

To see how this might work, I used an algorithm similar to that in Section \ref{MOND_governing_equations} to model the orbit of Proxima Cen. I treated it as a test particle orbiting the much more massive $\alpha$ Cen A and B, which I considered as a single point mass of ${2.043 M_\odot}$ given that they are in a tight orbit separated by only ${\sim 18}$ AU \citep{Kervella_2016}. The EF was taken to be towards the Galactic centre and of a magnitude sufficient to maintain the observed $v_{c, \odot}$ of 232.8 km/s, assuming the Sun is 8.2 kpc from the Galactic centre \citep{McMillan_2017}. As this is much larger than the distance to $\alpha$ Cen, it feels nearly the same $\bm{g}_{ext}$ as the Sun.

I used the gravitational field found in this way to integrate the orbit of Proxima Cen forwards, starting with the radial velocity and proper motion measurements in table 2 of \citet{Kervella_2017}. I also found the Proxima Cen trajectory in Newtonian gravity. In both cases, the observations are assumed to span a negligibly short fraction of the ${\sim 500}$ kyr orbital period, allowing the force to be approximated as constant and the trajectory as parabolic.

\begin{figure}
	\centering
		\includegraphics[width = 8.5cm] {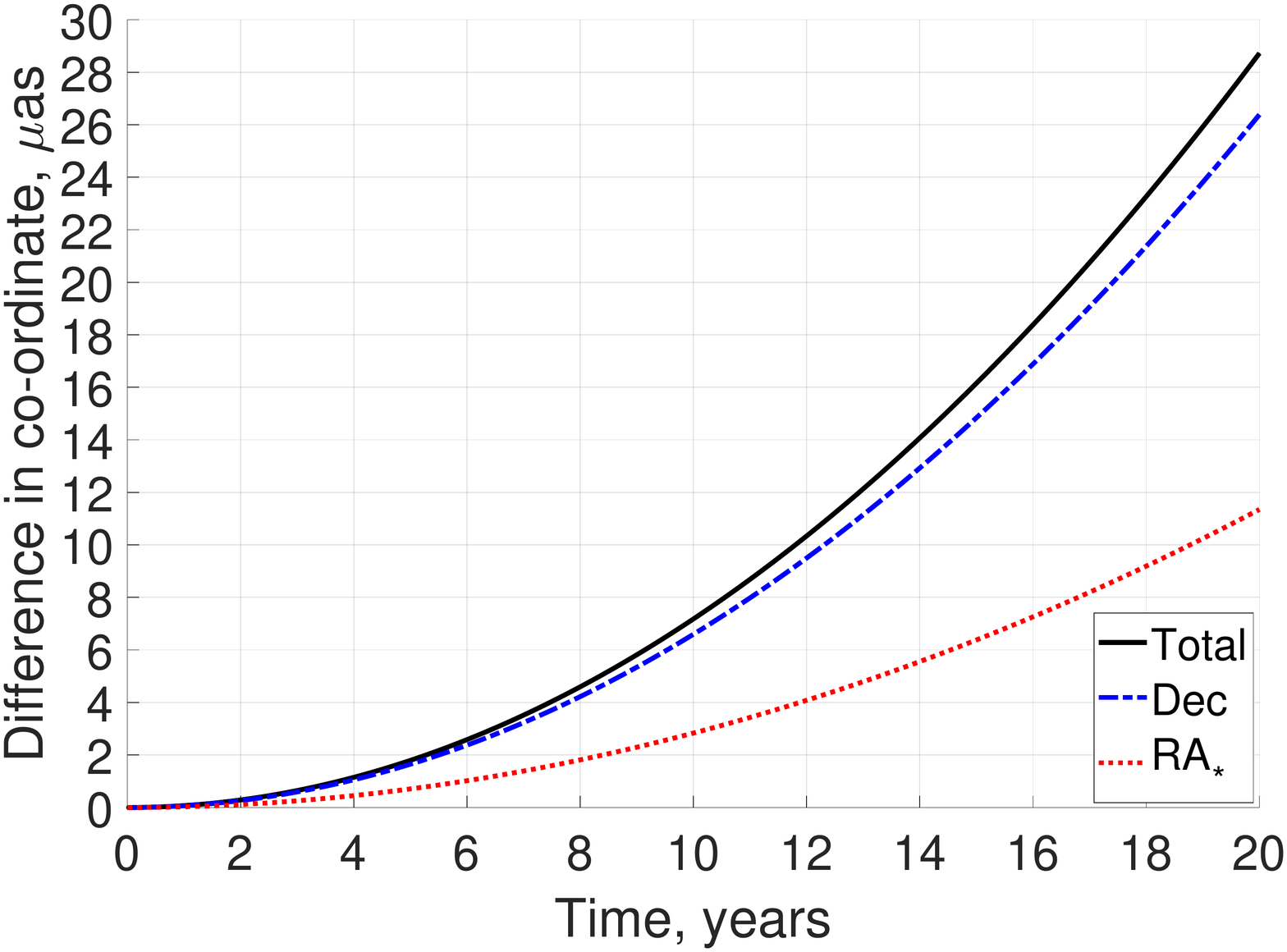}
		\caption{Difference in sky position of Proxima Cen depending on whether Newtonian gravity or MOND governs its orbit about $\alpha$ Cen A and B. The same initial conditions are used for both trajectories \citep[][table 2]{Kervella_2017}. The difference in right ascension has been scaled by the cosine of the declination so it corresponds to an actual angular difference. The total angular difference grows quadratically with time and is 7.18 $\mu$as after 10 years. Astronomers might try to fit the data by varying the initial conditions, in which case the angular differences would be ${\sim \frac{1}{4}}$ that shown here.\textsuperscript{\ref{A}}}
\label{Theia_test}
\end{figure}

The angular difference between the trajectories on our sky is shown in Figure \ref{Theia_test}. Unless the initial conditions are known exactly, the difference would actually be $\frac{1}{4}$ of that shown because astronomers would try to fit the data using different initial conditions.\footnote{The exact ratio will depend on spacecraft performance and other factors. I assumed the fit to data would be designed to minimise its $\chi^2$ with respect to observations taken at regular intervals with equal accuracy. In this case, the best linear fit to the parabola $y = x^2$ over the range ${0-1}$ is given by $y = \frac{3}{4}x$.\label{A}} Even so, a parabola can only be fit with a straight line for so long. Thus, if Theia is flown and achieves $\mu$as astrometric precision over a few years, it should be able to directly determine how much Proxima Cen accelerates towards $\alpha$ Cen A and B. This is because the acceleration is expected to be ${\sim 40\%}$ higher in MOND compared to Newtonian dynamics, though it is only ${\sim a_0}$ in either case and thus very small.

In principle, the radial velocity $v_r$ of Proxima Cen could also be used to distinguish these theories. However, a constant acceleration causes $v_r$ to change linearly with time, whereas the position would respond quadratically. Thus, $v_r$ would only differ by 0.5 cm/s between the models after a decade of observations. This would be very challenging to detect, making it a much less plausible test of MOND than using precise astrometry of Proxima Cen.

One possible complication with such tests is that an undetected exoplanet could also cause an extra acceleration. However, as perceived at Proxima Cen, the exoplanet is quite likely to be in a different direction than $\alpha$ Cen. Moreover, a short period exoplanet would show up in multi-epoch observations. This would not work with a sufficiently long period, but in this case the greater distance implies the exoplanet must be more massive and so more likely to be detected. This is especially true given our proximity to the system enlarging the angles involved, thus making it easier to achieve sufficient starlight suppression in the region of interest. If an anomalous acceleration was detected, then intensive observations could be taken in its direction from Proxima Cen.

In the long run, interstellar precursor missions should be able to test MOND directly as ${\nu_{ext} \approx 1.5}$ in the Solar neighbourhood due to the rest of the MW. Thus, a Cavendish-type experiment performed $\ga 10$ kAU from the Sun should yield rather different forces in MOND than for a similar experiment on the Earth. Alternatively, laser ranging measurements to a spacecraft at such a distance could be used to directly constrain the gravity exerted on it by the Sun \citep{Christian_2017}. The Breakthrough Starshot initiative plans to send spacecraft to much larger distances and thus explore the nearest stars \citep{Merali_2016}. Testing MOND may provide a valuable incentive for interstellar precursor missions that only reach a few percent of this distance.

For such tests to be accurate, the non-gravitational accelerations would need to be constrained as these were important in the case of the Pioneer anomaly $-$ this probably arose due to anisotropy of radiation emitted by the spacecraft \citep{Turyshev_2012}. As any on-board computations require energy to be radiated, such effects can't be completely avoided. Although they could perhaps be reduced, it is almost certainly necessary and quite feasible to have an on-board accelerometer that accurately measures such non-gravitational accelerations \citep{Lenoir_2011}. Combined with tracking data, this could allow rather sensitive tests of gravity.

\subsection{Extensions to the basic MOND paradigm}

The MOND paradigm may need further modification to satisfy all observational constraints. It has been suggested that an additional sterile neutrino species with an 11 eV mass could resolve outstanding issues of MOND at the galaxy cluster scale \citep{Angus_2010} and with the cosmic microwave background radiation \citep{Angus_2009}. This hot DM model gets the large scale structure of the Universe approximately correct, but has difficulty matching the detailed shape of the matter power spectrum at galaxy cluster scales \citep{Diaferio_2011}. Interestingly, recent structure growth measurements reveal some tension with $\Lambda$CDM that may be reduced if at least some of its DM was dynamically hot \citep{Nesseris_2017}.

Other hybrid MOND-DM approaches have also been considered, for instance the superfluid DM model \citep{Khoury_2016}. It suggests that galaxies are surrounded by DM halos which cause additional non-gravitational forces between the baryons mediated by phonons in the superfluid DM. This can naturally reproduce the RAR. Even purely baryonic satellites of the MW within its DM halo would behave as if they were governed by MOND. However, the large extent of the MW and M31 satellite planes means that the model faces similar issues to $\Lambda$CDM $-$ the more distant satellite plane members would still be Newtonian. Moreover, superfluid DM only creates an extra non-gravitational force on baryons, making it unclear how the theory can explain galaxy-galaxy weak lensing measurements that are consistent with light deflection governed by the RAR but not by standard gravity of the baryons alone \citep{Milgrom_2013}. Although the normal gravitational effect of the superfluid DM would make it act similarly to $\Lambda$CDM on large scales, its distribution would need to be rather finely tuned to satisfy Equation \ref{Light_deflection_MOND} and avoid radii where the deflection angle undergoes a Keplerian decline before rising again due to the DM halo. In fact, this may be a problem for $\Lambda$CDM \citep[e.g.][]{Faber_1979}, motivating some workers to consider adjusting the basic paradigm \citep[e.g.][]{Kamda_2017}.

Given these difficulties, it is entirely possible that the empirical MOND equations need further modification. As the cluster-scale issues faced by MOND \citep{Sanders_2003} arise in regions with a deep potential well, it has been suggested that the gravitational potential affects the acceleration parameter $a_{_0}$ \citep{Zhao_2012}. This theory of extended MOND (EMOND) can be tuned to match galaxy and cluster-scale observations fairly well \citep{Hodson_2017}. Further work could help determine if EMOND is consistent with the internal dynamics of galaxies in cluster environments, where the local value of $a_{_0}$ should be higher than in more isolated field galaxies.

MOND is an empirical theory whose more fundamental basis (if any) could help reconcile issues faced by its current formulation. One approach to understanding why MOND arises at all is the emergent gravity theory \citep{Verlinde_2016}. Although it is underpinned by some interesting theoretical ideas, the model faces difficulty explaining the observed RAR \citep{Verlinde_test_2017}. A decisive test of the theory should be possible in the near future because it predicts extra perihelion precession of Solar System planets, especially Mars \citep{Iorio_2017}.

Originally, MOND was formulated as a non-relativistic theory \citep{Bekenstein_Milgrom_1984}. It can be embedded within a relativistic framework \citep{Bekenstein_2004}, a very relevant exercise in light of recent GW detections \citep{LIGO_2016}. Expecting these results, GW propagation in MOND was briefly explored by \citet{Milgrom_2014}, though of course further work would be highly relevant. It is particularly important to check whether relativistic MOND theories are consistent with GWs propagating at speeds very close to the speed of light, as required to explain recent observations of GW170817 \citep{LIGO_Virgo_2017}. Perhaps MOND can't be reconciled with these observations \citep{Chesler_2017}. It is unclear whether their arguments rule out particular relativistic generalisations of MOND or the paradigm in its entirety. As MOND was originally formulated to deal with non-relativistic systems (galaxies), the former seems more likely. However, only time will tell if it is possible to formulate MOND in a way consistent with special relativity and the near-simultaneous arrival of GWs and their electromagnetic counterpart. Even if this is possible, the galaxy cluster and larger scale issues faced by MOND may yet prove its undoing.

\section{Conclusions}
\label{Conclusions}

In this thesis, I explored several situations within and beyond the Local Group (LG) that are likely to reveal behaviour characteristic of MOND if it is correct. I began by considering how galaxy cluster collision velocities could be measured more accurately to search for a high-velocity tail not expected in $\Lambda$CDM but expected if gravity is enhanced at long range (Section \ref{Bullet_Cluster}). Despite possible differences of several hundred km/s, the large distance to such systems makes it difficult to determine their kinematics. Thus, the remaining contributions in this portfolio focus on the LG.

In Section \ref{Local_Group_2D}, I described a $\Lambda$CDM dynamical analysis of the LG that treated it as axisymmetric about the MW-M31 line. This revealed several galaxies with much higher radial velocities than expected. I suggested that these HVGs were flung out by the MW/M31 around the time of their past close flyby, an event which would have happened in MOND \citep{Zhao_2013} but not $\Lambda$CDM due to dynamical friction between their DM halos \citep{Privon_2013}.

In Section \ref{Local_Group_3D}, I used an algorithm provided by one of the founders of the $\Lambda$CDM paradigm (Peebles) to address the issue of whether it really faces difficulty explaining the kinematics of LG dwarfs. The main difference was that the algorithm used a full 3D model, allowing rigorous consideration of tides raised by external galaxies and galaxy groups within 10 Mpc. Despite using a different algorithm written by different people in different programming languages, my conclusions remained broadly similar and were confirmed in a similar analysis by \citet{Peebles_2017}. Such HVGs do not easily arise in cosmological simulations due to interactions of LG dwarfs with analogues of MW or M31 satellites \citep[][figures 3 and 6]{Sales_2007}. Thus, although some real LG satellites may be missing from my model due to a lack of detectable baryons, it is unclear how this would explain my results as such satellites should still be present in $\Lambda$CDM simulations.

To test my proposed scenario regarding the HVGs, I focused on comparing their spatial distribution and other properties to what these ought to be if the HVGs really were flung out by three-body interactions with the MW and M31 (Section \ref{HVG_plane}). I found that the HVGs should preferentially lie within the plane of the MW-M31 orbit (Figure \ref{z_distribution}). The HVGs do indeed define a rather thin plane (Figure \ref{Plane_offset_Delta_GRV_GA}) oriented so the MW-M31 line is only $16^\circ$ out of this plane. Thus, the properties of the HVGs are broadly what they should be if their anomalous kinematics arose due to passing near the spacetime location of a past MW-M31 flyby.

To gain experience with detailed MW models of the sort likely to become testable in the GAIA era, I calculated the escape velocity curve of the MW in MOND (Section \ref{v_esc_article}) and compared it with recent measurements covering Galactocentric radii of 8$-$50 kpc \citep{Williams_2017}. I was able to account for both the amplitude and radial gradient of the Galactic escape velocity curve using a baryonic mass model consistent with the much more accurately known MW rotation curve \citep[][table 1 model Q4ZB]{McGaugh_2016_MW}. My results suggest that the MW has only a fairly low mass hot gas corona around it and is embedded in an EF of ${\sim 0.03a_{_0}}$.

In Section \ref{Future_prospects}, I briefly reviewed some possible avenues for future investigations to help determine if there really are substantial deviations from standard gravity at low acceleration. Although many of these ideas could be useful, the wide binary test of gravity seems the most promising near-term line of attack due to the impending release of data from the GAIA mission \citep{Perryman_2001}. This would be an almost direct test involving a very different type of low-acceleration system to the rotating disk galaxies which originally inspired MOND. Nonetheless, both systems are alike in that their overall density greatly exceeds the cosmic mean value. Therefore, the application of MOND to these non-relativistic systems is quite clear.


\section{Acknowledgements}
\label{Acknowledgements}

\subsection{General acknowledgements}
The work on the high-velocity galaxy plane (Section \ref{HVG_plane}) was suggested by Marcel Pawlowski and benefited from a visit to Princeton hosted by Nima Arkani-Hamed at the Institute for Advanced Studies. That work and the 3D $\Lambda$CDM model of the Local Group (Section \ref{Local_Group_3D}) were both based on a \textsc{fortran} algorithm kindly lent by P. J. E. Peebles, who also provided much advice regarding its operation. All the other algorithms were set up using \textsc{matlab}$^\text{\textregistered}$.

\subsection{Funding}
IB is supported by Science and Technology Facilities Council studentship 1506672. The visit to Princeton was funded by a Scottish Universities' Physics Alliance travel grant.

\bibliographystyle{mnras}
\bibliography{PHD_bbl}

\begin{thebibliography}{}
\makeatletter
\relax
\def\mn@urlcharsother{\let\do\@makeother \do\$\do\&\do\#\do\^\do\_\do\%\do\~}
\def\mn@doi{\begingroup\mn@urlcharsother \@ifnextchar [ {\mn@doi@}
  {\mn@doi@[]}}
\def\mn@doi@[#1]#2{\def\@tempa{#1}\ifx\@tempa\@empty \href
  {http://dx.doi.org/#2} {doi:#2}\else \href {http://dx.doi.org/#2} {#1}\fi
  \endgroup}
\def\mn@eprint#1#2{\mn@eprint@#1:#2::\@nil}
\def\mn@eprint@arXiv#1{\href {http://arxiv.org/abs/#1} {{\tt arXiv:#1}}}
\def\mn@eprint@dblp#1{\href {http://dblp.uni-trier.de/rec/bibtex/#1.xml}
  {dblp:#1}}
\def\mn@eprint@#1:#2:#3:#4\@nil{\def\@tempa {#1}\def\@tempb {#2}\def\@tempc
  {#3}\ifx \@tempc \@empty \let \@tempc \@tempb \let \@tempb \@tempa \fi \ifx
  \@tempb \@empty \def\@tempb {arXiv}\fi \@ifundefined
  {mn@eprint@\@tempb}{\@tempb:\@tempc}{\expandafter \expandafter \csname
  mn@eprint@\@tempb\endcsname \expandafter{\@tempc}}}

\bibitem[\protect\citeauthoryear{{ATLAS Collaboration}}{{ATLAS
  Collaboration}}{2015}]{ATLAS_Collaboration_2015}
{ATLAS Collaboration} 2015, \mn@doi [Journal of High Energy Physics]
  {10.1007/JHEP10(2015)134}, \href
  {http://adsabs.harvard.edu/abs/2015JHEP...10..134A} {10, 134}

\bibitem[\protect\citeauthoryear{{Ahmed}, {Brooks}  \& {Christensen}}{{Ahmed}
  et~al.}{2017}]{Ahmed_2017}
{Ahmed} S.~H.,  {Brooks} A.~M.,   {Christensen} C.~R.,  2017, \mn@doi [MNRAS]
  {10.1093/mnras/stw3271}, \href
  {http://adsabs.harvard.edu/abs/2017MNRAS.466.3119A} {466, 3119}

\bibitem[\protect\citeauthoryear{{Ahn} et~al.,}{{Ahn} et~al.}{2012}]{Ahn_2012}
{Ahn} C.~P.,  et~al., 2012, \mn@doi [ApJS] {10.1088/0067-0049/203/2/21}, \href
  {http://adsabs.harvard.edu/abs/2012ApJS..203...21A} {203, 21}

\bibitem[\protect\citeauthoryear{{Alcock} et~al.,}{{Alcock}
  et~al.}{2000}]{MACHO_2000}
{Alcock} C.,  et~al., 2000, \mn@doi [ApJ] {10.1086/309512}, \href
  {http://adsabs.harvard.edu/abs/2000ApJ...542..281A} {542, 281}

\bibitem[\protect\citeauthoryear{{Andrews}, {Chanam{\'e}}  \&
  {Ag{\"u}eros}}{{Andrews} et~al.}{2017}]{Andrews_2017}
{Andrews} J.~J.,  {Chanam{\'e}} J.,   {Ag{\"u}eros} M.~A.,  2017, \mn@doi
  [MNRAS] {10.1093/mnras/stx2000}, \href
  {http://adsabs.harvard.edu/abs/2017MNRAS.472..675A} {472, 675}

\bibitem[\protect\citeauthoryear{{Angus}}{{Angus}}{2009}]{Angus_2009}
{Angus} G.~W.,  2009, \mn@doi [MNRAS] {10.1111/j.1365-2966.2008.14341.x}, \href
  {http://adsabs.harvard.edu/abs/2009MNRAS.394..527A} {394, 527}

\bibitem[\protect\citeauthoryear{{Angus} \& {Diaferio}}{{Angus} \&
  {Diaferio}}{2011}]{Diaferio_2011}
{Angus} G.~W.,  {Diaferio} A.,  2011, \mn@doi [MNRAS]
  {10.1111/j.1365-2966.2011.19321.x}, \href
  {http://adsabs.harvard.edu/abs/2011MNRAS.417..941A} {417, 941}

\bibitem[\protect\citeauthoryear{{Angus}, {Famaey}  \& {Diaferio}}{{Angus}
  et~al.}{2010}]{Angus_2010}
{Angus} G.~W.,  {Famaey} B.,   {Diaferio} A.,  2010, \mn@doi [MNRAS]
  {10.1111/j.1365-2966.2009.15895.x}, \href
  {http://adsabs.harvard.edu/abs/2010MNRAS.402..395A} {402, 395}

\bibitem[\protect\citeauthoryear{{Angus}, {Diaferio}  \& {Kroupa}}{{Angus}
  et~al.}{2011}]{Angus_2011}
{Angus} G.~W.,  {Diaferio} A.,   {Kroupa} P.,  2011, \mn@doi [MNRAS]
  {10.1111/j.1365-2966.2011.19138.x}, \href
  {http://adsabs.harvard.edu/abs/2011MNRAS.416.1401A} {416, 1401}

\bibitem[\protect\citeauthoryear{{Aragon-Calvo}, {Silk}  \&
  {Szalay}}{{Aragon-Calvo} et~al.}{2011}]{Aragon_Calvo_2011}
{Aragon-Calvo} M.~A.,  {Silk} J.,   {Szalay} A.~S.,  2011, \mn@doi [MNRAS]
  {10.1111/j.1745-3933.2011.01071.x}, \href
  {http://adsabs.harvard.edu/abs/2011MNRAS.415L..16A} {415, L16}

\bibitem[\protect\citeauthoryear{{Babcock}}{{Babcock}}{1939}]{Babcock_1939}
{Babcock} H.~W.,  1939, \mn@doi [Lick Observatory Bulletin]
  {10.5479/ADS/bib/1939LicOB.19.41B}, \href
  {http://adsabs.harvard.edu/abs/1939LicOB..19...41B} {19, 41}

\bibitem[\protect\citeauthoryear{{Baer}, {Choi}, {Kim}  \& {Roszkowski}}{{Baer}
  et~al.}{2015}]{Baer_2015}
{Baer} H.,  {Choi} K.-Y.,  {Kim} J.~E.,   {Roszkowski} L.,  2015, \mn@doi
  [Physics Reports] {10.1016/j.physrep.2014.10.002}, \href
  {http://adsabs.harvard.edu/abs/2015PhR...555....1B} {555, 1}

\bibitem[\protect\citeauthoryear{{Banik}}{{Banik}}{2014}]{Banik_2014}
{Banik} I.,  2014, preprint, \href
  {http://adsabs.harvard.edu/abs/2014arXiv1406.4538B} {Arxiv} (\mn@eprint
  {arXiv} {1406.4538v2})

\bibitem[\protect\citeauthoryear{{Banik} \& {Zhao}}{{Banik} \&
  {Zhao}}{2015a}]{BANIK_2015_MCE}
{Banik} I.,  {Zhao} H.,  2015a, \mn@doi [MNRAS] {10.1093/mnras/stv802}, \href
  {http://adsabs.harvard.edu/abs/2015MNRAS.450.3155B} {450, 3155}

\bibitem[\protect\citeauthoryear{{Banik} \& {Zhao}}{{Banik} \&
  {Zhao}}{2015b}]{Banik_2015_analytic}
{Banik} I.,  {Zhao} H.,  2015b, preprint, \href
  {http://adsabs.harvard.edu/abs/2015arXiv150908457B} {Arxiv} (\mn@eprint
  {arXiv} {1509.08457})

\bibitem[\protect\citeauthoryear{{Banik} \& {Zhao}}{{Banik} \&
  {Zhao}}{2016}]{BANIK_ZHAO_2016}
{Banik} I.,  {Zhao} H.,  2016, \mn@doi [MNRAS] {10.1093/mnras/stw787}, \href
  {http://adsabs.harvard.edu/abs/2016MNRAS.459.2237B} {459, 2237}

\bibitem[\protect\citeauthoryear{{Banik} \& {Zhao}}{{Banik} \&
  {Zhao}}{2017}]{BANIK_ZHAO_2017}
{Banik} I.,  {Zhao} H.,  2017, \mn@doi [MNRAS] {10.1093/mnras/stx151}, \href
  {http://adsabs.harvard.edu/abs/2017MNRAS.467.2180B} {467, 2180}

\bibitem[\protect\citeauthoryear{{Banik} \& {Zhao}}{{Banik} \&
  {Zhao}}{2018a}]{BANIK_2017_ESCAPE}
{Banik} I.,  {Zhao} H.,  2018a, \mn@doi [MNRAS] {10.1093/mnras/stx2350}, \href
  {http://adsabs.harvard.edu/abs/2018MNRAS.473..419B} {473, 419}

\bibitem[\protect\citeauthoryear{{Banik} \& {Zhao}}{{Banik} \&
  {Zhao}}{2018b}]{BANIK_2017_ANISOTROPY}
{Banik} I.,  {Zhao} H.,  2018b, \mn@doi [MNRAS] {10.1093/mnras/stx2596}, \href
  {http://adsabs.harvard.edu/abs/2018MNRAS.473.4033B} {473, 4033}

\bibitem[\protect\citeauthoryear{{Banik}, {O'Ryan}  \& {Zhao}}{{Banik}
  et~al.}{2018}]{Banik_2018}
{Banik} I.,  {O'Ryan} D.,   {Zhao} H.,  2018, \mn@doi [MNRAS]
  {10.1093/mnras/sty919}, \href
  {http://adsabs.harvard.edu/abs/2018MNRAS.tmp..879B} {accepted}

\bibitem[\protect\citeauthoryear{{Barnes} \& {Hernquist}}{{Barnes} \&
  {Hernquist}}{1992}]{Barnes_1992}
{Barnes} J.~E.,  {Hernquist} L.,  1992, \mn@doi [Nature] {10.1038/360715a0},
  \href {http://adsabs.harvard.edu/abs/1992Natur.360..715B} {360, 715}

\bibitem[\protect\citeauthoryear{{Beech}}{{Beech}}{2009}]{Beech_2009}
{Beech} M.,  2009, \mn@doi [MNRAS] {10.1111/j.1745-3933.2009.00718.x}, \href
  {http://adsabs.harvard.edu/abs/2009MNRAS.399L..21B} {399, L21}

\bibitem[\protect\citeauthoryear{{Beech}}{{Beech}}{2011}]{Beech_2011}
{Beech} M.,  2011, \mn@doi [ApSS] {10.1007/s10509-011-0665-2}, \href
  {http://adsabs.harvard.edu/abs/2011Ap\%26SS.333..419B} {333, 419}

\bibitem[\protect\citeauthoryear{{Bekenstein}}{{Bekenstein}}{2004}]{Bekenstein_2004}
{Bekenstein} J.~D.,  2004, \mn@doi [Physical Review D]
  {10.1103/PhysRevD.70.083509}, \href
  {http://adsabs.harvard.edu/abs/2004PhRvD..70h3509B} {70, 083509}

\bibitem[\protect\citeauthoryear{{Bekenstein} \& {Milgrom}}{{Bekenstein} \&
  {Milgrom}}{1984}]{Bekenstein_Milgrom_1984}
{Bekenstein} J.,  {Milgrom} M.,  1984, \mn@doi [ApJ] {10.1086/162570}, \href
  {http://adsabs.harvard.edu/abs/1984ApJ...286....7B} {286, 7}

\bibitem[\protect\citeauthoryear{{Bell} \& {de Jong}}{{Bell} \& {de
  Jong}}{2001}]{Bell_de_Jong_2001}
{Bell} E.~F.,  {de Jong} R.~S.,  2001, \mn@doi [ApJ] {10.1086/319728}, \href
  {http://adsabs.harvard.edu/abs/2001ApJ...550..212B} {550, 212}

\bibitem[\protect\citeauthoryear{{Bellazzini}, {Oosterloo}, {Fraternali}  \&
  {Beccari}}{{Bellazzini} et~al.}{2013}]{Bellazzini_2013}
{Bellazzini} M.,  {Oosterloo} T.,  {Fraternali} F.,   {Beccari} G.,  2013,
  \mn@doi [A\&A] {10.1051/0004-6361/201322744}, \href
  {http://adsabs.harvard.edu/abs/2013A\%26A...559L..11B} {559, L11}

\bibitem[\protect\citeauthoryear{{Berenji}, {Gaskins}  \& {Meyer}}{{Berenji}
  et~al.}{2016}]{Berenji_2016}
{Berenji} B.,  {Gaskins} J.,   {Meyer} M.,  2016, \mn@doi [Physical Review D]
  {10.1103/PhysRevD.93.045019}, \href
  {http://adsabs.harvard.edu/abs/2016PhRvD..93d5019B} {93, 045019}

\bibitem[\protect\citeauthoryear{{Berezhiani} \& {Khoury}}{{Berezhiani} \&
  {Khoury}}{2016}]{Berezhiani_2015}
{Berezhiani} L.,  {Khoury} J.,  2016, \mn@doi [Physics Letters B]
  {10.1016/j.physletb.2015.12.054}, \href
  {http://adsabs.harvard.edu/abs/2016PhLB..753..639B} {753, 639}

\bibitem[\protect\citeauthoryear{{Bienaym{\'e}}, {Famaey}, {Wu}, {Zhao}  \&
  {Aubert}}{{Bienaym{\'e}} et~al.}{2009}]{Bienayme_2009}
{Bienaym{\'e}} O.,  {Famaey} B.,  {Wu} X.,  {Zhao} H.~S.,   {Aubert} D.,  2009,
  \mn@doi [A\&A] {10.1051/0004-6361/200809978}, \href
  {http://adsabs.harvard.edu/abs/2009A\%26A...500..801B} {500, 801}

\bibitem[\protect\citeauthoryear{{B{\'{\i}}lek}, {Thies}, {Kroupa}  \&
  {Famaey}}{{B{\'{\i}}lek} et~al.}{2017}]{Bilek_2017}
{B{\'{\i}}lek} M.,  {Thies} I.,  {Kroupa} P.,   {Famaey} B.,  2017, \mn@doi
  [A\&A] {10.1051/0004-6361/201731939}, \href
  {http://adsabs.harvard.edu/abs/2017arXiv171204938B} {accepted}

\bibitem[\protect\citeauthoryear{{Birkinshaw} \& {Gull}}{{Birkinshaw} \&
  {Gull}}{1983}]{Birkinshaw_Gull_1983}
{Birkinshaw} M.,  {Gull} S.~F.,  1983, \mn@doi [Nature] {10.1038/302315a0},
  \href {http://adsabs.harvard.edu/abs/1983Natur.302..315B} {302, 315}

\bibitem[\protect\citeauthoryear{{Blaksley} \& {Bonamente}}{{Blaksley} \&
  {Bonamente}}{2010}]{Blaksley_Bonamente_2009}
{Blaksley} C.,  {Bonamente} M.,  2010, \mn@doi [New Astronomy]
  {10.1016/j.newast.2009.07.004}, \href
  {http://adsabs.harvard.edu/abs/2010NewA...15..159B} {15, 159}

\bibitem[\protect\citeauthoryear{{Bolton}, {Burles}, {Koopmans}, {Treu},
  {Gavazzi}, {Moustakas}, {Wayth}  \& {Schlegel}}{{Bolton}
  et~al.}{2008}]{Bolton_2008}
{Bolton} A.~S.,  {Burles} S.,  {Koopmans} L.~V.~E.,  {Treu} T.,  {Gavazzi} R.,
  {Moustakas} L.~A.,  {Wayth} R.,   {Schlegel} D.~J.,  2008, \mn@doi [ApJ]
  {10.1086/589327}, \href {http://adsabs.harvard.edu/abs/2008ApJ...682..964B}
  {682, 964}

\bibitem[\protect\citeauthoryear{{Borsanyi} et~al.,}{{Borsanyi}
  et~al.}{2016}]{Borsanyi_2016}
{Borsanyi} S.,  et~al., 2016, \mn@doi [Nature] {10.1038/nature20115}, \href
  {http://adsabs.harvard.edu/abs/2016Natur.539...69B} {539, 69}

\bibitem[\protect\citeauthoryear{{Borucki}, {Koch}, {Dunham}  \&
  {Jenkins}}{{Borucki} et~al.}{1997}]{Borucki_1997}
{Borucki} W.~J.,  {Koch} D.~G.,  {Dunham} E.~W.,   {Jenkins} J.~M.,  1997, in
  {Soderblom} D.,  ed.,  Astronomical Society of the Pacific Conference Series
  Vol. 119, Planets Beyond the Solar System and the Next Generation of Space
  Missions. p.~153

\bibitem[\protect\citeauthoryear{{Bournaud} et~al.,}{{Bournaud}
  et~al.}{2007}]{Bournaud_2007}
{Bournaud} F.,  et~al., 2007, \mn@doi [Science] {10.1126/science.1142114},
  \href {http://adsabs.harvard.edu/abs/2007Sci...316.1166B} {316, 1166}

\bibitem[\protect\citeauthoryear{{Bovy} \& {Rix}}{{Bovy} \&
  {Rix}}{2013}]{Bovy_2013}
{Bovy} J.,  {Rix} H.-W.,  2013, \mn@doi [ApJ] {10.1088/0004-637X/779/2/115},
  \href {http://adsabs.harvard.edu/abs/2013ApJ...779..115B} {779, 115}

\bibitem[\protect\citeauthoryear{{Bowden}, {Evans}  \& {Belokurov}}{{Bowden}
  et~al.}{2013}]{Bowden_2013}
{Bowden} A.,  {Evans} N.~W.,   {Belokurov} V.,  2013, \mn@doi [MNRAS]
  {10.1093/mnras/stt1253}, \href
  {http://adsabs.harvard.edu/abs/2013MNRAS.435..928B} {435, 928}

\bibitem[\protect\citeauthoryear{{Brada} \& {Milgrom}}{{Brada} \&
  {Milgrom}}{1999}]{Brada_Milgrom_1998}
{Brada} R.,  {Milgrom} M.,  1999, \mn@doi [ApJ] {10.1086/307402}, \href
  {http://adsabs.harvard.edu/abs/1999ApJ...519..590B} {519, 590}

\bibitem[\protect\citeauthoryear{{Brimioulle}, {Seitz}, {Lerchster}, {Bender}
  \& {Snigula}}{{Brimioulle} et~al.}{2013}]{Brimioulle_2013}
{Brimioulle} F.,  {Seitz} S.,  {Lerchster} M.,  {Bender} R.,   {Snigula} J.,
  2013, \mn@doi [MNRAS] {10.1093/mnras/stt525}, \href
  {http://adsabs.harvard.edu/abs/2013MNRAS.432.1046B} {432, 1046}

\bibitem[\protect\citeauthoryear{{Broeils}}{{Broeils}}{1992}]{Broeils_1992}
{Broeils} A.~H.,  1992, A\&A, \href
  {http://adsabs.harvard.edu/abs/1992A%26A...256...19B} {256, 19}

\bibitem[\protect\citeauthoryear{{Brunthaler}, {Reid}, {Falcke}, {Greenhill}
  \& {Henkel}}{{Brunthaler} et~al.}{2005}]{Brunthaler_2005}
{Brunthaler} A.,  {Reid} M.~J.,  {Falcke} H.,  {Greenhill} L.~J.,   {Henkel}
  C.,  2005, \mn@doi [Science] {10.1126/science.1108342}, \href
  {http://adsabs.harvard.edu/abs/2005Sci...307.1440B} {307, 1440}

\bibitem[\protect\citeauthoryear{{Cai}, {Li}, {Cole}, {Frenk}  \&
  {Neyrinck}}{{Cai} et~al.}{2014}]{Cai_2014}
{Cai} Y.-C.,  {Li} B.,  {Cole} S.,  {Frenk} C.~S.,   {Neyrinck} M.,  2014,
  \mn@doi [MNRAS] {10.1093/mnras/stu154}, \href
  {http://adsabs.harvard.edu/abs/2014MNRAS.439.2978C} {439, 2978}

\bibitem[\protect\citeauthoryear{{Caldwell} et~al.,}{{Caldwell}
  et~al.}{2017}]{Caldwell_2017}
{Caldwell} N.,  et~al., 2017, \mn@doi [ApJ] {10.3847/1538-4357/aa688e}, \href
  {http://adsabs.harvard.edu/abs/2017ApJ...839...20C} {839, 20}

\bibitem[\protect\citeauthoryear{{Candlish}}{{Candlish}}{2016}]{Candlish_2016_velocity}
{Candlish} G.~N.,  2016, \mn@doi [MNRAS] {10.1093/mnras/stw1130}, \href
  {http://adsabs.harvard.edu/abs/2016MNRAS.460.2571C} {460, 2571}

\bibitem[\protect\citeauthoryear{{Candlish}, {Smith}  \&
  {Fellhauer}}{{Candlish} et~al.}{2016}]{Candlish_2016}
{Candlish} G.~N.,  {Smith} R.,   {Fellhauer} M.,  2016, in Journal of Physics
  Conference Series. p. 012012, \mn@doi{10.1088/1742-6596/720/1/012012}

\bibitem[\protect\citeauthoryear{{Carignan}, {Chemin}, {Huchtmeier}  \&
  {Lockman}}{{Carignan} et~al.}{2006}]{Carignan_2006}
{Carignan} C.,  {Chemin} L.,  {Huchtmeier} W.~K.,   {Lockman} F.~J.,  2006,
  \mn@doi [ApJL] {10.1086/503869}, \href
  {http://adsabs.harvard.edu/abs/2006ApJ...641L.109C} {641, L109}

\bibitem[\protect\citeauthoryear{{Carlip}}{{Carlip}}{2001}]{Carlip_2001}
{Carlip} S.,  2001, \mn@doi [Reports on Progress in Physics]
  {10.1088/0034-4885/64/8/301}, \href
  {http://adsabs.harvard.edu/abs/2001RPPh...64..885C} {64, 885}

\bibitem[\protect\citeauthoryear{{Carr}}{{Carr}}{1994}]{Carr_1994}
{Carr} B.,  1994, \mn@doi [ARA\&A] {10.1146/annurev.astro.32.1.531}, \href
  {http://adsabs.harvard.edu/abs/1994ARA\%26A..32..531C} {32, 531}

\bibitem[\protect\citeauthoryear{{Casetti-Dinescu}, {Girard}  \&
  {Schriefer}}{{Casetti-Dinescu} et~al.}{2018}]{Dinescu_2018}
{Casetti-Dinescu} D.~I.,  {Girard} T.~M.,   {Schriefer} M.,  2018, \mn@doi
  [MNRAS] {10.1093/mnras/stx2645}, \href
  {http://adsabs.harvard.edu/abs/2018MNRAS.473.4064C} {473, 4064}

\bibitem[\protect\citeauthoryear{{Cautun}, {Bose}, {Frenk}, {Guo}, {Han},
  {Hellwing}, {Sawala}  \& {Wang}}{{Cautun} et~al.}{2015}]{Cautun_2015}
{Cautun} M.,  {Bose} S.,  {Frenk} C.~S.,  {Guo} Q.,  {Han} J.,  {Hellwing}
  W.~A.,  {Sawala} T.,   {Wang} W.,  2015, \mn@doi [MNRAS]
  {10.1093/mnras/stv1557}, \href
  {http://adsabs.harvard.edu/abs/2015MNRAS.452.3838C} {452, 3838}

\bibitem[\protect\citeauthoryear{{Chae}, {Bernardi}  \& {Sheth}}{{Chae}
  et~al.}{2017}]{Chae_2017}
{Chae} K.-H.,  {Bernardi} M.,   {Sheth} R.~K.,  2017, preprint, \href
  {http://adsabs.harvard.edu/abs/2017arXiv170708280C} {Arxiv} (\mn@eprint
  {arXiv} {1707.08280})

\bibitem[\protect\citeauthoryear{{Chandrasekhar}}{{Chandrasekhar}}{1943}]{Chandrasekhar_1943}
{Chandrasekhar} S.,  1943, \mn@doi [ApJ] {10.1086/144517}, \href
  {http://adsabs.harvard.edu/abs/1943ApJ....97..255C} {97, 255}

\bibitem[\protect\citeauthoryear{{Chesler} \& {Loeb}}{{Chesler} \&
  {Loeb}}{2017}]{Chesler_2017}
{Chesler} P.~M.,  {Loeb} A.,  2017, \mn@doi [Physical Review Letters]
  {10.1103/PhysRevLett.119.031102}, \href
  {http://adsabs.harvard.edu/abs/2017PhRvL.119c1102C} {119, 031102}

\bibitem[\protect\citeauthoryear{{Chiu}, {Ko}  \& {Tian}}{{Chiu}
  et~al.}{2006}]{Chiu_2006}
{Chiu} M.-C.,  {Ko} C.-M.,   {Tian} Y.,  2006, \mn@doi [ApJ] {10.1086/498241},
  \href {http://adsabs.harvard.edu/abs/2006ApJ...636..565C} {636, 565}

\bibitem[\protect\citeauthoryear{{Christian} \& {Loeb}}{{Christian} \&
  {Loeb}}{2017}]{Christian_2017}
{Christian} P.,  {Loeb} A.,  2017, \mn@doi [ApJL]
  {10.3847/2041-8213/834/2/L20}, \href
  {http://adsabs.harvard.edu/abs/2017ApJ...834L..20C} {834, L20}

\bibitem[\protect\citeauthoryear{{Clark}, {Lewis}  \& {Scott}}{{Clark}
  et~al.}{2016}]{Clark_2016}
{Clark} H.~A.,  {Lewis} G.~F.,   {Scott} P.,  2016, \mn@doi [MNRAS]
  {10.1093/mnras/stv2743}, \href
  {http://adsabs.harvard.edu/abs/2016MNRAS.456.1394C} {456, 1394}

\bibitem[\protect\citeauthoryear{{Clowe}, {Gonzalez}  \& {Markevitch}}{{Clowe}
  et~al.}{2004}]{Clowe_2004}
{Clowe} D.,  {Gonzalez} A.,   {Markevitch} M.,  2004, \mn@doi [ApJ]
  {10.1086/381970}, \href {http://adsabs.harvard.edu/abs/2004ApJ...604..596C}
  {604, 596}

\bibitem[\protect\citeauthoryear{{Clowe}, {Brada{\v c}}, {Gonzalez},
  {Markevitch}, {Randall}, {Jones}  \& {Zaritsky}}{{Clowe}
  et~al.}{2006}]{Dark_Matter_Proof}
{Clowe} D.,  {Brada{\v c}} M.,  {Gonzalez} A.~H.,  {Markevitch} M.,  {Randall}
  S.~W.,  {Jones} C.,   {Zaritsky} D.,  2006, \mn@doi [ApJ] {10.1086/508162},
  \href {http://adsabs.harvard.edu/abs/2006ApJ...648L.109C} {648, L109}

\bibitem[\protect\citeauthoryear{{Collins} et~al.,}{{Collins}
  et~al.}{2015}]{Collins_2015}
{Collins} M.~L.~M.,  et~al., 2015, \mn@doi [ApJL]
  {10.1088/2041-8205/799/1/L13}, \href
  {http://adsabs.harvard.edu/abs/2015ApJ...799L..13C} {799, L13}

\bibitem[\protect\citeauthoryear{{Combes}}{{Combes}}{2014}]{Combes_2014}
{Combes} F.,  2014, \mn@doi [A\&A] {10.1051/0004-6361/201424990}, \href
  {http://adsabs.harvard.edu/abs/2014A\%26A...571A..82C} {571, A82}

\bibitem[\protect\citeauthoryear{Cyburt, Fields, Olive  \& Yeh}{Cyburt
  et~al.}{2016}]{Cyburt_2016}
Cyburt R.~H.,  Fields B.~D.,  Olive K.~A.,   Yeh T.-H.,  2016, \mn@doi [Rev.
  Mod. Phys.] {10.1103/RevModPhys.88.015004}, \href
  {https://link.aps.org/doi/10.1103/RevModPhys.88.015004} {88, 015004}

\bibitem[\protect\citeauthoryear{{Dalcanton} et~al.,}{{Dalcanton}
  et~al.}{2009}]{Dalcanton_2009}
{Dalcanton} J.~J.,  et~al., 2009, \mn@doi [ApJS] {10.1088/0067-0049/183/1/67},
  \href {http://adsabs.harvard.edu/abs/2009ApJS..183...67D} {183, 67}

\bibitem[\protect\citeauthoryear{{Desmond}}{{Desmond}}{2017a}]{Desmond_2016}
{Desmond} H.,  2017a, \mn@doi [MNRAS] {10.1093/mnras/stw2571}, \href
  {http://adsabs.harvard.edu/abs/2017MNRAS.464.4160D} {464, 4160}

\bibitem[\protect\citeauthoryear{{Desmond}}{{Desmond}}{2017b}]{Desmond_2017}
{Desmond} H.,  2017b, \mn@doi [MNRAS] {10.1093/mnrasl/slx134}, \href
  {http://adsabs.harvard.edu/abs/2017MNRAS.472L..35D} {472, L35}

\bibitem[\protect\citeauthoryear{{Duc}, {Paudel}, {McDermid}, {Cuillandre},
  {Serra}, {Bournaud}, {Cappellari}  \& {Emsellem}}{{Duc}
  et~al.}{2014}]{Duc_2014}
{Duc} P.-A.,  {Paudel} S.,  {McDermid} R.~M.,  {Cuillandre} J.-C.,  {Serra} P.,
   {Bournaud} F.,  {Cappellari} M.,   {Emsellem} E.,  2014, \mn@doi [MNRAS]
  {10.1093/mnras/stu330}, \href
  {http://adsabs.harvard.edu/abs/2014MNRAS.440.1458D} {440, 1458}

\bibitem[\protect\citeauthoryear{{Duffy} et~al.,}{{Duffy}
  et~al.}{2006}]{Duffy_2006}
{Duffy} L.~D.,  et~al., 2006, \mn@doi [Physical Review D]
  {10.1103/PhysRevD.74.012006}, \href
  {http://adsabs.harvard.edu/abs/2006PhRvD..74a2006D} {74, 012006}

\bibitem[\protect\citeauthoryear{{Dyson}, {Eddington}  \& {Davidson}}{{Dyson}
  et~al.}{1920}]{Eddington_1919}
{Dyson} F.~W.,  {Eddington} A.~S.,   {Davidson} C.,  1920, \mn@doi [Royal
  Society of London Philosophical Transactions Series A]
  {10.1098/rsta.1920.0009}, \href
  {http://adsabs.harvard.edu/abs/1920RSPTA.220..291D} {220, 291}

\bibitem[\protect\citeauthoryear{{Einstein}}{{Einstein}}{1905}]{Einstein_1905}
{Einstein} A.,  1905, \mn@doi [Annalen der Physik] {10.1002/andp.19053221004},
  \href {http://adsabs.harvard.edu/abs/1905AnP...322..891E} {322, 891}

\bibitem[\protect\citeauthoryear{{Einstein}}{{Einstein}}{1915}]{Einstein_1915}
{Einstein} A.,  1915, Sitzungsberichte der K{\"o}niglich Preu{\ss}ischen
  Akademie der Wissenschaften (Berlin), Seite 844-847., \href
  {http://adsabs.harvard.edu/abs/1915SPAW.......844E} {pp 844--847}

\bibitem[\protect\citeauthoryear{{Faber} \& {Gallagher}}{{Faber} \&
  {Gallagher}}{1979}]{Faber_1979}
{Faber} S.~M.,  {Gallagher} J.~S.,  1979, \mn@doi [ARA\&A]
  {10.1146/annurev.aa.17.090179.001031}, \href
  {http://adsabs.harvard.edu/abs/1979ARA%26A..17..135F} {17, 135}

\bibitem[\protect\citeauthoryear{{Famaey} \& {Binney}}{{Famaey} \&
  {Binney}}{2005}]{Famaey_Binney_2005}
{Famaey} B.,  {Binney} J.,  2005, \mn@doi [MNRAS]
  {10.1111/j.1365-2966.2005.09474.x}, \href
  {http://adsabs.harvard.edu/abs/2005MNRAS.363..603F} {363, 603}

\bibitem[\protect\citeauthoryear{{Famaey} \& {McGaugh}}{{Famaey} \&
  {McGaugh}}{2012}]{Famaey_McGaugh_2012}
{Famaey} B.,  {McGaugh} S.~S.,  2012, \mn@doi [Living Reviews in Relativity]
  {10.12942/lrr-2012-10}, \href
  {http://adsabs.harvard.edu/abs/2012LRR....15...10F} {15, 10}

\bibitem[\protect\citeauthoryear{{Famaey}, {Bruneton}  \& {Zhao}}{{Famaey}
  et~al.}{2007}]{Famaey_2007}
{Famaey} B.,  {Bruneton} J.-P.,   {Zhao} H.,  2007, \mn@doi [MNRAS]
  {10.1111/j.1745-3933.2007.00308.x}, \href
  {http://adsabs.harvard.edu/abs/2007MNRAS.377L..79F} {377, L79}

\bibitem[\protect\citeauthoryear{{Famaey}, {Khoury}  \& {Penco}}{{Famaey}
  et~al.}{2018}]{Famaey_2017}
{Famaey} B.,  {Khoury} J.,   {Penco} R.,  2018, \mn@doi [JCAP]
  {10.1088/1475-7516/2018/03/038}, \href
  {http://adsabs.harvard.edu/abs/2018JCAP...03..038F} {3, 038}

\bibitem[\protect\citeauthoryear{{Fattahi} et~al.,}{{Fattahi}
  et~al.}{2016}]{Fattahi_2016}
{Fattahi} A.,  et~al., 2016, \mn@doi [MNRAS] {10.1093/mnras/stv2970}, \href
  {http://adsabs.harvard.edu/abs/2016MNRAS.457..844F} {457, 844}

\bibitem[\protect\citeauthoryear{{Fattahi}, {Navarro}, {Frenk}, {Oman},
  {Sawala}  \& {Schaller}}{{Fattahi} et~al.}{2017}]{Fattahi_2018}
{Fattahi} A.,  {Navarro} J.~F.,  {Frenk} C.~S.,  {Oman} K.,  {Sawala} T.,
  {Schaller} M.,  2017, preprint, \href
  {http://adsabs.harvard.edu/abs/2017arXiv170703898F} {Arxiv} (\mn@eprint
  {arXiv} {1707.03898})

\bibitem[\protect\citeauthoryear{{Fermi-LAT Collaboration}}{{Fermi-LAT
  Collaboration}}{2015}]{Ackerman_2015}
{Fermi-LAT Collaboration} 2015, \mn@doi [Physical Review Letters]
  {10.1103/PhysRevLett.115.231301}, \href
  {http://adsabs.harvard.edu/abs/2015PhRvL.115w1301A} {115, 231301}

\bibitem[\protect\citeauthoryear{{Fernando}, {Arias}, {Guglielmo}, {Lewis},
  {Ibata}  \& {Power}}{{Fernando} et~al.}{2017}]{Fernando_2016}
{Fernando} N.,  {Arias} V.,  {Guglielmo} M.,  {Lewis} G.~F.,  {Ibata} R.~A.,
  {Power} C.,  2017, \mn@doi [MNRAS] {10.1093/mnras/stw2694}, \href
  {http://adsabs.harvard.edu/abs/2017MNRAS.465..641F} {465, 641}

\bibitem[\protect\citeauthoryear{{Fernando}, {Arias}, {Lewis}, {Ibata}  \&
  {Power}}{{Fernando} et~al.}{2018}]{Fernando_2018}
{Fernando} N.,  {Arias} V.,  {Lewis} G.~F.,  {Ibata} R.~A.,   {Power} C.,
  2018, \mn@doi [MNRAS] {10.1093/mnras/stx2483}, \href
  {http://adsabs.harvard.edu/abs/2018MNRAS.473.2212F} {473, 2212}

\bibitem[\protect\citeauthoryear{Fletcher \& Powell}{Fletcher \&
  Powell}{1963}]{Fletcher_1963}
Fletcher R.,  Powell M. J.~D.,  1963, \mn@doi [The Computer Journal]
  {10.1093/comjnl/6.2.163}, \href {http://dx.doi.org/10.1093/comjnl/6.2.163}
  {6, 163}

\bibitem[\protect\citeauthoryear{{Flores}, {Hammer}, {Fouquet}, {Puech},
  {Kroupa}, {Yang}  \& {Pawlowski}}{{Flores} et~al.}{2016}]{Flores_2016}
{Flores} H.,  {Hammer} F.,  {Fouquet} S.,  {Puech} M.,  {Kroupa} P.,  {Yang}
  Y.,   {Pawlowski} M.,  2016, \mn@doi [MNRAS] {10.1093/mnrasl/slv189}, \href
  {http://adsabs.harvard.edu/abs/2016MNRAS.457L..14F} {457, L14}

\bibitem[\protect\citeauthoryear{{Francis} \& {Anderson}}{{Francis} \&
  {Anderson}}{2014}]{Francis_2014}
{Francis} C.,  {Anderson} E.,  2014, \mn@doi [Celestial Mechanics and Dynamical
  Astronomy] {10.1007/s10569-014-9541-z}, \href
  {http://adsabs.harvard.edu/abs/2014CeMDA.118..399F} {118, 399}

\bibitem[\protect\citeauthoryear{{Freeman}}{{Freeman}}{1970}]{Freeman_1970}
{Freeman} K.~C.,  1970, \mn@doi [ApJ] {10.1086/150474}, \href
  {http://adsabs.harvard.edu/abs/1970ApJ...160..811F} {160, 811}

\bibitem[\protect\citeauthoryear{{Garaldi}, {Romano-D{\'{\i}}az},
  {Borzyszkowski}  \& {Porciani}}{{Garaldi} et~al.}{2018}]{Garaldi_2017}
{Garaldi} E.,  {Romano-D{\'{\i}}az} E.,  {Borzyszkowski} M.,   {Porciani} C.,
  2018, \mn@doi [MNRAS] {10.1093/mnras/stx2489}, \href
  {http://adsabs.harvard.edu/abs/2018MNRAS.473.2234G} {473, 2234}

\bibitem[\protect\citeauthoryear{{Gentile}, {Famaey}, {Combes}, {Kroupa},
  {Zhao}  \& {Tiret}}{{Gentile} et~al.}{2007}]{Gentile_2007}
{Gentile} G.,  {Famaey} B.,  {Combes} F.,  {Kroupa} P.,  {Zhao} H.~S.,
  {Tiret} O.,  2007, \mn@doi [A\&A] {10.1051/0004-6361:20078081}, \href
  {http://adsabs.harvard.edu/abs/2007A\%26A...472L..25G} {472, L25}

\bibitem[\protect\citeauthoryear{{Gentile}, {Baes}, {Famaey}  \& {van
  Acoleyen}}{{Gentile} et~al.}{2010}]{Gentile_2010}
{Gentile} G.,  {Baes} M.,  {Famaey} B.,   {van Acoleyen} K.,  2010, \mn@doi
  [MNRAS] {10.1111/j.1365-2966.2010.16838.x}, \href
  {http://adsabs.harvard.edu/abs/2010MNRAS.406.2493G} {406, 2493}

\bibitem[\protect\citeauthoryear{{Gilmore} \& {Reid}}{{Gilmore} \&
  {Reid}}{1983}]{Gilmore_1983}
{Gilmore} G.,  {Reid} N.,  1983, \mn@doi [MNRAS] {10.1093/mnras/202.4.1025},
  \href {http://adsabs.harvard.edu/abs/1983MNRAS.202.1025G} {202, 1025}

\bibitem[\protect\citeauthoryear{{G{\'o}mez} et~al.,}{{G{\'o}mez}
  et~al.}{2012}]{Gomez_2012}
{G{\'o}mez} P.~L.,  et~al., 2012, \mn@doi [AJ] {10.1088/0004-6256/144/3/79},
  \href {http://adsabs.harvard.edu/abs/2012AJ....144...79G} {144, 79}

\bibitem[\protect\citeauthoryear{{Gonzalez}, {Clowe}, {Brada{\v c}},
  {Zaritsky}, {Jones}  \& {Markevitch}}{{Gonzalez}
  et~al.}{2009}]{Gonzalez_2009}
{Gonzalez} A.~H.,  {Clowe} D.,  {Brada{\v c}} M.,  {Zaritsky} D.,  {Jones} C.,
   {Markevitch} M.,  2009, \mn@doi [ApJ] {10.1088/0004-637X/691/1/525}, \href
  {http://adsabs.harvard.edu/abs/2009ApJ...691..525G} {691, 525}

\bibitem[\protect\citeauthoryear{{Gregersen} et~al.,}{{Gregersen}
  et~al.}{2015}]{Gregersen_2015}
{Gregersen} D.,  et~al., 2015, \mn@doi [AJ] {10.1088/0004-6256/150/6/189},
  \href {http://adsabs.harvard.edu/abs/2015AJ....150..189G} {150, 189}

\bibitem[\protect\citeauthoryear{{Griest}}{{Griest}}{1993}]{Griest_1993}
{Griest} K.,  1993, in {Akerlof} C.~W.,  {Srednicki} M.~A.,  eds,  Annals of
  the New York Academy of Sciences Vol. 688, Texas/PASCOS '92: Relativistic
  Astrophysics and Particle Cosmology. p.~390 (\mn@eprint {} {hep-ph/9303253}),
  \mn@doi{10.1111/j.1749-6632.1993.tb43912.x}

\bibitem[\protect\citeauthoryear{{Hammer}, {Yang}, {Flores}, {Puech}  \&
  {Fouquet}}{{Hammer} et~al.}{2015}]{Hammer_2015}
{Hammer} F.,  {Yang} Y.~B.,  {Flores} H.,  {Puech} M.,   {Fouquet} S.,  2015,
  \mn@doi [ApJ] {10.1088/0004-637X/813/2/110}, \href
  {http://adsabs.harvard.edu/abs/2015ApJ...813..110H} {813, 110}

\bibitem[\protect\citeauthoryear{{Hayden} et~al.,}{{Hayden}
  et~al.}{2015}]{Hayden_2015}
{Hayden} M.~R.,  et~al., 2015, \mn@doi [ApJ] {10.1088/0004-637X/808/2/132},
  \href {http://adsabs.harvard.edu/abs/2015ApJ...808..132H} {808, 132}

\bibitem[\protect\citeauthoryear{{Hernandez}, {Jim{\'e}nez}  \&
  {Allen}}{{Hernandez} et~al.}{2012}]{Hernandez_2012}
{Hernandez} X.,  {Jim{\'e}nez} M.~A.,   {Allen} C.,  2012, \mn@doi [European
  Physical Journal C] {10.1140/epjc/s10052-012-1884-6}, \href
  {http://adsabs.harvard.edu/abs/2012EPJC...72.1884H} {72, 1884}

\bibitem[\protect\citeauthoryear{{Hodson} \& {Zhao}}{{Hodson} \&
  {Zhao}}{2017}]{Hodson_2017}
{Hodson} A.~O.,  {Zhao} H.,  2017, \mn@doi [A\&A]
  {10.1051/0004-6361/201629358}, \href
  {http://adsabs.harvard.edu/abs/2017A\%26A...598A.127H} {598, A127}

\bibitem[\protect\citeauthoryear{{Hohl}}{{Hohl}}{1971}]{Hohl_1971}
{Hohl} F.,  1971, \mn@doi [ApJ] {10.1086/151091}, \href
  {http://adsabs.harvard.edu/abs/1971ApJ...168..343H} {168, 343}

\bibitem[\protect\citeauthoryear{{Hubble}}{{Hubble}}{1929}]{Hubble_1929}
{Hubble} E.,  1929, \mn@doi [Proceedings of the National Academy of Science]
  {10.1073/pnas.15.3.168}, \href
  {http://adsabs.harvard.edu/abs/1929PNAS...15..168H} {15, 168}

\bibitem[\protect\citeauthoryear{{Ibata} et~al.,}{{Ibata}
  et~al.}{2013}]{Ibata_2013}
{Ibata} R.~A.,  et~al., 2013, \mn@doi [Nature] {10.1038/nature11717}, \href
  {http://adsabs.harvard.edu/abs/2013Natur.493...62I} {493, 62}

\bibitem[\protect\citeauthoryear{{Ibata}, {Ibata}, {Famaey}  \&
  {Lewis}}{{Ibata} et~al.}{2014a}]{Ibata_2014_Nature}
{Ibata} N.~G.,  {Ibata} R.~A.,  {Famaey} B.,   {Lewis} G.~F.,  2014a, \mn@doi
  [Nature] {10.1038/nature13481}, \href
  {http://adsabs.harvard.edu/abs/2014Natur.511..563I} {511, 563}

\bibitem[\protect\citeauthoryear{{Ibata}, {Ibata}, {Lewis}, {Martin}, {Conn},
  {Elahi}, {Arias}  \& {Fernando}}{{Ibata} et~al.}{2014b}]{Ibata_2014}
{Ibata} R.~A.,  {Ibata} N.~G.,  {Lewis} G.~F.,  {Martin} N.~F.,  {Conn} A.,
  {Elahi} P.,  {Arias} V.,   {Fernando} N.,  2014b, \mn@doi [ApJL]
  {10.1088/2041-8205/784/1/L6}, \href
  {http://adsabs.harvard.edu/abs/2014ApJ...784L...6I} {784, L6}

\bibitem[\protect\citeauthoryear{{Ibata}, {Famaey}, {Lewis}, {Ibata}  \&
  {Martin}}{{Ibata} et~al.}{2015}]{Ibata_2015}
{Ibata} R.~A.,  {Famaey} B.,  {Lewis} G.~F.,  {Ibata} N.~G.,   {Martin} N.,
  2015, \mn@doi [ApJ] {10.1088/0004-637X/805/1/67}, \href
  {http://adsabs.harvard.edu/abs/2015ApJ...805...67I} {805, 67}

\bibitem[\protect\citeauthoryear{{Iocco}, {Pato}  \& {Bertone}}{{Iocco}
  et~al.}{2015}]{Iocco_Bertone_2015}
{Iocco} F.,  {Pato} M.,   {Bertone} G.,  2015, \mn@doi [Physical Review D]
  {10.1103/PhysRevD.92.084046}, \href
  {http://adsabs.harvard.edu/abs/2015PhRvD..92h4046I} {92, 084046}

\bibitem[\protect\citeauthoryear{{Iorio}}{{Iorio}}{2017}]{Iorio_2017}
{Iorio} L.,  2017, \mn@doi [European Physical Journal C]
  {10.1140/epjc/s10052-017-4722-z}, \href
  {http://adsabs.harvard.edu/abs/2017EPJC...77..149I} {77, 149}

\bibitem[\protect\citeauthoryear{{Ir{\v s}i{\v c}} et~al.,}{{Ir{\v s}i{\v c}}
  et~al.}{2017}]{Irsic_2017}
{Ir{\v s}i{\v c}} V.,  et~al., 2017, \mn@doi [Physical Review D]
  {10.1103/PhysRevD.96.023522}, \href
  {http://adsabs.harvard.edu/abs/2017PhRvD..96b3522I} {96, 023522}

\bibitem[\protect\citeauthoryear{{Jacobs}, {Rizzi}, {Tully}, {Shaya}, {Makarov}
   \& {Makarova}}{{Jacobs} et~al.}{2009}]{Jacobs_2009}
{Jacobs} B.~A.,  {Rizzi} L.,  {Tully} R.~B.,  {Shaya} E.~J.,  {Makarov} D.~I.,
   {Makarova} L.,  2009, \mn@doi [AJ] {10.1088/0004-6256/138/2/332}, \href
  {http://adsabs.harvard.edu/abs/2009AJ....138..332J} {138, 332}

\bibitem[\protect\citeauthoryear{{Jansen} et~al.,}{{Jansen}
  et~al.}{2001}]{Jansen_2001}
{Jansen} F.,  et~al., 2001, \mn@doi [A\&A] {10.1051/0004-6361:20000036}, \href
  {http://adsabs.harvard.edu/abs/2001A\%26A...365L...1J} {365, L1}

\bibitem[\protect\citeauthoryear{{Jayaraman}, {Gilmore}, {Wyse}, {Norris}  \&
  {Belokurov}}{{Jayaraman} et~al.}{2013}]{Jayaraman_2013}
{Jayaraman} A.,  {Gilmore} G.,  {Wyse} R.~F.~G.,  {Norris} J.~E.,   {Belokurov}
  V.,  2013, \mn@doi [MNRAS] {10.1093/mnras/stt221}, \href
  {http://adsabs.harvard.edu/abs/2013MNRAS.431..930J} {431, 930}

\bibitem[\protect\citeauthoryear{{Jee}, {Hughes}, {Menanteau}, {Sif{\'o}n},
  {Mandelbaum}, {Barrientos}, {Infante}  \& {Ng}}{{Jee}
  et~al.}{2014}]{Jee_2014}
{Jee} M.~J.,  {Hughes} J.~P.,  {Menanteau} F.,  {Sif{\'o}n} C.,  {Mandelbaum}
  R.,  {Barrientos} L.~F.,  {Infante} L.,   {Ng} K.~Y.,  2014, \mn@doi [ApJ]
  {10.1088/0004-637X/785/1/20}, \href
  {http://adsabs.harvard.edu/abs/2014ApJ...785...20J} {785, 20}

\bibitem[\protect\citeauthoryear{{Jungman}, {Kamionkowski}  \&
  {Griest}}{{Jungman} et~al.}{1996}]{Jungman_1996}
{Jungman} G.,  {Kamionkowski} M.,   {Griest} K.,  1996, \mn@doi [Physics
  Reports] {10.1016/0370-1573(95)00058-5}, \href
  {http://adsabs.harvard.edu/abs/1996PhR...267..195J} {267, 195}

\bibitem[\protect\citeauthoryear{{Juri{\'c}} et~al.,}{{Juri{\'c}}
  et~al.}{2008}]{Juric_2008}
{Juri{\'c}} M.,  et~al., 2008, \mn@doi [ApJ] {10.1086/523619}, \href
  {http://adsabs.harvard.edu/abs/2008ApJ...673..864J} {673, 864}

\bibitem[\protect\citeauthoryear{{Kafle}, {Sharma}, {Lewis}  \&
  {Bland-Hawthorn}}{{Kafle} et~al.}{2012}]{Kafle_2012}
{Kafle} P.~R.,  {Sharma} S.,  {Lewis} G.~F.,   {Bland-Hawthorn} J.,  2012,
  \mn@doi [ApJ] {10.1088/0004-637X/761/2/98}, \href
  {http://adsabs.harvard.edu/abs/2012ApJ...761...98K} {761, 98}

\bibitem[\protect\citeauthoryear{{Kahn} \& {Woltjer}}{{Kahn} \&
  {Woltjer}}{1959}]{Kahn_Woltjer_1959}
{Kahn} F.~D.,  {Woltjer} L.,  1959, \mn@doi [ApJ] {10.1086/146762}, \href
  {http://adsabs.harvard.edu/abs/1959ApJ...130..705K} {130, 705}

\bibitem[\protect\citeauthoryear{{Kamada}, {Kaplinghat}, {Pace}  \&
  {Yu}}{{Kamada} et~al.}{2017}]{Kamda_2017}
{Kamada} A.,  {Kaplinghat} M.,  {Pace} A.~B.,   {Yu} H.-B.,  2017, \mn@doi
  [Physical Review Letters] {10.1103/PhysRevLett.119.111102}, \href
  {http://adsabs.harvard.edu/abs/2017PhRvL.119k1102K} {119, 111102}

\bibitem[\protect\citeauthoryear{{Kamionkowski}}{{Kamionkowski}}{1998}]{Kamionkowski_1998}
{Kamionkowski} M.,  1998, in {Gava} E.,  {Masiero} A.,  {Narain} K.~S.,
  {Randjbar-Daemi} S.,  {Senjanovic} G.,  {Smirnov} A.,   {Shafi} Q.,  eds,
  High Energy Physics and Cosmology, 1997 Summer School. p.~394 (\mn@eprint {}
  {hep-ph/9710467})

\bibitem[\protect\citeauthoryear{{Katz}, {McGaugh}, {Teuben}  \&
  {Angus}}{{Katz} et~al.}{2013}]{Katz_2013}
{Katz} H.,  {McGaugh} S.,  {Teuben} P.,   {Angus} G.~W.,  2013, \mn@doi [ApJ]
  {10.1088/0004-637X/772/1/10}, \href
  {http://adsabs.harvard.edu/abs/2013ApJ...772...10K} {772, 10}

\bibitem[\protect\citeauthoryear{{Keller} \& {Wadsley}}{{Keller} \&
  {Wadsley}}{2017}]{Keller_Wadsley_2017}
{Keller} B.~W.,  {Wadsley} J.~W.,  2017, \mn@doi [ApJL]
  {10.3847/2041-8213/835/1/L17}, \href
  {http://adsabs.harvard.edu/abs/2017ApJ...835L..17K} {835, L17}

\bibitem[\protect\citeauthoryear{{Keller}, {Wadsley}  \& {Couchman}}{{Keller}
  et~al.}{2016}]{Keller_Wadsley_2016}
{Keller} B.~W.,  {Wadsley} J.,   {Couchman} H.~M.~P.,  2016, \mn@doi [MNRAS]
  {10.1093/mnras/stw2029}, \href
  {http://adsabs.harvard.edu/abs/2016MNRAS.463.1431K} {463, 1431}

\bibitem[\protect\citeauthoryear{{Kerins} \& {Carr}}{{Kerins} \&
  {Carr}}{1995}]{Kerins_1995}
{Kerins} E.~J.,  {Carr} B.~J.,  1995, \mn@doi [Nuclear Physics B: Proceedings
  Supplements] {10.1016/0920-5632(95)00469-P}, \href
  {http://adsabs.harvard.edu/abs/1995NuPhS..43..157K} {43, 157}

\bibitem[\protect\citeauthoryear{{Kervella}, {Mignard}, {M{\'e}rand}  \&
  {Th{\'e}venin}}{{Kervella} et~al.}{2016}]{Kervella_2016}
{Kervella} P.,  {Mignard} F.,  {M{\'e}rand} A.,   {Th{\'e}venin} F.,  2016,
  \mn@doi [A\&A] {10.1051/0004-6361/201629201}, \href
  {http://adsabs.harvard.edu/abs/2016A\%26A...594A.107K} {594, A107}

\bibitem[\protect\citeauthoryear{{Kervella}, {Th{\'e}venin}  \&
  {Lovis}}{{Kervella} et~al.}{2017}]{Kervella_2017}
{Kervella} P.,  {Th{\'e}venin} F.,   {Lovis} C.,  2017, \mn@doi [A\&A]
  {10.1051/0004-6361/201629930}, \href
  {http://adsabs.harvard.edu/abs/2017A\%26A...598L...7K} {598, L7}

\bibitem[\protect\citeauthoryear{{Khoury}}{{Khoury}}{2016}]{Khoury_2016}
{Khoury} J.,  2016, \mn@doi [Physical Review D] {10.1103/PhysRevD.93.103533},
  \href {http://adsabs.harvard.edu/abs/2016PhRvD..93j3533K} {93, 103533}

\bibitem[\protect\citeauthoryear{{Kirby}, {Bullock}, {Boylan-Kolchin},
  {Kaplinghat}  \& {Cohen}}{{Kirby} et~al.}{2014}]{Kirby_2014}
{Kirby} E.~N.,  {Bullock} J.~S.,  {Boylan-Kolchin} M.,  {Kaplinghat} M.,
  {Cohen} J.~G.,  2014, \mn@doi [MNRAS] {10.1093/mnras/stu025}, \href
  {http://adsabs.harvard.edu/abs/2014MNRAS.439.1015K} {439, 1015}

\bibitem[\protect\citeauthoryear{{Klimentowski}, {{\L}okas}, {Knebe},
  {Gottl{\"o}ber}, {Martinez-Vaquero}, {Yepes}  \& {Hoffman}}{{Klimentowski}
  et~al.}{2010}]{Klimentowski_2010}
{Klimentowski} J.,  {{\L}okas} E.~L.,  {Knebe} A.,  {Gottl{\"o}ber} S.,
  {Martinez-Vaquero} L.~A.,  {Yepes} G.,   {Hoffman} Y.,  2010, \mn@doi [MNRAS]
  {10.1111/j.1365-2966.2009.16024.x}, \href
  {http://adsabs.harvard.edu/abs/2010MNRAS.402.1899K} {402, 1899}

\bibitem[\protect\citeauthoryear{{Klypin}, {Kravtsov}, {Valenzuela}  \&
  {Prada}}{{Klypin} et~al.}{1999}]{Klypin_1999}
{Klypin} A.,  {Kravtsov} A.~V.,  {Valenzuela} O.,   {Prada} F.,  1999, \mn@doi
  [ApJ] {10.1086/307643}, \href
  {http://adsabs.harvard.edu/abs/1999ApJ...522...82K} {522, 82}

\bibitem[\protect\citeauthoryear{{Komatsu} et~al.,}{{Komatsu}
  et~al.}{2011}]{Komatsu_2011}
{Komatsu} E.,  et~al., 2011, \mn@doi [ApJS] {10.1088/0067-0049/192/2/18}, \href
  {http://adsabs.harvard.edu/abs/2011ApJS..192...18K} {192, 18}

\bibitem[\protect\citeauthoryear{{Kormendy}, {Drory}, {Bender}  \&
  {Cornell}}{{Kormendy} et~al.}{2010}]{Kormendy_2010}
{Kormendy} J.,  {Drory} N.,  {Bender} R.,   {Cornell} M.~E.,  2010, \mn@doi
  [ApJ] {10.1088/0004-637X/723/1/54}, \href
  {http://adsabs.harvard.edu/abs/2010ApJ...723...54K} {723, 54}

\bibitem[\protect\citeauthoryear{{Kraljic} \& {Sarkar}}{{Kraljic} \&
  {Sarkar}}{2015}]{Kraljic_2015}
{Kraljic} D.,  {Sarkar} S.,  2015, \mn@doi [JCAP]
  {10.1088/1475-7516/2015/04/050}, \href
  {http://adsabs.harvard.edu/abs/2015JCAP...04..050K} {4, 050}

\bibitem[\protect\citeauthoryear{{Kroupa}}{{Kroupa}}{2015}]{Kroupa_2015}
{Kroupa} P.,  2015, \mn@doi [Canadian Journal of Physics]
  {10.1139/cjp-2014-0179}, \href
  {http://adsabs.harvard.edu/abs/2015CaJPh..93..169K} {93, 169}

\bibitem[\protect\citeauthoryear{{Kroupa}, {Theis}  \& {Boily}}{{Kroupa}
  et~al.}{2005}]{Kroupa_2005}
{Kroupa} P.,  {Theis} C.,   {Boily} C.~M.,  2005, \mn@doi [A\&A]
  {10.1051/0004-6361:20041122}, \href
  {http://adsabs.harvard.edu/abs/2005A\%26A...431..517K} {431, 517}

\bibitem[\protect\citeauthoryear{{Kunder} et~al.,}{{Kunder}
  et~al.}{2017}]{Kunder_2017}
{Kunder} A.,  et~al., 2017, \mn@doi [AJ] {10.3847/1538-3881/153/2/75}, \href
  {http://adsabs.harvard.edu/abs/2017AJ....153...75K} {153, 75}

\bibitem[\protect\citeauthoryear{{LIGO Collaboration}}{{LIGO
  Collaboration}}{2016}]{LIGO_2016}
{LIGO Collaboration} 2016, \mn@doi [Physical Review Letters]
  {10.1103/PhysRevLett.116.061102}, \href
  {http://adsabs.harvard.edu/abs/2016PhRvL.116f1102A} {116, 061102}

\bibitem[\protect\citeauthoryear{{LUX Collaboration}}{{LUX
  Collaboration}}{2017}]{LUX_2016}
{LUX Collaboration} 2017, \mn@doi [Physical Review Letters]
  {10.1103/PhysRevLett.118.021303}, \href
  {http://adsabs.harvard.edu/abs/2017PhRvL.118b1303A} {118, 021303}

\bibitem[\protect\citeauthoryear{{Lage} \& {Farrar}}{{Lage} \&
  {Farrar}}{2014}]{Lage_Farrar_2014}
{Lage} C.,  {Farrar} G.,  2014, \mn@doi [ApJ] {10.1088/0004-637X/787/2/144},
  \href {http://adsabs.harvard.edu/abs/2014ApJ...787..144L} {787, 144}

\bibitem[\protect\citeauthoryear{{Lelli}, {McGaugh}  \& {Schombert}}{{Lelli}
  et~al.}{2016}]{SPARC}
{Lelli} F.,  {McGaugh} S.~S.,   {Schombert} J.~M.,  2016, \mn@doi [AJ]
  {10.3847/0004-6256/152/6/157}, \href
  {http://adsabs.harvard.edu/abs/2016AJ....152..157L} {152, 157}

\bibitem[\protect\citeauthoryear{{Lelli}, {McGaugh}  \& {Schombert}}{{Lelli}
  et~al.}{2017a}]{Verlinde_test_2017}
{Lelli} F.,  {McGaugh} S.~S.,   {Schombert} J.~M.,  2017a, \mn@doi [MNRAS]
  {10.1093/mnrasl/slx031}, \href
  {http://adsabs.harvard.edu/abs/2017MNRAS.468L..68L} {468, L68}

\bibitem[\protect\citeauthoryear{{Lelli}, {McGaugh}, {Schombert}  \&
  {Pawlowski}}{{Lelli} et~al.}{2017b}]{Lelli_2017}
{Lelli} F.,  {McGaugh} S.~S.,  {Schombert} J.~M.,   {Pawlowski} M.~S.,  2017b,
  \mn@doi [ApJ] {10.3847/1538-4357/836/2/152}, \href
  {http://adsabs.harvard.edu/abs/2017ApJ...836..152L} {836, 152}

\bibitem[\protect\citeauthoryear{{Lena}, {Robinson}, {Marconi}, {Axon},
  {Capetti}, {Merritt}  \& {Batcheldor}}{{Lena} et~al.}{2014}]{Lena_2014}
{Lena} D.,  {Robinson} A.,  {Marconi} A.,  {Axon} D.~J.,  {Capetti} A.,
  {Merritt} D.,   {Batcheldor} D.,  2014, \mn@doi [ApJ]
  {10.1088/0004-637X/795/2/146}, \href
  {http://adsabs.harvard.edu/abs/2014ApJ...795..146L} {795, 146}

\bibitem[\protect\citeauthoryear{{Lenoir}, {Christophe}  \& {Reynaud}}{{Lenoir}
  et~al.}{2011}]{Lenoir_2011}
{Lenoir} B.,  {Christophe} B.,   {Reynaud} S.,  2011, preprint, \href
  {http://adsabs.harvard.edu/abs/2011arXiv1107.0861L} {Arxiv} (\mn@eprint
  {arXiv} {1107.0861})

\bibitem[\protect\citeauthoryear{{Leonard} \& {Tremaine}}{{Leonard} \&
  {Tremaine}}{1990}]{Leonard_1990}
{Leonard} P.~J.~T.,  {Tremaine} S.,  1990, \mn@doi [ApJ] {10.1086/168638},
  \href {http://adsabs.harvard.edu/abs/1990ApJ...353..486L} {353, 486}

\bibitem[\protect\citeauthoryear{{Lin} \& {Shu}}{{Lin} \&
  {Shu}}{1964}]{Lin_Shu_1964}
{Lin} C.~C.,  {Shu} F.~H.,  1964, \mn@doi [ApJ] {10.1086/147955}, \href
  {http://adsabs.harvard.edu/abs/1964ApJ...140..646L} {140, 646}

\bibitem[\protect\citeauthoryear{{L{\'o}pez-Corredoira} \&
  {Kroupa}}{{L{\'o}pez-Corredoira} \& {Kroupa}}{2016}]{Lopez_Corredoira_2016}
{L{\'o}pez-Corredoira} M.,  {Kroupa} P.,  2016, \mn@doi [ApJ]
  {10.3847/0004-637X/817/1/75}, \href
  {http://adsabs.harvard.edu/abs/2016ApJ...817...75L} {817, 75}

\bibitem[\protect\citeauthoryear{{L{\"u}ghausen}, {Famaey}  \&
  {Kroupa}}{{L{\"u}ghausen} et~al.}{2015}]{PoR}
{L{\"u}ghausen} F.,  {Famaey} B.,   {Kroupa} P.,  2015, \mn@doi [Canadian
  Journal of Physics] {10.1139/cjp-2014-0168}, \href
  {http://adsabs.harvard.edu/abs/2015CaJPh..93..232L} {93, 232}

\bibitem[\protect\citeauthoryear{{Ma} et~al.,}{{Ma} et~al.}{1998}]{Ma_1998}
{Ma} C.,  et~al., 1998, \mn@doi [AJ] {10.1086/300408}, \href
  {http://adsabs.harvard.edu/abs/1998AJ....116..516M} {116, 516}

\bibitem[\protect\citeauthoryear{{Ma}, {Wu}, {Wang}, {Fan}, {Zhou}, {Wu},
  {Jiang}  \& {Chen}}{{Ma} et~al.}{2010}]{Ma_2010}
{Ma} J.,  {Wu} Z.,  {Wang} S.,  {Fan} Z.,  {Zhou} X.,  {Wu} J.,  {Jiang} Z.,
  {Chen} J.,  2010, \mn@doi [PASP] {10.1086/656567}, \href
  {http://adsabs.harvard.edu/abs/2010PASP..122.1164M} {122, 1164}

\bibitem[\protect\citeauthoryear{{Maji}, {Zhu}, {Marinacci}  \& {Li}}{{Maji}
  et~al.}{2017}]{Maji_2017}
{Maji} M.,  {Zhu} Q.,  {Marinacci} F.,   {Li} Y.,  2017, preprint, \href
  {http://adsabs.harvard.edu/abs/2017arXiv170200497M} {Arxiv} (\mn@eprint
  {arXiv} {1702.00497})

\bibitem[\protect\citeauthoryear{{Makarov}, {Makarova}  \& {Uklein}}{{Makarov}
  et~al.}{2013}]{Makarov_2013}
{Makarov} D.~I.,  {Makarova} L.~N.,   {Uklein} R.~I.,  2013, \mn@doi
  [Astrophysical Bulletin] {10.1134/S1990341313020016}, \href
  {http://adsabs.harvard.edu/abs/2013AstBu..68..125M} {68, 125}

\bibitem[\protect\citeauthoryear{{Martig}, {Bournaud}, {Croton}, {Dekel}  \&
  {Teyssier}}{{Martig} et~al.}{2012}]{Martig_2012}
{Martig} M.,  {Bournaud} F.,  {Croton} D.~J.,  {Dekel} A.,   {Teyssier} R.,
  2012, \mn@doi [ApJ] {10.1088/0004-637X/756/1/26}, \href
  {http://adsabs.harvard.edu/abs/2012ApJ...756...26M} {756, 26}

\bibitem[\protect\citeauthoryear{{Massey}, {Henning}  \&
  {Kraan-Korteweg}}{{Massey} et~al.}{2003}]{Massey_2003}
{Massey} P.,  {Henning} P.~A.,   {Kraan-Korteweg} R.~C.,  2003, \mn@doi [AJ]
  {10.1086/378908}, \href {http://adsabs.harvard.edu/abs/2003AJ....126.2362M}
  {126, 2362}

\bibitem[\protect\citeauthoryear{{Mastropietro} \& {Burkert}}{{Mastropietro} \&
  {Burkert}}{2008}]{Mastropietro_Burkert_2008}
{Mastropietro} C.,  {Burkert} A.,  2008, \mn@doi [MNRAS]
  {10.1111/j.1365-2966.2008.13626.x}, \href
  {http://adsabs.harvard.edu/abs/2008MNRAS.389..967M} {389, 967}

\bibitem[\protect\citeauthoryear{{McConnachie}}{{McConnachie}}{2012}]{McConnachie_2012}
{McConnachie} A.~W.,  2012, \mn@doi [AJ] {10.1088/0004-6256/144/1/4}, \href
  {http://adsabs.harvard.edu/abs/2012AJ....144....4M} {144, 4}

\bibitem[\protect\citeauthoryear{{McConnachie} et~al.,}{{McConnachie}
  et~al.}{2009}]{PANDAS}
{McConnachie} A.~W.,  et~al., 2009, \mn@doi [Nature] {10.1038/nature08327},
  \href {http://adsabs.harvard.edu/abs/2009Natur.461...66M} {461, 66}

\bibitem[\protect\citeauthoryear{{McGaugh}}{{McGaugh}}{1996}]{McGaugh_1996}
{McGaugh} S.~S.,  1996, MNRAS, \href
  {http://adsabs.harvard.edu/abs/1996MNRAS.280..337M} {280, 337}

\bibitem[\protect\citeauthoryear{{McGaugh}}{{McGaugh}}{2008}]{McGaugh_2008}
{McGaugh} S.~S.,  2008, \mn@doi [ApJ] {10.1086/589148}, \href
  {http://adsabs.harvard.edu/abs/2008ApJ...683..137M} {683, 137}

\bibitem[\protect\citeauthoryear{{McGaugh}}{{McGaugh}}{2011}]{McGaugh_2011}
{McGaugh} S.~S.,  2011, \mn@doi [Physical Review Letters]
  {10.1103/PhysRevLett.106.121303}, \href
  {http://adsabs.harvard.edu/abs/2011PhRvL.106l1303M} {106, 121303}

\bibitem[\protect\citeauthoryear{{McGaugh}}{{McGaugh}}{2016a}]{McGaugh_2016_MW}
{McGaugh} S.~S.,  2016a, \mn@doi [ApJ] {10.3847/0004-637X/816/1/42}, \href
  {http://adsabs.harvard.edu/abs/2016ApJ...816...42M} {816, 42}

\bibitem[\protect\citeauthoryear{{McGaugh}}{{McGaugh}}{2016b}]{McGaugh_2016}
{McGaugh} S.~S.,  2016b, \mn@doi [ApJL] {10.3847/2041-8205/832/1/L8}, \href
  {http://adsabs.harvard.edu/abs/2016ApJ...832L...8M} {832, L8}

\bibitem[\protect\citeauthoryear{{McGaugh} \& {Milgrom}}{{McGaugh} \&
  {Milgrom}}{2013}]{McGaugh_2013}
{McGaugh} S.,  {Milgrom} M.,  2013, \mn@doi [ApJ]
  {10.1088/0004-637X/775/2/139}, \href
  {http://adsabs.harvard.edu/abs/2013ApJ...775..139M} {775, 139}

\bibitem[\protect\citeauthoryear{{McGaugh} \& {Wolf}}{{McGaugh} \&
  {Wolf}}{2010}]{McGaugh_2010}
{McGaugh} S.~S.,  {Wolf} J.,  2010, \mn@doi [ApJ]
  {10.1088/0004-637X/722/1/248}, \href
  {http://adsabs.harvard.edu/abs/2010ApJ...722..248M} {722, 248}

\bibitem[\protect\citeauthoryear{{McGaugh}, {Schombert}  \& {Bothun}}{{McGaugh}
  et~al.}{1995}]{McGaugh_1995}
{McGaugh} S.~S.,  {Schombert} J.~M.,   {Bothun} G.~D.,  1995, \mn@doi [AJ]
  {10.1086/117427}, \href {http://adsabs.harvard.edu/abs/1995AJ....109.2019M}
  {109, 2019}

\bibitem[\protect\citeauthoryear{{McGaugh}, {Lelli}  \& {Schombert}}{{McGaugh}
  et~al.}{2016}]{McGaugh_Lelli_2016}
{McGaugh} S.,  {Lelli} F.,   {Schombert} J.,  2016, \mn@doi [Phys. Rev. Lett.]
  {10.1103/PhysRevLett.117.201101}, \href
  {http://adsabs.harvard.edu/abs/2016arXiv160905917M} {117, 201101}

\bibitem[\protect\citeauthoryear{{McMillan}}{{McMillan}}{2011}]{McMillan_2011}
{McMillan} P.~J.,  2011, \mn@doi [MNRAS] {10.1111/j.1365-2966.2011.18564.x},
  \href {http://adsabs.harvard.edu/abs/2011MNRAS.414.2446M} {414, 2446}

\bibitem[\protect\citeauthoryear{{McMillan}}{{McMillan}}{2017}]{McMillan_2017}
{McMillan} P.~J.,  2017, \mn@doi [MNRAS] {10.1093/mnras/stw2759}, \href
  {http://adsabs.harvard.edu/abs/2017MNRAS.465...76M} {465, 76}

\bibitem[\protect\citeauthoryear{{McQuinn} et~al.,}{{McQuinn}
  et~al.}{2015}]{McQuinn_2015}
{McQuinn} K.~B.~W.,  et~al., 2015, \mn@doi [ApJ] {10.1088/0004-637X/812/2/158},
  \href {http://adsabs.harvard.edu/abs/2015ApJ...812..158M} {812, 158}

\bibitem[\protect\citeauthoryear{{Menanteau} et~al.,}{{Menanteau}
  et~al.}{2012}]{Menanteau_2012}
{Menanteau} F.,  et~al., 2012, \mn@doi [ApJ] {10.1088/0004-637X/748/1/7}, \href
  {http://adsabs.harvard.edu/abs/2012ApJ...748....7M} {748, 7}

\bibitem[\protect\citeauthoryear{{Merali}}{{Merali}}{2016}]{Merali_2016}
{Merali} Z.,  2016, Science, \href
  {http://adsabs.harvard.edu/abs/2016Sci...352.1040M} {352, 1040}

\bibitem[\protect\citeauthoryear{{Milgrom}}{{Milgrom}}{1983}]{Milgrom_1983}
{Milgrom} M.,  1983, \mn@doi [ApJ] {10.1086/161130}, \href
  {http://adsabs.harvard.edu/abs/1983ApJ...270..365M} {270, 365}

\bibitem[\protect\citeauthoryear{{Milgrom}}{{Milgrom}}{1986}]{Milgrom_1986}
{Milgrom} M.,  1986, \mn@doi [ApJ] {10.1086/164021}, \href
  {http://adsabs.harvard.edu/abs/1986ApJ...302..617M} {302, 617}

\bibitem[\protect\citeauthoryear{{Milgrom}}{{Milgrom}}{1999}]{Milgrom_1999}
{Milgrom} M.,  1999, \mn@doi [Phys. Lett. A] {10.1016/S0375-9601(99)00077-8},
  \href {http://adsabs.harvard.edu/abs/1999PhLA..253..273M} {253, 273}

\bibitem[\protect\citeauthoryear{{Milgrom}}{{Milgrom}}{2010}]{QUMOND}
{Milgrom} M.,  2010, \mn@doi [MNRAS] {10.1111/j.1365-2966.2009.16184.x}, \href
  {http://adsabs.harvard.edu/abs/2010MNRAS.403..886M} {403, 886}

\bibitem[\protect\citeauthoryear{{Milgrom}}{{Milgrom}}{2013}]{Milgrom_2013}
{Milgrom} M.,  2013, \mn@doi [Physical Review Letters]
  {10.1103/PhysRevLett.111.041105}, \href
  {http://adsabs.harvard.edu/abs/2013PhRvL.111d1105M} {111, 041105}

\bibitem[\protect\citeauthoryear{{Milgrom}}{{Milgrom}}{2014}]{Milgrom_2014}
{Milgrom} M.,  2014, \mn@doi [Physical Review D] {10.1103/PhysRevD.89.024027},
  \href {http://adsabs.harvard.edu/abs/2014PhRvD..89b4027M} {89, 024027}

\bibitem[\protect\citeauthoryear{{Milgrom}}{{Milgrom}}{2016}]{Milgrom_2016}
{Milgrom} M.,  2016, preprint, \href
  {http://adsabs.harvard.edu/abs/2016arXiv161007538M} {Arxiv} (\mn@eprint
  {arXiv} {1610.07538})

\bibitem[\protect\citeauthoryear{{Mirabel}, {Dottori}  \& {Lutz}}{{Mirabel}
  et~al.}{1992}]{Mirabel_1992}
{Mirabel} I.~F.,  {Dottori} H.,   {Lutz} D.,  1992, A\&A, \href
  {http://adsabs.harvard.edu/abs/1992A\%26A...256L..19M} {256, L19}

\bibitem[\protect\citeauthoryear{{Molnar} \& {Broadhurst}}{{Molnar} \&
  {Broadhurst}}{2015}]{Molnar_2014}
{Molnar} S.~M.,  {Broadhurst} T.,  2015, \mn@doi [ApJ]
  {10.1088/0004-637X/800/1/37}, \href
  {http://adsabs.harvard.edu/abs/2015ApJ...800...37M} {800, 37}

\bibitem[\protect\citeauthoryear{{Molnar}, {Chiu}, {Broadhurst}  \&
  {Stadel}}{{Molnar} et~al.}{2013a}]{Molnar_2013}
{Molnar} S.~M.,  {Chiu} I.-N.~T.,  {Broadhurst} T.,   {Stadel} J.~G.,  2013a,
  \mn@doi [ApJ] {10.1088/0004-637X/779/1/63}, \href
  {http://adsabs.harvard.edu/abs/2013ApJ...779...63M} {779, 63}

\bibitem[\protect\citeauthoryear{{Molnar}, {Chiu}, {Broadhurst}  \&
  {Stadel}}{{Molnar} et~al.}{2013b}]{Molnar_2013_Abell}
{Molnar} S.~M.,  {Chiu} I.-N.~T.,  {Broadhurst} T.,   {Stadel} J.~G.,  2013b,
  \mn@doi [ApJ] {10.1088/0004-637X/779/1/63}, \href
  {http://adsabs.harvard.edu/abs/2013ApJ...779...63M} {779, 63}

\bibitem[\protect\citeauthoryear{{M{\"u}ller}, {Pawlowski}, {Jerjen}  \&
  {Lelli}}{{M{\"u}ller} et~al.}{2018}]{Muller_2018}
{M{\"u}ller} O.,  {Pawlowski} M.~S.,  {Jerjen} H.,   {Lelli} F.,  2018, \mn@doi
  [Science] {10.1126/science.aao1858}, \href
  {http://adsabs.harvard.edu/abs/2018arXiv180200081M} {359, 534}

\bibitem[\protect\citeauthoryear{{NANOGrav Collaboration}}{{NANOGrav
  Collaboration}}{2018}]{NANOGrav_2018}
{NANOGrav Collaboration} 2018, preprint, \href
  {http://adsabs.harvard.edu/abs/2018arXiv180102617A} {Arxiv} (\mn@eprint
  {arXiv} {1801.02617})

\bibitem[\protect\citeauthoryear{{Nesseris}, {Pantazis}  \&
  {Perivolaropoulos}}{{Nesseris} et~al.}{2017}]{Nesseris_2017}
{Nesseris} S.,  {Pantazis} G.,   {Perivolaropoulos} L.,  2017, \mn@doi
  [Physical Review D] {10.1103/PhysRevD.96.023542}, \href
  {http://adsabs.harvard.edu/abs/2017PhRvD..96b3542N} {96, 023542}

\bibitem[\protect\citeauthoryear{{Nicastro}, {Senatore}, {Krongold}, {Mathur}
  \& {Elvis}}{{Nicastro} et~al.}{2016}]{Nicastro_2016}
{Nicastro} F.,  {Senatore} F.,  {Krongold} Y.,  {Mathur} S.,   {Elvis} M.,
  2016, \mn@doi [ApJL] {10.3847/2041-8205/828/1/L12}, \href
  {http://adsabs.harvard.edu/abs/2016ApJ...828L..12N} {828, L12}

\bibitem[\protect\citeauthoryear{{Noguchi}}{{Noguchi}}{1999}]{Noguchi_1999}
{Noguchi} M.,  1999, \mn@doi [ApJ] {10.1086/306932}, \href
  {http://adsabs.harvard.edu/abs/1999ApJ...514...77N} {514, 77}

\bibitem[\protect\citeauthoryear{{Norris} et~al.,}{{Norris}
  et~al.}{2016}]{Norris_2016}
{Norris} M.~A.,  et~al., 2016, \mn@doi [ApJ] {10.3847/0004-637X/832/2/198},
  \href {http://adsabs.harvard.edu/abs/2016ApJ...832..198N} {832, 198}

\bibitem[\protect\citeauthoryear{{Okazaki} \& {Taniguchi}}{{Okazaki} \&
  {Taniguchi}}{2000}]{Okazaki_2000}
{Okazaki} T.,  {Taniguchi} Y.,  2000, \mn@doi [ApJ] {10.1086/317109}, \href
  {http://adsabs.harvard.edu/abs/2000ApJ...543..149O} {543, 149}

\bibitem[\protect\citeauthoryear{{Olling} \& {Merrifield}}{{Olling} \&
  {Merrifield}}{2001}]{Olling_Merrifield_2001}
{Olling} R.~P.,  {Merrifield} M.~R.,  2001, \mn@doi [MNRAS]
  {10.1046/j.1365-8711.2001.04581.x}, \href
  {http://adsabs.harvard.edu/abs/2001MNRAS.326..164O} {326, 164}

\bibitem[\protect\citeauthoryear{{Ostriker} \& {Peebles}}{{Ostriker} \&
  {Peebles}}{1973}]{Ostriker_Peebles_1973}
{Ostriker} J.~P.,  {Peebles} P.~J.~E.,  1973, \mn@doi [ApJ] {10.1086/152513},
  \href {http://adsabs.harvard.edu/abs/1973ApJ...186..467O} {186, 467}

\bibitem[\protect\citeauthoryear{{Ostriker} \& {Steinhardt}}{{Ostriker} \&
  {Steinhardt}}{1995}]{Ostriker_Steinhardt_1995}
{Ostriker} J.~P.,  {Steinhardt} P.~J.,  1995, \mn@doi [Nature]
  {10.1038/377600a0}, \href {http://adsabs.harvard.edu/abs/1995Natur.377..600O}
  {377, 600}

\bibitem[\protect\citeauthoryear{{Paczynski}}{{Paczynski}}{1986}]{Paczynski_1986}
{Paczynski} B.,  1986, \mn@doi [ApJ] {10.1086/164140}, \href
  {http://adsabs.harvard.edu/abs/1986ApJ...304....1P} {304, 1}

\bibitem[\protect\citeauthoryear{{PandaX-II Collaboration}}{{PandaX-II
  Collaboration}}{2016}]{PandaX_2016}
{PandaX-II Collaboration} 2016, \mn@doi [Phys. Rev. Lett.]
  {10.1103/PhysRevLett.117.121303}, \href
  {http://adsabs.harvard.edu/abs/2016arXiv160707400P} {117, 121303}

\bibitem[\protect\citeauthoryear{{Papastergis}, {Adams}  \& {van der
  Hulst}}{{Papastergis} et~al.}{2016}]{Papastergis_2016}
{Papastergis} E.,  {Adams} E.~A.~K.,   {van der Hulst} J.~M.,  2016, \mn@doi
  [A\&A] {10.1051/0004-6361/201628410}, \href
  {http://adsabs.harvard.edu/abs/2016A\%26A...593A..39P} {593, A39}

\bibitem[\protect\citeauthoryear{{Pawlowski}}{{Pawlowski}}{2016}]{Pawlowski_2016}
{Pawlowski} M.~S.,  2016, \mn@doi [MNRAS] {10.1093/mnras/stv2673}, \href
  {http://adsabs.harvard.edu/abs/2016MNRAS.456..448P} {456, 448}

\bibitem[\protect\citeauthoryear{{Pawlowski} \& {Kroupa}}{{Pawlowski} \&
  {Kroupa}}{2013}]{Kroupa_2013}
{Pawlowski} M.~S.,  {Kroupa} P.,  2013, \mn@doi [MNRAS]
  {10.1093/mnras/stt1429}, \href
  {http://adsabs.harvard.edu/abs/2013MNRAS.435.2116P} {435, 2116}

\bibitem[\protect\citeauthoryear{{Pawlowski} et~al.,}{{Pawlowski}
  et~al.}{2014}]{Pawlowski_2014}
{Pawlowski} M.~S.,  et~al., 2014, \mn@doi [MNRAS] {10.1093/mnras/stu1005},
  \href {http://adsabs.harvard.edu/abs/2014MNRAS.442.2362P} {442, 2362}

\bibitem[\protect\citeauthoryear{{Pawlowski}, {Famaey}, {Merritt}  \&
  {Kroupa}}{{Pawlowski} et~al.}{2015}]{Pawlowski_2015}
{Pawlowski} M.~S.,  {Famaey} B.,  {Merritt} D.,   {Kroupa} P.,  2015, \mn@doi
  [ApJ] {10.1088/0004-637X/815/1/19}, \href
  {http://adsabs.harvard.edu/abs/2015ApJ...815...19P} {815, 19}

\bibitem[\protect\citeauthoryear{{Pawlowski} et~al.,}{{Pawlowski}
  et~al.}{2017}]{Pawlowski_2017}
{Pawlowski} M.~S.,  et~al., 2017, \mn@doi [Astronomische Nachrichten]
  {10.1002/asna.201713366}, \href
  {http://adsabs.harvard.edu/abs/2017AN....338..854P} {338, 854}

\bibitem[\protect\citeauthoryear{{Pazy}}{{Pazy}}{2013}]{Pazy_2013}
{Pazy} E.,  2013, \mn@doi [Phys. Rev. D] {10.1103/PhysRevD.87.084063}, \href
  {http://adsabs.harvard.edu/abs/2013PhRvD..87h4063P} {87, 084063}

\bibitem[\protect\citeauthoryear{{Pe{\~n}arrubia} \&
  {Fattahi}}{{Pe{\~n}arrubia} \& {Fattahi}}{2017}]{Fattahi_2017}
{Pe{\~n}arrubia} J.,  {Fattahi} A.,  2017, \mn@doi [MNRAS]
  {10.1093/mnras/stx323}, \href
  {http://adsabs.harvard.edu/abs/2017MNRAS.468.1300P} {468, 1300}

\bibitem[\protect\citeauthoryear{{Pe{\~n}arrubia}, {Ma}, {Walker}  \&
  {McConnachie}}{{Pe{\~n}arrubia} et~al.}{2014}]{Jorge_2014}
{Pe{\~n}arrubia} J.,  {Ma} Y.-Z.,  {Walker} M.~G.,   {McConnachie} A.,  2014,
  \mn@doi [MNRAS] {10.1093/mnras/stu879}, \href
  {http://adsabs.harvard.edu/abs/2014MNRAS.443.2204P} {443, 2204}

\bibitem[\protect\citeauthoryear{{Peebles}}{{Peebles}}{2017}]{Peebles_2017}
{Peebles} P.~J.~E.,  2017, preprint, \href
  {http://adsabs.harvard.edu/abs/2017arXiv170510683P} {Arxiv} (\mn@eprint
  {arXiv} {1705.10683})

\bibitem[\protect\citeauthoryear{{Peebles} \& {Tully}}{{Peebles} \&
  {Tully}}{2013}]{Peebles_2013}
{Peebles} P.~J.~E.,  {Tully} R.~B.,  2013, preprint, \href
  {http://adsabs.harvard.edu/abs/2013arXiv1302.6982P} {Arxiv} (\mn@eprint
  {arXiv} {1302.6982})

\bibitem[\protect\citeauthoryear{{Peebles}, {Tully}  \& {Shaya}}{{Peebles}
  et~al.}{2011}]{Shaya_2011}
{Peebles} P.~J.~E.,  {Tully} R.~B.,   {Shaya} E.~J.,  2011, preprint, \href
  {http://adsabs.harvard.edu/abs/2011arXiv1105.5596P} {Arxiv} (\mn@eprint
  {arXiv} {1105.5596})

\bibitem[\protect\citeauthoryear{{Peng}, {Maiolino}  \& {Cochrane}}{{Peng}
  et~al.}{2015}]{Rachel_2015}
{Peng} Y.,  {Maiolino} R.,   {Cochrane} R.,  2015, \mn@doi [Nature]
  {10.1038/nature14439}, \href
  {http://adsabs.harvard.edu/abs/2015arXiv150503143P} {521, 192}

\bibitem[\protect\citeauthoryear{{Perryman} et~al.,}{{Perryman}
  et~al.}{2001}]{Perryman_2001}
{Perryman} M.~A.~C.,  et~al., 2001, \mn@doi [A\&A]
  {10.1051/0004-6361:20010085}, \href
  {http://adsabs.harvard.edu/abs/2001A\%26A...369..339P} {369, 339}

\bibitem[\protect\citeauthoryear{{Phelps}, {Nusser}  \& {Desjacques}}{{Phelps}
  et~al.}{2013}]{Phelps_2013}
{Phelps} S.,  {Nusser} A.,   {Desjacques} V.,  2013, \mn@doi [ApJ]
  {10.1088/0004-637X/775/2/102}, \href
  {http://adsabs.harvard.edu/abs/2013ApJ...775..102P} {775, 102}

\bibitem[\protect\citeauthoryear{{Piffl} et~al.,}{{Piffl}
  et~al.}{2014}]{Piffl_2014}
{Piffl} T.,  et~al., 2014, \mn@doi [A\&A] {10.1051/0004-6361/201322531}, \href
  {http://adsabs.harvard.edu/abs/2014A\%26A...562A..91P} {562, A91}

\bibitem[\protect\citeauthoryear{{Pittordis} \& {Sutherland}}{{Pittordis} \&
  {Sutherland}}{2017}]{Pittordis_2017}
{Pittordis} C.,  {Sutherland} W.,  2017, preprint, \href
  {http://adsabs.harvard.edu/abs/2017arXiv171110867P} {Arxiv} (\mn@eprint
  {arXiv} {1711.10867})

\bibitem[\protect\citeauthoryear{{Planck Collaboration XIII}}{{Planck
  Collaboration XIII}}{2016}]{Planck_2015}
{Planck Collaboration XIII} 2016, \mn@doi [A\&A] {10.1051/0004-6361/201525830},
  \href {http://adsabs.harvard.edu/abs/2016A\%26A...594A..13P} {594, A13}

\bibitem[\protect\citeauthoryear{{Planck Collaboration XVI}}{{Planck
  Collaboration XVI}}{2014}]{Planck_Cosmological_Parameters}
{Planck Collaboration XVI} 2014, \mn@doi [A\&A] {10.1051/0004-6361/201321591},
  \href {http://adsabs.harvard.edu/abs/2014A\%26A...571A..16P} {571, A16}

\bibitem[\protect\citeauthoryear{{Planck Collaboration XXVII}}{{Planck
  Collaboration XXVII}}{2014}]{Planck_2013}
{Planck Collaboration XXVII} 2014, \mn@doi [A\&A]
  {10.1051/0004-6361/201321556}, \href
  {http://adsabs.harvard.edu/abs/2014A\%26A...571A..27P} {571, A27}

\bibitem[\protect\citeauthoryear{{Plummer}}{{Plummer}}{1911}]{Plummer_1911}
{Plummer} H.~C.,  1911, \mn@doi [MNRAS] {10.1093/mnras/71.5.460}, \href
  {http://adsabs.harvard.edu/abs/1911MNRAS..71..460P} {71, 460}

\bibitem[\protect\citeauthoryear{{Privon}, {Barnes}, {Evans}, {Hibbard}, {Yun},
  {Mazzarella}, {Armus}  \& {Surace}}{{Privon} et~al.}{2013}]{Privon_2013}
{Privon} G.~C.,  {Barnes} J.~E.,  {Evans} A.~S.,  {Hibbard} J.~E.,  {Yun}
  M.~S.,  {Mazzarella} J.~M.,  {Armus} L.,   {Surace} J.,  2013, \mn@doi [ApJ]
  {10.1088/0004-637X/771/2/120}, \href
  {http://adsabs.harvard.edu/abs/2013ApJ...771..120P} {771, 120}

\bibitem[\protect\citeauthoryear{{Quillen} \& {Garnett}}{{Quillen} \&
  {Garnett}}{2001}]{Quillen_2001}
{Quillen} A.~C.,  {Garnett} D.~R.,  2001, in {Funes} J.~G.,  {Corsini} E.~M.,
  eds,  Astronomical Society of the Pacific Conference Series Vol. 230, Galaxy
  Disks and Disk Galaxies. pp 87--88 (\mn@eprint {arXiv} {astro-ph/0004210})

\bibitem[\protect\citeauthoryear{{Refsdal}}{{Refsdal}}{1966}]{Refsdal_1966}
{Refsdal} S.,  1966, MNRAS, \href
  {http://adsabs.harvard.edu/abs/1966MNRAS.134..315R} {134, 315}

\bibitem[\protect\citeauthoryear{{Riess} et~al.,}{{Riess}
  et~al.}{1998}]{Riess_1998}
{Riess} A.~G.,  et~al., 1998, \mn@doi [AJ] {10.1086/300499}, \href
  {http://adsabs.harvard.edu/abs/1998AJ....116.1009R} {116, 1009}

\bibitem[\protect\citeauthoryear{{Roberts} \& {Whitehurst}}{{Roberts} \&
  {Whitehurst}}{1975}]{Roberts_1975}
{Roberts} M.~S.,  {Whitehurst} R.~N.,  1975, \mn@doi [ApJ] {10.1086/153889},
  \href {http://adsabs.harvard.edu/abs/1975ApJ...201..327R} {201, 327}

\bibitem[\protect\citeauthoryear{{Rogstad} \& {Shostak}}{{Rogstad} \&
  {Shostak}}{1972}]{Rogstad_1972}
{Rogstad} D.~H.,  {Shostak} G.~S.,  1972, \mn@doi [ApJ] {10.1086/151636}, \href
  {http://adsabs.harvard.edu/abs/1972ApJ...176..315R} {176, 315}

\bibitem[\protect\citeauthoryear{{Rubin} \& {Ford}}{{Rubin} \&
  {Ford}}{1970}]{Rubin_Ford_1970}
{Rubin} V.~C.,  {Ford} Jr. W.~K.,  1970, \mn@doi [ApJ] {10.1086/150317}, \href
  {http://adsabs.harvard.edu/abs/1970ApJ...159..379R} {159, 379}

\bibitem[\protect\citeauthoryear{{Ruskin}}{{Ruskin}}{2017}]{Ruskin_2017}
{Ruskin} S.,  2017, \mn@doi [American Journal of Physics] {10.1119/1.4968558},
  \href {http://adsabs.harvard.edu/abs/2017AmJPh..85..159R} {85, 159}

\bibitem[\protect\citeauthoryear{{Ryu}, {Perna}, {Haiman}, {Ostriker}  \&
  {Stone}}{{Ryu} et~al.}{2018}]{Ryu_2018}
{Ryu} T.,  {Perna} R.,  {Haiman} Z.,  {Ostriker} J.~P.,   {Stone} N.~C.,  2018,
  \mn@doi [MNRAS] {10.1093/mnras/stx2524}, \href
  {http://adsabs.harvard.edu/abs/2018MNRAS.473.3410R} {473, 3410}

\bibitem[\protect\citeauthoryear{{Salem}, {Besla}, {Bryan}, {Putman}, {van der
  Marel}  \& {Tonnesen}}{{Salem} et~al.}{2015}]{Salem_2015}
{Salem} M.,  {Besla} G.,  {Bryan} G.,  {Putman} M.,  {van der Marel} R.~P.,
  {Tonnesen} S.,  2015, \mn@doi [ApJ] {10.1088/0004-637X/815/1/77}, \href
  {http://adsabs.harvard.edu/abs/2015ApJ...815...77S} {815, 77}

\bibitem[\protect\citeauthoryear{{Sales}, {Navarro}, {Abadi}  \&
  {Steinmetz}}{{Sales} et~al.}{2007}]{Sales_2007}
{Sales} L.~V.,  {Navarro} J.~F.,  {Abadi} M.~G.,   {Steinmetz} M.,  2007,
  \mn@doi [MNRAS] {10.1111/j.1365-2966.2007.12026.x}, \href
  {http://adsabs.harvard.edu/abs/2007MNRAS.379.1475S} {379, 1475}

\bibitem[\protect\citeauthoryear{{Salucci} \& {Turini}}{{Salucci} \&
  {Turini}}{2017}]{Salucci_2017}
{Salucci} P.,  {Turini} N.,  2017, preprint, \href
  {http://adsabs.harvard.edu/abs/2017arXiv170701059S} {Arxiv} (\mn@eprint
  {arXiv} {1707.01059})

\bibitem[\protect\citeauthoryear{{Salucci}, {Lapi}, {Tonini}, {Gentile},
  {Yegorova}  \& {Klein}}{{Salucci} et~al.}{2007}]{Salucci_2007}
{Salucci} P.,  {Lapi} A.,  {Tonini} C.,  {Gentile} G.,  {Yegorova} I.,
  {Klein} U.,  2007, \mn@doi [MNRAS] {10.1111/j.1365-2966.2007.11696.x}, \href
  {http://adsabs.harvard.edu/abs/2007MNRAS.378...41S} {378, 41}

\bibitem[\protect\citeauthoryear{{Sancisi}}{{Sancisi}}{2004}]{Sancisi_2004}
{Sancisi} R.,  2004, in {Ryder} S.,  {Pisano} D.,  {Walker} M.,   {Freeman} K.,
   eds,  IAU Symposium Vol. 220, Dark Matter in Galaxies. p.~233 (\mn@eprint {}
  {astro-ph/0311348})

\bibitem[\protect\citeauthoryear{{Sandage}}{{Sandage}}{1986}]{Sandage_1986}
{Sandage} A.,  1986, \mn@doi [ApJ] {10.1086/164387}, \href
  {http://adsabs.harvard.edu/abs/1986ApJ...307....1S} {307, 1}

\bibitem[\protect\citeauthoryear{{Sanders}}{{Sanders}}{2003}]{Sanders_2003}
{Sanders} R.~H.,  2003, \mn@doi [MNRAS] {10.1046/j.1365-8711.2003.06596.x},
  \href {http://adsabs.harvard.edu/abs/2003MNRAS.342..901S} {342, 901}

\bibitem[\protect\citeauthoryear{{Sarazin}}{{Sarazin}}{1986}]{Sarazin_1986}
{Sarazin} C.~L.,  1986, \mn@doi [Reviews of Modern Physics]
  {10.1103/RevModPhys.58.1}, \href
  {http://adsabs.harvard.edu/abs/1986RvMP...58....1S} {58, 1}

\bibitem[\protect\citeauthoryear{{Scarpa}, {Ottolina}, {Falomo}  \&
  {Treves}}{{Scarpa} et~al.}{2017}]{Scarpa_2017}
{Scarpa} R.,  {Ottolina} R.,  {Falomo} R.,   {Treves} A.,  2017, \mn@doi
  [International Journal of Modern Physics D] {10.1142/S0218271817500675},
  \href {http://adsabs.harvard.edu/abs/2017IJMPD..2650067S} {26, 1750067}

\bibitem[\protect\citeauthoryear{{Schechter}}{{Schechter}}{1976}]{Schechter_1976}
{Schechter} P.,  1976, \mn@doi [ApJ] {10.1086/154079}, \href
  {http://adsabs.harvard.edu/abs/1976ApJ...203..297S} {203, 297}

\bibitem[\protect\citeauthoryear{{Schmidt}}{{Schmidt}}{1958}]{Schmidt_1958}
{Schmidt} K.~H.,  1958, Astronomische Nachrichten, \href
  {http://adsabs.harvard.edu/abs/1958AN....284...76S} {284, 76}

\bibitem[\protect\citeauthoryear{{Sch{\"o}nrich}}{{Sch{\"o}nrich}}{2012}]{Schonrich_2012}
{Sch{\"o}nrich} R.,  2012, \mn@doi [MNRAS] {10.1111/j.1365-2966.2012.21631.x},
  \href {http://adsabs.harvard.edu/abs/2012MNRAS.427..274S} {427, 274}

\bibitem[\protect\citeauthoryear{{Shannon} et~al.,}{{Shannon}
  et~al.}{2015}]{Shannon_2015}
{Shannon} R.~M.,  et~al., 2015, \mn@doi [Science] {10.1126/science.aab1910},
  \href {http://adsabs.harvard.edu/abs/2015Sci...349.1522S} {349, 1522}

\bibitem[\protect\citeauthoryear{{Shappee} \& {Stanek}}{{Shappee} \&
  {Stanek}}{2011}]{Shappee_2011}
{Shappee} B.~J.,  {Stanek} K.~Z.,  2011, \mn@doi [ApJ]
  {10.1088/0004-637X/733/2/124}, \href
  {http://adsabs.harvard.edu/abs/2011ApJ...733..124S} {733, 124}

\bibitem[\protect\citeauthoryear{{Sikivie}}{{Sikivie}}{1983}]{Sikivie_1983}
{Sikivie} P.,  1983, \mn@doi [Physical Review Letters]
  {10.1103/PhysRevLett.51.1415}, \href
  {http://adsabs.harvard.edu/abs/1983PhRvL..51.1415S} {51, 1415}

\bibitem[\protect\citeauthoryear{{Slipher}}{{Slipher}}{1913}]{Slipher_1913}
{Slipher} V.~M.,  1913, Lowell Observatory Bulletin, \href
  {http://adsabs.harvard.edu/abs/1913LowOB...2...56S} {2, 56}

\bibitem[\protect\citeauthoryear{{Smolin}}{{Smolin}}{2017}]{Smolin_2017}
{Smolin} L.,  2017, \mn@doi [Physical Review D] {10.1103/PhysRevD.96.083523},
  \href {http://adsabs.harvard.edu/abs/2017PhRvD..96h3523S} {96, 083523}

\bibitem[\protect\citeauthoryear{{Snaith}, {Haywood}, {Di Matteo}, {Lehnert},
  {Combes}, {Katz}  \& {G{\'o}mez}}{{Snaith} et~al.}{2014}]{Snaith_2014}
{Snaith} O.~N.,  {Haywood} M.,  {Di Matteo} P.,  {Lehnert} M.~D.,  {Combes} F.,
   {Katz} D.,   {G{\'o}mez} A.,  2014, \mn@doi [ApJL]
  {10.1088/2041-8205/781/2/L31}, \href
  {http://adsabs.harvard.edu/abs/2014ApJ...781L..31S} {781, L31}

\bibitem[\protect\citeauthoryear{{Sohn}}{{Sohn}}{2016}]{Sohn_2016}
{Sohn} S.,  2016, {Proper Motions of the Crater-Leo Group: Testing the Group
  Infall Scenario}, HST Proposal

\bibitem[\protect\citeauthoryear{{Sohn} et~al.,}{{Sohn}
  et~al.}{2017}]{Sohn_2017}
{Sohn} S.~T.,  et~al., 2017, \mn@doi [ApJ] {10.3847/1538-4357/aa917b}, \href
  {http://adsabs.harvard.edu/abs/2017ApJ...849...93S} {849, 93}

\bibitem[\protect\citeauthoryear{{Steigman} \& {Turner}}{{Steigman} \&
  {Turner}}{1985}]{Steigman_1985}
{Steigman} G.,  {Turner} M.~S.,  1985, \mn@doi [Nuclear Physics B]
  {10.1016/0550-3213(85)90537-1}, \href
  {http://adsabs.harvard.edu/abs/1985NuPhB.253..375S} {253, 375}

\bibitem[\protect\citeauthoryear{{Swaters}, {Sancisi}, {van Albada}  \& {van
  der Hulst}}{{Swaters} et~al.}{2009}]{Swaters_2009}
{Swaters} R.~A.,  {Sancisi} R.,  {van Albada} T.~S.,   {van der Hulst} J.~M.,
  2009, \mn@doi [A\&A] {10.1051/0004-6361:200810516}, \href
  {http://adsabs.harvard.edu/abs/2009A\%26A...493..871S} {493, 871}

\bibitem[\protect\citeauthoryear{{Tenneti}, {Mao}, {Croft}, {Di Matteo},
  {Kosowsky}, {Zago}  \& {Zentner}}{{Tenneti} et~al.}{2018}]{Tenneti_2018}
{Tenneti} A.,  {Mao} Y.-Y.,  {Croft} R.~A.~C.,  {Di Matteo} T.,  {Kosowsky} A.,
   {Zago} F.,   {Zentner} A.~R.,  2018, \mn@doi [MNRAS]
  {10.1093/mnras/stx3010}, \href
  {http://adsabs.harvard.edu/abs/2018MNRAS.474.3125T} {474, 3125}

\bibitem[\protect\citeauthoryear{{Teyssier}}{{Teyssier}}{2002}]{Teyssier_2002}
{Teyssier} R.,  2002, \mn@doi [A\&A] {10.1051/0004-6361:20011817}, \href
  {http://adsabs.harvard.edu/abs/2002A\%26A...385..337T} {385, 337}

\bibitem[\protect\citeauthoryear{{Theia Collaboration}}{{Theia
  Collaboration}}{2017}]{Theia_2017}
{Theia Collaboration} 2017, preprint, \href
  {http://adsabs.harvard.edu/abs/2017arXiv170701348T} {Arxiv} (\mn@eprint
  {arXiv} {1707.01348})

\bibitem[\protect\citeauthoryear{{Thompson} \& {Nagamine}}{{Thompson} \&
  {Nagamine}}{2012}]{Thompson_Nagamine_2012}
{Thompson} R.,  {Nagamine} K.,  2012, \mn@doi [MNRAS]
  {10.1111/j.1365-2966.2011.20000.x}, \href
  {http://adsabs.harvard.edu/abs/2012MNRAS.419.3560T} {419, 3560}

\bibitem[\protect\citeauthoryear{{Tian} \& {Ko}}{{Tian} \&
  {Ko}}{2017}]{Tian_2017}
{Tian} Y.,  {Ko} C.-M.,  2017, \mn@doi [MNRAS] {10.1093/mnras/stx2056}, \href
  {http://adsabs.harvard.edu/abs/2017MNRAS.472..765T} {472, 765}

\bibitem[\protect\citeauthoryear{{Tisserand} et~al.,}{{Tisserand}
  et~al.}{2007}]{EROS_2007}
{Tisserand} P.,  et~al., 2007, \mn@doi [A\&A] {10.1051/0004-6361:20066017},
  \href {http://adsabs.harvard.edu/abs/2007A\%26A...469..387T} {469, 387}

\bibitem[\protect\citeauthoryear{{Torrealba}, {Koposov}, {Belokurov}  \&
  {Irwin}}{{Torrealba} et~al.}{2016}]{Torrealba_2016}
{Torrealba} G.,  {Koposov} S.~E.,  {Belokurov} V.,   {Irwin} M.,  2016, \mn@doi
  [MNRAS] {10.1093/mnras/stw733}, \href
  {http://adsabs.harvard.edu/abs/2016MNRAS.459.2370T} {459, 2370}

\bibitem[\protect\citeauthoryear{{Tucker}, {Tananbaum}  \&
  {Remillard}}{{Tucker} et~al.}{1995}]{Tucker_1995}
{Tucker} W.~H.,  {Tananbaum} H.,   {Remillard} R.~A.,  1995, \mn@doi [ApJ]
  {10.1086/175627}, \href {http://adsabs.harvard.edu/abs/1995ApJ...444..532T}
  {444, 532}

\bibitem[\protect\citeauthoryear{{Tucker} et~al.,}{{Tucker}
  et~al.}{1998}]{Tucker_1998}
{Tucker} W.,  et~al., 1998, \mn@doi [ApJL] {10.1086/311234}, \href
  {http://adsabs.harvard.edu/abs/1998ApJ...496L...5T} {496, L5}

\bibitem[\protect\citeauthoryear{{Tully}}{{Tully}}{2015}]{Tully_2015}
{Tully} R.~B.,  2015, \mn@doi [AJ] {10.1088/0004-6256/149/2/54}, \href
  {http://adsabs.harvard.edu/abs/2015AJ....149...54T} {149, 54}

\bibitem[\protect\citeauthoryear{{Tully} \& {Fisher}}{{Tully} \&
  {Fisher}}{1977}]{Tully_Fisher_1977}
{Tully} R.~B.,  {Fisher} J.~R.,  1977, A\&A, \href
  {http://adsabs.harvard.edu/abs/1977A\%26A....54..661T} {54, 661}

\bibitem[\protect\citeauthoryear{{Tully} et~al.,}{{Tully}
  et~al.}{2013}]{Tully_2013}
{Tully} R.~B.,  et~al., 2013, \mn@doi [AJ] {10.1088/0004-6256/146/4/86}, \href
  {http://adsabs.harvard.edu/abs/2013AJ....146...86T} {146, 86}

\bibitem[\protect\citeauthoryear{{Turyshev}, {Toth}, {Kinsella}, {Lee}, {Lok}
  \& {Ellis}}{{Turyshev} et~al.}{2012}]{Turyshev_2012}
{Turyshev} S.~G.,  {Toth} V.~T.,  {Kinsella} G.,  {Lee} S.-C.,  {Lok} S.~M.,
  {Ellis} J.,  2012, \mn@doi [Physical Review Letters]
  {10.1103/PhysRevLett.108.241101}, \href
  {http://adsabs.harvard.edu/abs/2012PhRvL.108x1101T} {108, 241101}

\bibitem[\protect\citeauthoryear{{Verlinde}}{{Verlinde}}{2017}]{Verlinde_2016}
{Verlinde} E.~P.,  2017, \mn@doi [SciPost Physics]
  {10.21468/SciPostPhys.2.3.016}, \href
  {http://adsabs.harvard.edu/abs/2016arXiv161102269V} {2, 016}

\bibitem[\protect\citeauthoryear{{Viel}, {Becker}, {Bolton}  \&
  {Haehnelt}}{{Viel} et~al.}{2013}]{Viel_2013}
{Viel} M.,  {Becker} G.~D.,  {Bolton} J.~S.,   {Haehnelt} M.~G.,  2013, \mn@doi
  [Physical Review D] {10.1103/PhysRevD.88.043502}, \href
  {http://adsabs.harvard.edu/abs/2013PhRvD..88d3502V} {88, 043502}

\bibitem[\protect\citeauthoryear{{Virgo \& LIGO Collaborations}}{{Virgo \& LIGO
  Collaborations}}{2017}]{LIGO_Virgo_2017}
{Virgo \& LIGO Collaborations} 2017, \mn@doi [Physical Review Letters]
  {10.1103/PhysRevLett.119.161101}, \href
  {http://adsabs.harvard.edu/abs/2017PhRvL.119p1101A} {119, 161101}

\bibitem[\protect\citeauthoryear{{Wang} \& {Mohanty}}{{Wang} \&
  {Mohanty}}{2017}]{Wang_2017}
{Wang} Y.,  {Mohanty} S.~D.,  2017, \mn@doi [Physical Review Letters]
  {10.1103/PhysRevLett.118.151104}, \href
  {http://adsabs.harvard.edu/abs/2017PhRvL.118o1104W} {118, 151104}

\bibitem[\protect\citeauthoryear{{Wang}, {Greene}, {Ju}, {Rafikov}, {Ruan}  \&
  {Schneider}}{{Wang} et~al.}{2017}]{Wang_2016}
{Wang} L.,  {Greene} J.~E.,  {Ju} W.,  {Rafikov} R.~R.,  {Ruan} J.~J.,
  {Schneider} D.~P.,  2017, \mn@doi [ApJ] {10.3847/1538-4357/834/2/129}, \href
  {http://adsabs.harvard.edu/abs/2017ApJ...834..129W} {834, 129}

\bibitem[\protect\citeauthoryear{{Wetzstein}, {Naab}  \& {Burkert}}{{Wetzstein}
  et~al.}{2007}]{Wetzstein_2007}
{Wetzstein} M.,  {Naab} T.,   {Burkert} A.,  2007, \mn@doi [MNRAS]
  {10.1111/j.1365-2966.2006.11360.x}, \href
  {http://adsabs.harvard.edu/abs/2007MNRAS.375..805W} {375, 805}

\bibitem[\protect\citeauthoryear{{White} \& {Frenk}}{{White} \&
  {Frenk}}{1991}]{LCDM_1991}
{White} S.~D.~M.,  {Frenk} C.~S.,  1991, \mn@doi [ApJ] {10.1086/170483}, \href
  {http://adsabs.harvard.edu/abs/1991ApJ...379...52W} {379, 52}

\bibitem[\protect\citeauthoryear{{Williams}, {Belokurov}, {Casey}  \&
  {Evans}}{{Williams} et~al.}{2017a}]{Williams_2017}
{Williams} A.~A.,  {Belokurov} V.,  {Casey} A.~R.,   {Evans} N.~W.,  2017a,
  \mn@doi [MNRAS] {10.1093/mnras/stx508}, \href
  {http://adsabs.harvard.edu/abs/2017MNRAS.468.2359W} {468, 2359}

\bibitem[\protect\citeauthoryear{{Williams} et~al.,}{{Williams}
  et~al.}{2017b}]{Dolphin_2017}
{Williams} B.~F.,  et~al., 2017b, \mn@doi [ApJ] {10.3847/1538-4357/aa862a},
  \href {http://adsabs.harvard.edu/abs/2017arXiv170802617W} {846, 145}

\bibitem[\protect\citeauthoryear{{Wu}, {Famaey}, {Gentile}, {Perets}  \&
  {Zhao}}{{Wu} et~al.}{2008}]{Wu_2008}
{Wu} X.,  {Famaey} B.,  {Gentile} G.,  {Perets} H.,   {Zhao} H.,  2008, \mn@doi
  [MNRAS] {10.1111/j.1365-2966.2008.13198.x}, \href
  {http://adsabs.harvard.edu/abs/2008MNRAS.386.2199W} {386, 2199}

\bibitem[\protect\citeauthoryear{{York} et~al.,}{{York} et~al.}{2000}]{SDSS}
{York} D.~G.,  et~al., 2000, \mn@doi [AJ] {10.1086/301513}, \href
  {http://adsabs.harvard.edu/abs/2000AJ....120.1579Y} {120, 1579}

\bibitem[\protect\citeauthoryear{{Zacharias}, {Finch}  \&
  {Frouard}}{{Zacharias} et~al.}{2017}]{Zacharias_2017}
{Zacharias} N.,  {Finch} C.,   {Frouard} J.,  2017, \mn@doi [AJ]
  {10.3847/1538-3881/aa6196}, \href
  {http://adsabs.harvard.edu/abs/2017AJ....153..166Z} {153, 166}

\bibitem[\protect\citeauthoryear{{Zanella} et~al.,}{{Zanella}
  et~al.}{2015}]{Zanella_2015}
{Zanella} A.,  et~al., 2015, \mn@doi [Nature] {10.1038/nature14409}, \href
  {http://adsabs.harvard.edu/abs/2015Natur.521...54Z} {521, 54}

\bibitem[\protect\citeauthoryear{{Zhao} \& {Famaey}}{{Zhao} \&
  {Famaey}}{2012}]{Zhao_2012}
{Zhao} H.,  {Famaey} B.,  2012, \mn@doi [Physical Review D]
  {10.1103/PhysRevD.86.067301}, \href
  {http://adsabs.harvard.edu/abs/2012PhRvD..86f7301Z} {86, 067301}

\bibitem[\protect\citeauthoryear{{Zhao}, {Famaey}, {L{\"u}ghausen}  \&
  {Kroupa}}{{Zhao} et~al.}{2013}]{Zhao_2013}
{Zhao} H.,  {Famaey} B.,  {L{\"u}ghausen} F.,   {Kroupa} P.,  2013, \mn@doi
  [A\&A] {10.1051/0004-6361/201321879}, \href
  {http://adsabs.harvard.edu/abs/2013A\%26A...557L...3Z} {557, L3}

\bibitem[\protect\citeauthoryear{{Zwicky}}{{Zwicky}}{1937}]{Zwicky_1937}
{Zwicky} F.,  1937, \mn@doi [ApJ] {10.1086/143864}, \href
  {http://adsabs.harvard.edu/abs/1937ApJ....86..217Z} {86, 217}

\bibitem[\protect\citeauthoryear{{van den Bergh}}{{van den
  Bergh}}{1999}]{Van_den_Bergh_1999}
{van den Bergh} S.,  1999, \mn@doi [ApJL] {10.1086/312044}, \href
  {http://adsabs.harvard.edu/abs/1999ApJ...517L..97V} {517, L97}

\bibitem[\protect\citeauthoryear{{van der Marel} \& {Kallivayalil}}{{van der
  Marel} \& {Kallivayalil}}{2014}]{Van_der_Marel_2014}
{van der Marel} R.~P.,  {Kallivayalil} N.,  2014, \mn@doi [ApJ]
  {10.1088/0004-637X/781/2/121}, \href
  {http://adsabs.harvard.edu/abs/2014ApJ...781..121V} {781, 121}

\bibitem[\protect\citeauthoryear{{van der Marel}, {Fardal}, {Besla}, {Beaton},
  {Sohn}, {Anderson}, {Brown}  \& {Guhathakurta}}{{van der Marel}
  et~al.}{2012a}]{M31_motion}
{van der Marel} R.~P.,  {Fardal} M.,  {Besla} G.,  {Beaton} R.~L.,  {Sohn}
  S.~T.,  {Anderson} J.,  {Brown} T.,   {Guhathakurta} P.,  2012a, \mn@doi
  [ApJ] {10.1088/0004-637X/753/1/8}, \href
  {http://adsabs.harvard.edu/abs/2012ApJ...753....8V} {753, 8}

\bibitem[\protect\citeauthoryear{{van der Marel}, {Besla}, {Cox}, {Sohn}  \&
  {Anderson}}{{van der Marel} et~al.}{2012b}]{Van_der_Marel_2012}
{van der Marel} R.~P.,  {Besla} G.,  {Cox} T.~J.,  {Sohn} S.~T.,   {Anderson}
  J.,  2012b, \mn@doi [ApJ] {10.1088/0004-637X/753/1/9}, \href
  {http://adsabs.harvard.edu/abs/2012ApJ...753....9V} {753, 9}

\makeatother
\end{thebibliography}
\bsp
\label{lastpage}
\end{document}